\documentclass{aa}  
\usepackage{graphicx}
\usepackage{txfonts}
\usepackage{float}
\usepackage{mathtools}
\usepackage{subcaption}
\usepackage[version=4]{mhchem}
\usepackage{gensymb}
\usepackage{tikzsymbols}
\usepackage{rotating}
\usepackage{multirow}
\usepackage{mleftright}
\begin{document} 
   \title{Cloud property trends in hot and ultra-hot giant gas planets (WASP-43b, WASP-103b, WASP-121b, HAT-P-7b, and WASP-18b)}
   
   \titlerunning{Trending clouds in hot  giant gas planets and ultra-hot Jupiters}
   
   \author{
   Ch. Helling\inst{1,2,3}, 
   D. Lewis\inst{1,2},
   D. Samra\inst{1,2}, 
   L. Carone\inst{4},
   V. Graham\inst{6},
   O. Herbort\inst{1,2,5},
   K. L. Chubb\inst{3},
   M. Min\inst{3},
   R. Waters\inst{3,7},
   V. Parmentier\inst{8}, 
   N. Mayne\inst{9}}

 \authorrunning{Ch. Helling et. al.}
 \institute{
         Centre for Exoplanet Science, University of St Andrews, North Haugh, St Andrews, KY169SS, UK  \\ \email{ch80@st-andrews.ac.uk}
 \and
     SUPA, School of Physics \& Astronomy, University of St Andrews, North Haugh, St Andrews, KY169SS, UK 
              \and
              SRON Netherlands Institute for Space Research, Sorbonnelaan 2, 3584 CA Utrecht, NL
         \and  
         Max Planck Institute for Astronomy, Königstuhl 17, 69117 Heidelberg, Germany
\and
    School of Earth \& Environmental Studies, University of St Andrews, Irvine Building, St Andrews, KY16 9AL, UK
\and
    School of Physics and Astronomy, University of Glasgow, University Avenue, Glasgow, G12 8QQ, UK
         \and
         Institute for Mathematics, Astrophysics \& Particle Physics, Department of Astrophysics, Radboud University, P.O. Box 9010, MC 62, NL-6500 GL Nijmegen, The Netherlands
         \and
    Department of Physics, University of Oxford, Parks Rd, Oxford, OX1 3PU, UK
     \and
     Physics and Astronomy, College of Engineering, Mathematics and Physical Sciences, University of Exeter, EX4 4QL, UK
}
\date{\today}
     \date{Received September 15, 2996; accepted March 16, 2997}

\abstract
{Ultra-hot Jupiters are the hottest exoplanets discovered so far. Observations begin to provide insight into the composition of their extended atmospheres and their chemical day/night asymmetries. Both are strongly affected by cloud formation.}
{We explore trends in cloud properties for a sample of five giant gas planets: the  hot  gas giant WASP-43b and  the four ultra-hot Jupiters WASP-18b, HAT-P-7b, WASP-103b, and WASP-121b. This provides a reference frame for cloud properties for the JWST targets WASP-43b and  WASP-121b. We further explore chemically inert tracers  to observe geometrical asymmetries of ultra-hot Jupiters, and  if the location of inner boundary of a 3D GCM matters for the clouds that form. }
{A homogeneous set of 3D GCM results is used as input for a kinetic cloud formation code to evaluate the cloud opacity and gas parameters like C/O,  mean molecular weight, and degree of ionisation. We cast our results in terms of integrated quantities to enable a global comparison between the sample  planets.}
{The large day/night temperature differences of ultra-hot Jupiters cause large chemical asymmetries: cloud-free days but cloudy nights, atomic vs. molecular gases and respectively different mean molecular weights, deep thermal ionospheres vs.  low-ionised atmospheres, undepleted vs enhanced C/O. WASP-18b, as the heaviest planet in the sample, has the lowest global C/O.}
{The global climate may be considered as similar amongst ultra-hot Jupiters, but different to that of hot gas giants.  The local weather, however, is individual for each planet since the local thermodynamic conditions, and hence the local cloud and gas properties, differ. The morning and the evening terminator of ultra-hot Jupiters will carry signatures of their strong chemical asymmetry such that ingress/egress asymmetries can be expected. An increased C/O ratio is a clear sign of cloud formation, making cloud modelling a necessity  when utilizing C/O (or other mineral ratios)  as tracer for planet formation. The changing geometrical extension of the atmosphere from the day to the nightside  may be probed through chemically inert species like helium. Ultra-hot Jupiters are likely to develop deep atmospheric ionospheres which may impact the atmosphere dynamics through MHD processes. }


\maketitle



\section{Introduction}\label{intro}

Ultra-hot Jupiters provide the unique opportunity to study a wide range of  atmospheric regimes in one object due to their strong day-night temperature difference of $> 2000$K resulting from the close proximity to their host stars in combination with an inefficient heat redistribution from the day- to the nighside  (\citealt{2013ApJ...776..134P}). For one of the hottest known ultra-hot Jupiters, KELT-9b, several indicators for a very high-temperature dayside have been found: Ca~II (\citealt{2020ApJ...888L..13T}) as well as Fe and Fe$^+$ (\citealt{2018Natur.560..453H}) at the limb, in addition to an extended hydrogen-envelope (\citealt{2018NatAs...2..714Y}) prompting investigations into the atmospheric mass loss of KELT-9b (e.g. \citealt{2020A&A...638A..87W}). \cite{2020arXiv201000997F} discuss the differences in the H$\alpha$ and H$\beta$ lines as observed by different groups with different instruments and different reduction pipelines. Employing  non-LTE radiative transfer, a temperature of $10^4\,\ldots\,1.1\times 10^4$K is retrieved to be consistent with the PEPSI transmission spectra. The first CHEOPS science paper by \cite{2020arXiv200913403L} presents WASP-189b with an global dayside  temperature of $\approx 3400$K for a nonreflective atmosphere with inefficient heat redistribution.
Also WASP-121b shows less extreme temperatures than KELT-9b such that the presence of neutral atoms like Mg, Na, Ca, Cr, Fe and also Ni and V have been inferred from observations of the terminator regions (\citealt{2020arXiv200611308H}).  However, \cite{2020arXiv200611308H} did not detect 
 Ti and TiO  on WASP-121b which may be an indication for  cloud-depleted Ti abundances at the cold evening terminator similar to WASP-18b or HAT-P-7b (\citealt{2019A&A...626A.133H,2019A&A...631A..79H}). \cite{2020MNRAS.493.2215G} detect Fe~I in WASP-121b applying VLT/UVES transit observations.  Partial coverage of WASP-121b by aerosols was suggest along the terminator for WASP-121b by \cite{2017ApJ...845L..20K} based on phase equilibrium considerations. 
 \cite{2020Natur.580..597E} present ESPRESSO/VLT observations of the ultra-hot Jupiter WASP-78b showing a day/night terminator asymmetry as suggested for  WASP-18b or HAT-P-7b. Neutral iron has been suggested to be present on the hot morning terminator but not on the colder evening terminator, leading the authors to claim that iron rain should form on the nightside of WASP-78b.
Based on detailed cloud modelling, \cite{2019A&A...626A.133H,2019A&A...631A..79H} show that the dayside of the ultra-hot Jupiters WASP-18b and HAT-P7b are likely to be cloud free while the nightside will be covered in clouds.  In comparison, hot  giant gas planets, like HD\,189733b, HD\,209458b and WASP-43b, have a global day/night cloud coverage over a large pressure range (\citealt{Lee2015,Lee2016,2018A&A...615A..97L,helling2020mineral}) due to their smaller day-night temperature differences.


\begin{table*}
 \centering
 \begin{tabular}{c | c c c c}
 \hline\hline
 \textbf{Stellar Parameters}  & \textbf{WASP-103} & \textbf{WASP-18} & \textbf{WASP-121} & \textbf{HAT-P-7}\\

 T$_{\text{eff}}$[K] & 6100$\pm$100$^{4}$  & 6400$\pm$100$^{5}$ & 6500$\pm$100$^{1}$ & 6300$\pm$100$^{2}$\\

 Mass [M$_{\text{Sun}}$] & 1.220$\pm$0.039$^{4}$ & 1.25$\pm$0.13$^{5}$ &  1.353$\pm$0.080$^{1}$ & 1.361$\pm$0.021$^{2}$ \\

 Radius [R$_{\text{Sun}}$]  & 1.436$\pm$0.052$^{4}$  & 1.216$\pm$0.067$^{5}$ & 1.458$\pm$0.030$^{1}$ & 1.094$\pm$0.01$^{2}$ \\

 Spectral Type  & F8V & F6V & F6V & F8\\  

 Metallicity [Fe/H][dex]  &  0.06$\pm$0.13$^{4}$ & 0.11$\pm$0.08$^{5}$ & 0.13$\pm$0.09$^{1}$ & 0.26$\pm$0.08$^{2}$ \\

 \hline
 \textbf{Orbital Parameters} & \textbf{WASP-103b} & \textbf{WASP-18b} & \textbf{WASP-121b}  & \textbf{HAT-P-7b} \\ [0.5ex]
 Semi-Major Axis [AU] & 0.01985$\pm$0.00021$^{4}$ & 0.020266$\pm$0.00068$^{5}$ & 0.02544$\pm$0.00050$^{1}$ & 0.0379$\pm$0.0004$^{3}$ \\ 

 Orbital Period [days] & 0.92554$\pm$0.000019 $^{4}$ & 0.94145$\pm$0.00000134$^{6}$ & 1.2749255$\pm$0.00000025$^{1}$ & 2.20474$^{2}$  \\ [0.5ex]

 \hline
 \textbf{Planetary Parameters} \\ [0.5ex] 
 T$_{\text{eq}}$[K]  & 2500$\pm$100$^{4}$ & 2400$\pm$100 $^{7}$ & 2400$\pm$100$^{1}$  & 2200$\pm$100$^{2}$  \\


 Mass [M$_{\text{Jup}}]$  & 1.49$\pm$0.088$^{4}$ & 10.30$\pm$0.69$^{5}$ & 1.183$\pm$0.064$^{1}$ & 1.74$\pm$0.03$^{2}$ \\

 Radius [R$_{\text{Jup}}$] & 1.528$\pm$ 0.073$^{4}$  & 1.106$\pm$0.072$^{5}$ & 1.807$\pm$0.039$^{1}$ & 1.431$\pm$0.011$^{2}$ \\

 Density [g cm$^{-3}$] & 0.554$\pm$0.056$^{10}$ & 10.096$\pm$1.324$^{10}$ & 0.266$\pm$0.017$^{10}$ & 0.787$\pm$0.017$^{10}$\\

 g [g$_{\text{Jup}}$] & 0.638$\pm$ 0.057$^{10}$ & 8.420$\pm$0.959$^{10}$ & 0.362$\pm$0.023$^{10}$ & 0.850$\pm$ 0.017$^{10}$ \\
 
 log$_{10}$ (g [cm s$^{-2}$]) & 3.219$\pm$2.172$^{10}$ & 4.339$\pm$3.395$^{10}$  & 2.973$\pm$1.767$^{10}$ & 3.343$\pm$1.652$^{10}$ \\ [0.5ex]
 \hline
 \end{tabular}
 \begin{tabular}{c | c c c c }
 \hline
 \textbf{Stellar Parameters}  & \textcolor{gray}{\textbf{HD 209458}}& \textbf{WASP-43} & \textcolor{gray}{\textbf{HD 189733}} & \textcolor{gray}{\textbf{HD 86226}} \\
 T$_{\text{eff}}$[K]& \textcolor{gray}{6100$\pm$100$^{8}$} & 4500$\pm$100$^{9}$ & \textcolor{gray}{5000$\pm$ 100$^{8}$} & \textcolor{gray}{5900$\pm$100$^{11}$} \\
 Mass [M$_{\text{Sun}}$] &  \textcolor{gray}{1.119$\pm$ 0.033$^{8}$}  & 0.717$\pm$0.025$^{9}$ & \textcolor{gray}{0.806 $\pm$ 0.048$^{8}$} & \textcolor{gray}{1.019$\pm$0.066$^{11}$}\\
 Radius [R$_{\text{Sun}}$] & \textcolor{gray}{1.155 $\pm$ 0.016$^{8}$} & 0.667$\pm$0.01$^{9}$ & \textcolor{gray}{0.756 $\pm$ 0.018$^{8}$} & \textcolor{gray}{1.053$\pm$0.026$^{11}$}\\
 Spectral Type & \textcolor{gray}{G0V} & K7V & \textcolor{gray}{K1.5V} & \textcolor{gray}{G2V}\\
 Metallicity [Fe/H][dex] & \textcolor{gray}{0.00 $\pm$ 0.05$^{8}$} & -0.01$\pm$0.012$^{9}$ & \textcolor{gray}{-0.03 $\pm$0.08$^{8}$} & \textcolor{gray}{0.018$\pm$0.057$^{11}$} \\
 \hline
 \textbf{Orbital Parameters} & \textcolor{gray}{\textbf{HD 209458b}}  & \textbf{WASP-43b} & \textcolor{gray}{\textbf{HD 189733b}} & \textcolor{gray}{\textbf{HD 86226c}}\\ [0.5ex]
 Semi-Major Axis [AU] & \textcolor{gray}{0.04707 $\pm$0.00047$^{8}$} & 0.01526$\pm$0.00018$^{9}$ & \textcolor{gray}{0.03099 $\pm$0.00063$^{8}$}&\textcolor{gray}{0.049$\pm$0.001$^{11}$} \\ 
 Orbital Period [days] & \textcolor{gray}{3.524746$^{8}$} & 0.81347753$\pm$0.00000071$^{9}$  & \textcolor{gray}{2.218573$^{8}$} &\textcolor{gray}{3.98442$\pm$0.00018$^{11}$} \\ [0.5ex] 
 \hline
 \textbf{Planetary Parameters} \\ [0.5ex] 
 T$_{\text{eq}}$[K] & \textcolor{gray}{1500$\pm$100$^{8}$} & 1400$\pm$100$^{9}$  & \textcolor{gray}{1200$\pm$100$^{8}$} &\textcolor{gray}{1300$\pm$100$^{11}$} \\
 
 
 Mass [M$_{\text{Jup}}]$  & \textcolor{gray}{0.685$\pm$0.015$^{8}$} & 2.034$\pm$0.052$^{9}$ & \textcolor{gray}{1.144$\pm$0.057$^{8}$}&\textcolor{gray}{7.25$\pm$1.19$^{11}$ [M$_{\rm E}$] or 0.023 M$_{\text{Jup}}$} \\
 
 Radius [R$_{\text{Jup}}$] & \textcolor{gray}{1.359$\pm$0.019$^{8}$} & 1.036$\pm$0.019$^{9}$  &  \textcolor{gray}{1.138$\pm$0.027$^{8}$}&\textcolor{gray}{2.16$\pm$0.08$^{11}$ [R$_{\rm E}$] or 0.193 R$_{\text{Jup}}$} \\

 Density [g cm$^{-3}$] & \textcolor{gray}{0.362$\pm$0.012$^{10}$} & 2.426$\pm$0.100$^{10}$ & \textcolor{gray}{1.029$\pm$0.066$^{10}$}&\textcolor{gray}{3.97$\pm$0.78$^{11}$}\\
 
 g [g$_{\text{Jup}}$] & \textcolor{gray}{0.371$\pm$0.017$^{10}$} & 1.900$\pm$ 0.069$^{10}$ & \textcolor{gray}{0.883$\pm$0.053$^{10}$} &\textcolor{gray}{0.614$\pm$0.106$^{10}$} \\
 
 log$_{10}$ (g [cm s$^{-2}$]) & \textcolor{gray}{2.983$\pm$1.453$^{10}$}  & 3.691$\pm$2.253$^{10}$  & \textcolor{gray}{3.360$\pm$2.138$^{10}$}&\textcolor{gray}{3.202$\pm$2.438$^{10}$} \\ [0.5ex]
 \hline
 \end{tabular}
 \caption[]{Physical parameters of the exoplanets, their host star and of the orbit. The equilibrium temperature given in the literature have been rounded to the next 100K as it in unrealistic to expect that it can be determined to a precision of 1K or 10K.  \textbf{Top:} Ultra-hot Jupiters. \textbf{Bottom:} hot  giant gas planets and HD 86226c, a close-in sub-Neptune. Grayed planets are used for comparison only, black coloured planets are composing the gas giant sample studied here.
 
 \textbf{References}: $^{1}$\citet{10.1093/mnras/stw522}, $^{2}$\citet{vaneylen13}, $^{3}$\citet{pal09}, $^{4}$\citet{Gillon14}, $^{5}$\citet{Hellier2009}, $^{6}$\citet{Pearson_2019}, $^{7}$\citet{Sheppard17}, $^{8}$\citet{2008ApJ...677.1324T}, $^{9}$\citet{2012A&A...542A...4G},$^{10}$This Work (calculated using the referenced data),$^{11}$\citet{teske2020tess}.}
 \label{table:stpl} 
 \end{table*}

\begin{figure*}[ht]
    \includegraphics[width=20pc]{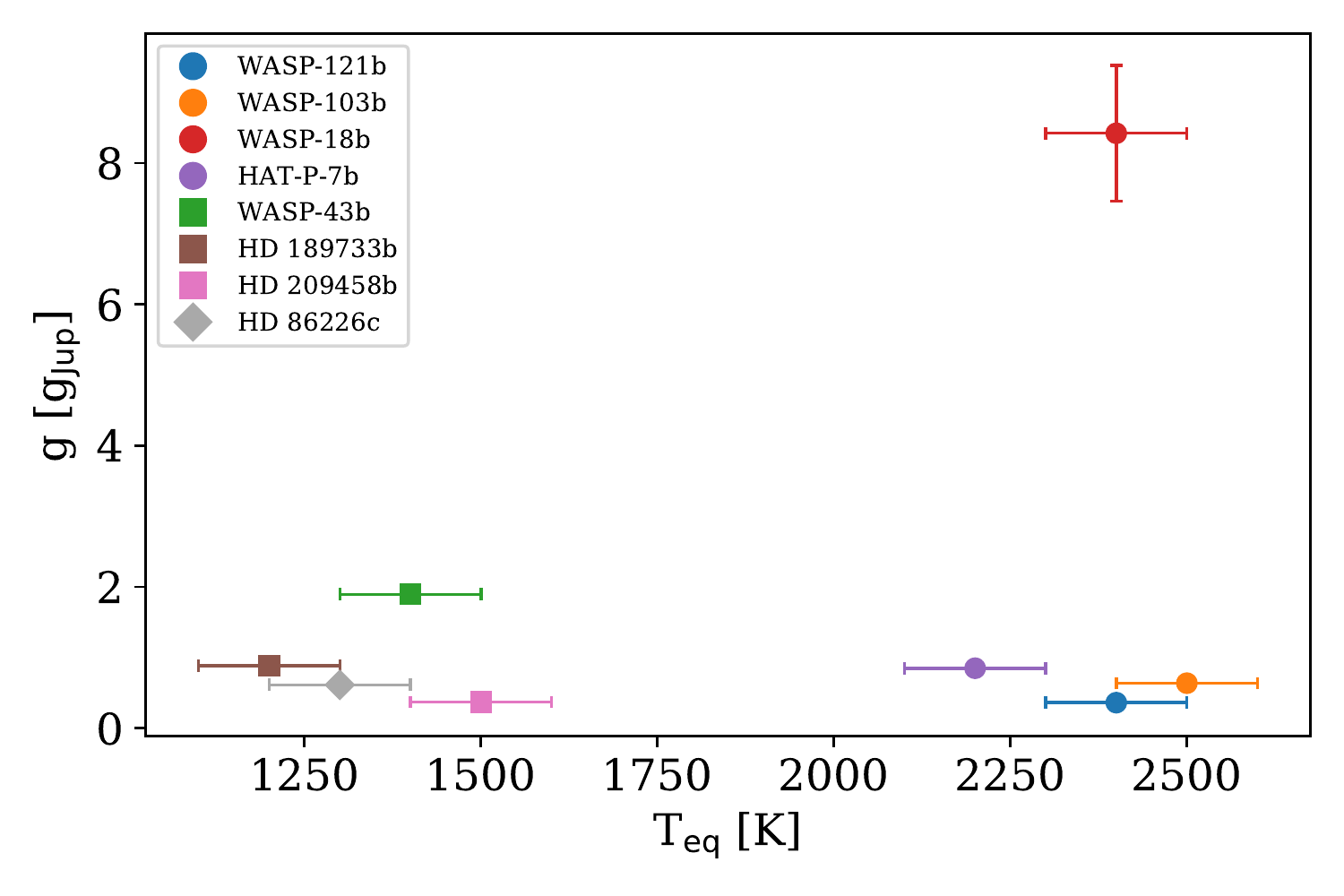}
    \includegraphics[width=20pc]{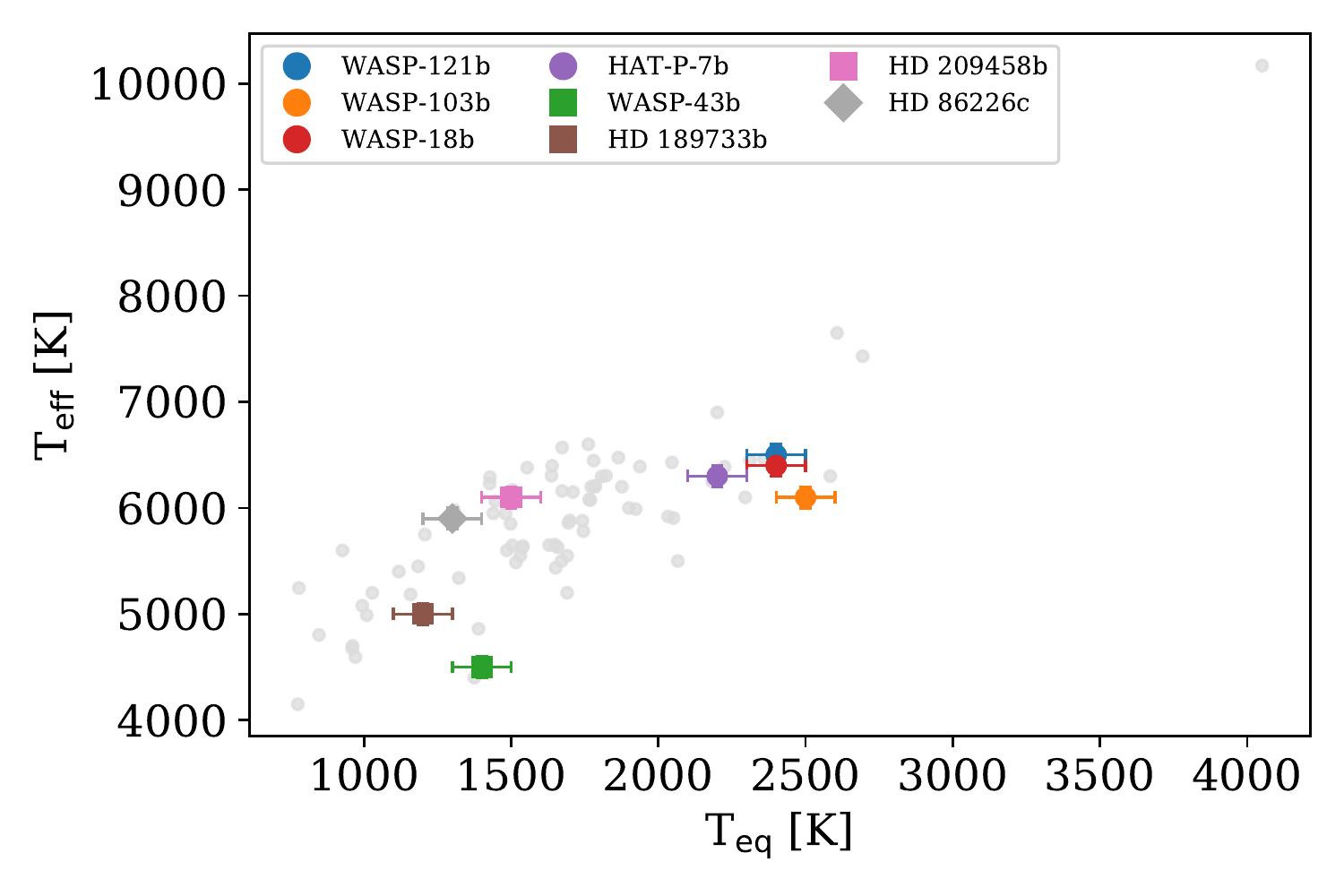}\\
    \includegraphics[width=20pc]{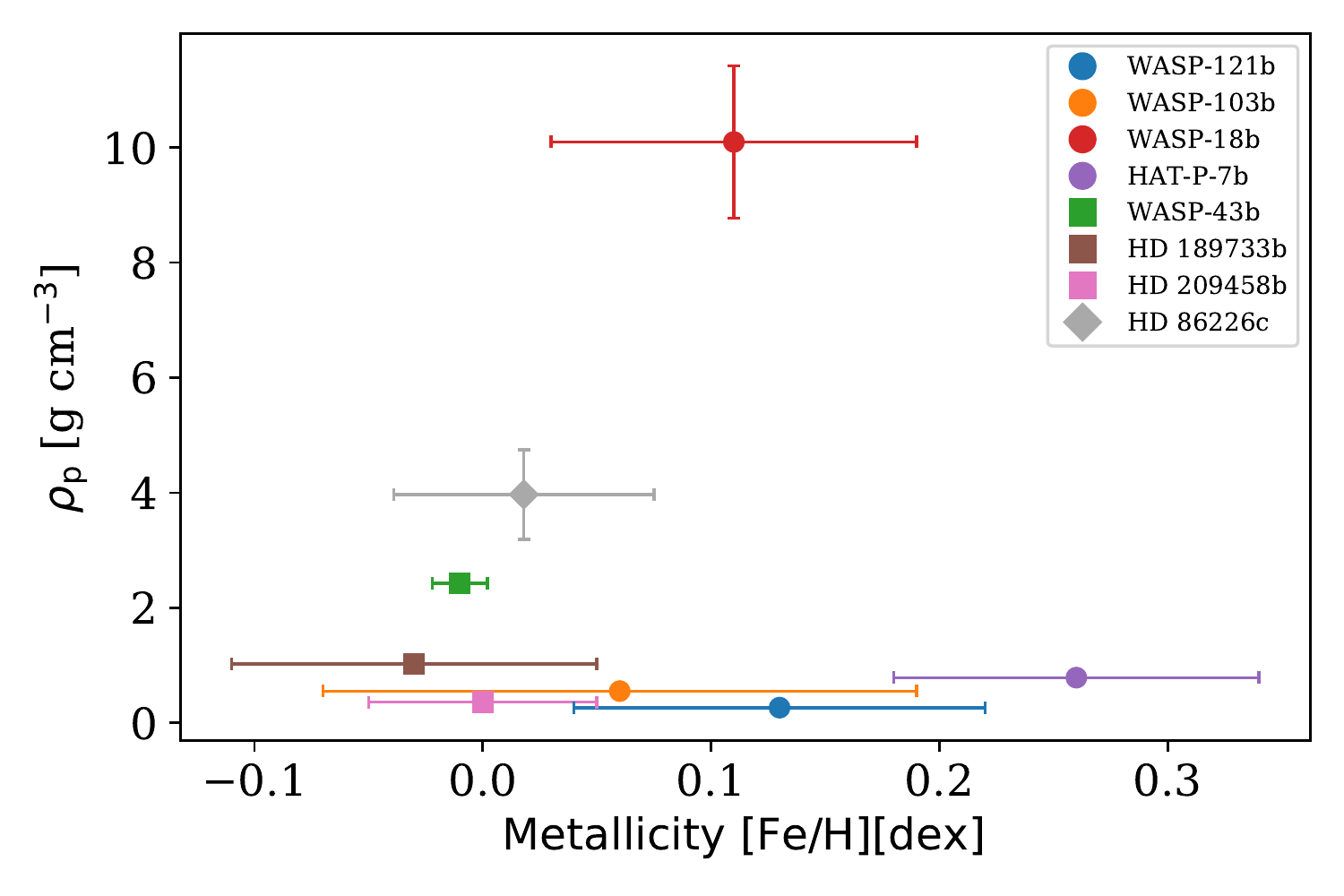}
    \includegraphics[width=20pc]{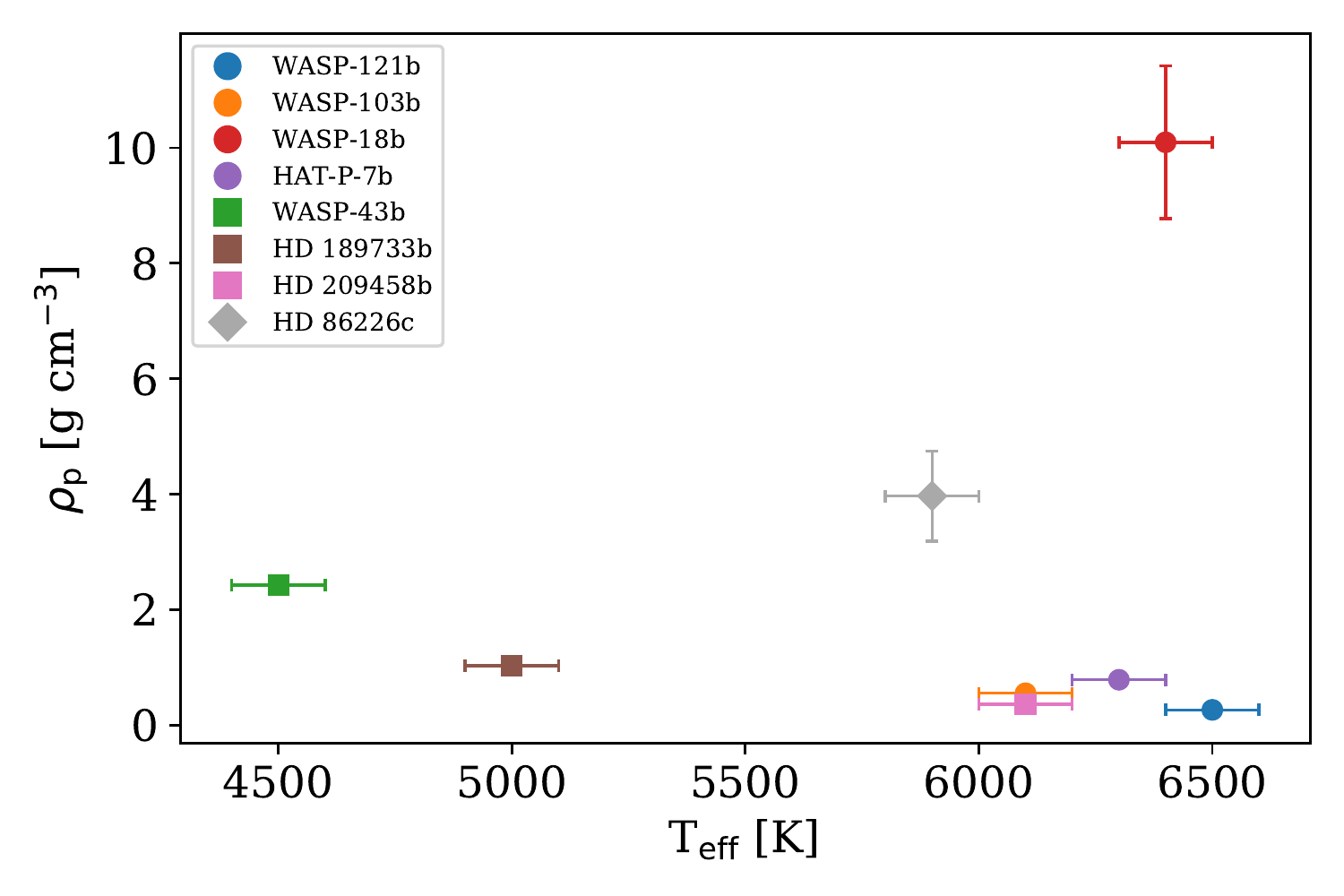}
    \caption{ Relationship between selected system properties for  hot  giant Jupiters (WASP-43b, HD 189733b, HD 209458b) and ultra-hot Jupiters (HAT-P-7b, WASP-18b, WASP-103b, WASP-121b). The two classes of giant gas planets are clearly separated  with respect to their global planetary temperature, $T_{\rm eq}$ [K] and the planets' surface gravity, $g$ [g$_{\rm Jup}$].
    There is not as clear a separation between the hot giant gas planets and ultra-hot Jupiters regarding their host star's effective temperature, $T_{\rm eff}$ [K], bulk planetary density $\rho_{\rm p}$,  and stellar metallicity [Fe/H]. The ultra-hot Jupiters are shown by circle markers, the giant-gas planets by the square markers and a sub-Neptune (HD 86226c) by the diamond marker. We also include the sample of hot and the ultra-hot Jupiters from Table 1 in \citet{2020arXiv200715287B} in the $T_{\rm eff}$ vs $T_{\rm eq}$ plot (top right) as smaller light gray points. }  
   \label{fig:Teq_Teff_all}
\end{figure*}

The theoretical modelling of exoplanet atmospheres has reached a level that may allow to study a set of ultra-hot Jupiters and  hot  giant gas planets with respect to their potential trends in cloud properties.  In this paper, we consider the ultra-hot Jupiters WASP-103b, WASP-121b, HAT-P-7b, and WASP-18b, and the hot  giant gas planet WASP-43b. WASP-43b and WASP-18b are ERS JWST targets (program 1366).
WASP-43b (program 1224), and  WASP-121b (program 1201) are GTO JWST targets\footnote{https://www.stsci.edu/jwst/observing-programs/program-information}.
The ultra-hot Jupiters in our sample have a global temperature of  $T_{\rm eq}> 2000$K in comparison to the hot  giant gas planet $T_{\rm eq}<1500$K. Table~\ref{table:stpl} shows that this distinction holds also when including HD\,189733b and HD\,209458b. Figure~\ref{fig:Teq_Teff_all} demonstrates that these  sub-classes of giant gas planets separate well in the g$_{\rm P}$/T$_{\rm eq}$ and the T$_{\rm eff}$/T$_{\rm eq}$ diagrams. As a comparison, we add the close-in sub-Neptune HD~86226c (\citealt{teske2020tess}) which has a global temperature similar to those of hot  gas giants, and falls into the giant gas planet  g$_{\rm P}$/T$_{\rm eq}$ corner 
of Fig.~\ref{fig:Teq_Teff_all} (top left). \cite{2020arXiv201012745C} show that the IRAC2 (4.5$\mu$m) irradiation temperature of the hot Neptune LTT~9779b is roughly comparable to that of HAT-P-7b, and generally consistent with that of giant gas planets.

For our comparison study, we use results from 3D GCMs as input for our kinetic cloud formation code in order to study differences and similarities of the cloud coverage of these planets. The atmosphere dynamics enters our simulation only indirectly due to its effect on the local thermodynamic properties. Starting from the global temperature/pressure/velocity structure, cloud properties like cloud particle formation rate, mean particle size, material compositions, the dust-to-gas ratio are investigated, and also characteristic gas-phase properties like C/O, mean molecular weight and degree of ionisation which all provide insight into various processes beyond cloud formation. C/O (and also other mineral ratios like Si/O, Fe/O, \citealt{2019A&A...626A.133H}) 
are used as potential markers for planet formation scenarios
(e.g. \citealt{2014Life....4..142H, 2019A&A...632A..63C}), the mean molecular weight is important to guide our understanding of atmospheric extensions, and the degree of ionisation shows in how far electrostatic or electromagnetic processes require attention in exoplanet atmospheres (e.g. \citealt{2015MNRAS.454.3977R}). The latter is an essential step towards magnetosphere studies of exoplanets (for example, \citealt{2018A&A...616A.182V,2020ApJ...895...62S})

The paper is structured as follows.
After the introduction (Sect.~\ref{intro}), Sect.~\ref{s:ap} outlines our modelling approach to study the cloud properties and some key gas characteristic for the five  gas planets in our sample listed in black colour in  Table~\ref{table:stpl}. Section~\ref{s:input} summarises the input.  Section~\ref{s:clouds} compares the cloud properties of the different planets. 
Section~\ref{s:gas} compares the planets with respect to the cloud feedback on the atmospheric gas (C/O), the  mean molecular weight, and also the degree of ionisation.  Section~\ref{s:wasp_gcm_comp} looks at how the inner boundary of the General Circulation Model (GCM) affects the (T$_{\rm
  gas}$(z), p$_{\rm gas}$(z), v$_{\rm z}(x,y,z)$)-profiles and consequently the possible cloud properties for WASP-43b. Observational implications are explored in Sect.~\ref{s:obs}, Sect.~\ref{s:disc} presents our discussion and Sect.~\ref{s:concl} presents our conclusions.
  
     \begin{figure*}
   \centering
   \includegraphics[width=21pc]{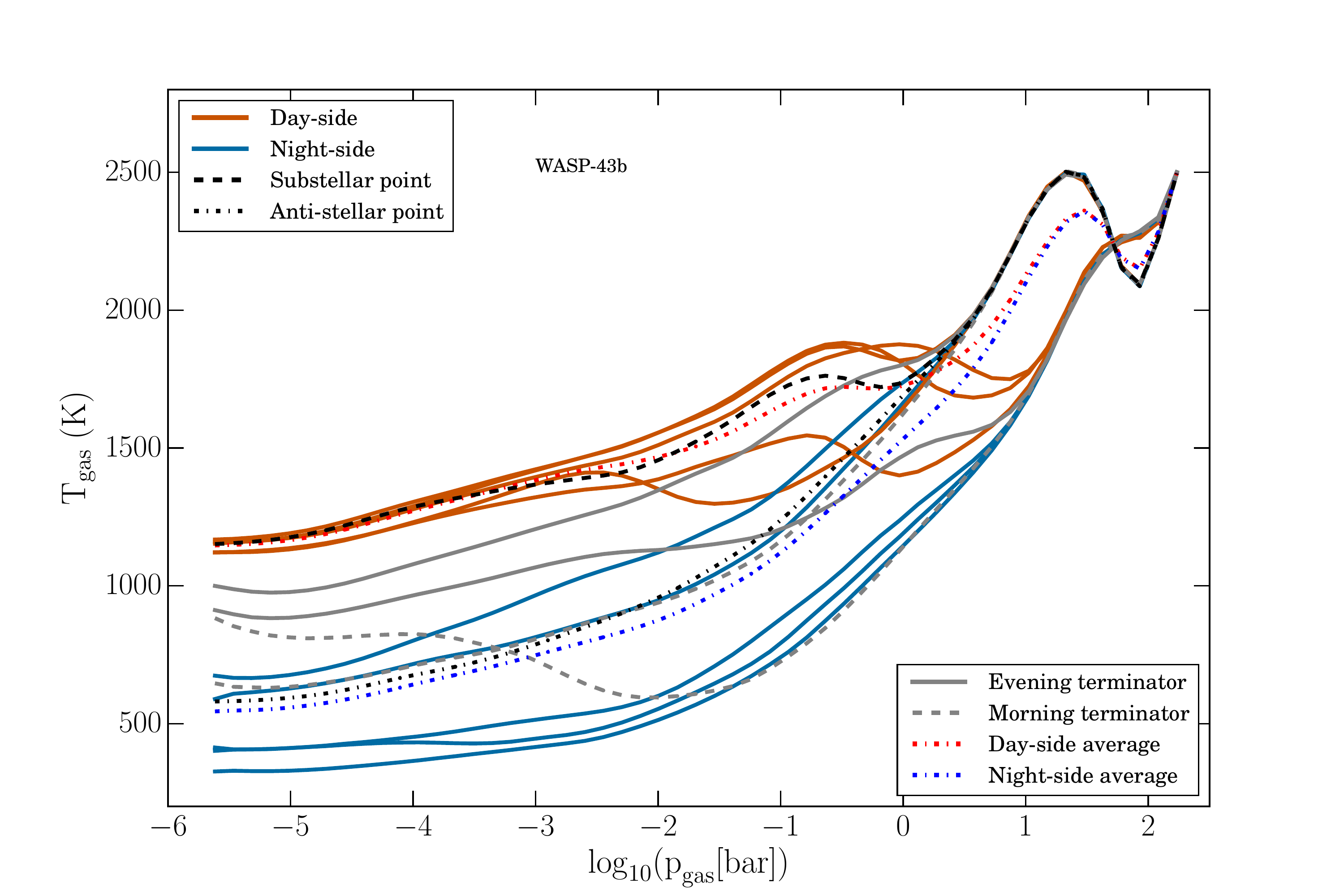}
   \includegraphics[width=21pc]{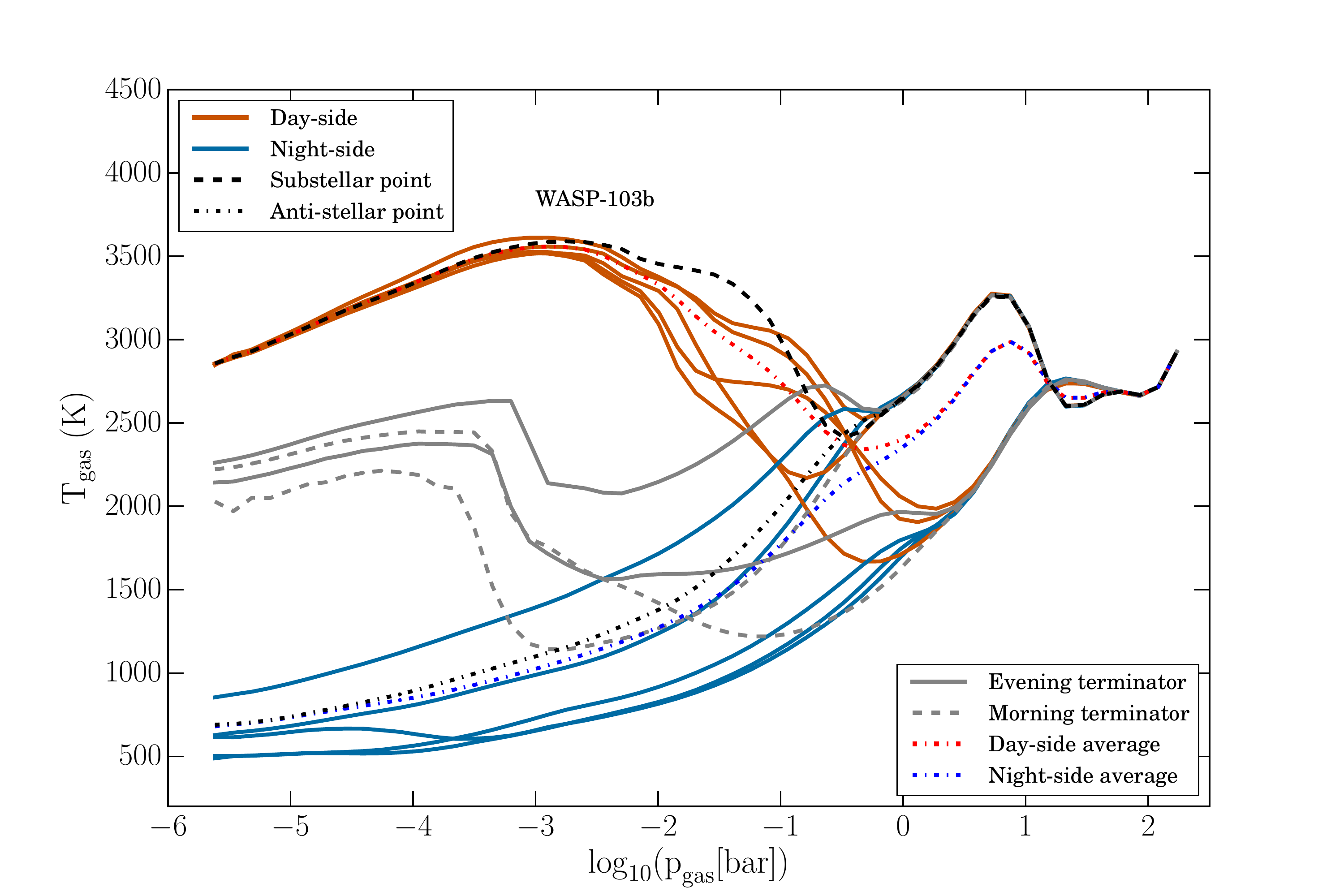}\\
   \includegraphics[width=21pc]{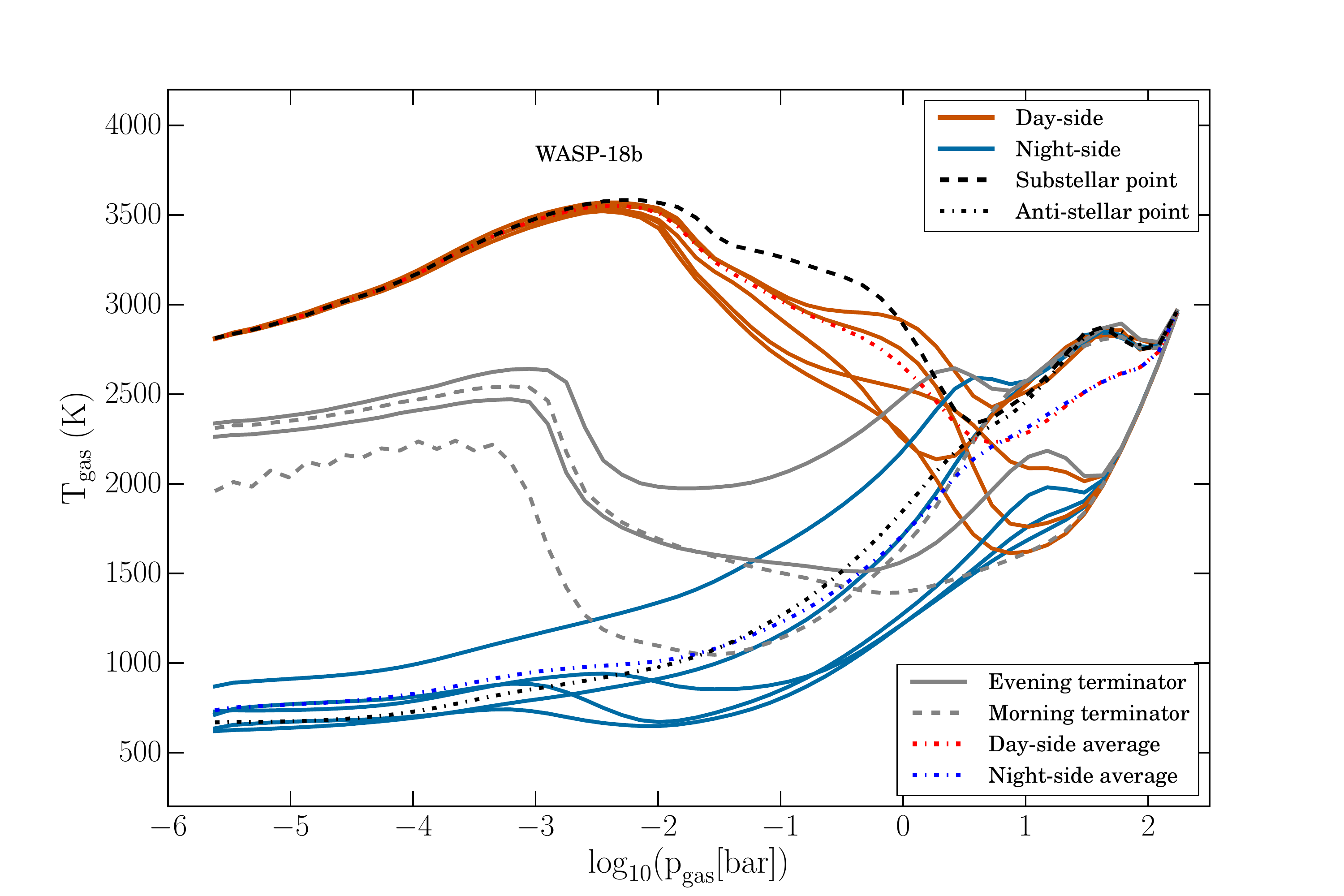}
   \includegraphics[width=21pc]{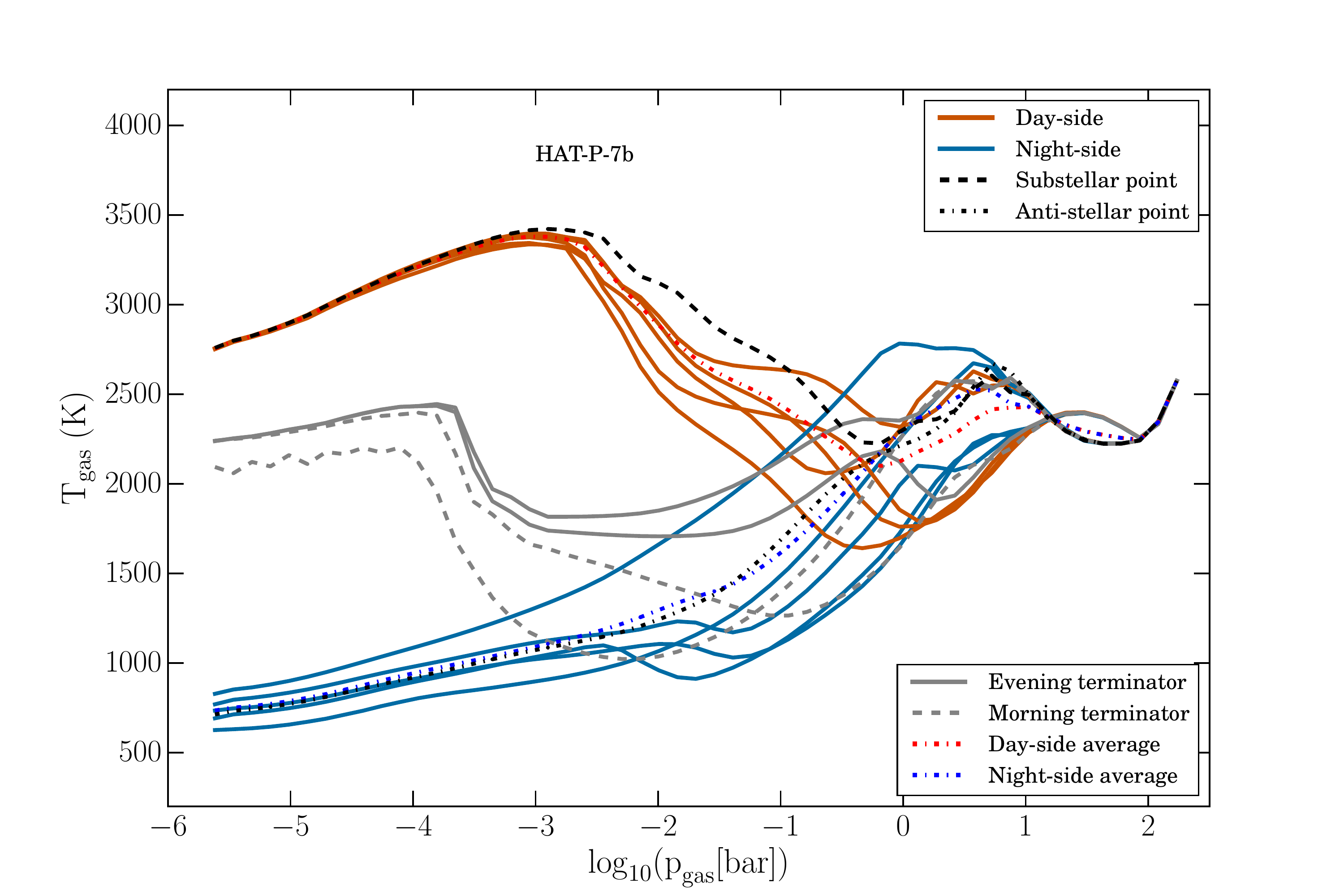}\\
   \includegraphics[width=21pc]{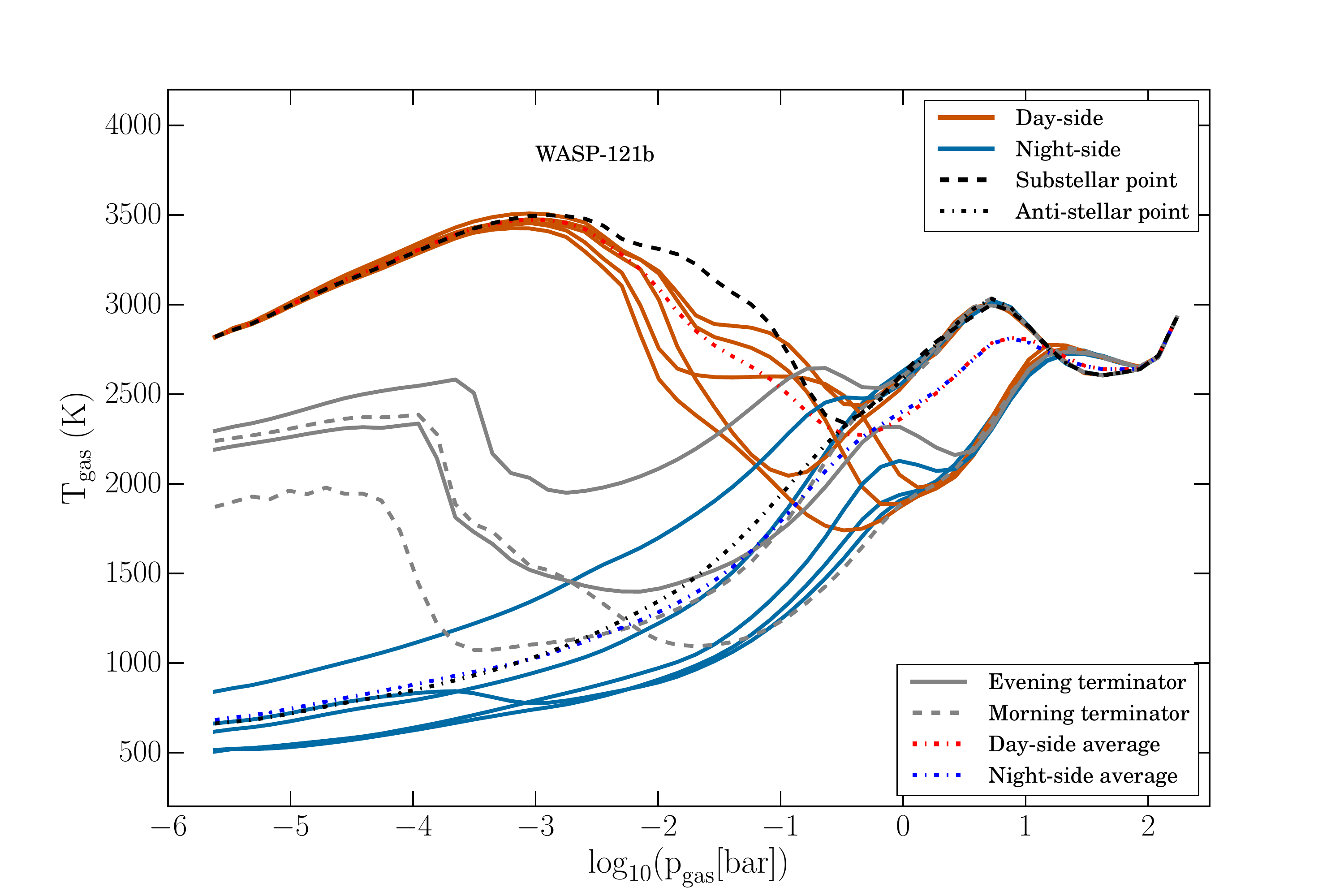}
   \includegraphics[width=21pc]{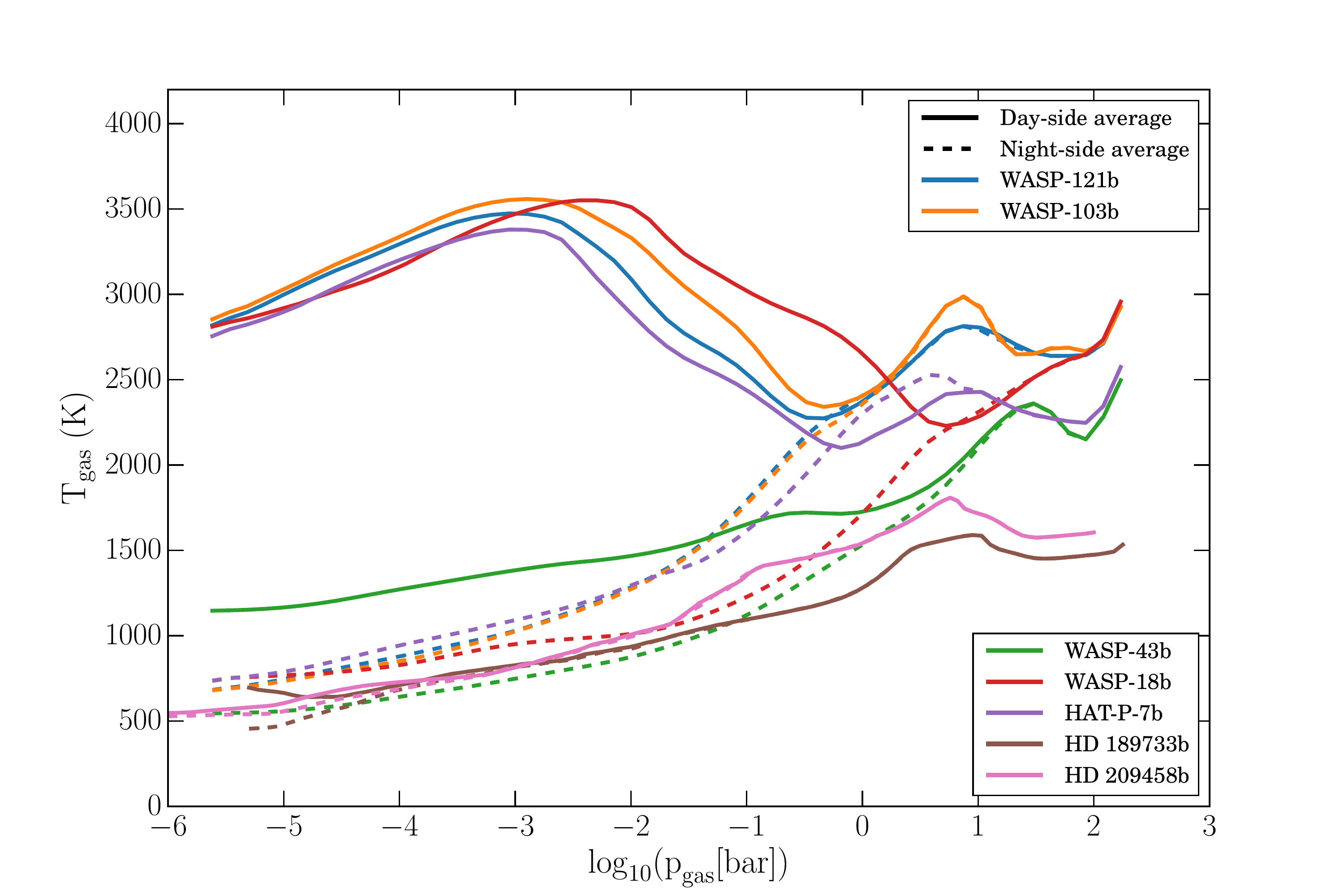}
      \caption{The 16 1D $(T_{\rm gas}, p_{\rm gas})$-profiles of the hot  giant gas planet WASP-43b, and the ultra-hot Jupiters WASP-18b, HAT-P-7b, WASP-103b, and WASP-121b. The day- and nightside average profiles (lower right panel) exclude the terminators.  The same longitudes, $\phi$, and latitudes, $\theta$, are sample for each planets. The sampled longitudes are $\phi=0^o,45^o,90^o,135^o,-180^o,-135^o, -90^o$ ($\phi<0$ nightside), the latitudes are the equator $\theta=0^o$ and $\theta=45^o$ in the northern hemisphere. The substellar point is $(\theta,\phi)=(0^o,0^o)$ (black dashed), the antistellar point is $(\theta,\phi)=(0^o,-180^o)$ (black dash-dot), the terminators are at $\phi = 90^o , -90^o$ (grey lines). Day/night temperature differences  of $\approx 2500\,\ldots\,3000$K occur in the atmospheres of ultra-hot Jupiters.}      
      \label{Tp_all_new}
   \end{figure*}

\section{Approach}\label{s:ap}
We adopted a two-step approach in order to examine the cloud structures of five gas planets in comparisons, similar to works on the hot Jupiters HD\,189733b, HD\,209458b and WASP-43b (\citealt{Lee2015,2016MNRAS.460..855H}), and the ultra-hot Jupiters WASP-18b (\citealt{2019A&A...626A.133H}) and HAT-P-7b (\citealt{2019A&A...631A..79H,Mol2019}): The first modelling step produced pre-calculated 3D GCM results. These results were used as input for the second modelling step which was a kinetic cloud formation model consistently combined with equilibrium gas-chemistry calculations. We utilised 16 1D (T$_{\rm
  gas}$(z), p$_{\rm gas}$(z), v$_{\rm z}(x,y,z)$)-profiles for all planets in our ensemble (Fig.~\ref{Tp_all_new}). These profiles probe specific locations (incl. morning and evening terminators, substellar and antistellar point) on the planetary globe and were distributed as depicted in Fig.~1 in \cite{2019A&A...626A.133H}. The same longitudes, $\phi$, and latitudes, $\theta$, are sampled for all ensemble planets studied here. The sampled longitudes are $\phi=0^o,45^o,90^o,135^o,-180^o,-135^o, -90^o$ with $\phi<0$ on the nightside. The latitudes are at the equator $\theta=0^o$ and $\theta=45^o$ in the northern hemisphere. The 3D simulations assume that the southern hemisphere is similar to the northern hemisphere. The substellar point is $(\theta,\phi)=(0^o,0^o)$ (black dashed), the antistellar point is $(\theta,\phi)=(0^o,-180^o)$ (black dash-dot), the terminators are at $\phi = 90^o , -90^o$ (grey lines in Fig.~\ref{Tp_all_new}). T$_{\rm gas}$(z) is the local gas temperature [K], p$_{\rm gas}$(z) is the local gas pressure [bar], and v$_{\rm z}$(x,y,z) is the local vertical velocity component [cm s$^{-1}$]. We use the solar element abundances from \cite{2009ARA&A..47..481A}
  for the  undepleted element abundances.

  This two-step approach has the limitation of not explicitly taking into account the potential effect of horizontal winds on cloud formation,  nor the opacity of the cloud particles on the atmospheric structure.  However, processes governing the formation of mineral clouds are determined by local thermodynamic properties which are the result of 3D dynamic atmosphere  simulations. 
  Cloud particle properties such as particle size or particle composition should be smeared out in longitude compared to the results shown here. We note that comparing~\citet{Lee2015} (without horizontal advection) and ~\citet{Lee2016} (including horizontal advection), the non-coupled problem is both more computationally feasible, easier to interpret and provides reasonable first order insights into the expected  atmospheric cloud properties. The situation is somewhat different for photochemically triggered cloud formation. Photochemical hydrocarbon-haze production, for example, is determined by the external radiation field. We did, however, show that in the case of efficient mineral cloud formation, that hydrocarbon hazes play no role for the cloud opacity on the nightside and the terminator (\citealt{2020arXiv200514595H}). The dayside of all the ultra-hot Jupiters will be too hot for hydrocarbon hazes to be thermally stable.

\smallskip\noindent
{\it 3D GCM input:} We utilize the 3D thermal structures of WASP-43b, HAT-P-7b, WASP-18b, WASP-103b, and WASP-121b as described in 
\cite{2018A&A...617A.110P,2018AJ....156...10M} as input for our kinetic cloud-formation simulation. These 3D GCM structures were obtained with the cloud-free SPARC/MITgcm~\citep{Showman2009}  and was run for 300 Earth days, with the last 100 days used to calculate time averaged quantities. The pressure range covered is $2\cdot10^{-5}$bar$\,\ldots\,100$bar. Please refer to previous paper for more details of these models.

\smallskip
\noindent
{\it Kinetic cloud formation:} To preserve consistency for WASP-43b, HAT-P-7b, WASP-18b, WASP-103b, and WASP-121b, we apply the same set-up of our kinetic cloud formation model (nucleation, growth, evaporation, gravitational settling, element consumption and replenishment)  and equilibrium gas-phase calculations  as in Sect 2.1 in \cite{2019A&A...631A..79H}. Insight into numerical aspects of the solution can be found in Sect.~2.4 in \cite{Woitke2004}. Cloud particle formation  depletes the local gas phase, and gravitational settling causes these
elements to be deposited, for example, in the inner (high pressure) atmosphere where the cloud particles evaporate.  We use the local vertical velocity to calculate the necessary mixing timescale, $\tau_{\rm mix}\propto v_z(r)^{-1}$ as outlined in Sect. 2.4 in  \cite{Lee2016}. Hence, $\tau_{\rm mix}\propto K_{\rm zz}^{-1} \not= const$ along the probed atmospheric profiles. We acknowledge that this approach may introduce limitations which, however, will affect all planets in our sample similarly.   Our comparative study therefore remains valid.

Seed forming species (TiO$_2$, SiO) also need to be considered as surface growth material, since both processes (nucleation and growth) compete for the participating elements (Ti, Si, O, C, K, and Cl in this work). We consider the formation of 16 bulk materials ([s]=\ce{TiO2}[s], \ce{Mg2SiO4}[s], \ce{MgSiO3}[s], MgO[s], SiO[s], \ce{SiO2}[s], Fe[s], FeO[s], FeS[s], \ce{Fe2O3}[s], \ce{Fe2SiO4}[s], \ce{Al2O3}[s], \ce{CaTiO3}[s], \ce{CaSiO3}[s], C[s], KCl[s]) that form from 11 elements (Mg, Si, Ti, O, Fe, Al, Ca, S, C, K and Cl) by 128 surface reactions. The abundance of these 11 elements will decrease if cloud particles are forming (nucleation, growth) and increase if cloud particles evaporate. Sulfur has not been included in our present mineral cloud model.  Sulfuric materials in form of S[s], FeS[s],  MgS[s] would contribute by less than 10\% in volume fraction in a solar element abundance gas \citep[see Fig. 6 in][]{2019AREPS..47..583H}.

Other kinetic cloud models emphasise the importance of the Kelvin effect on cloud formation (e.g. \citealt{2018ApJ...860...18P,2020RAA....20...99Z}). The Kelvin effect refers to the decreasing thermal stability with increasing surface curvature, hence, with decreasing particle sizes. The Figures 3 in \cite{1996ASPC...96...69G} and in \cite{helling2013RSPTA} visualise the need for a super-cooling below thermal stability as result of the decreased surface binding energy with increasing surface curvature for small particles.  The effect of the resulting changing thermal stability  is taken into account in our nucleation model by determining the stable molecular clusters and deriving a surface tension as outlined in \cite{2015A&A...575A..11L}.


\begin{figure*}
\centering
  \includegraphics[width=21pc]{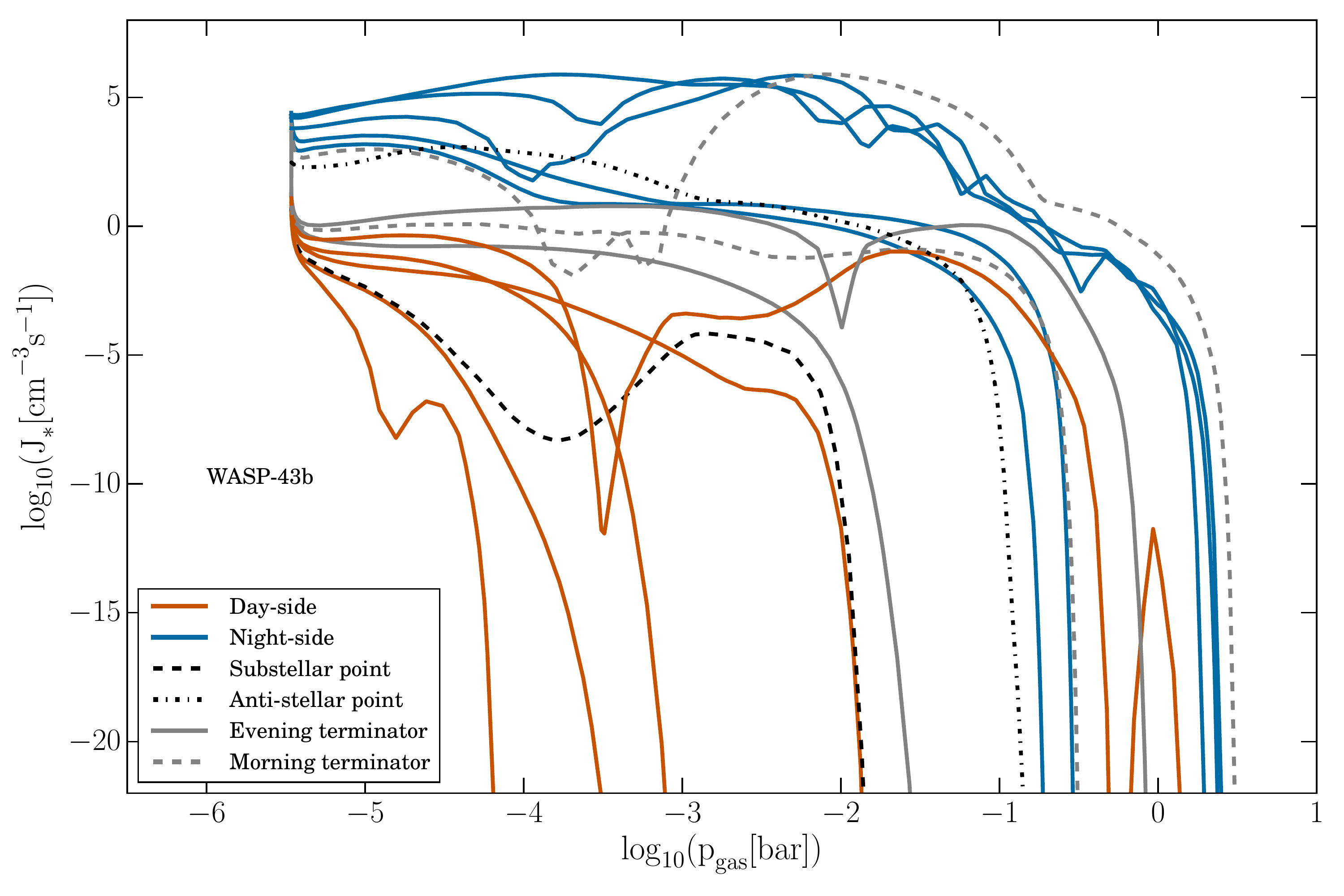}
   \includegraphics[width=21pc]{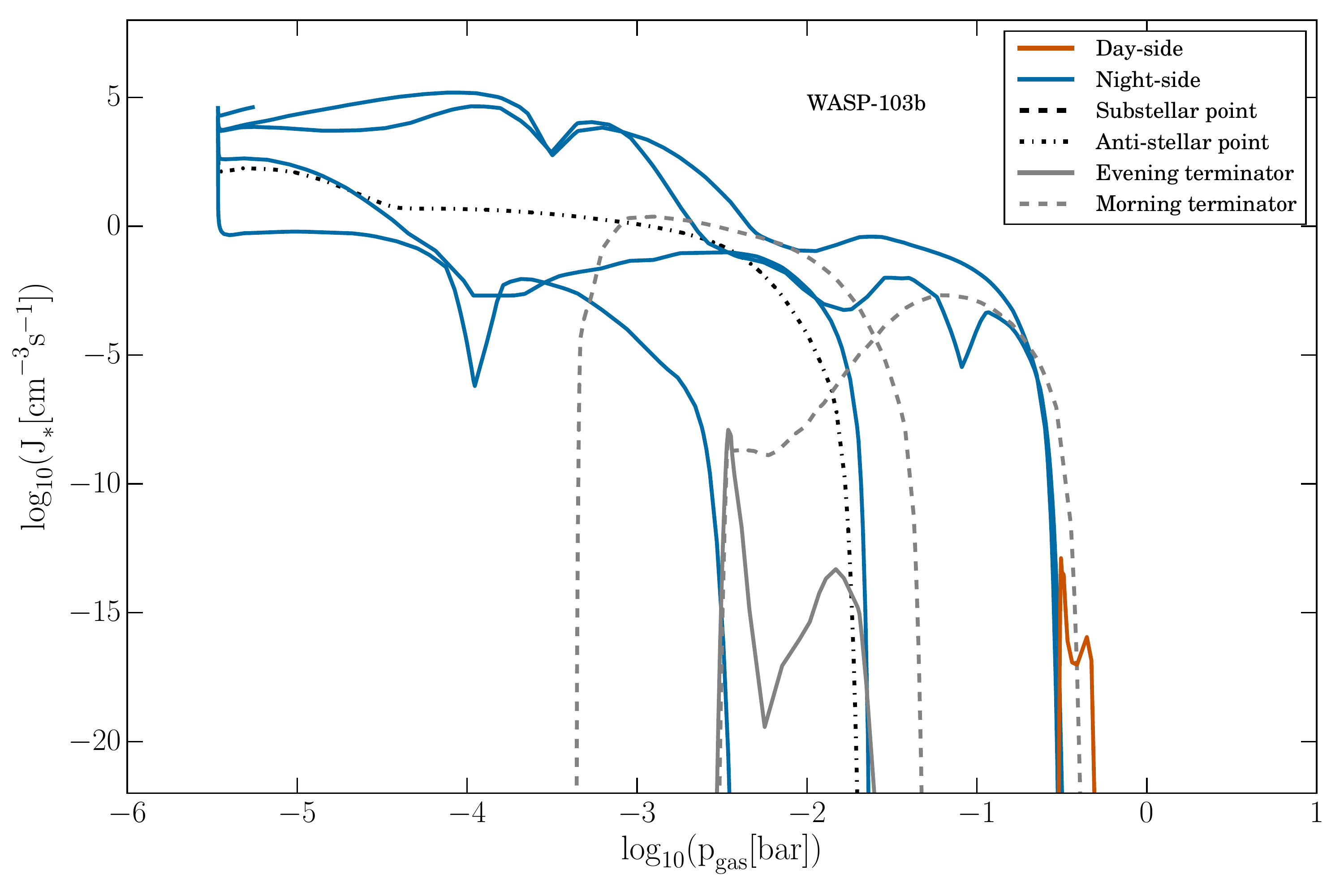}\\
   \includegraphics[width=21pc]{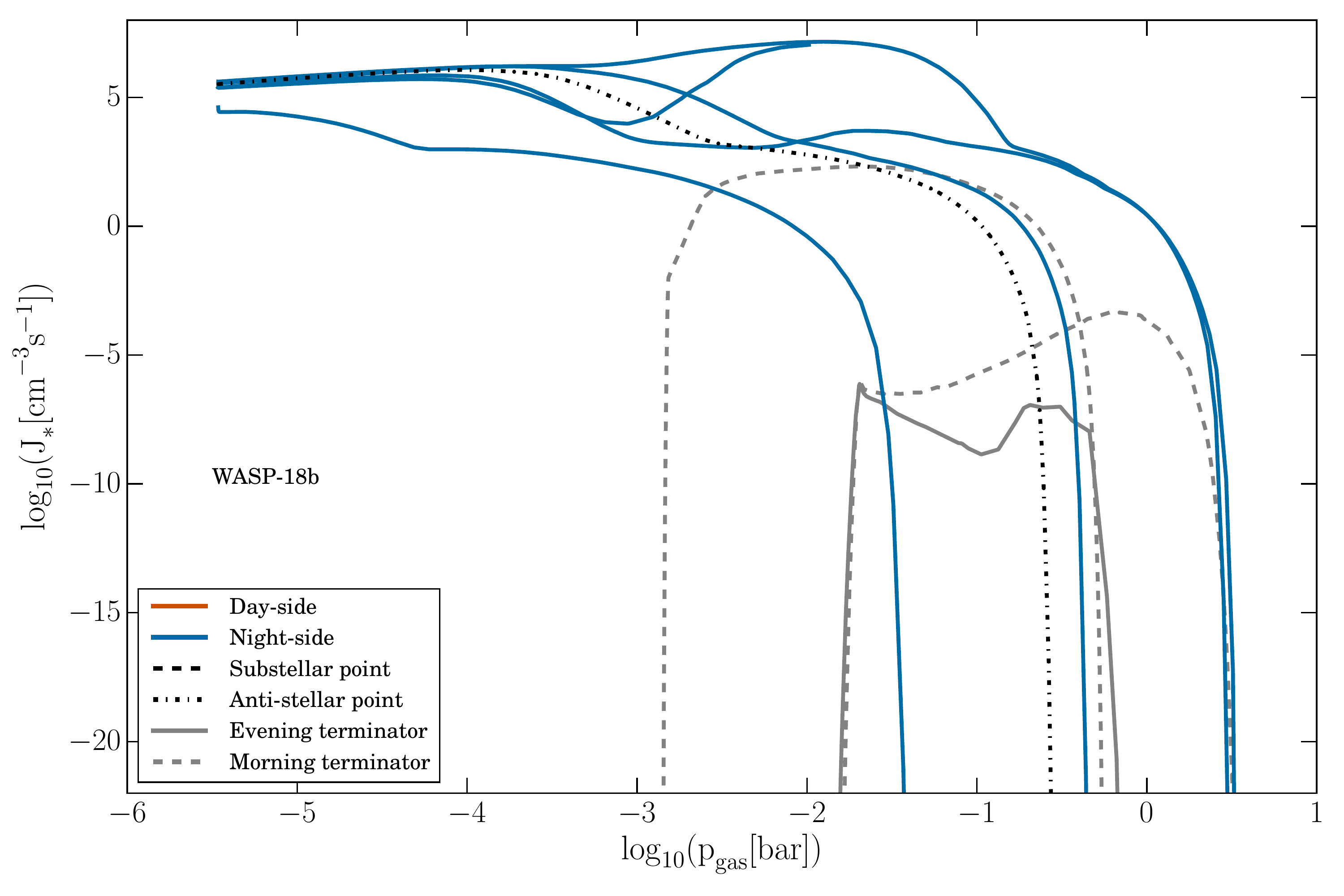}
   \includegraphics[width=21pc]{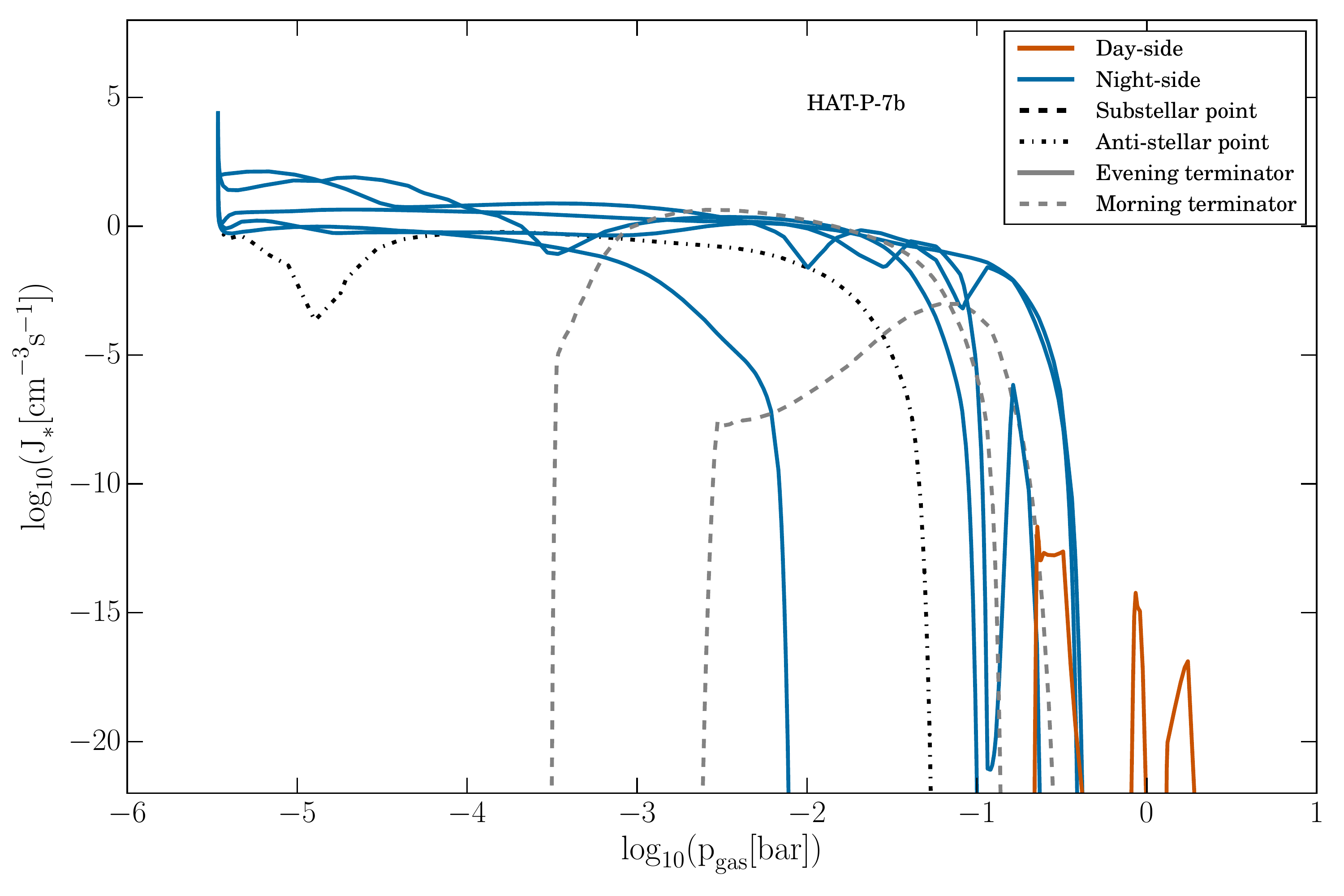}\\
   \includegraphics[width=21pc]{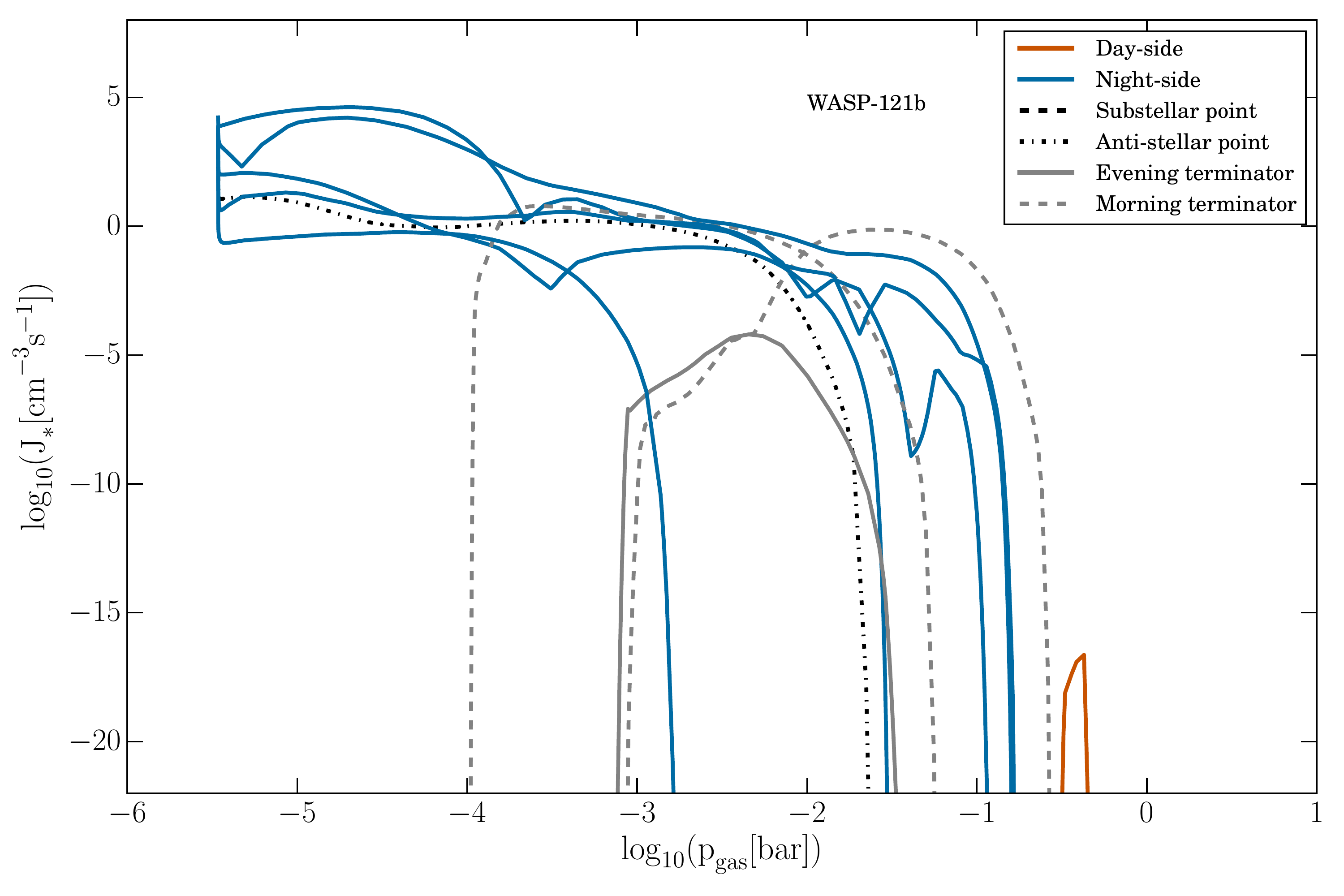}
   \includegraphics[width=21pc]{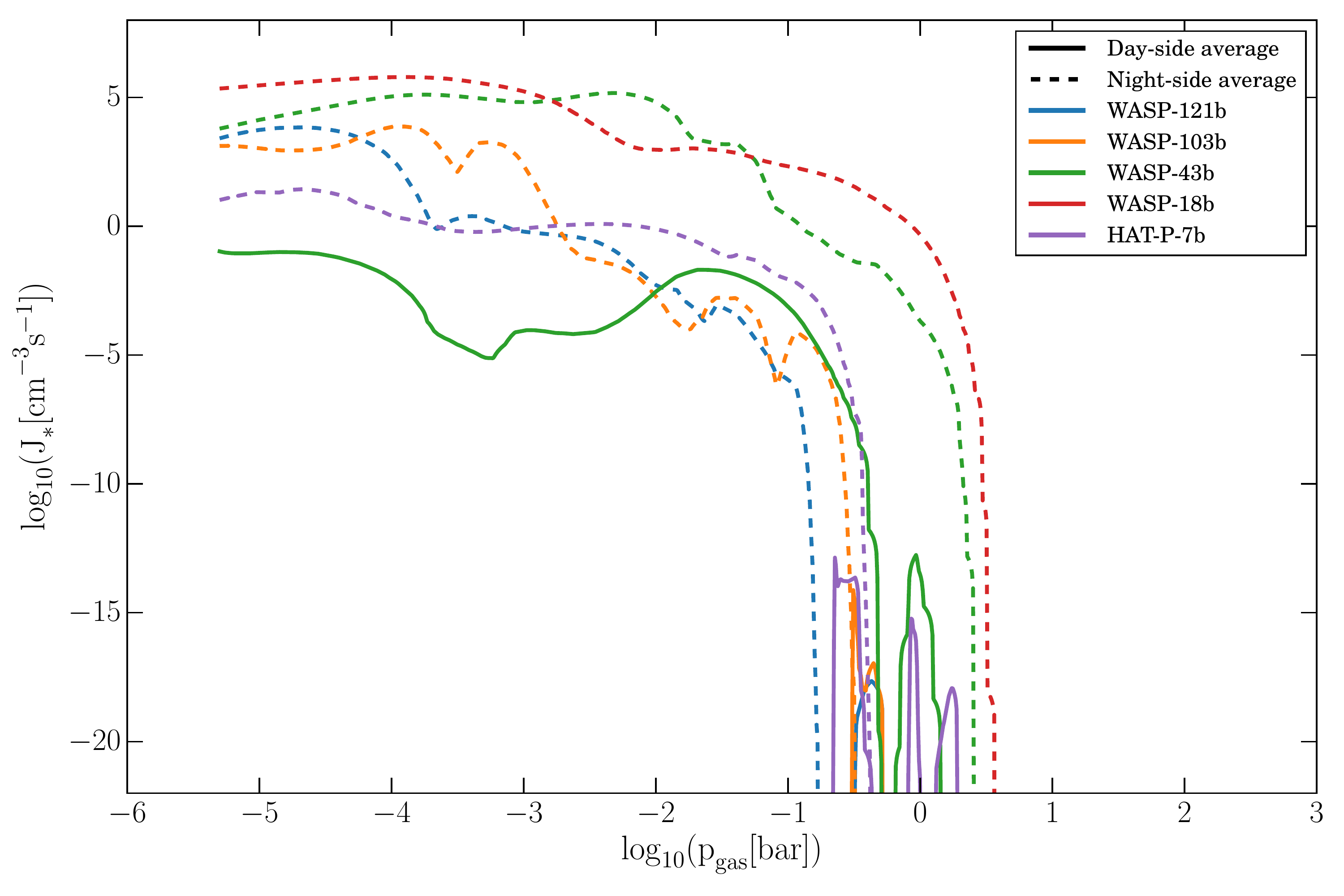}
  \caption{Total nucleation rate $J_{*}=\sum J_{i}$ [cm$^{-3}$ s$^{-1}$] for the 1D $(T_{\rm gas}, p_{\rm gas})$-profiles (Fig.~\ref{Tp_all_new}) for the hot  giant gas planets WASP-43b, 
    and the ultra-hot Jupiters WASP-18b, HAT-P-7b, WASP-103b, and WASP-121bs. The colour code is similar to Fig.~\ref{Tp_all_new}. The lower right panel shows the day (solid lines) and nightside (dashed lines) averaged seed formation rates, excluding the terminator profiles. All depicted planets show seed formation on the nightside. Only WASP-43b enables the nucleation process on the dayside efficiently. }
  \label{fig:Nuc_new}
\end{figure*}

\begin{figure*}
    \centering
    \includegraphics[width=21pc]{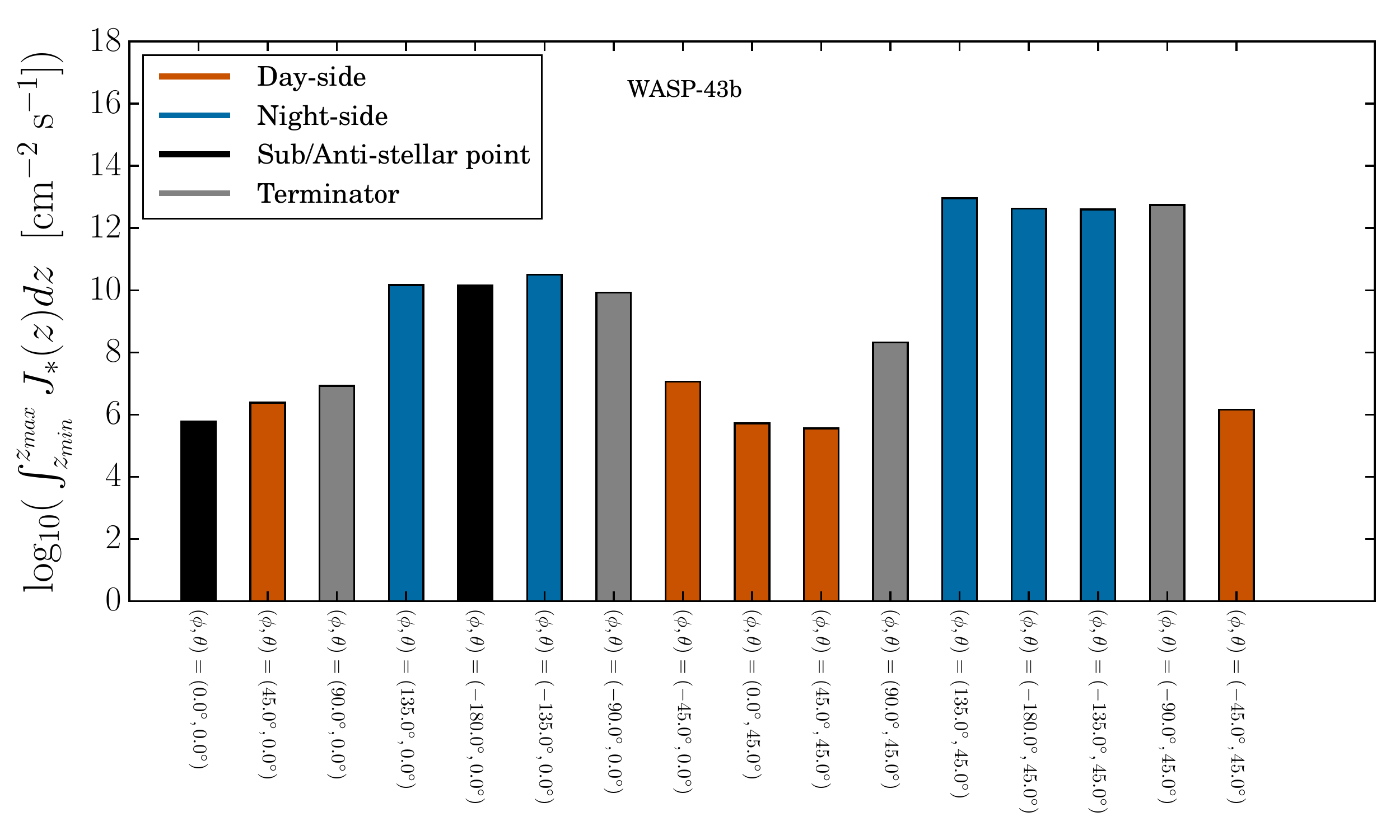}
    \includegraphics[width=21pc]{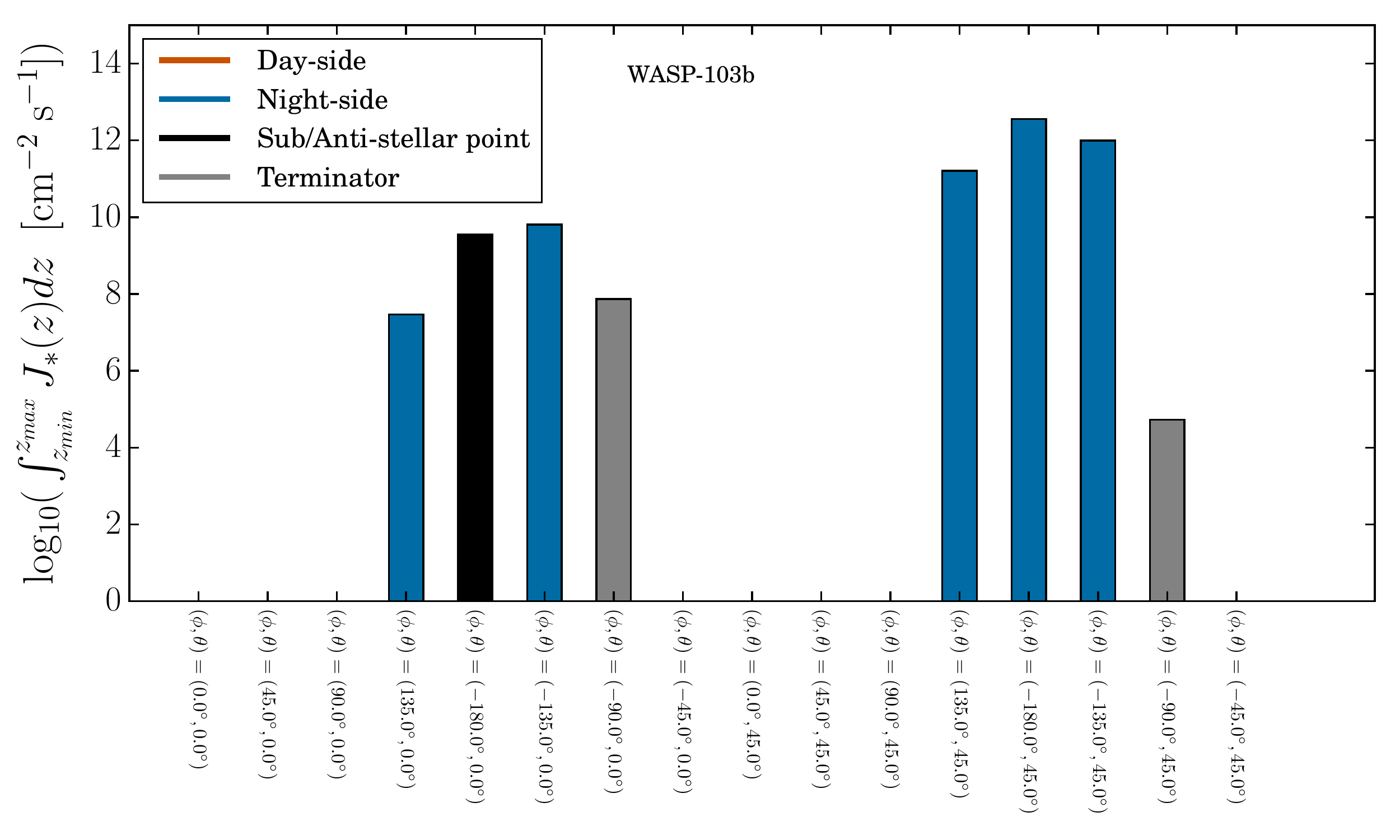}\\
    \includegraphics[width=21pc]{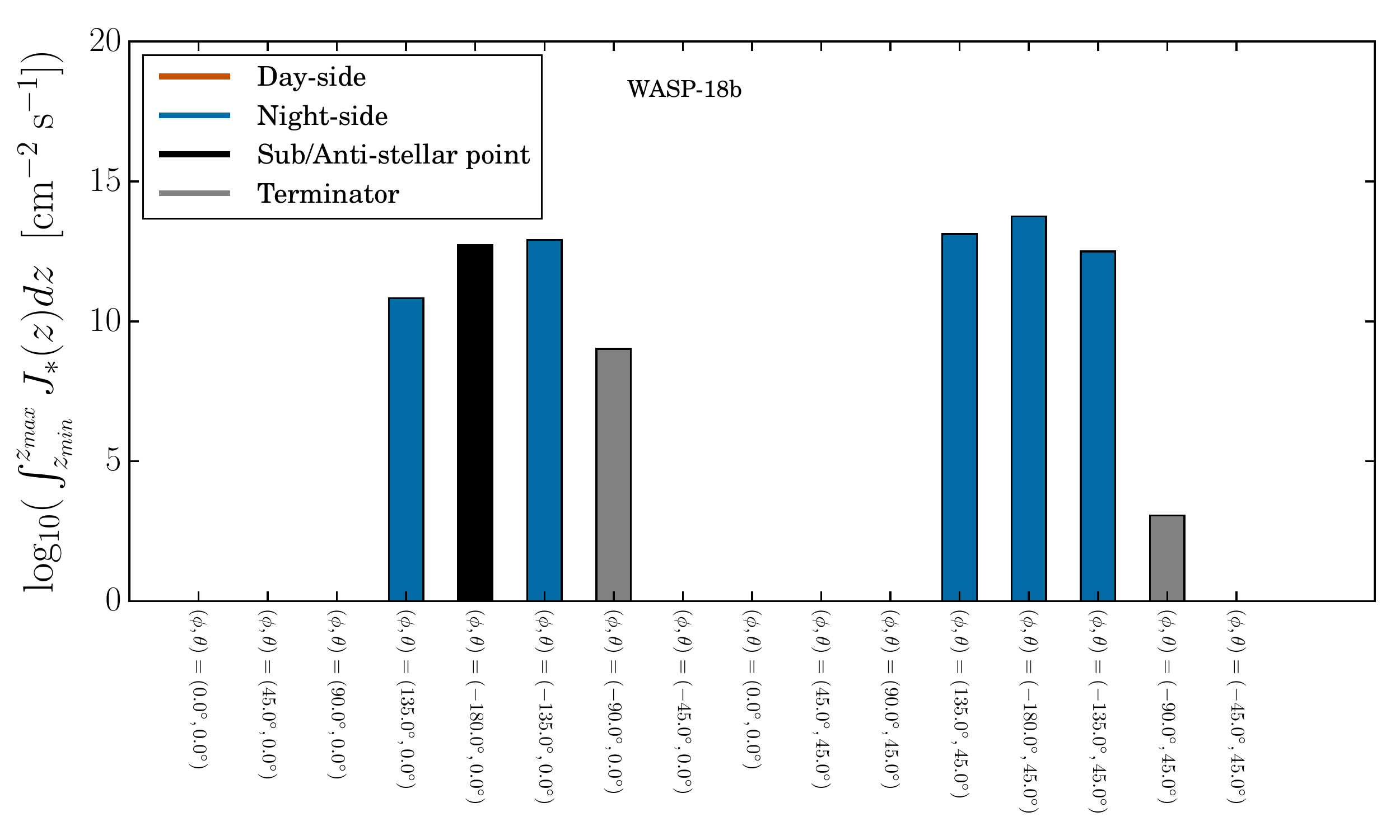}
    \includegraphics[width=21pc]{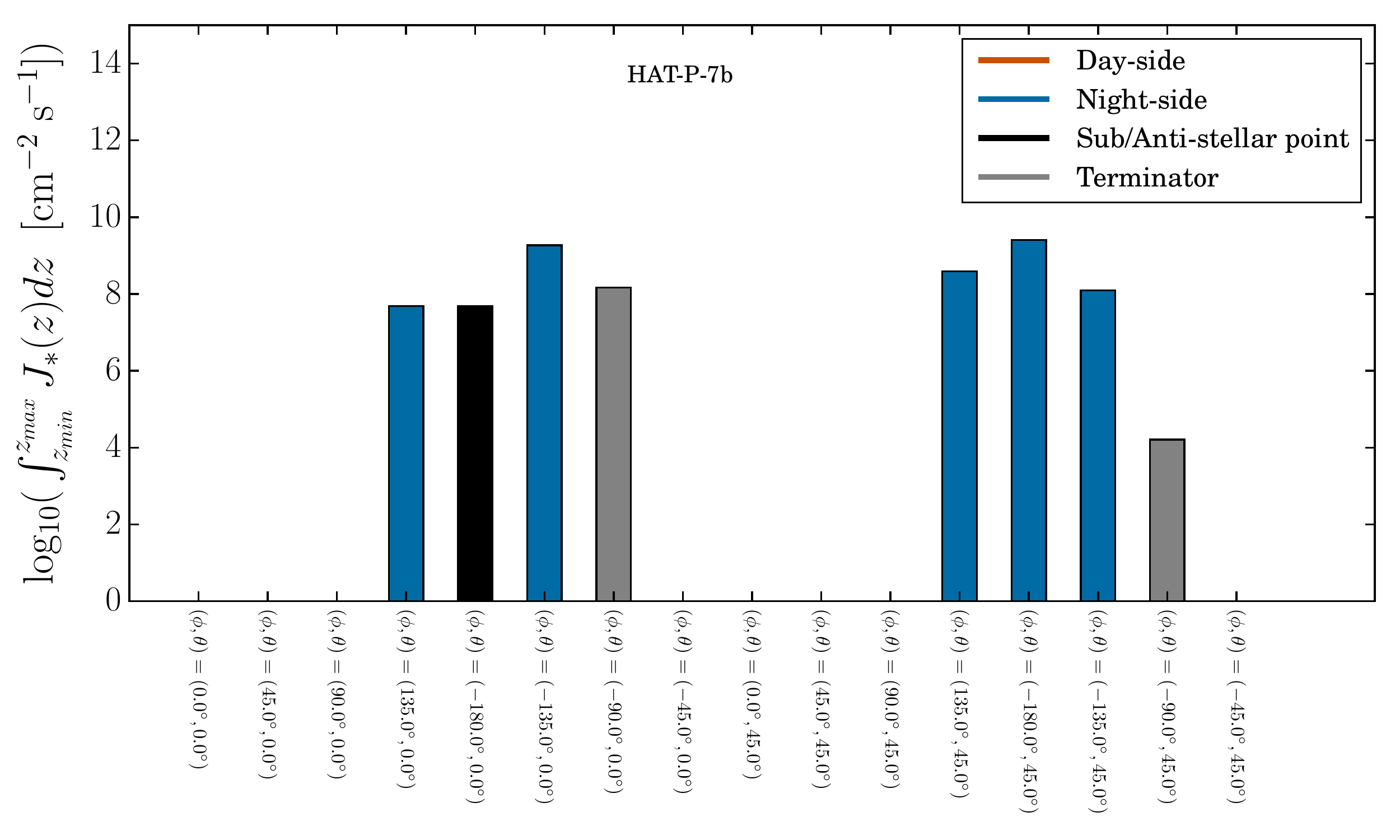}\\
    \includegraphics[width=21pc]{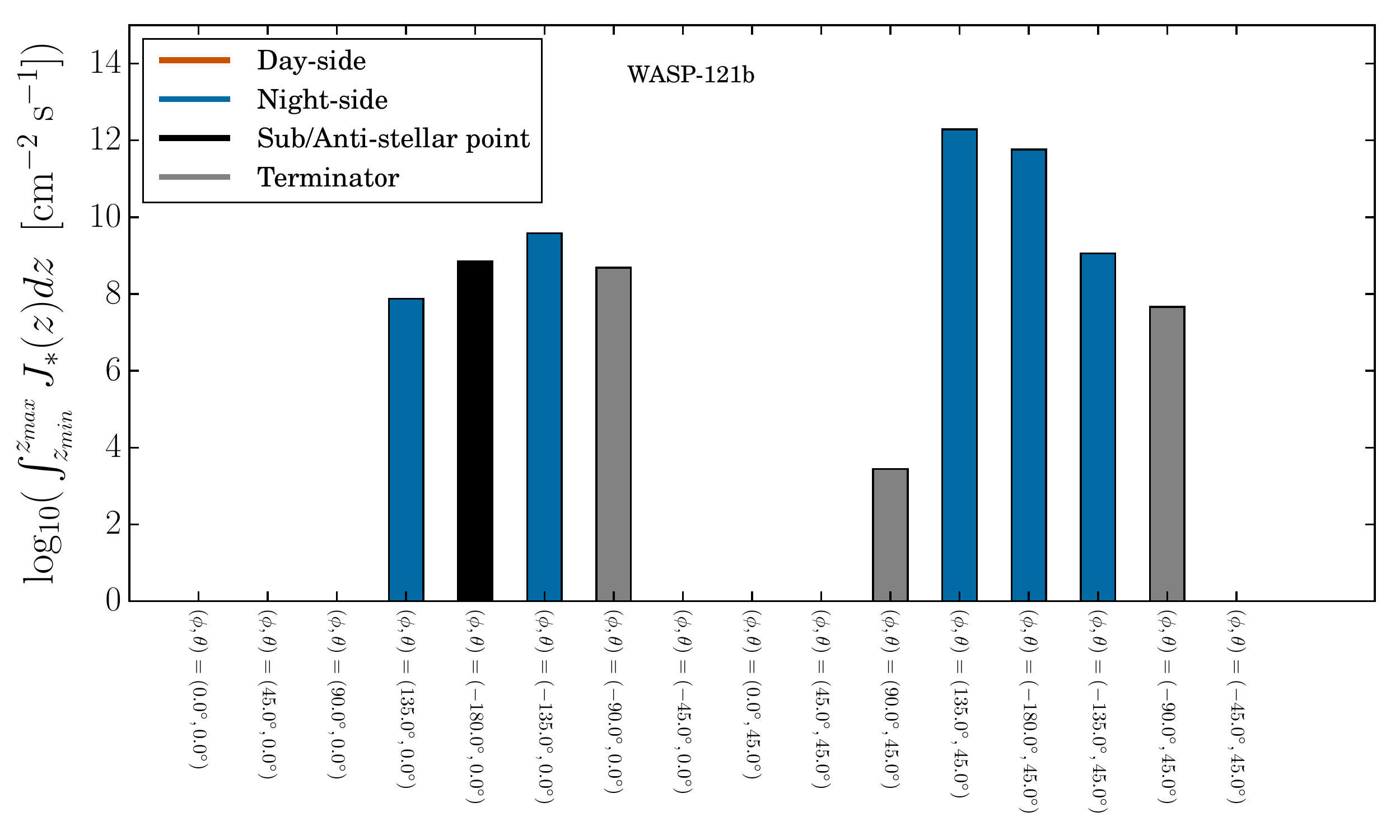}
    \caption{Column-integrated total nucleation rates $\int_{z_{min}}^{z_{max}} J_{*}(z) dz ~~ \mathrm{[cm^{-2}~s^{-1}]}$ for the sample planets  WASP-43b, WASP-18b, HAT-P-7b, WASP-103b, and WASP-121b. 
    WASP-103b has the highest integrated nucleation efficiency, HAT-P-7b the lowest in the nightside.}
    \label{fig:nuc_integrate}
\end{figure*}
\begin{figure*}
    \centering
    \includegraphics[width=21pc]{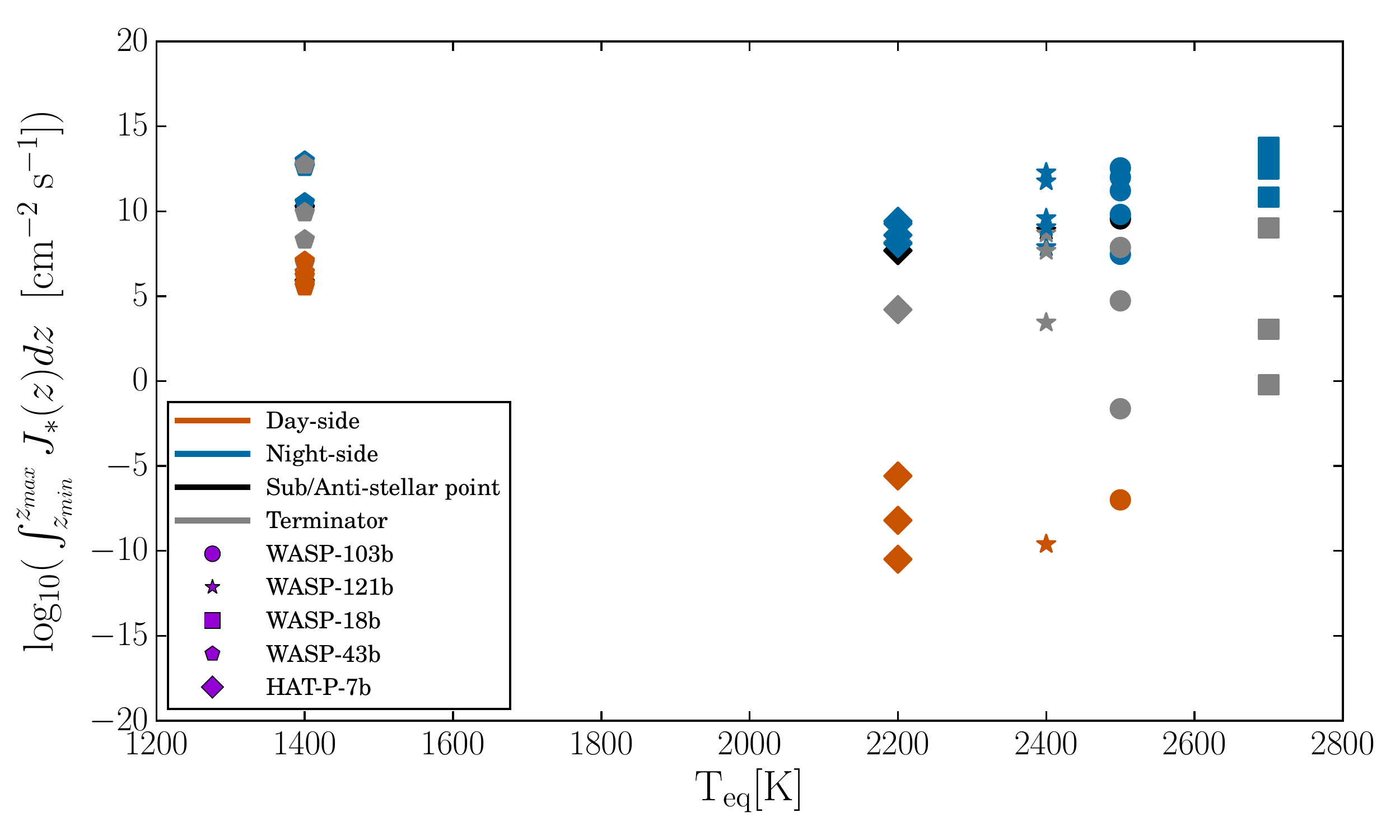}
    \includegraphics[width=21pc]{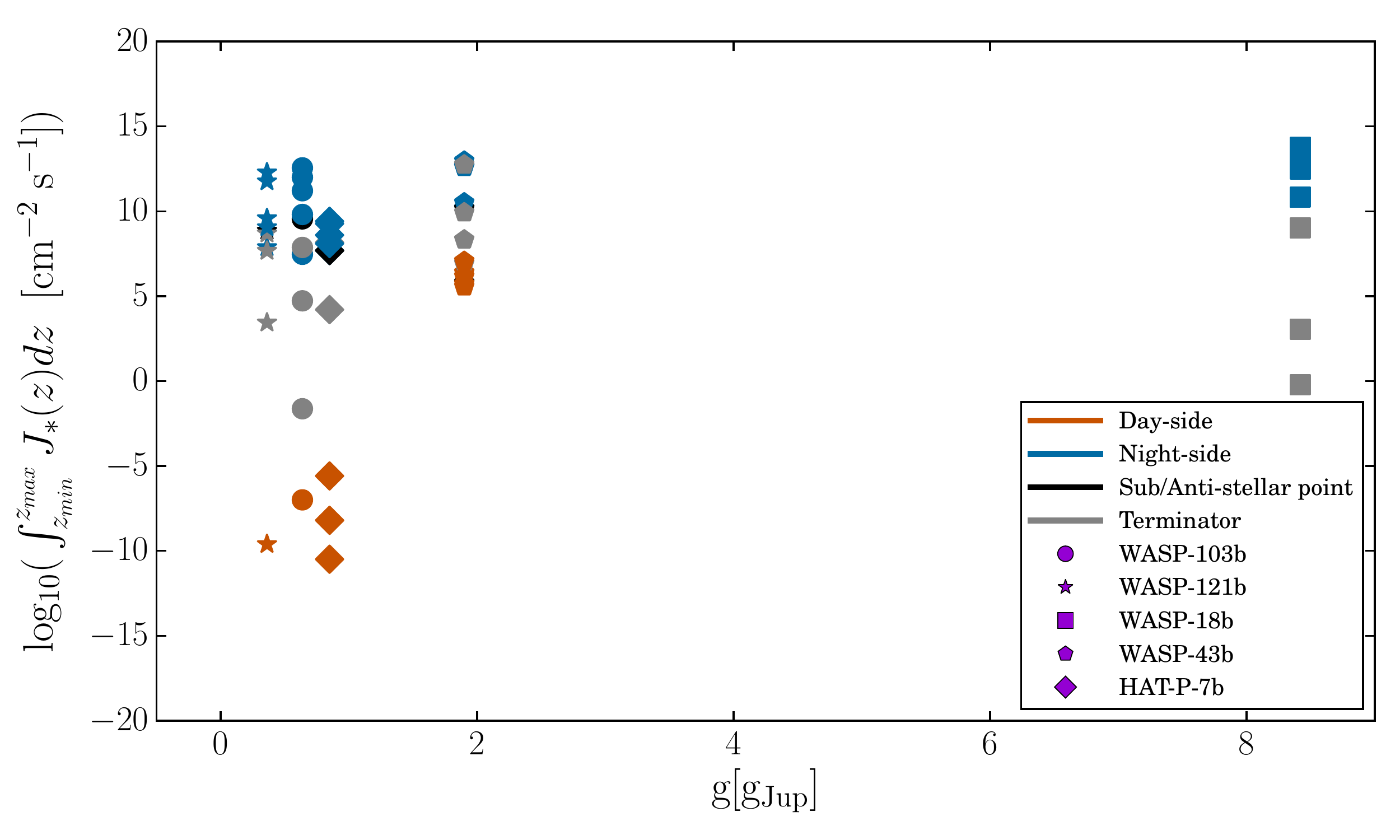} 
    \caption{The range of column-integrated nucleation rates from Fig.~\ref{fig:nuc_integrate} shown for $T_{\rm eq}$ [K] (left) and g [g$_{\rm Jup}$] (right) for the giant gas planet WASP-43b, and the ultra-hot Jupiters HAT-P-7b, WASP-18b, and WASP-103b, WASP-121b. The WASP-18b $T_{\rm eq}$ is offset by +200 K to avoid overlap. No one value suffices to describe the rate at which cloud particles form.}
    \label{fig:nuc_integrate_teq}
\end{figure*}


\section{The different atmosphere structures} 
\label{s:input}


Figure~\ref{Tp_all_new} summarises the 1D atmospheric $(T_{\rm gas}, p_{\rm gas}$)-profiles for the  hot giant gas planet WASP-43b, 
and the four ultra-hot Jupiters WASP-18b, HAT-P-7b, WASP-103b, WASP-121b. 
This sample of giant gas planets is homogeneous in that all  $(T_{\rm gas}, p_{\rm gas}$)-profiles result from the same 3D GCM code. All differences that will be explored in this paper will therefore be caused by the local thermodynamic conditions, and will not be caused by differences in numerical methods or other assumptions made in different hydrodynamic simulations. Section~\ref{s:wasp_gcm_comp} will, however, address the effect of the inner boundary for the example of WASP-43b based on results from different 3D GCMs.

Figure~\ref{Tp_all_new} shows that hot  giant gas planets and the ultra-hot Jupiters have substantially  different  day/night $(T_{\rm gas}, p_{\rm gas}$)-structures. The largest difference occurs on the dayside between these two sub-classes of gas giants. The nightsides appear more similar.
All ultra-hot Jupiters sampled reach maximum gas temperatures of $\approx 3400$K on the dayside which suggest that no clouds will form here. This value resembles the dayside value proposed from CHEOPS observation in \cite{2020arXiv200913403L}. The details of the individual temperature profiles differ, but all ultra-hot Jupiters have a comparably low nightside temperature.

Figure~\ref{Tp_all_new} (right lower panel) provides a comparison of the day- and the nightside averaged profiles (excluding the terminator regions).
All sampled planets have a hotter dayside with a temperature inversion occurring at $p_{\rm gas}\sim 10^{-3}\,\ldots\,10^{-2}$bar for the ultra-hot Jupiters. These gas temperature inversions typically display a change of 1500-2000K. Hot  giant gas planets have a much less pronounced temperature inversion happening deeper in the atmosphere than the ultra-hot Jupiters, at about 1 bar, with a range of only less than 500K. 
In the low pressure regimes, the dayside profiles from our simulations of the ultra-hot Jupiters lie at 2700K, whereas the dayside profiles of the hot  giant gas planets are much cooler at 1100K. The hot  giant gas planet (WASP-43b)
is the only planet in our sample that shows a net increase in temperature on the dayside with atmospheric depth. HAT-P-7b, WASP-18b, WASP-121b and WASP-103b  have a roughly equivalent temperature at $10^{-6}$ bar and 100 bar within the GCM modelling framework utilised here. An average dayside profile of  hot  giant gas planets (now including WASP-43b, HD\,189733b and HD\,209458b for comparison)  are a lot cooler than those of the ultra-hot Jupiters in our sample but exact differences vary across pressure ranges (Fig.~\ref{Tp_all_new}, lower right panel). The average nightside temperatures of the ultra-hot Jupiters are in the temperature and pressure ranges of the average dayside profiles of the hot  giant gas planets in our sample for $p_{\rm gas}<0.1$bar.



\section{Comparing cloud properties} \label{s:clouds}

\subsection{Nucleation Rate}\label{ss:nuc}

The nucleation rate, $J_*$ [cm$^{-3}$ s$^{-1}$], is an essential measure for the efficiency with which cloud formation occurs, hence, with which efficiency the gas is depleted and is undergoing a phase transition leading to the formation of cloud particles. Here we consider the formation of mineral cloud particles which is triggered by the nucleation of mainly TiO$_2$ and SiO.  We analyse the cloud formation efficiency for individual profiles first (Fig.~\ref{fig:Nuc_new}), before we proceed to integrated properties (Fig.~\ref{fig:nuc_integrate}) as the base for comparing column integrated nucleation rates for different planets according to their $T_{\rm eq}$ and g$_{P}$ (Fig.~\ref{fig:nuc_integrate_teq}).

Figure~\ref{fig:Nuc_new} demonstrates that the nightside gas temperatures  are low enough that nucleation takes place for all exoplanets of our sample, hot  giant gas planets and ultra-hot Jupiters. 
{\it We may conclude that most if not all ultra-hot Jupiters will have clouds forming on their nightsides.} Averaging over all nightside profiles (Fig.~\ref{fig:Nuc_new}, lower right panel) suggests the most efficient formation of cloud condensation nuclei  occurring on  WASP-18b and WASP-43b with average values of $J_*\approx 10^4\,\ldots\,10^6$ cm$^{-3}$s$^{-1}$, and the least efficient nucleation on the nightside of WASP-43b with values of $J_*\approx 10^{-4}\,\ldots\,10^1$ cm$^{-3}$s$^{-1}$ in the upper atmosphere. The least efficient nightside nucleation in the upper atmosphere occurs in HAT-P7b. The details of the nucleation profiles depend on the local thermodynamic conditions. The nightside averaged values are  higher than those retrieved with  ARCiS by \cite{2020A&A...642A..28M} for the 10 hot  giant gas planets published in \cite{2016Natur.529...59S}. Both samples contain different planets, however.

Therefore, the only planet forming a substantial amount of cloud particles on the dayside within our sample is the hot  giant gas planet WASP-43b,
and a tiny nucleation peak occurs in deeper, high-pressure atmospheric regions on the dayside of WASP-121b (Fig.~\ref{fig:Nuc_new}). This will have implications for the cloud particle sizes in Sect.~\ref{ss:amean}. We have demonstrated  (\citealt{Lee2015,Lee2016,2016MNRAS.460..855H,2018A&A...615A..97L}) that both the hot  giant gas planets HD\,189733b and HD\,209458b  form clouds on the day and on the nightside. 
{\it It might  therefore be reasonable to conclude that hot  giant gas planets can form clouds that cover the entire globe and that occupy a substantial pressure range.}

To be able to compare the cloud formation efficiency in the global atmospheres of our sample planets, column integrated nucleation rates $\int_{z_{min}}^{z_{max}} J_{*}(z) dz ~~ \mathrm{[cm^{-2}~s^{-1}]}$ are considered (Fig.~\ref{fig:nuc_integrate}). The left half of the plot shows the equatorial values ($\theta=0^o$) and the right half showing northern hemisphere values ($\theta=45^o$). The  order of profiles is starting at sub-stellar point ($\phi, \theta$)=($0^o, 0^o$), move around from East to West at the equator $\theta = 0^o$ (left half of the plot) and then the same for $\theta = 45^o$. We note that these values average out all local information as discussed before (Figure~\ref{fig:Nuc_new}) and should be considered as guiding rather then absolute values. The colour code is the same as in Fig.~\ref{fig:Nuc_new}. Nucleation is generally less efficient (on hot  giant gas planets) or completely absent (ultra-hot Jupiters) on the dayside. Nucleation is more efficient in the non-equatorial hemispheres for some planets (WASP-43b, WASP-103b, WASP-121b).
Nucleation is generally less efficient in the terminator regions for the ultra-hot Jupiters. The maximum nucleation efficiency is very individual for every planet.  Figure~\ref{fig:nuc_integrate_teq} (left) shows the much larger spread of nucleation values for globally hotter ultra-hot Jupiters compared to the shown hot  giant gas planet. This suggests that the cloud particle population will be more diverse in size on ultra-hot Jupiters than on hot  giant gas planets. 


\begin{figure*}
   \centering
   \includegraphics[width=21pc]{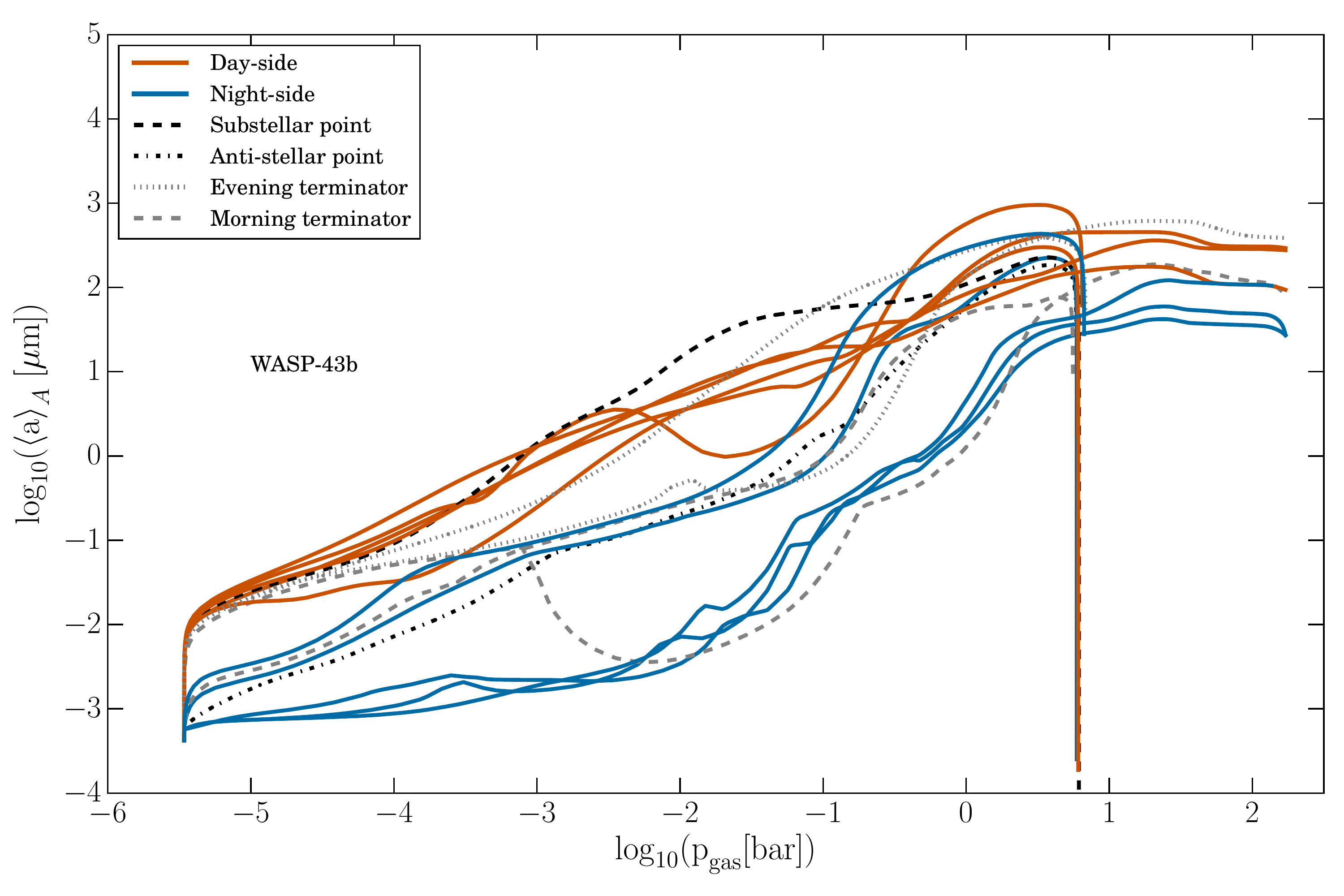}
   \includegraphics[width=21pc]{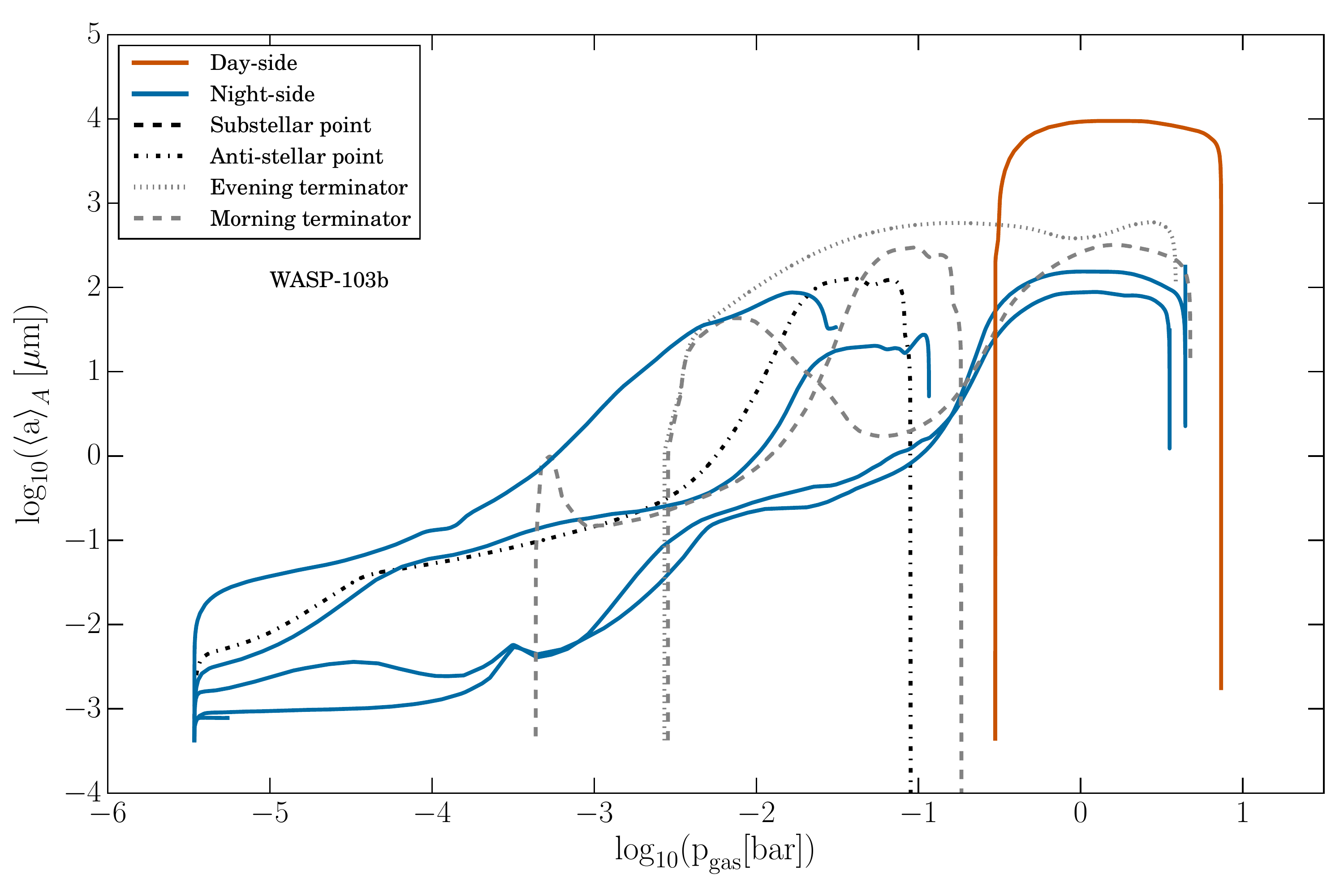}\\
   \includegraphics[width=21pc]{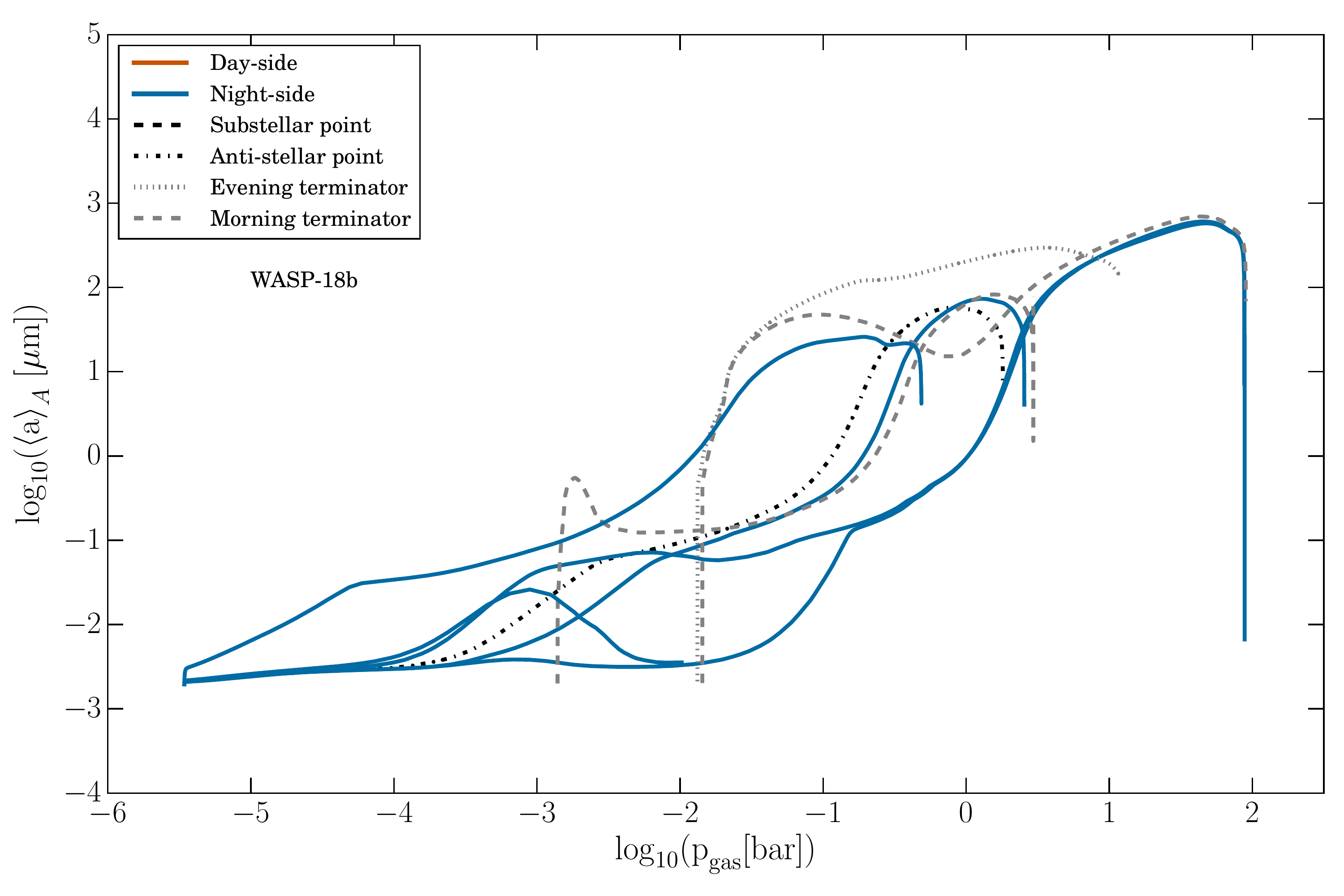}
   \includegraphics[width=21pc]{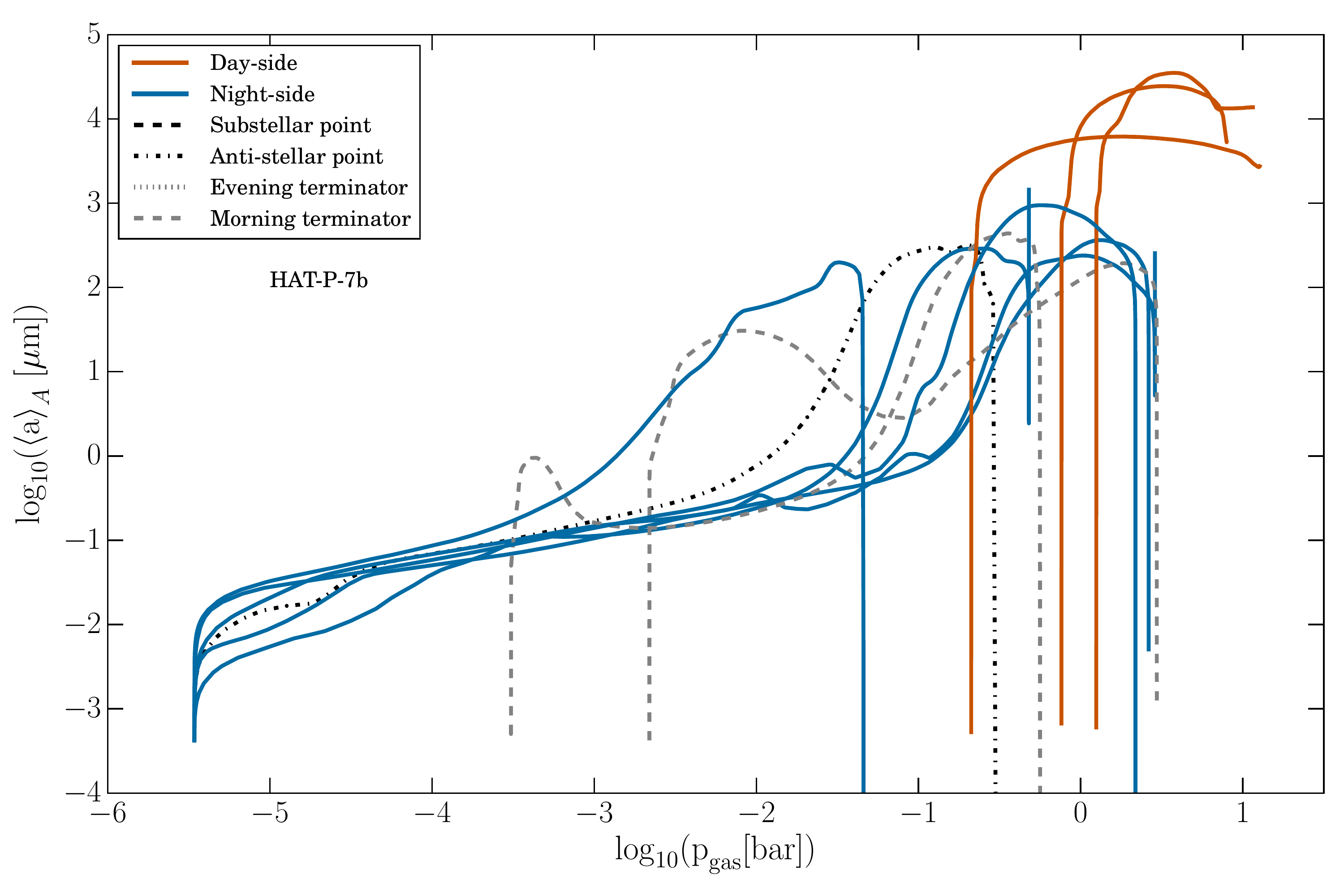}\\
   \includegraphics[width=21pc]{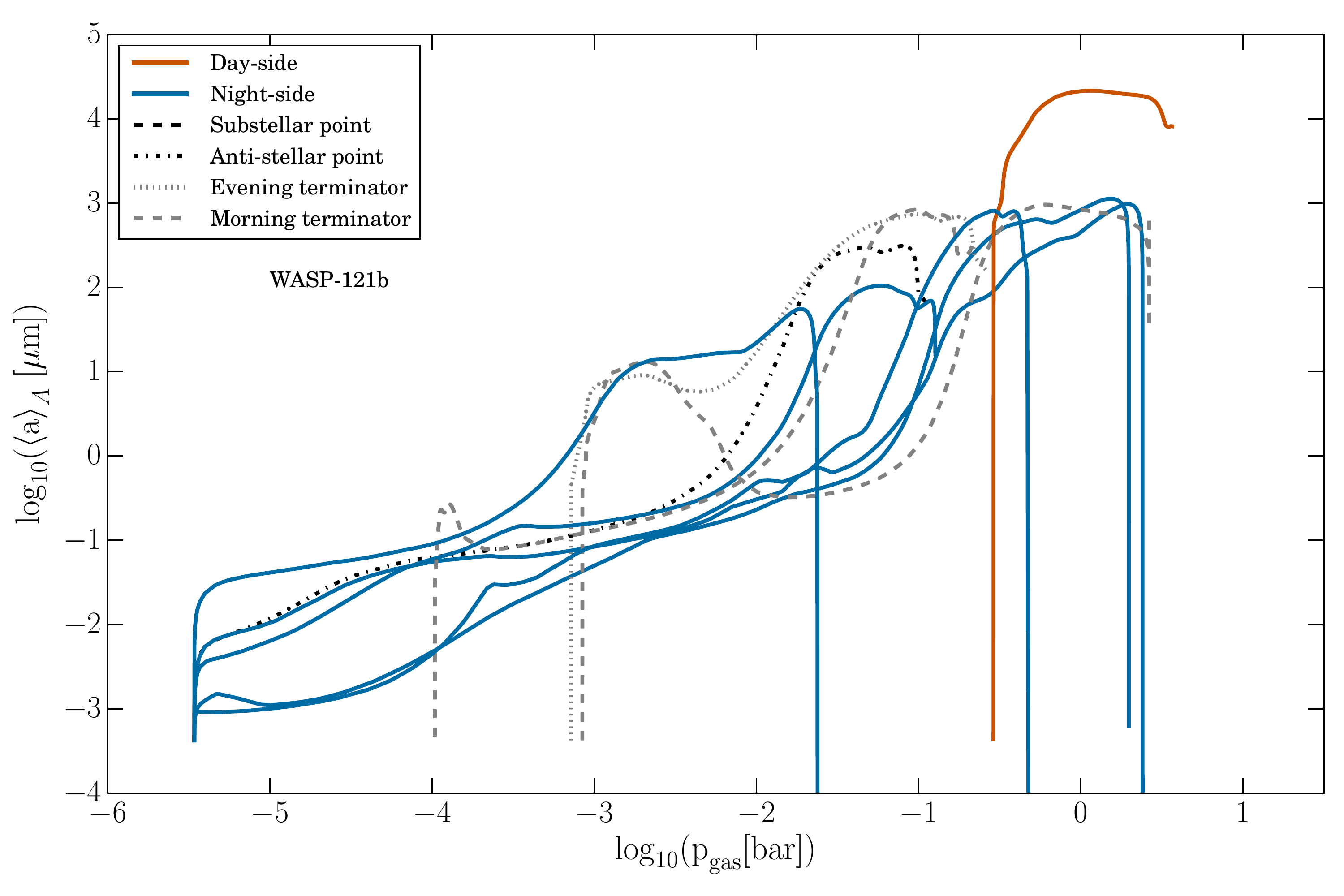}
   \includegraphics[width=21pc]{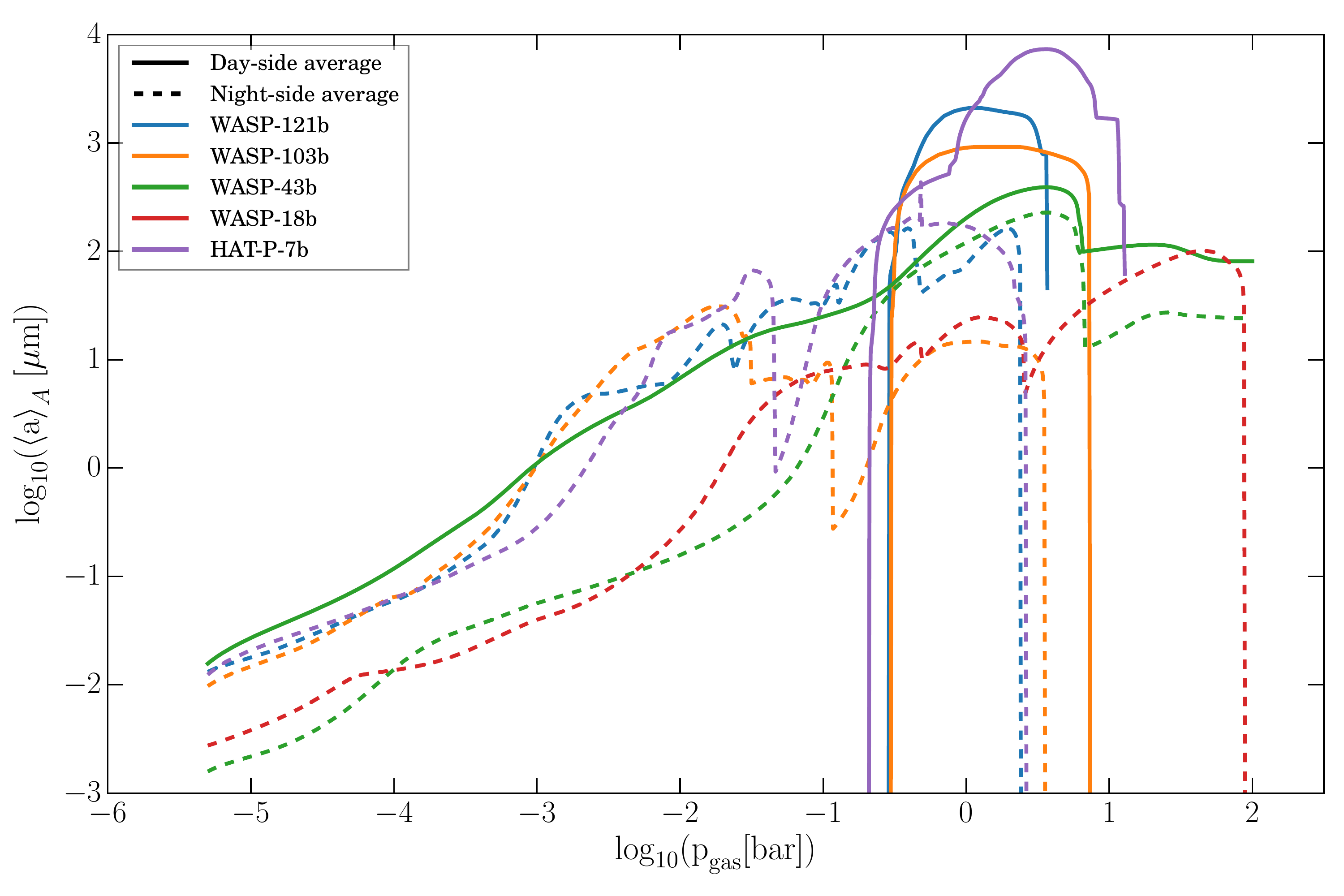}
    \caption{Surface averaged mean particle size, $ \langle a \rangle_{\rm A}$ [$\mu$m] (Eq.~\ref{eq:aa}), for the hot  giant gas planet WASP-43b,
    and the ultra-hot Jupiters WASP-18b, HAT-P-7b, WASP-103b, and WASP-121b. The lower right panel shows the day (solid lines) and nightside (dashed lines) averaged surface averaged mean particle size. The cloud particles sizes vary throughout the atmospheres with the largest particles occurring in the high-pressure, inner cloud layers. Cloud particles as large as 1cm may occur in low numbers in the inner dayside cloud layers of ultra-hot Jupiters  where nucleation is very inefficient, but present.}
    \label{fig:mean_particle_size_A}
\end{figure*}

\begin{figure*}
   \centering
   \includegraphics[width=21pc]{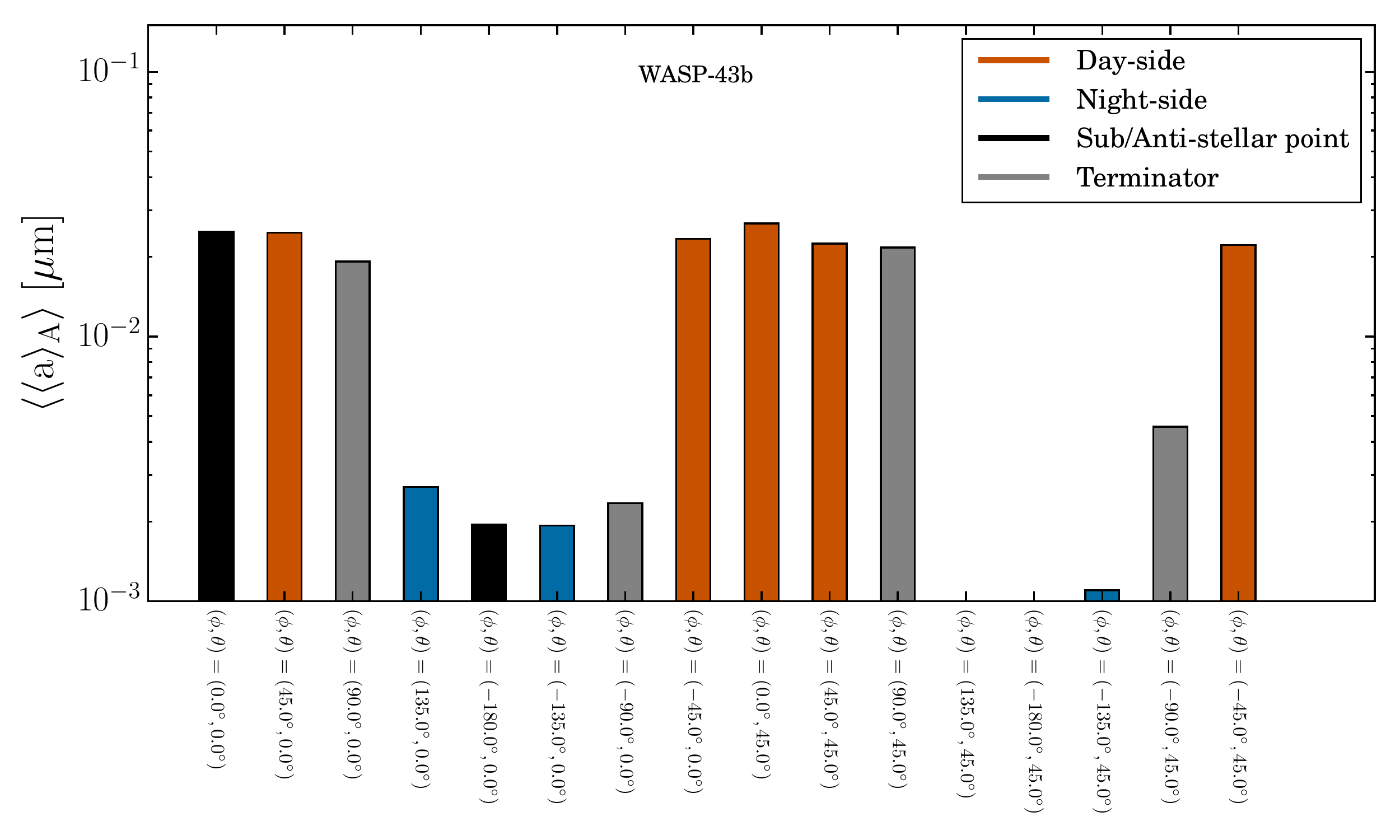}
   \includegraphics[width=21pc]{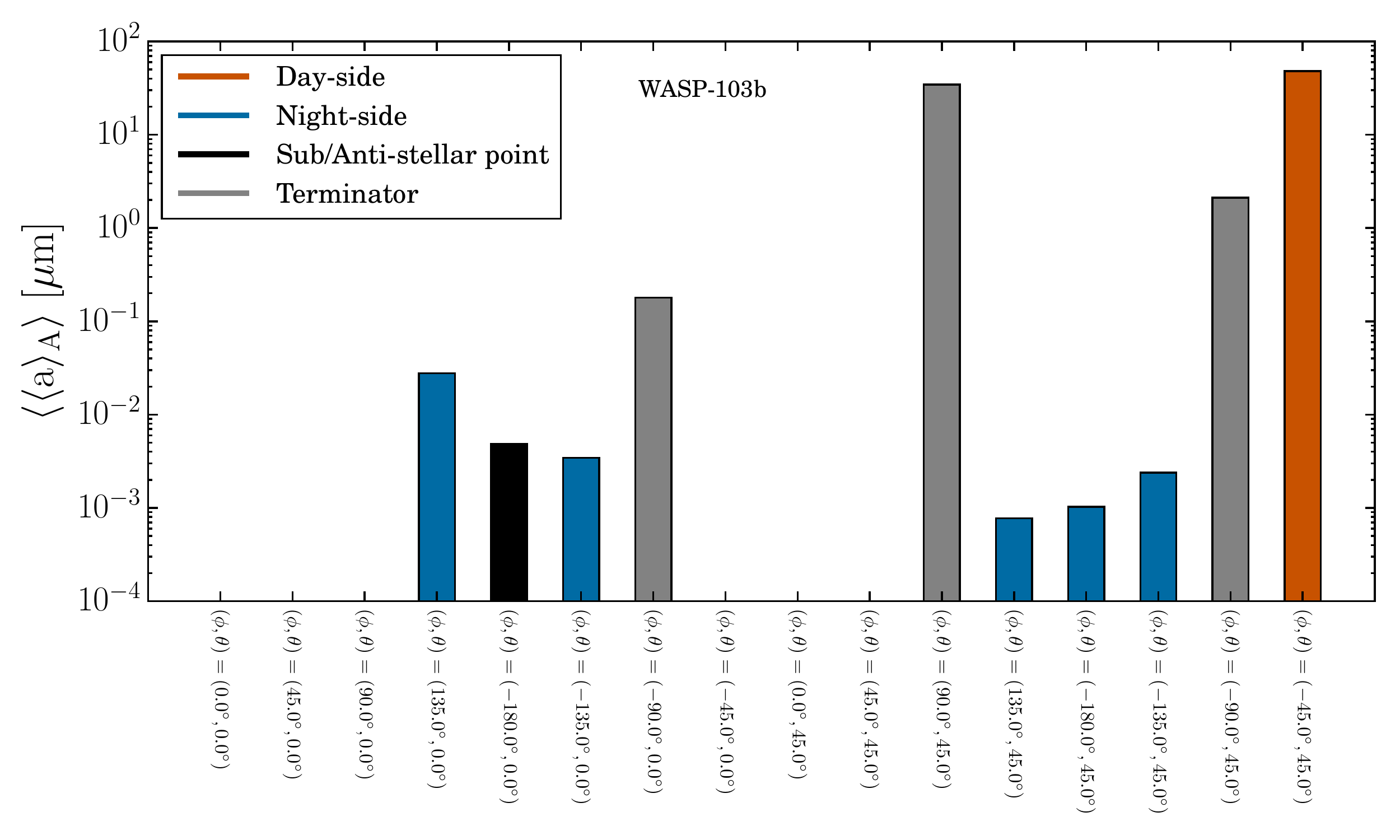}\\
   \includegraphics[width=21pc]{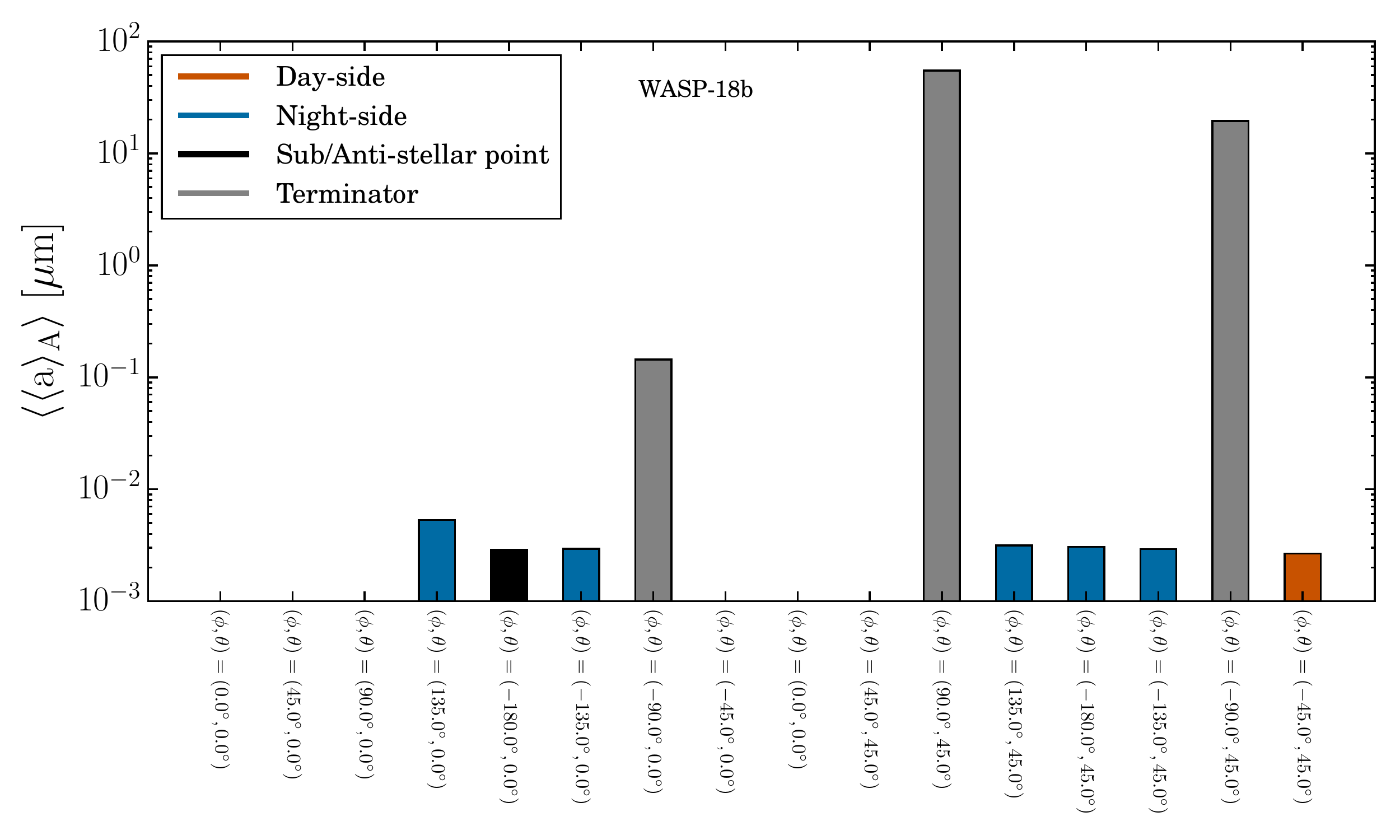}
   \includegraphics[width=21pc]{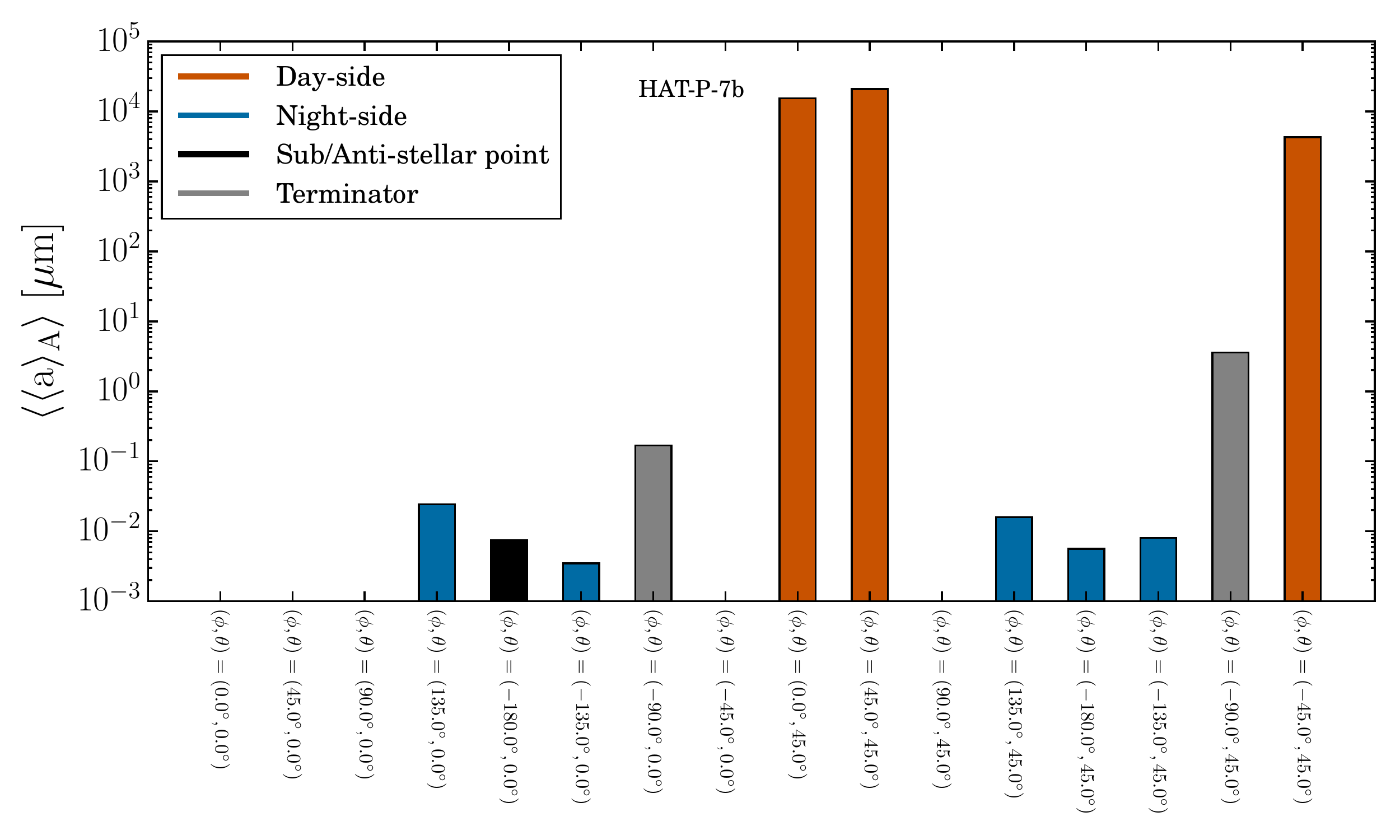}\\
   \includegraphics[width=21pc]{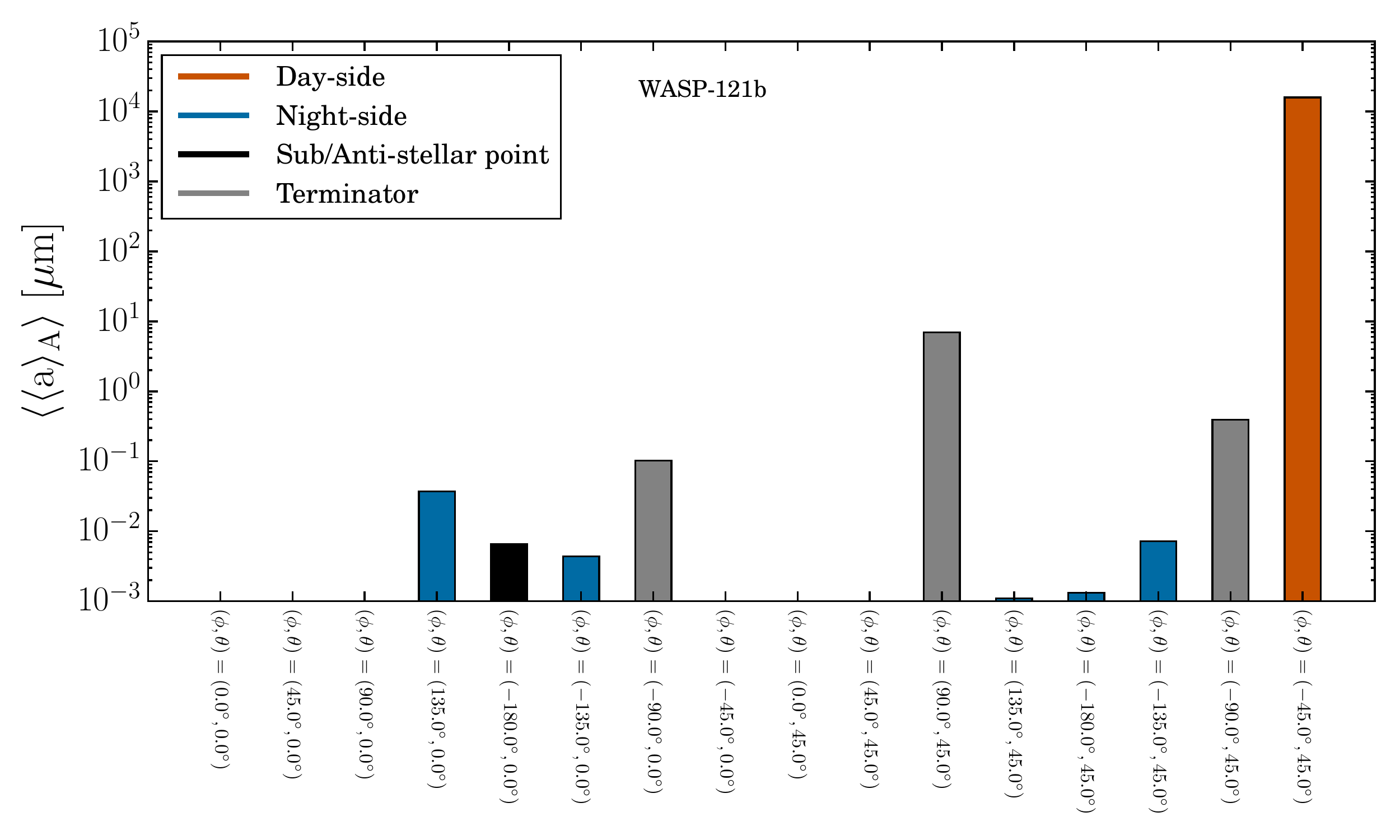}
   
    \caption{Integrated, number density weighted surface averaged mean particle size, $ \langle \langle a \rangle_{\rm A} \rangle = \int_{z_{min}}^{z_{max}} n_{d}(z)\langle a \rangle_{\rm A}(z) dz ~\big/ \int_{z_{min}}^{z_{max}} n_{d}(z) dz$, for the giant gas planet WASP-43b and the ultra-hot Jupiters WASP-18b, HAT-P-7b, WASP-103b, and WASP-121b.}
    \label{fig:integrated_mean_particle_size_A}
\end{figure*}
\begin{figure*}
    \centering
    \includegraphics[width=21pc]{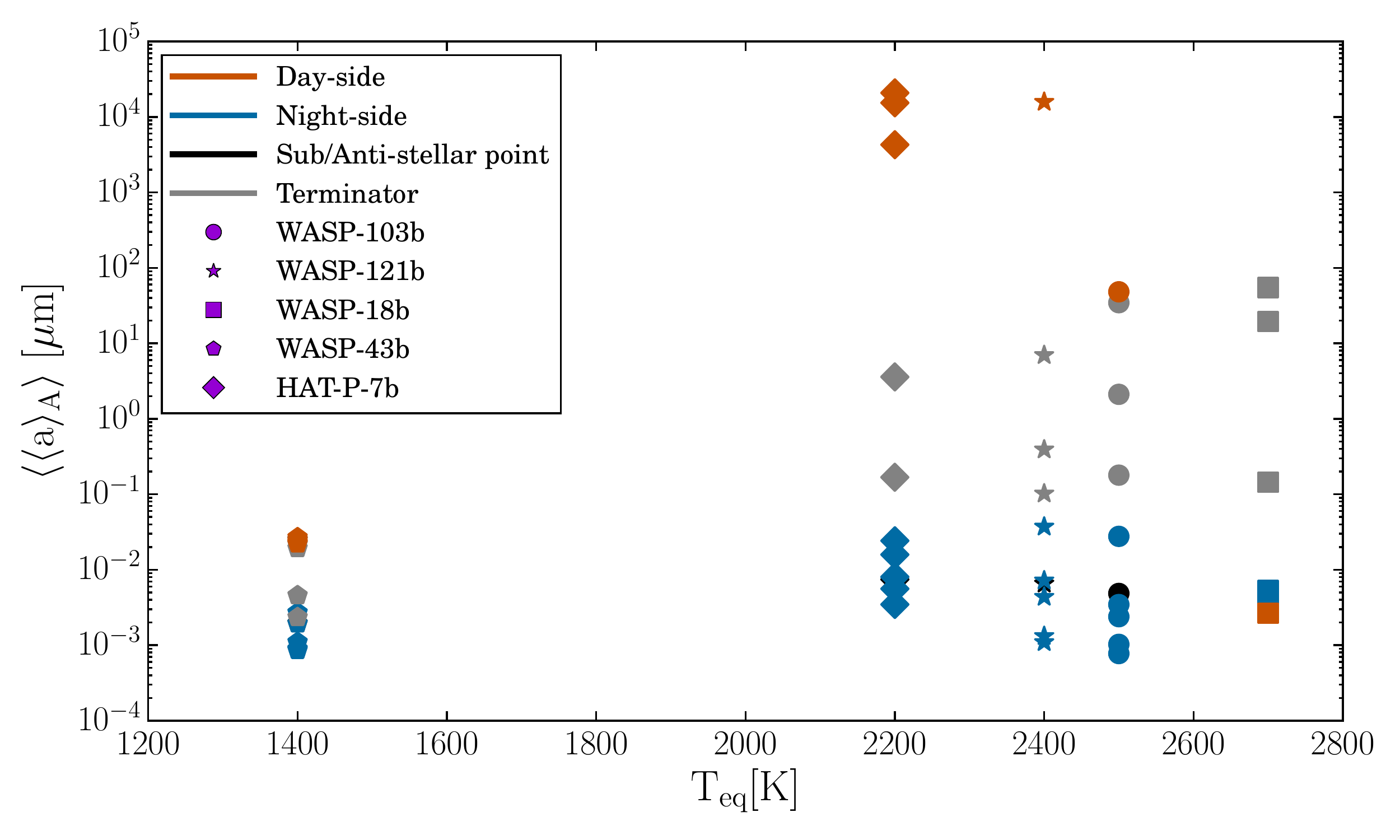}
    \includegraphics[width=21pc]{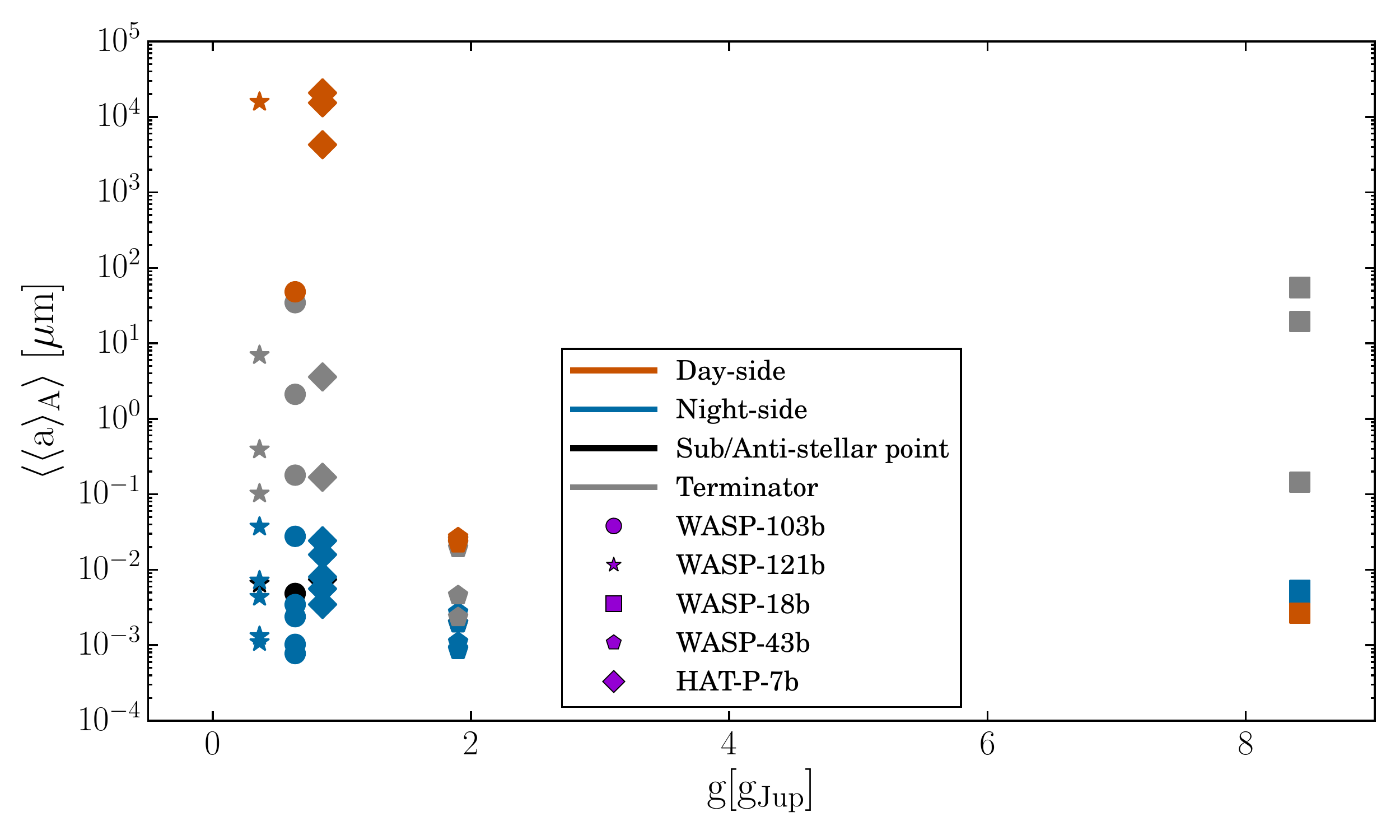}
    \caption{The range of integrated, number density weighted surface averaged mean particle size from Fig.~\ref{fig:integrated_mean_particle_size_A} shown for $T_{\rm eq}$ [K] (left) and g [g$_{\rm Jup}$] (right). The largest range of integrated cloud particles sizes occurs for ultra-hot Jupiters with small surface gravity, one being the JWST target WASP-121b. The WASP-18b $T_{\rm eq}$ is offest by +200 K to avoid overlap.}
    \label{fig:mean_particle_size_integrate_scatter}
\end{figure*}

\subsection{Mean Particle Size}\label{ss:amean}

The cloud particle size, $a$ [cm],  is an essential value entering the opacity calculation and shows how efficient surface growth depletes the gas phase through the growth of Mg/Si/O/Fe/Ti/Al/ containing minerals. For our purpose, we show the surface averaged mean particle size, $\langle a\rangle_{\rm A}$ [cm], which we will use for our opacity calculation in Sect.~\ref{s:obs},
\begin{equation}
    \centering
    \langle a\rangle_{\rm A} = \sqrt[3]{\frac{3}{4\pi}}\, \frac{L_3}{L_2},
    \label{eq:surf_size}
\end{equation}
with the dust moments $L_{2}$ and $L_{3}$,  \citep[Eq.A.1 in][]{helling2020mineral}. Further discussion of the different definitions of the mean particle size can be found in Appendix A of \citet{helling2020mineral}.


Similar to Sect.~\ref{ss:nuc}, we first present details of surface averaged mean particle size, $\langle a\rangle_{\rm A}$, in Fig.~\ref{fig:mean_particle_size_A}, before we proceed to integrated properties (Fig.~\ref{fig:integrated_mean_particle_size_A}), namely the integrated number density weighted surface averaged mean particle size, 
\begin{equation}
\label{eq:aa}
\langle \langle a \rangle_{\rm A} \rangle = \frac{\int_{z_{min}}^{z_{max}} n_{d}(z)\langle a \rangle_{\rm A}(z) dz }
{ \int_{z_{min}}^{z_{max}} n_{d}(z) dz} \quad \mbox{with}\quad n_{\rm d}(z) = \frac{\rho(z) L_3(z)}{4\pi  \langle a(z)\rangle_{\rm A}^3/3}
\end{equation}

as the base for comparing column integrated mean particle sizes for different planet parameter, $T_{\rm eq}$ and g$_{P}$ (Fig.~\ref{fig:mean_particle_size_integrate_scatter}).

Across both sides of the planets, the mean particle size, $ \langle a \rangle_{\rm A}$,  increases with pressure, as surface growth efficiency increases with increasing gas density (Fig.~\ref{fig:mean_particle_size_A}). Small cloud condensation seeds nucleate in the cool upper atmosphere. Due to gravitational settling these fall through the atmosphere, growing faster the deeper they fall. Both night and day show an increase in mean particle size with increasing pressure with dayside profiles always showing a larger particle radius  than the nightside. Cloud particles of the size of $\langle a\rangle\approx 0.001\mu$m reside about 2h in the atmosphere where $p_{\rm gas}\approx 10^{-5}$bar in the terminator region of WASP-43b, but remains for 170h on the nightside which has a somewhat less extended atmosphere of higher density.

\medskip
The mean particles sizes (plotted as number density weighted surface averaged, column integrated mean particle sizes, $\langle \langle a \rangle_{\rm A} \rangle$,  in  Fig,~\ref{fig:integrated_mean_particle_size_A}) are biggest in atmospheric regions of low nucleation efficiency. This results in a factor of 10 difference in size between day and nightside on the hot  giant gas planets WASP-43b, but causes the formation of cm-sized cloud particles at some terminator and dayside locations on the ultra-hot Jupiters. {\it Figure~\ref{fig:mean_particle_size_integrate_scatter} suggests that low-mass ultra-hot Jupiters have the widest range of cloud particles sizes across their atmospheres, indicating a strong spatial in-homogeneity of these atmospheres. This includes the JWST target WASP-121b, but not the much heavier WASP-18b.}

\begin{figure}
   \includegraphics[width=21pc]{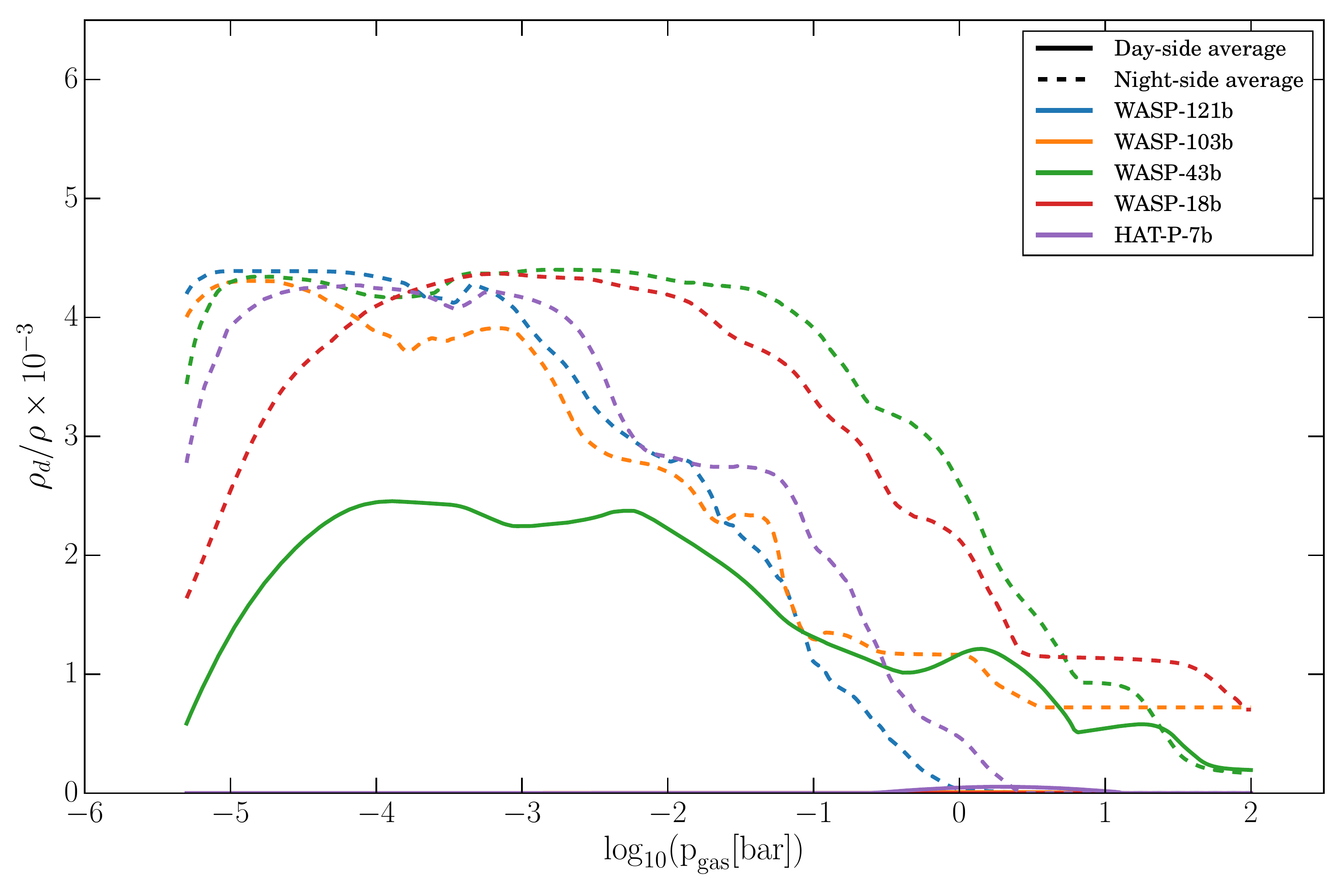}
        \caption{ The day (solid lines) and nighside (dashed lines) averaged dust-to-gas ratios, $\rho_{d}/\rho$, for the giant gas planet WASP-43b, and the ultra-hot Jupiters WASP-18b, HAT-P-7b, WASP-103b, and WASP-121b. The detailed results are in Fig.~\ref{fig:rhod_rho_all}. }
       \label{fig:rhod_rho_all1}
\end{figure}


\begin{figure*}
    \centering
    \includegraphics[width=21pc]{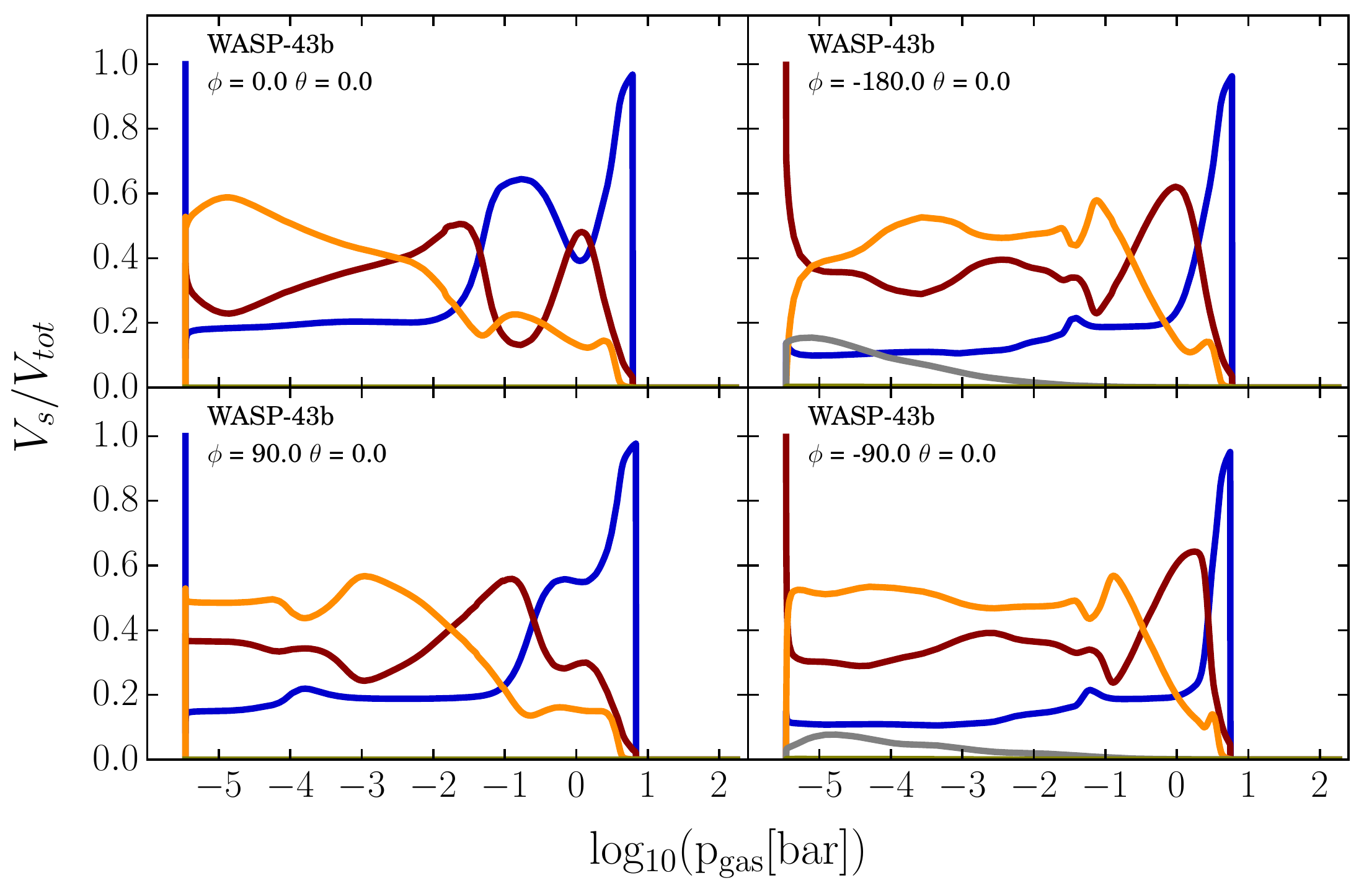}
    \includegraphics[width=21pc]{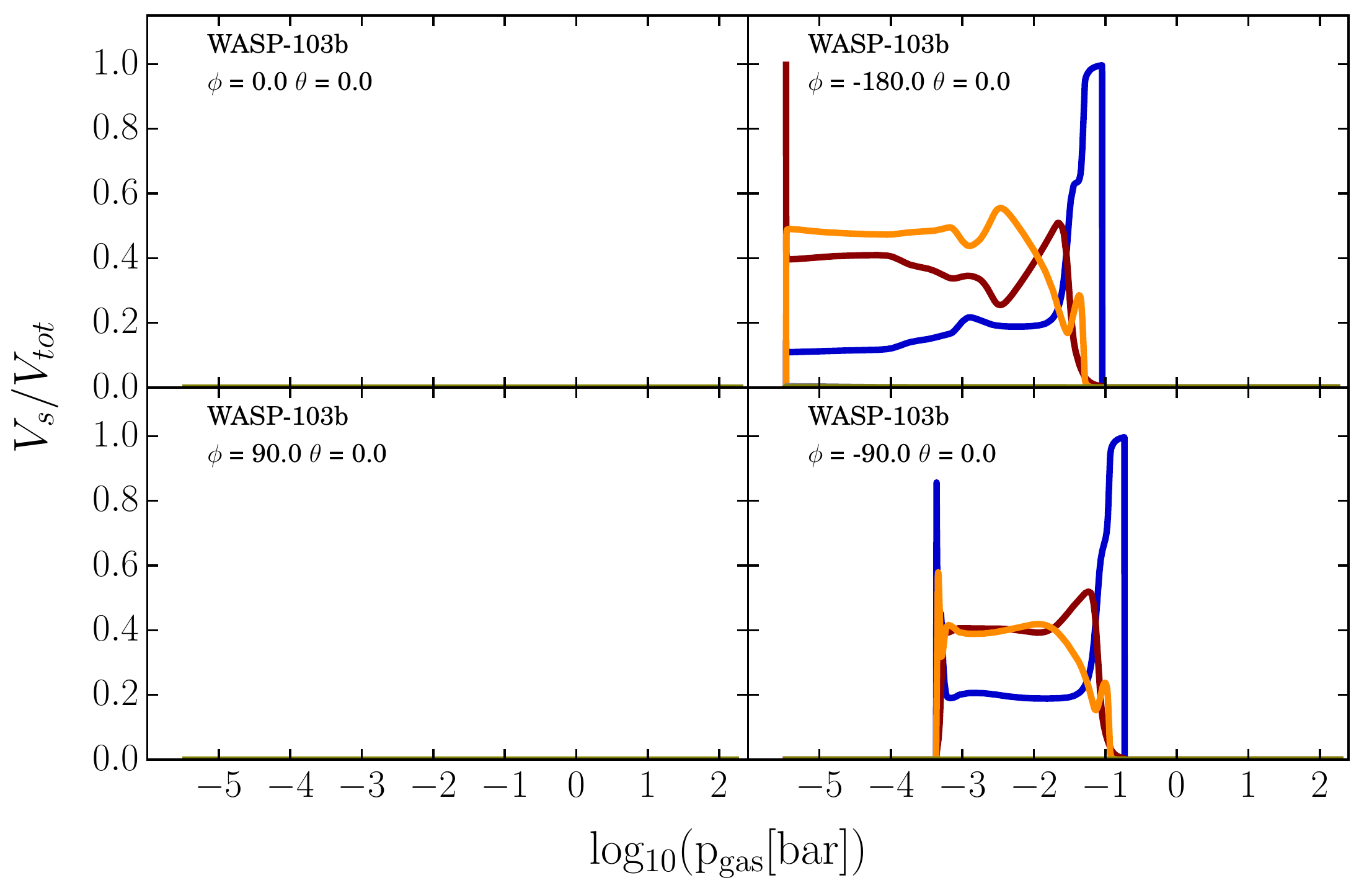}\\
    \includegraphics[width=21pc]{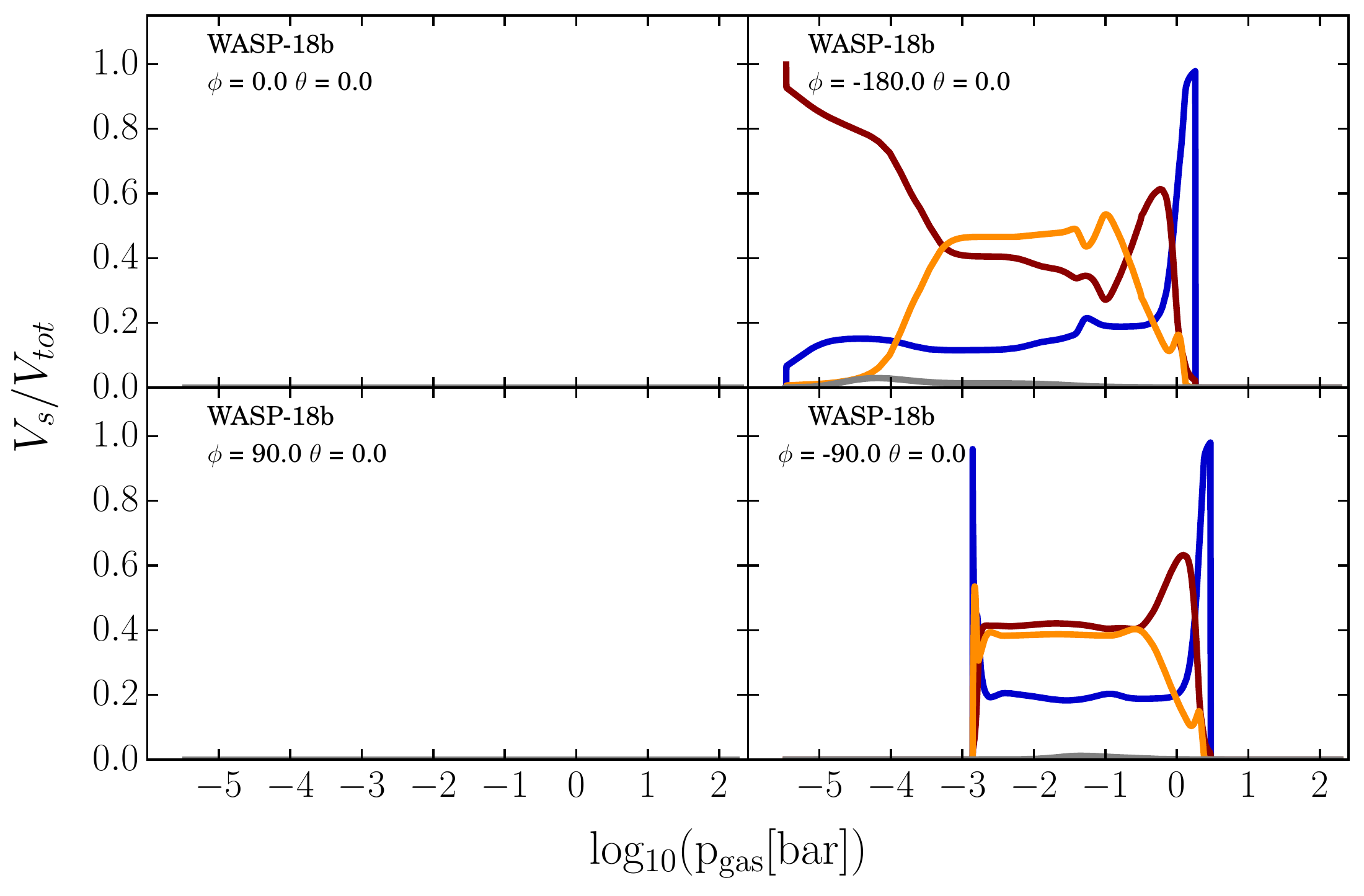}
    \includegraphics[width=21pc]{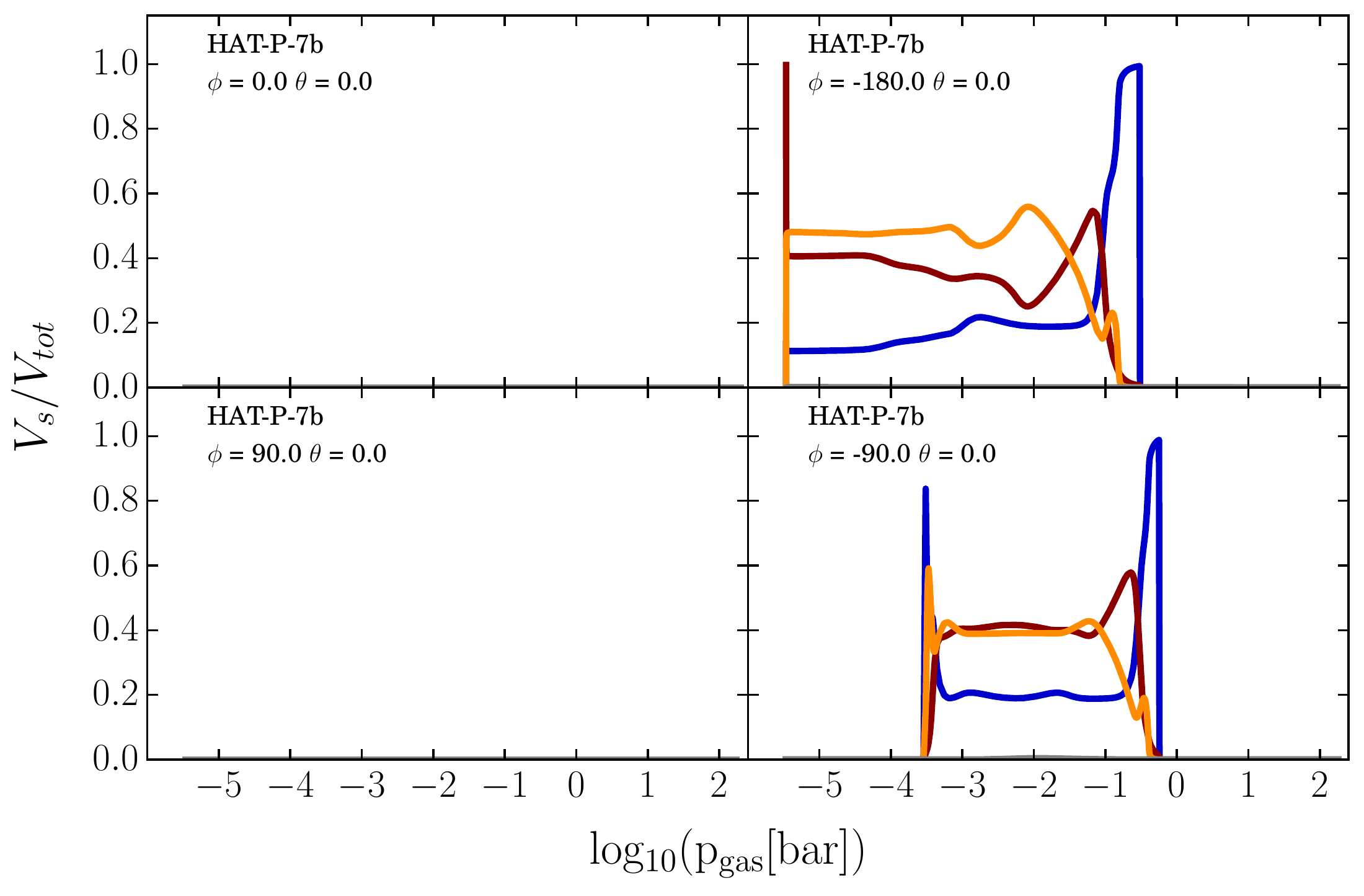}\\
    \includegraphics[width=21pc]{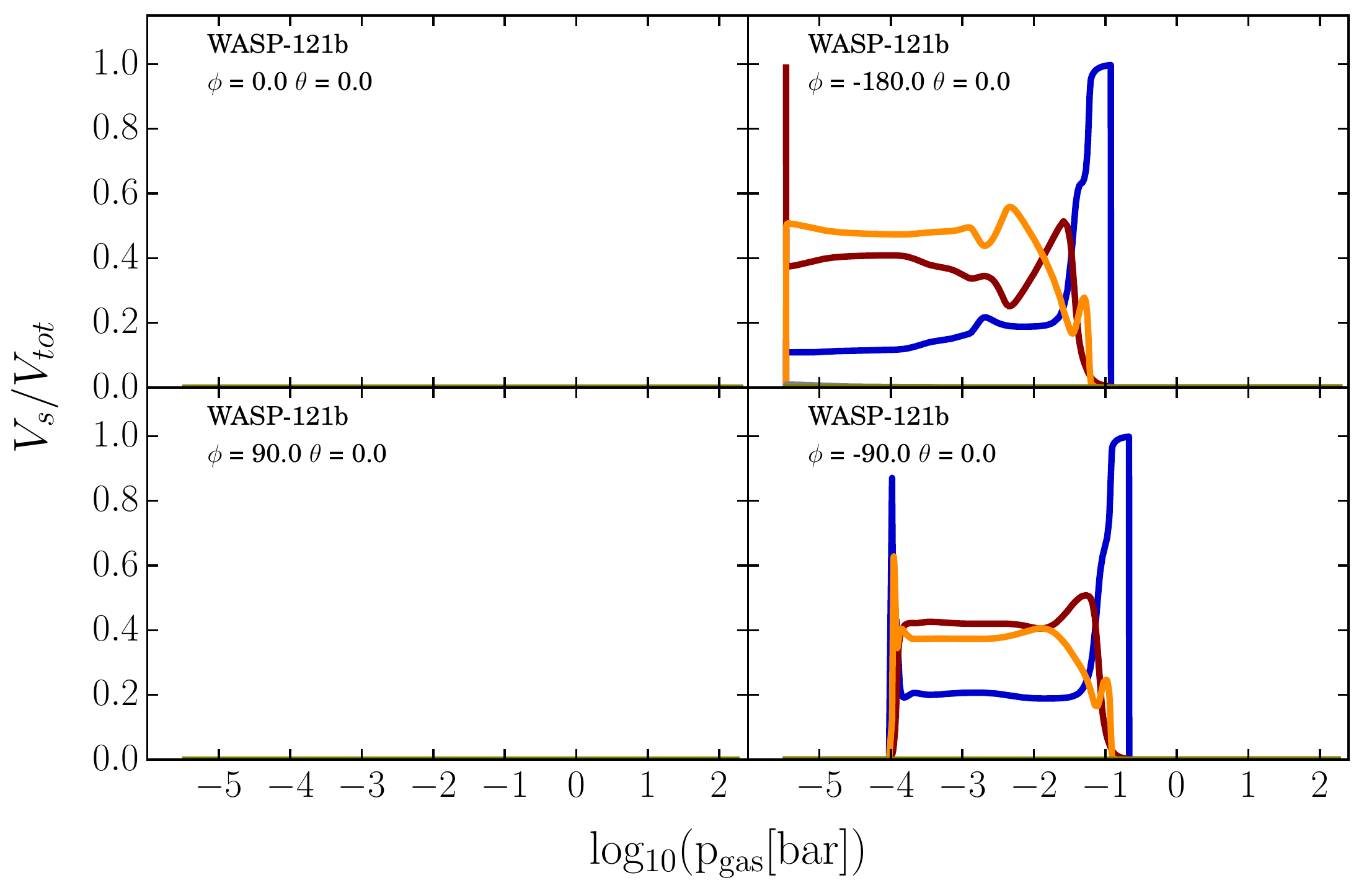}
    
    \caption{The volume fractions $V_{\rm s} / V_{\rm tot}$ of the material groups as defined in Table~\ref{tab:vol_frac_type_table} (blue: high temperature condensates, red: metal oxides, orange: silicates, grey: carbon, olive: salts). 
    Sub-stellar point: $(\phi, \theta)=(0.0^o, 0.0^o)$, Anti-stellar point: $(\phi, \theta)=(-180.0^o, 0.0^o)$, Equatorial Morning Terminator: $(\phi, \theta)=(-90.0^o, 0.0^o)$, Equatorial Evening Terminator: $(\phi, \theta)=(90.0^o, 0.0^o)$. There are no salt condensate species included for WASP-18b and HAT-P-7b due to their very low abundance. Empty panels represent profiles without cloud formation.}
    \label{fig:category_condensates}
\end{figure*}

\begin{figure*}
    \centering
    \includegraphics[width=21pc]{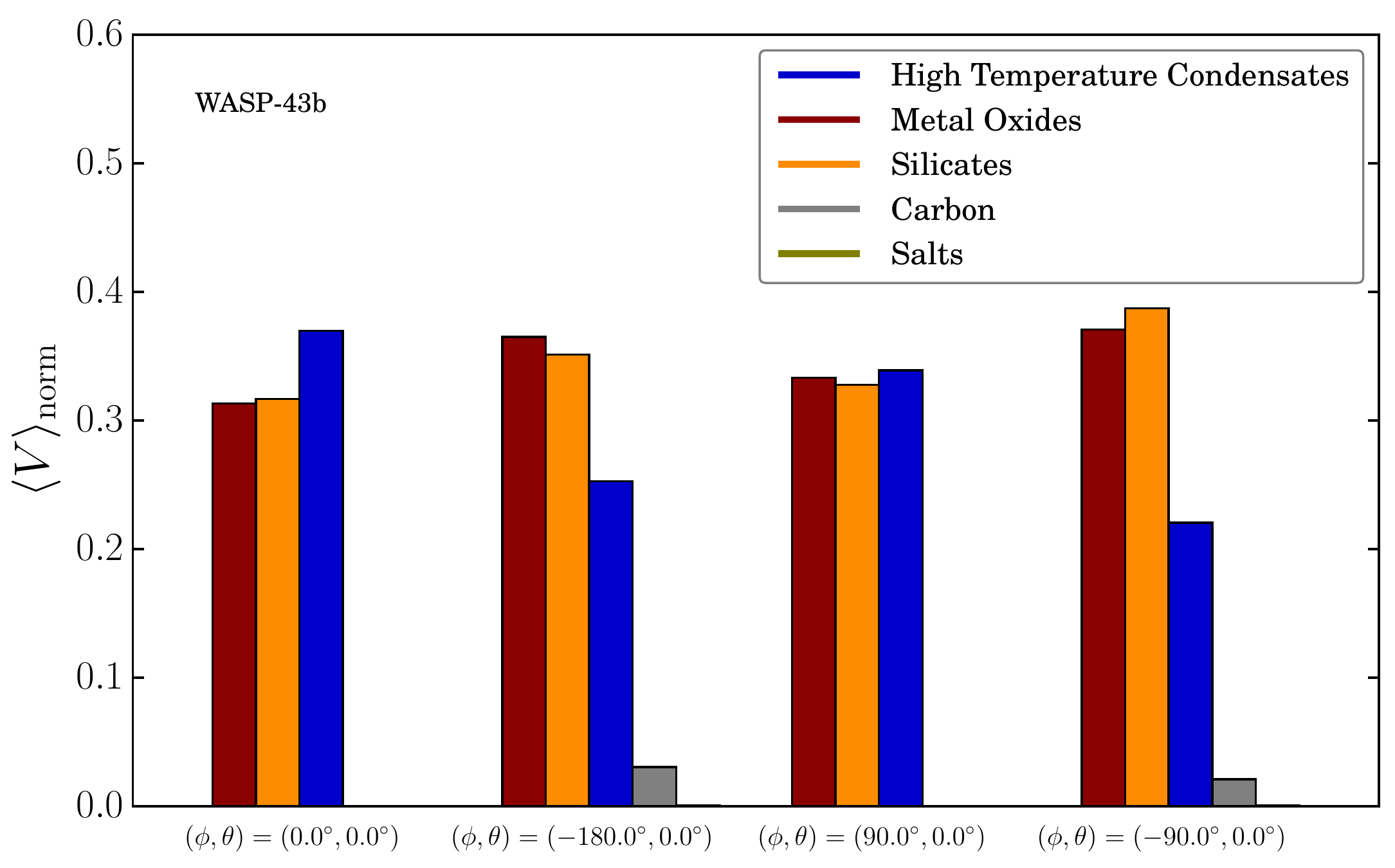}
    \includegraphics[width=21pc]{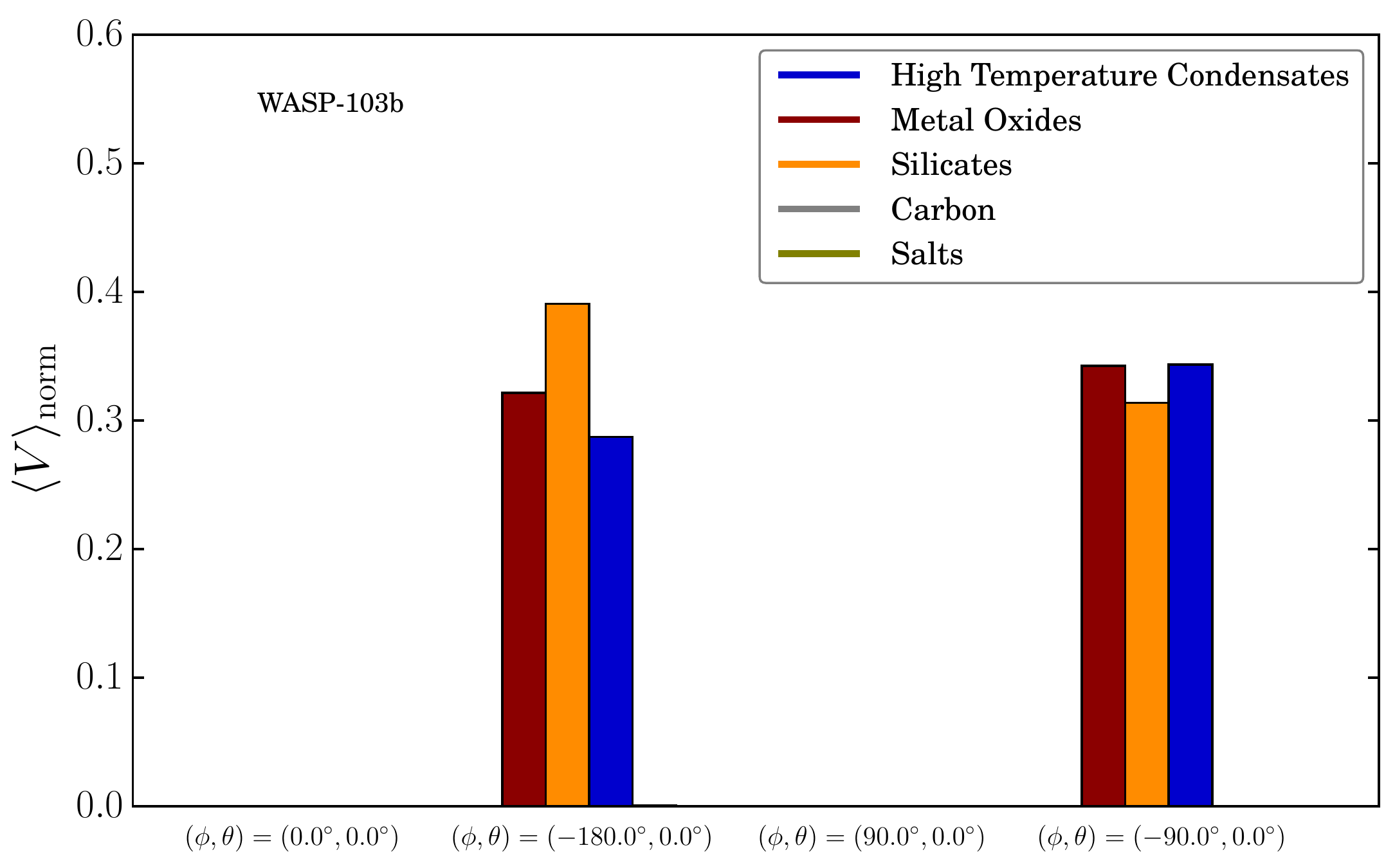}\\
    \includegraphics[width=21pc]{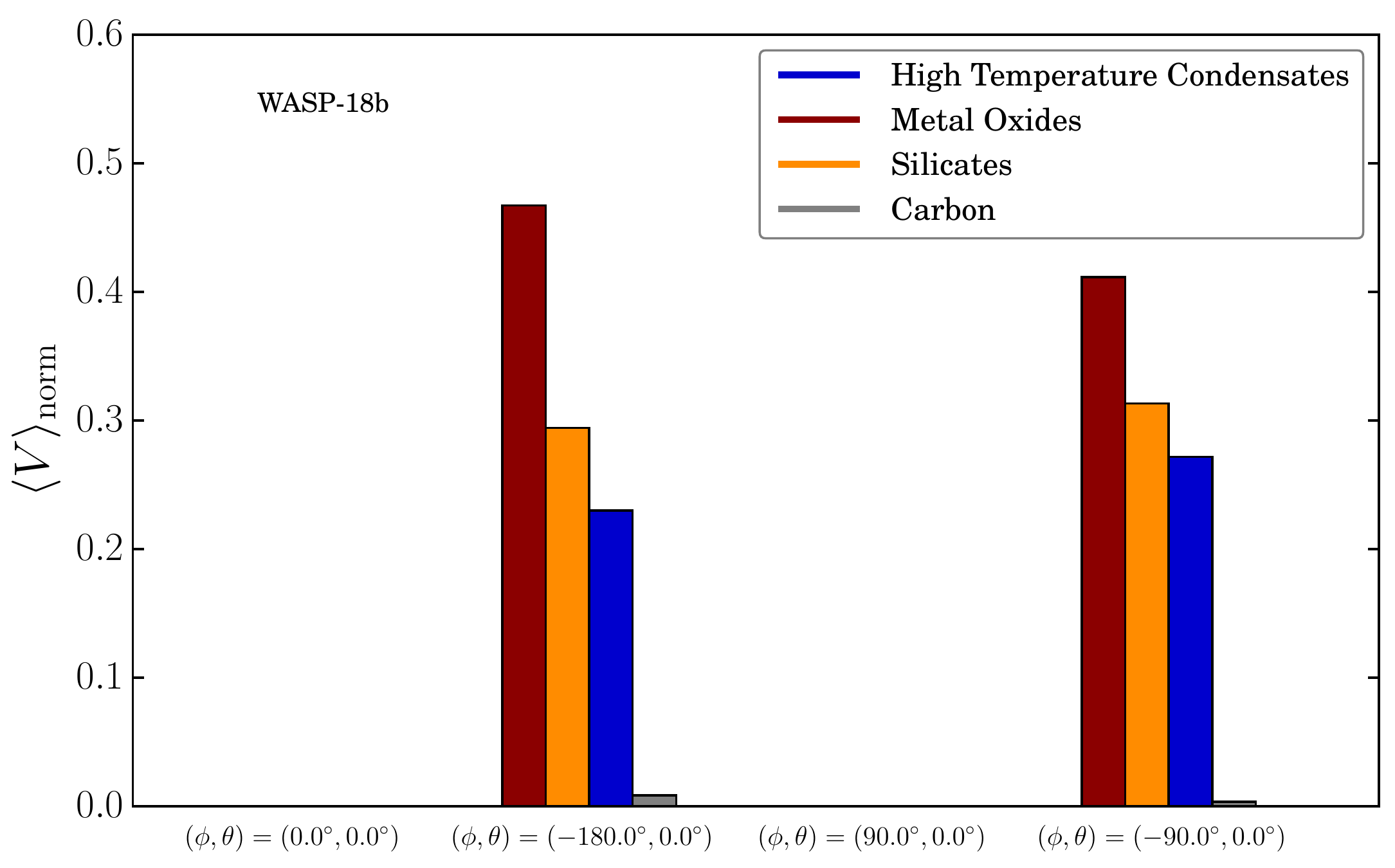}
    \includegraphics[width=21pc]{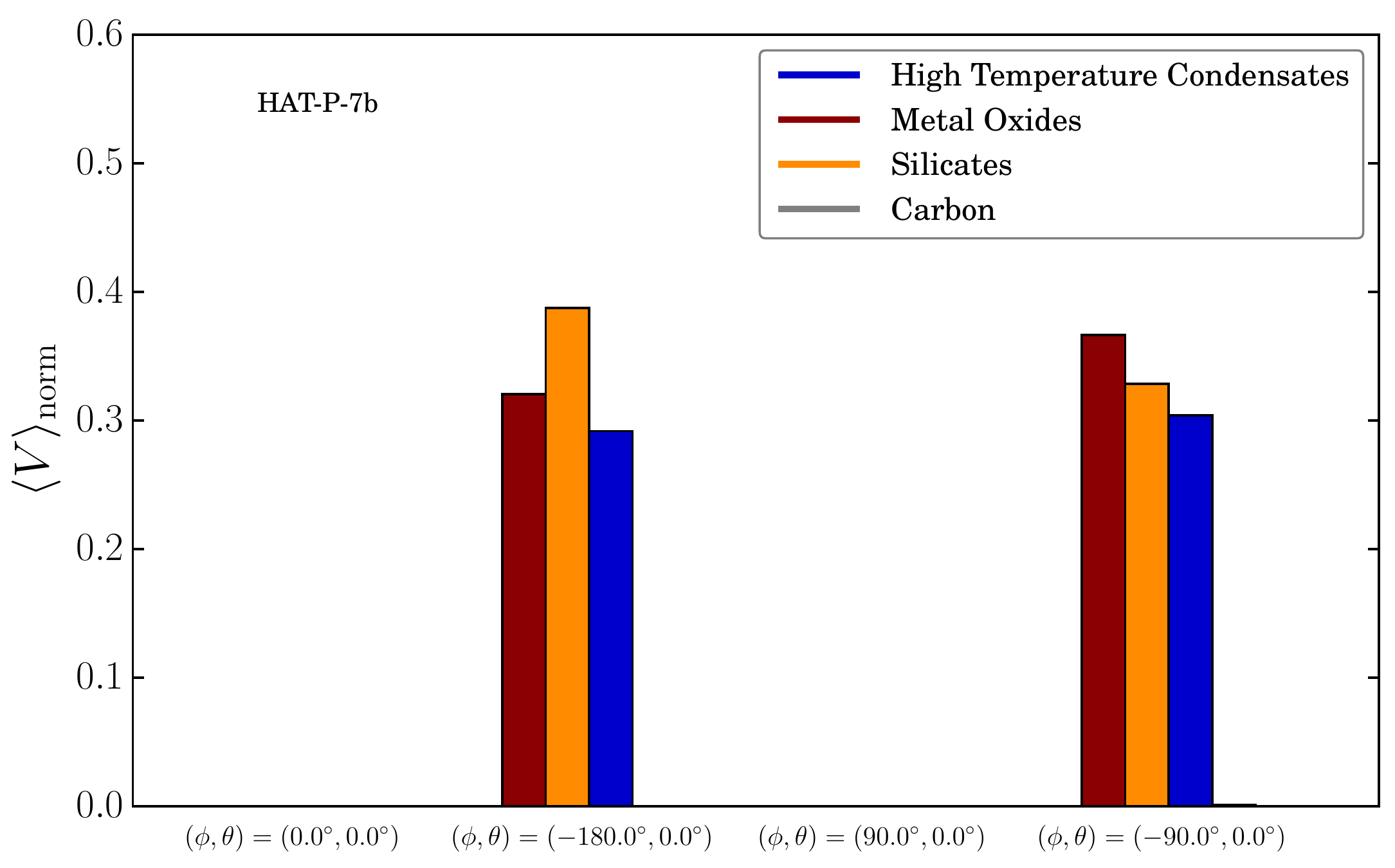}\\
    \includegraphics[width=21pc]{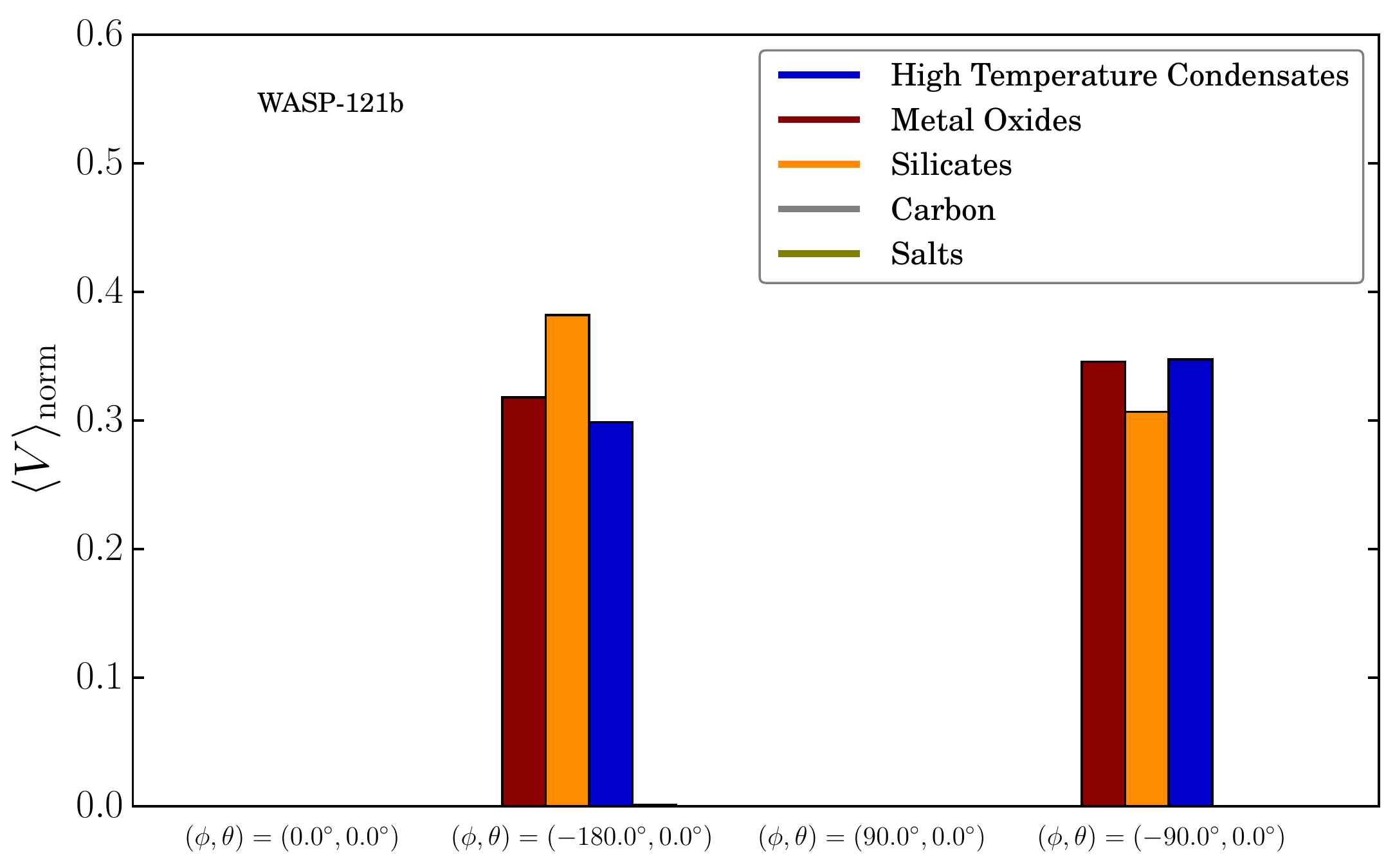}
    
    \caption{The normalised column integrated volume fractions $\langle V \rangle_{\rm norm}  = \int_{z_{min}}^{z_{max}} \frac{V_{s}(z)}{V_{tot}(z)} dz \mathrel{\Big/} \sum_{i}  \int_{z_{min}}^{z_{max}} \frac{V_{i}(z)}{V_{tot}(z)} dz$ where $s$ is the given material group and $i$ runs over all the material groups as defined in Table~\ref{tab:vol_frac_type_table} (blue: high temperature condensates, red: metal oxides, orange: silicates, grey: carbon, olive: salts). 
    Sub-stellar point: $(\phi, \theta)=(0.0^o, 0.0^o)$, Anti-stellar point: $(\phi, \theta)=(-180.0^o, 0.0^o)$, Equatorial Morning Terminator: $(\phi, \theta)=(-90.0^o, 0.0^o)$, Equatorial Evening Terminator: $(\phi, \theta)=(90.0^o, 0.0^o)$. The relative abundance of the metal oxides, silicates and high temperature condensates is comparable for the substellar point of the ultra-hot Jupiters WASP-103b, HAT-P-7b and WASP-121b.    }
    \label{fig:category_condensates_integrate_bars}
\end{figure*}

\subsection{Dust-to-Gas Ratio}


The dust-to-gas ratio, $\rho_{\rm d}/\rho$ ($\rho_{\rm d}$ -- cloud particle mass density, $\rho$ -- gas mass density), shows where the largest cloud particle mass is located in the atmosphere, or in broader terms, where most of the condensed mass is located. It therefore makes little sense to discuss column integrated values here. In other areas of astrophysics, $\rho_{\rm d}/\rho$, is used to measure the enrichment of gaseous environments with solid particles, like in planet forming disks, commentary tails, the ISM, AGB star winds, super novae ejects.
    
The general shape of the day- and nightside $\rho_{\rm d}/\rho$-profiles in our sample of giant exoplanets (Figs.~\ref{fig:rhod_rho_all1},~\ref{fig:rhod_rho_all}) are similar, all increasing to a maximum before falling back to zero at the inner atmosphere. The nightside of all low-mass ultra-hot Jupiters in our sample (WASP-103b, WASP-121b, HAT-P7b) have a steep increase of the atmospheric cloud particle mass load at the cloud top,  reaching the maximum $\rho_{\rm d}/\rho$ within a very narrow pressure interval. Amongst the ultra-hot Jupiters, WASP-18b stands out with a shallower increase of the cloud particle mass load at the cloud top more comparable to the dayside of the gas giant  WASP-43b.  Figure~\ref{fig:rhod_rho_all1} provides day- (solid line) and nightside (dashed lines) averaged $\rho_{\rm d}/\rho$ values  without the terminators for the planet ensemble considered here.

For the hot  giant gas planet WASP-43b in our sample, the nightside always displays a higher dust-to-gas ratio than the dayside. This is consistent with the lower nightside mean particle size. The dust-to-gas ratio sharply increases up to $4.5\cdot 10^{-3}$ where it stays fairly level before beginning to decrease at $\sim 0.1\,\ldots\,1$ bar. This  $\rho_{\rm d}/\rho$ values is reached if all elements like Mg/Si/O/Fe have condensed (\citealt{2018A&A...614A...1W}), indicating that these atmospheric parts have achieved thermal equilibrium. Figure~\ref{fig:rhod_rho_all} in comparison to Fig~\ref{fig:Nuc_new}
shows that, after a brief period of nucleation (which does not occur under thermal equilibrium) in the very upper atmosphere, the nightside is almost completely covered by a thick cloud at all lat/long profiles, with the cloud continuing deep into the atmosphere. Most cloud particles have evaporated at 1bar for the ultra-hot Jupiters, except for WASP-18b and the hot  gas giant WASP-43b.

The northern morning terminator point, $(\phi, \theta)=(-90^o, 45^o)$, of WASP-43b (Fig.~\ref{fig:rhod_rho_all}) displays a short, sharp peak in dust-to-gas ratio of almost  $6\cdot 10^{-3}$  at $10^{-3}$bar, being considerably higher than the other profiles. This peak coincides with an influx of cold gas at the terminator which boosts cloud particle formation.
Similar peaks of often lower 'amplitude' occur for all other planets of our sample for the morning terminator.  We note that substantial cloud particle mass is present at the evening terminator (grey dotted lines in the detailed plots of Fig.~\ref{fig:rhod_rho_all}) of some planets in our sample. 

There are generally less extended (for hot  giant gas planets) or no (for ultra-hot Jupiters) clouds on the dayside, shown by a lower dust-to-gas ratio, but this thinner cloud is still present across all dayside profiles and even continues deep into the atmosphere for WASP-43b. Dips in  $\rho_{\rm d}/\rho$ are consistent with the dayside temperature inversions.

\subsection{Material composition of cloud particles}

The material composition of the cloud particles gives insight into the changing chemical composition of the atmosphere in which the cloud particles form, and through which they fall while they continue to grow. The 16 bulk materials considered for bulk growth are listed in Section~\ref{s:ap}. Here we split these materials into five categories of condensates: high temperature condensates, metal oxides, silicates, carbon and salts. The chemical species contained within each group are listed in Table~\ref{tab:vol_frac_type_table}. The models for WASP-18b and HAT-P-7b do not include KCl[s].
The individual material volume fractions can be found in Section~\ref{s:ind_mat_fracs}, the main text focuses on the material groups only.

Figure~\ref{fig:category_condensates} shows the variation of the volume fraction for each of the material categories for each of the planets at four key points: the sub-stellar point $(\phi, \theta)=(0^o, 0^o)$, the anti-stellar point $(\phi, \theta)=(-180^o, 0^o)$, the equatorial morning terminator $(\phi, \theta)=(-90^o, 0^o)$, and the equatorial evening terminator $(\phi, \theta)=(90^o, 0^o)$. 
For the giant-gas planet WASP-43b, the upper atmosphere is dominated by silicates, making up $\sim$50\% of the cloud particle volume. The next most common are metal oxides at $\sim$30\%, and high-temperature condensates with $\sim$20\% at the hotter sub-stellar and morning terminator points. High-temperature condensates are closer to 10\% at the cooler anti-stellar and evening terminator points, with the extra $\sim$10\% being made up by carbon materials. In the very upper atmosphere at the anti-stellar point, carbon material volume fractions are higher than high-temperature condensate volume, but decrease steadily as pressure increases from 0.01 mbar. Deeper in the atmosphere, between 10 mbar and 1 bar the fraction of metal oxides increases, and the fraction of silicates decreases as silicates evaporate. After a small increase the high temperature condensates remain constant comprising $\sim$20\% of the total material volume. At 10 bar the material composition becomes dominated by high temperature condensates, with the remaining groups comprising around 5\% of the composition in total. 

\begin{table}[h]
    \centering
    \begin{tabular}{p{3cm}|p{4cm}}
    \hline
         Condensate group & Materials included\\
    \hline\hline
         Metal Oxides & SiO[s], SiO$_{2}$[s], MgO[s], FeO[s], Fe$_{2}$O$_{3}$[s] \\
         Silicates & MgSiO$_{3}$[s], Mg$_{2}$SiO$_{4}$[s], CaSiO$_{3}$[s], Fe$_{2}$SiO$_{4}$[s]\\
         Carbon & C[s] \\
         High Temperature\newline Condensates & TiO$_{2}$[s], Fe[s], Al$_{2}$O$_{3}$[s], CaTi$_{3}$[s], FeS[s]\\
         Salts & KCl[s]\\
    \hline
    \end{tabular}
    \caption{The 16 bulk materials considered in our model are grouped in 6 categories. [s] indicates condensate materials. } 
    \label{tab:vol_frac_type_table}
\end{table}

The material volume fractions of the anti-stellar point (WASP-103b, WASP-121b and HAT-P-7b) and the equatorial evening terminator (WASP-103b, WASP-121b, HAT-P-7b and WASP-18b) follow similar variations throughout the atmosphere as seen for the same points on WASP-43b. The upper atmosphere, above 1 mbar, of the anti-stellar point of WASP-18b is dominated by metal oxides and there is a small fraction of carbon between 0.1 mbar and 1 bar.
At the evening terminator there are condensates from approximately at 1 mbar (10 mbar for WASP-18b) to pressures of 0.1 bar for WASP-121b and WASP-103b, 1 bar for HAT-P-7b and WASP-18b, and 10 bar for WASP-43b at which the temperature inversion occurs. Salt species are negligible for all planets in our sample. No clouds are forming at the sub-stellar point and at the equatorial morning terminator for all the ultra-hot Jupiters (hence, the respective panels are empty).

Figure~\ref{fig:category_condensates} also allows  to see the pressure range, and thus how much of the atmosphere, over which cloud condensates are forming in the frame of our model. There are cloud particles forming throughout most of the atmosphere of WASP-43b, ranging from $p \approx 0.001-0.01 ~\rm{mbar} \dots 10 ~\rm{bar}$, for all four atmosphere profiles. For the ultra-hot Jupiters WASP-121b, WASP-103b and HAT-P-7b the anti-stellar point shows clouds forming between $p \approx 0.001-0.01 ~\rm{mbar} \dots 0.1 ~\rm{bar}$ however the morning terminator shows clouds forming in a narrower region of the atmosphere between $p \approx 0.1-1 ~\rm{mbar} \dots 0.1 ~\rm{bar}$. WASP-18b shows a similar pattern to the other ultra-hot Jupiters with the exception of cloud particles at the morning terminator forming deeper in the atmosphere between $p \approx 1-10~\rm{bar}$.

The material volume fractions for all planets 'oscillate' between groups as pressure is increased. This is caused by a  'switching' back and forth associated with the evaporation of individual species, which frees up elements that are then consumed by thermally stable species in other condensate groups: The transition between \ce{Fe2SiO4}[s] and Fe[s] as can be seen more clearly in Figs.~\ref{fig:mat_vol_WASP121b} to~\ref{fig:mat_vol_HATP7b}.
As a general rule, the volume fractions of silicates and carbon decrease whereas volume fractions of metal oxides and high temperature condensates increase as pressure increases as result of their thermal stability. {\it In all profiles, metal oxides become more common than silicates at higher pressures. }

Figure~\ref{fig:category_condensates_integrate_bars} shows the normalised column integrated volume fractions,
\begin{equation}
 \langle V \rangle_{\rm norm}  = \frac{\int_{z_{min}}^{z_{max}} \frac{V_{s}(z)}{V_{tot}(z)} dz}{\sum_{i}  \int_{z_{min}}^{z_{max}} \frac{V_{i}(z)}{V_{tot}(z)} dz},
\label{eqn:V_norm}
\end{equation}
where $i$ runs through each of the condensate groups listed in Table~\ref{tab:vol_frac_type_table}, for the same four points as shown in Fig.\ref{fig:category_condensates}. These values provide an average cloud composition at this particular point, however it does not contain the details on the local pressure and material composition variation and so should be used only as a guiding value. Both the anti-stellar point and the morning terminator of all planets show that metal oxides and silicates together dominate the cloud composition making up between $\sim$60-70\% of the total volume of the cloud particles. The remaining volume is comprised predominantly of high temperature condensates and small fractions of carbon. The sub-stellar point of WASP-43b shows metal oxides and silicates comprise approximately $\sim$30\% of the material composition each, with the remaining $\sim$40\% being high temperature condensates. The high temperature condensates, metal oxides and silicates are almost equal in their contribution to the volume of WASP-43b's evening terminator clouds. 


\begin{figure}
    \includegraphics[width=21pc]{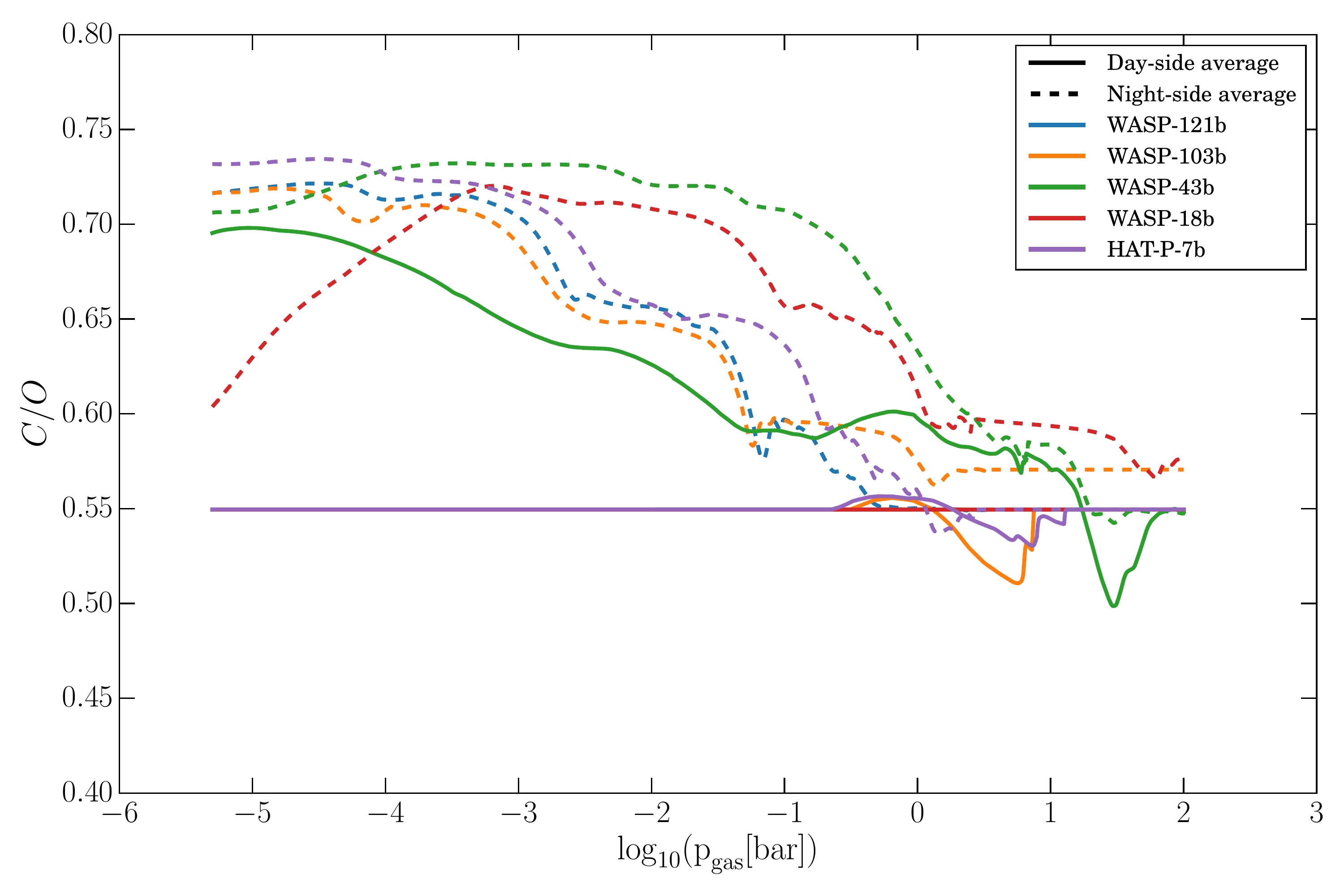}\\
    \includegraphics[width=21pc]{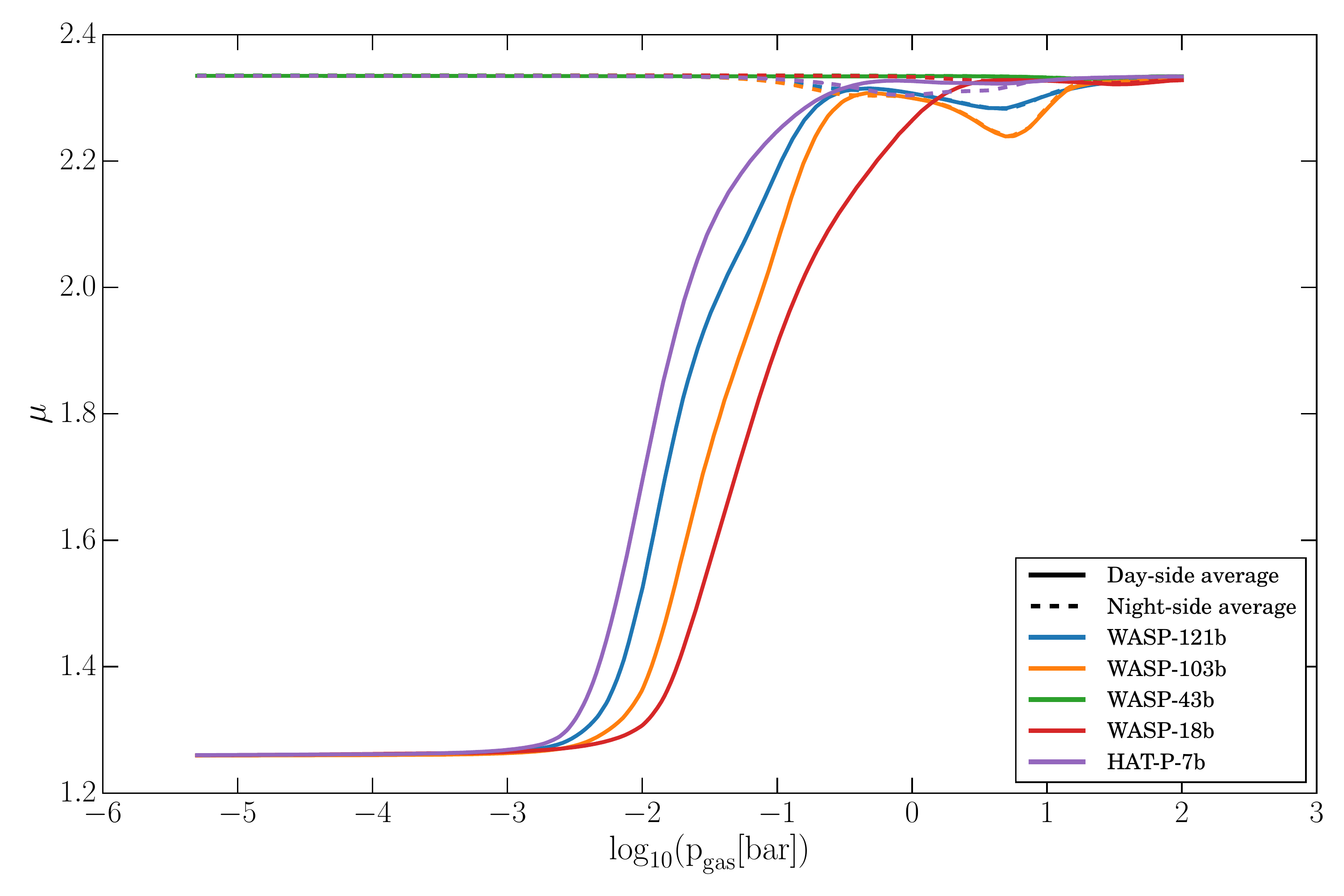}\\
       \includegraphics[width=21pc]{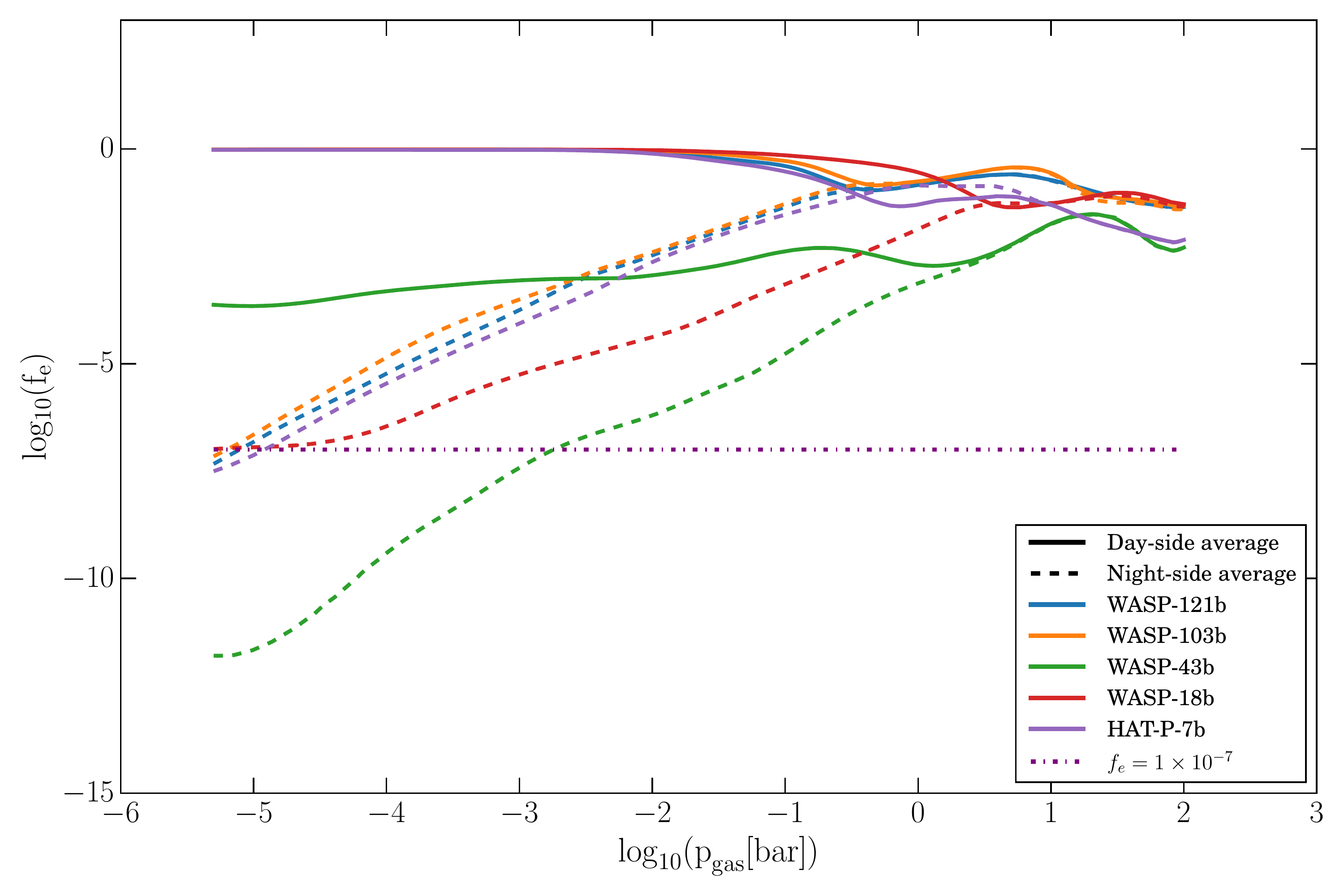}
    \caption{The carbon-to-oxygen ratio (C/O, top), atmospheric mean molecular weight, $\mu$ (middle), the degree of thermal ionisation, $f_{\rm e} = p_{\rm e}/\left(p_{\rm gas}+p_{e}\right)$ (bottom) for the hot  giant gas planet WASP-43b, and the ultra-hot Jupiters WASP-18b, HAT-P-7b, WASP-103b, and WASP-121b. The solar value C/O = 0.54 is plotted in dashed purple (top). The dash-dot purple line shows $f_{e} = 1\times 10^{-7}$ as a threshold for plasma behaviour (bottom). All ultra-hot Jupiters have dayside  thermal ionisation $f_{\rm e}>10^{-4}$ suggesting an extended dayside ionosphere. The detailed results for the individual planets are provided in Figs.~\ref{fig:rhod_rho_all} --~\ref{fig:deg_ion}.}
    \label{fig:CO_all1}
\end{figure}

\section{The comparison of characteristic global gas phase properties}\label{s:gas}

\subsection{The Carbon-to-Oxygen Ratio}

The carbon-to-oxygen ratio is often used in astrophysics to decide if an object is carbon rich, i.e. has more carbon than oxygen, or oxygen rich. Most of the exoplanet host stars will be oxygen-rich as the majority of stars in the universe are main sequence stars today. Once low-mass stars develop into  AGB stars, the star will become carbon-rich. The measurement of the stellar carbon and the oxygen abundance relies on high-resolution spectra, a technique which has only recently begun to be available in exoplanet research through instruments like CARMENES, PEPSI, and CRIRES+. So far, however, only the mere presence of atoms has been shown (e.g., \citealt{2019A&A...628A...9C,2020arXiv200702716Y}) and detailed abundance measurements are compromised by atmospheric clouds (e.g., \citealt{2018Natur.557..526N,2020MNRAS.494.5449C}). Here we focus on the local C/O which is determined by how much oxygen is locked in cloud particle materials like MgSiO3[s], MgO[s], Al$_2$O$_3$[s] etc. A similar exercise can be conducted for other element (or mineral) ratios as for example shown in \cite{2019A&A...626A.133H} (their Figs. A.3. ff.). 

As demonstrated in Figure~\ref{fig:CO_all1} (top) all cloud-forming profiles have on average a C/O ratio larger than the undepleted, solar value of 0.55 in the upper atmosphere indicating oxygen depletion by cloud particle condensation. This result holds also for the individual 1D profiles of the  3D atmospheres (Figure~\ref{fig:CO_all}).  The atmosphere turns more oxygen-rich at $\sim 1$ bar (0.1 bar for WASP-121b) in atmospheric regions where cloud particles efficiently evaporate and, hence, enrich the local gas phase with all elements previously locked within the cloud materials, including oxygen. Hence, cloud particle transport elements into the deeper atmosphere. The strongest enrichment occurs for WASP-43b, WASP-103b and HAT-P-7b for the dayside profiles with a small number of big cloud particles evaporating.

All planets in our sample can be expected to have a large range of C/O  values  from 0.54 to ~0.75  on the nightside (blue lines in Fig.~\ref{fig:CO_all}). The average nightside C/O values (Fig.~\ref{fig:CO_all1}, top) are similar among the ultra-hot Jupiters, except for WASP-18b which is the most massive planet with a larger bulk density than any of the other gas planets in our sample. 

Transmission spectra probe the terminator regions of the atmosphere which, according to our study, are very likely to differ in the C/O (grey dashed and grey dotted lines in Fig.~\ref{fig:CO_all}).
{\it Especially when considering the asymmetry in clouds, thus comparing the C/O for just above the opaque cloud level shows that 1D retrievals will fail to capture the global C/O ratio.}

Our computations confirm the conclusion by \cite{2020arXiv200715287B} that both, hot and ultra-hot Jupiters are likely to have C/O<0.8, but we can not confirm that these planets have a global solar or near solar C/O. The solar (or original) C/O can only be expected for the cloud free parts of the dayside. Lower than solar C/O would either point to an effective mass transport through the atmosphere (see C/O spikes < 0.55 on Fig ~\ref{fig:CO_all}) or to an originally higher oxygen or lower carbon content of the atmospheric gas. \cite{2020A&A...642A.229C} suggest that young giant gas planets should be expected to be more oxygen rich than older planets that have migrated through the disk already. This would be caused by accretion of  more icy material from high above the mid-plan of the planet-forming disk than originally assumed.

\begin{figure*}
{\ }\\*[-1cm]A
    \centering
    \includegraphics[width=19pc]{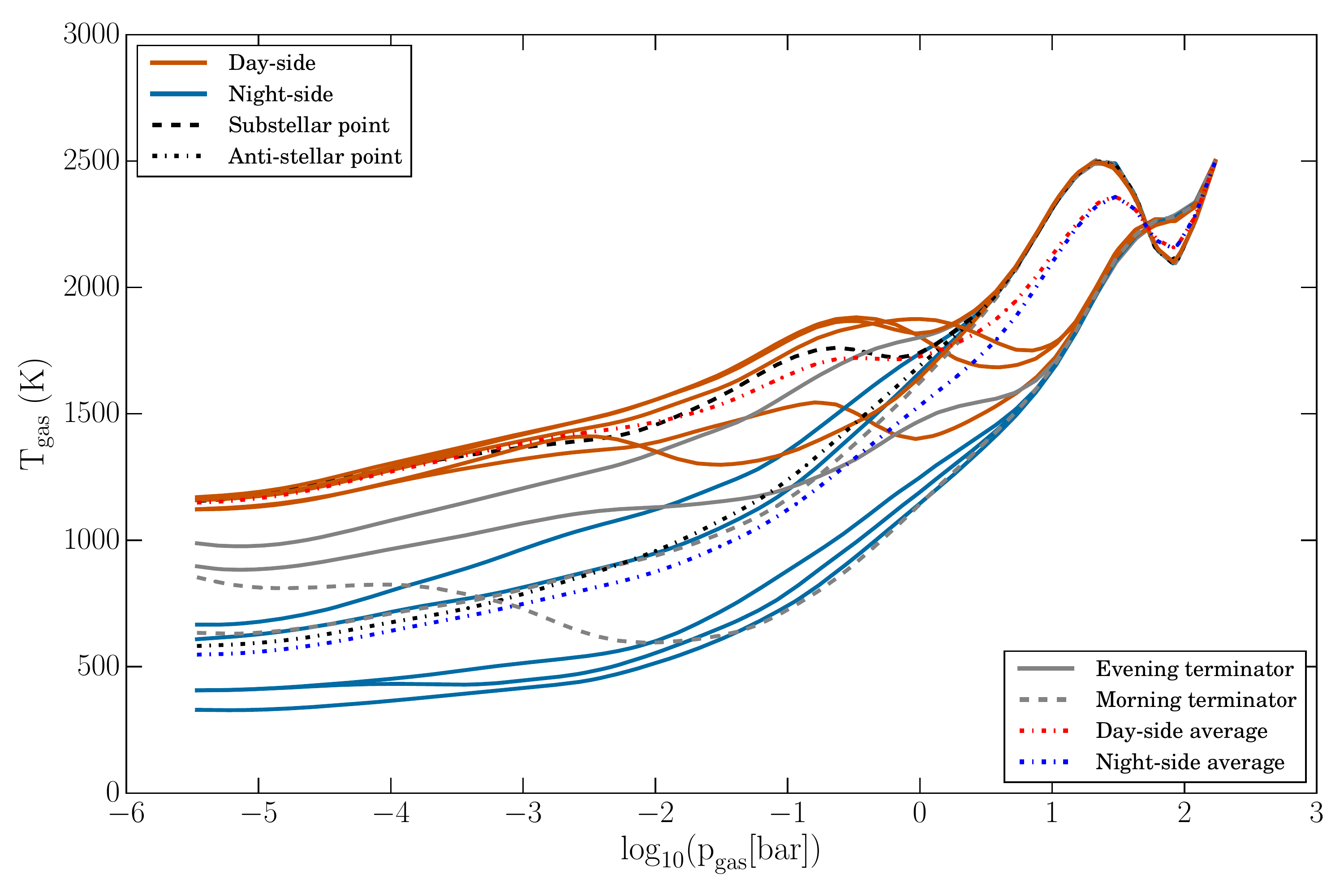}
    \includegraphics[width=19pc]{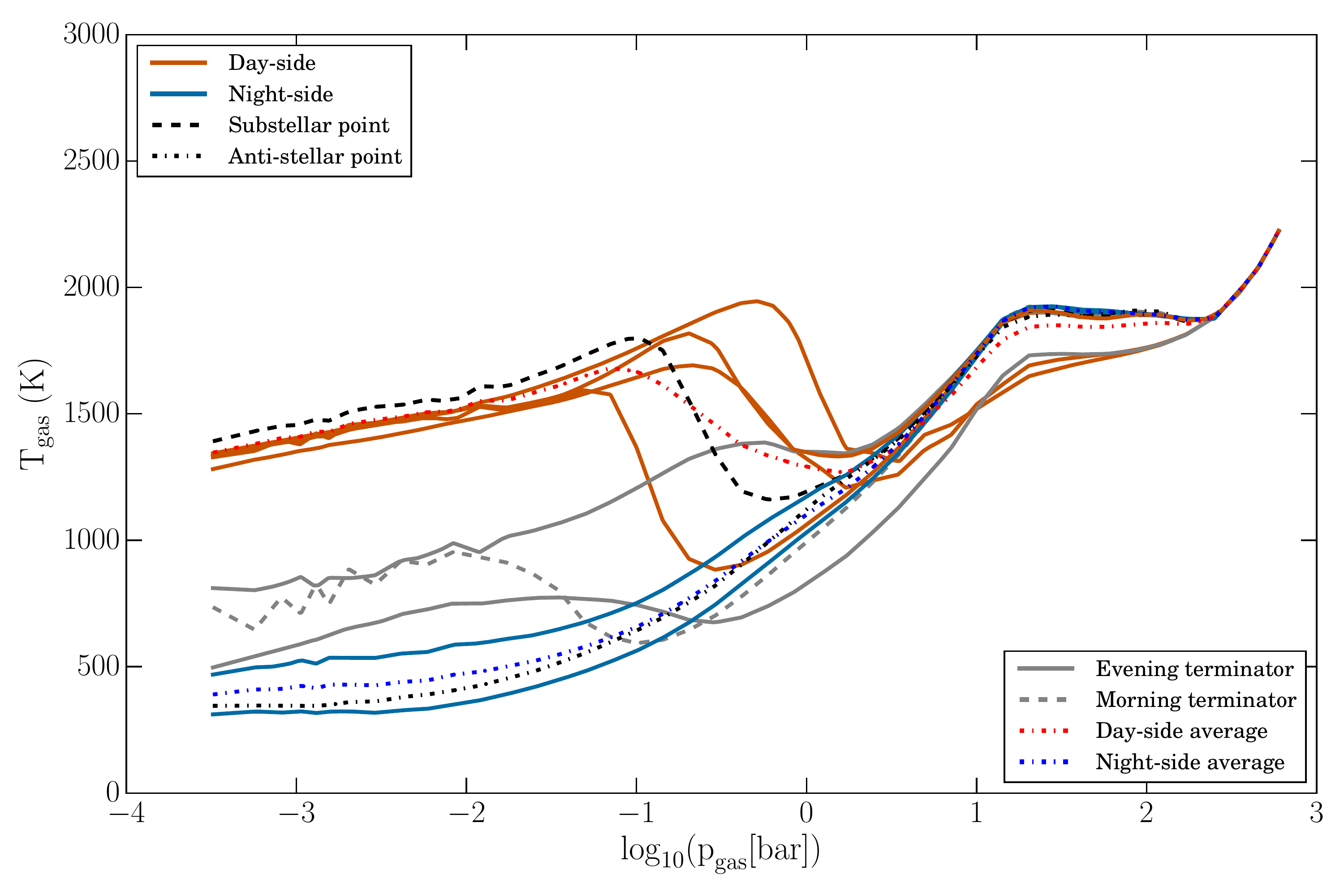}\\
    \includegraphics[width=19pc]{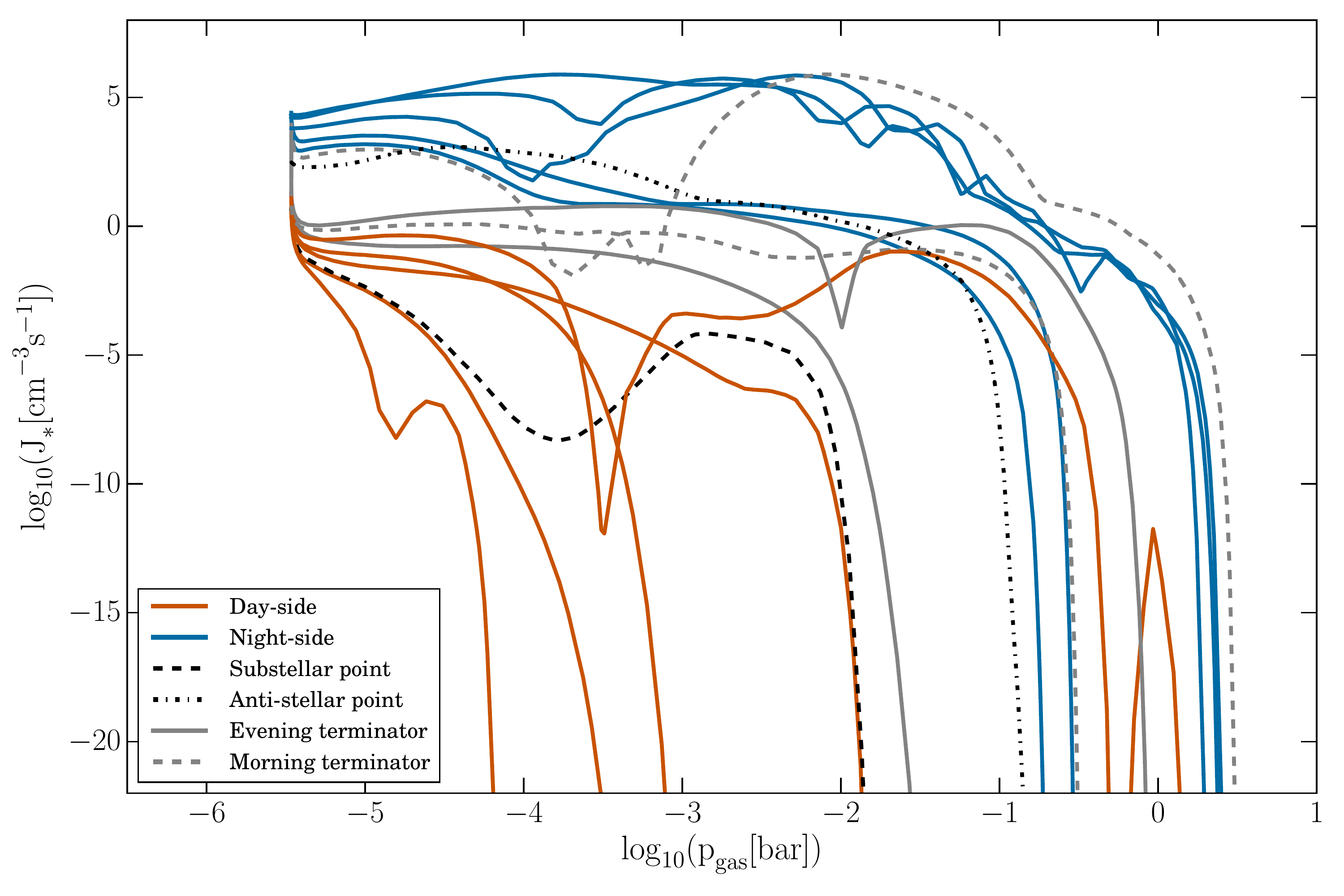}
    \includegraphics[width=19pc]{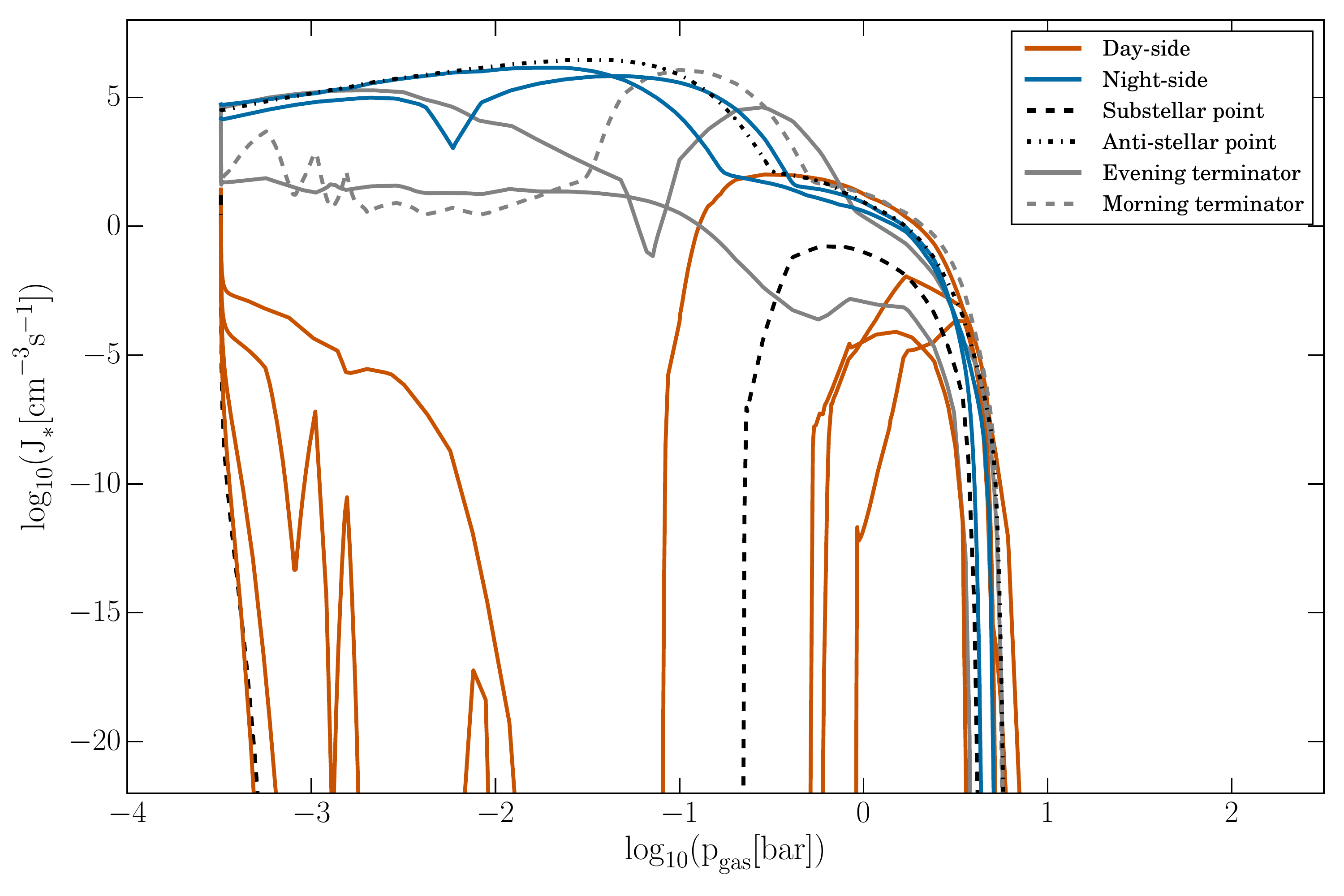}\\
    \includegraphics[width=19pc]{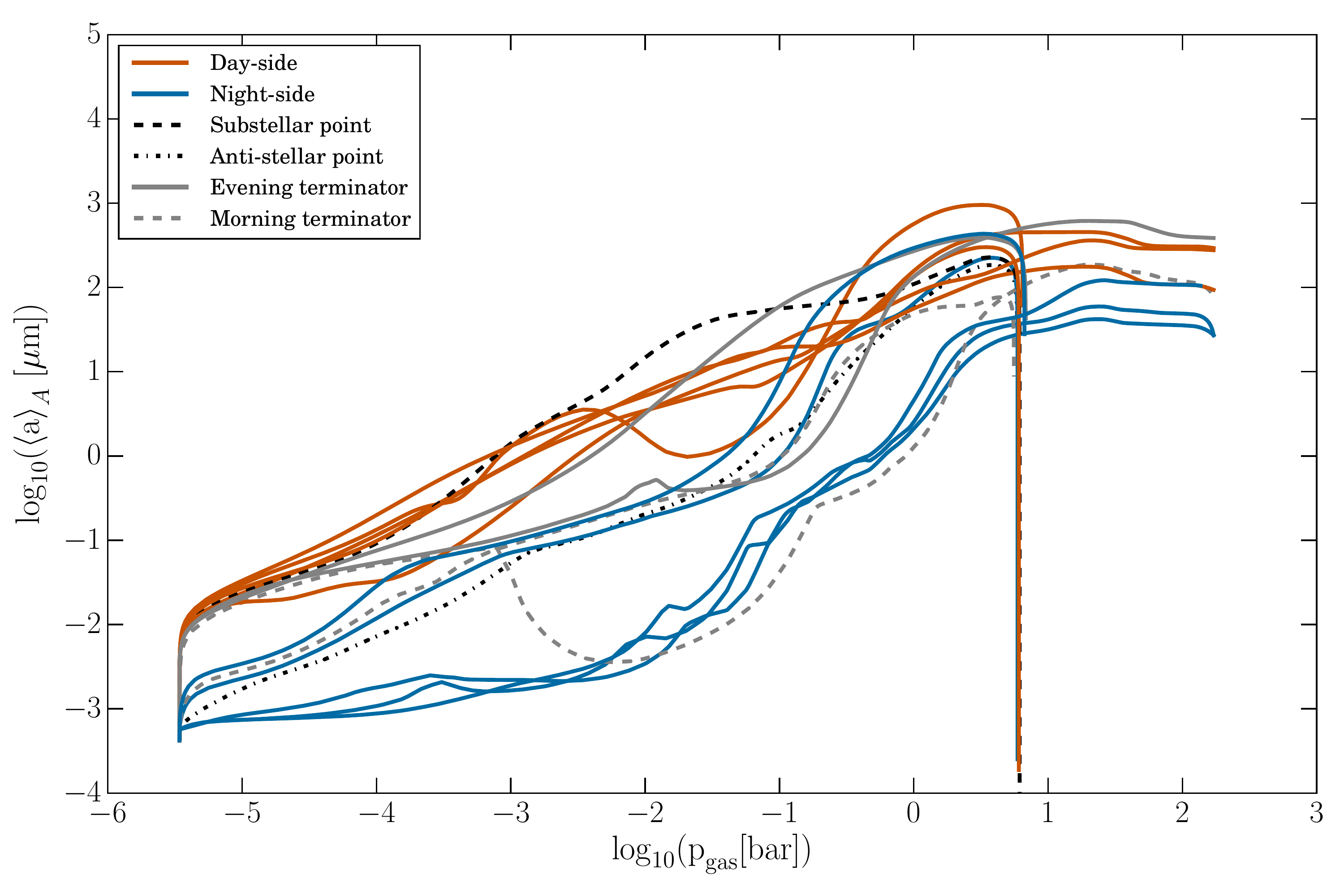}
    \includegraphics[width=19pc]{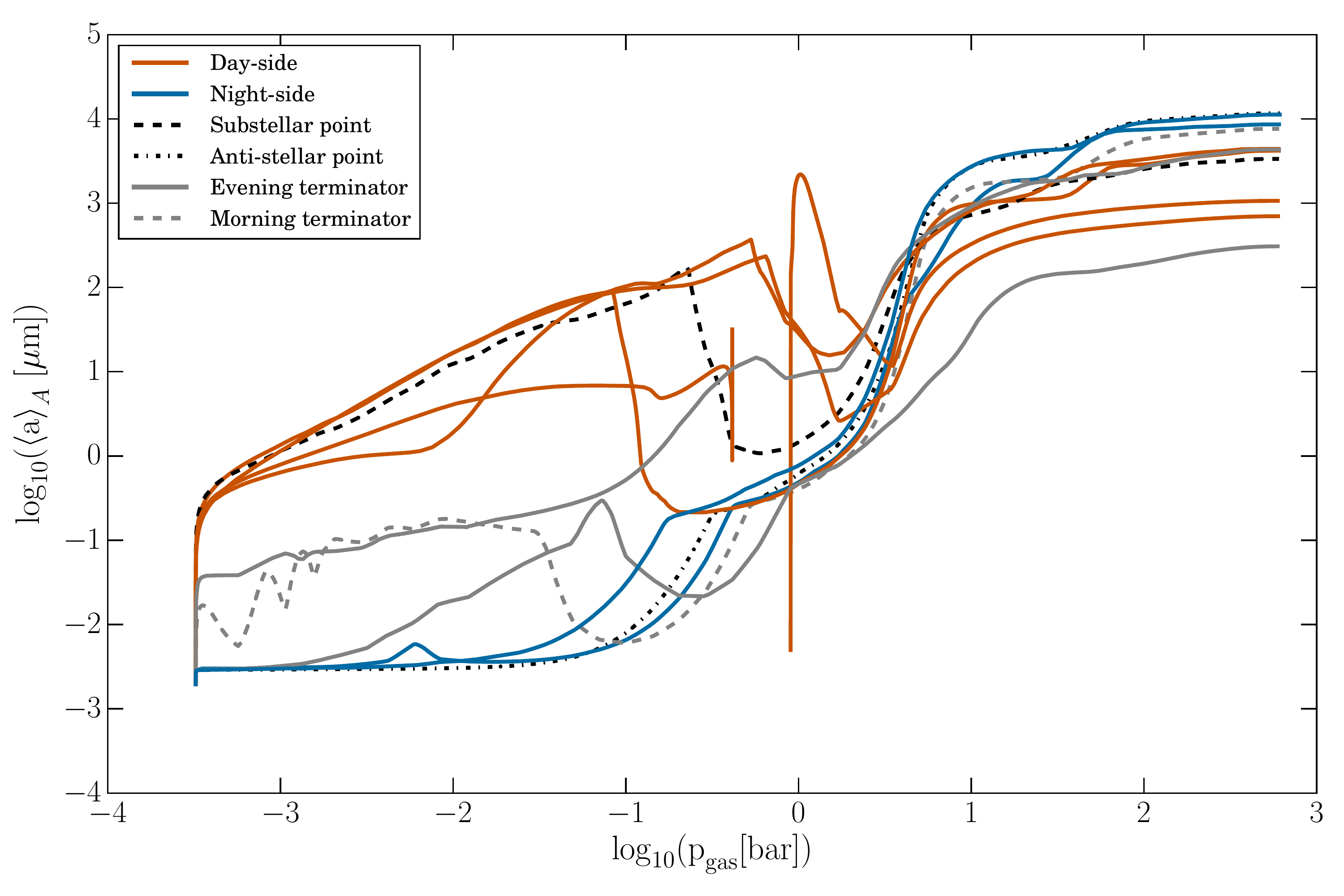}\\
    \includegraphics[width=19pc]{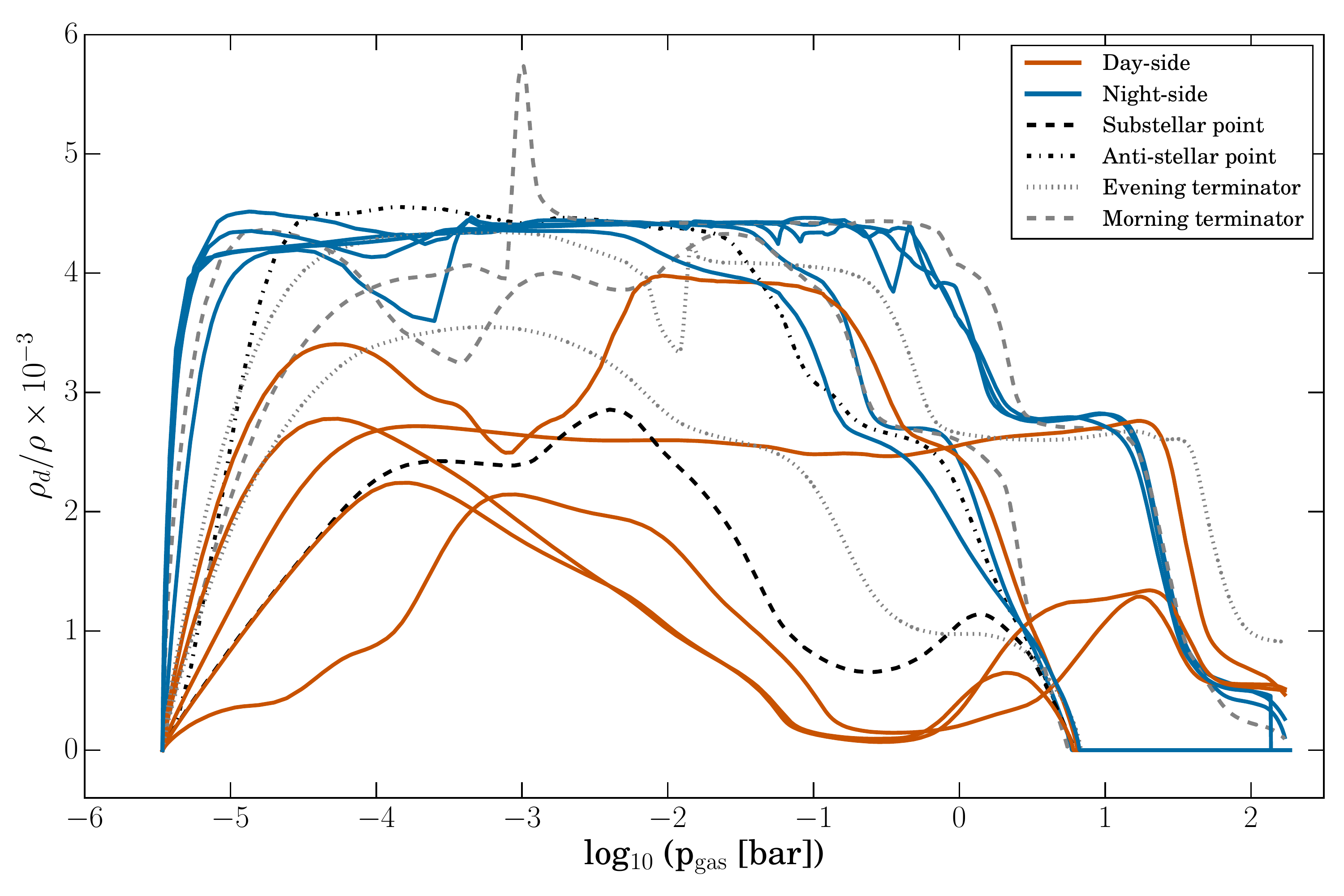}
    \includegraphics[width=19pc]{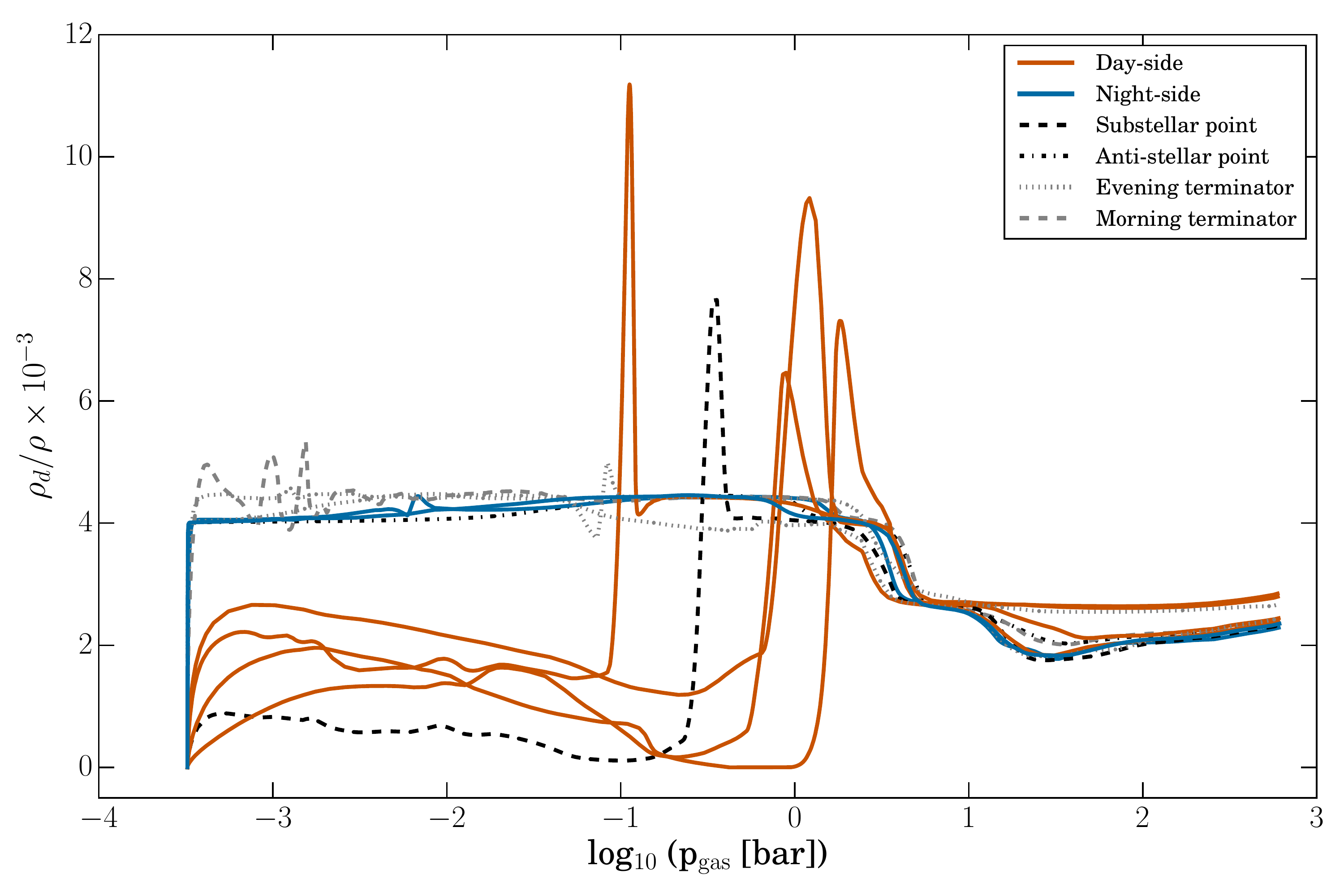}
    \caption{The effect of the inner boundary of the 3D GCM models on the $(T_{\rm gas}, p_{\rm gas}$)-profiles and the local cloud properties $J_{*}$, $\langle a \rangle_{A}$,  and $\rho_{d}/\rho$ for the giant gas planet example WASP-43b. \textbf{Left: } based on the 1D thermodynamic profiles from Parmentier et al. \textbf{Right: } based on the 1D thermodynamic profiles from  Carone et al. 
    }
    \label{fig:wasp43b_LC_plots}
\end{figure*}

\begin{figure*}
    \centering
    \includegraphics[width=19pc]{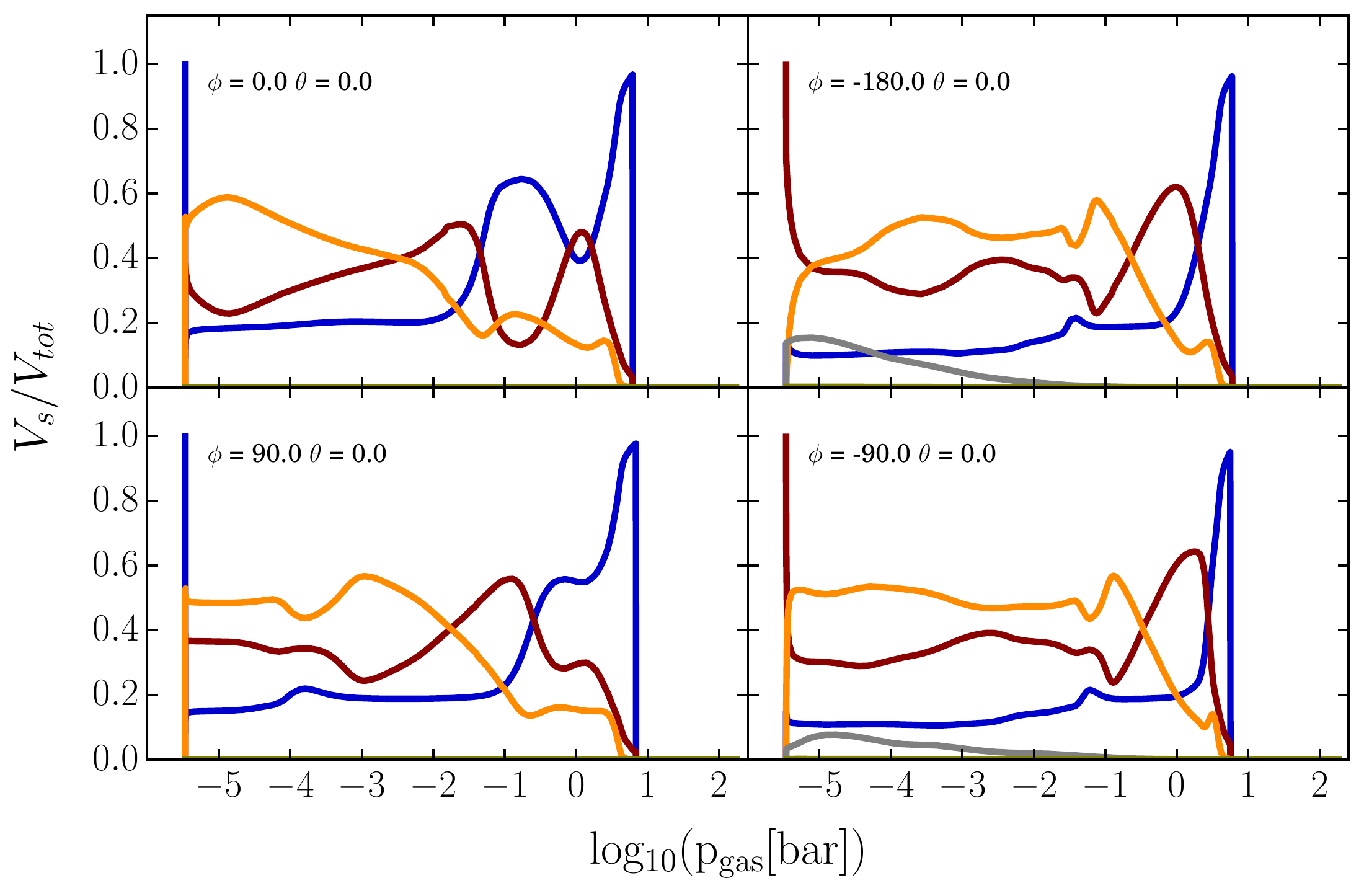}
    \includegraphics[width=19pc]{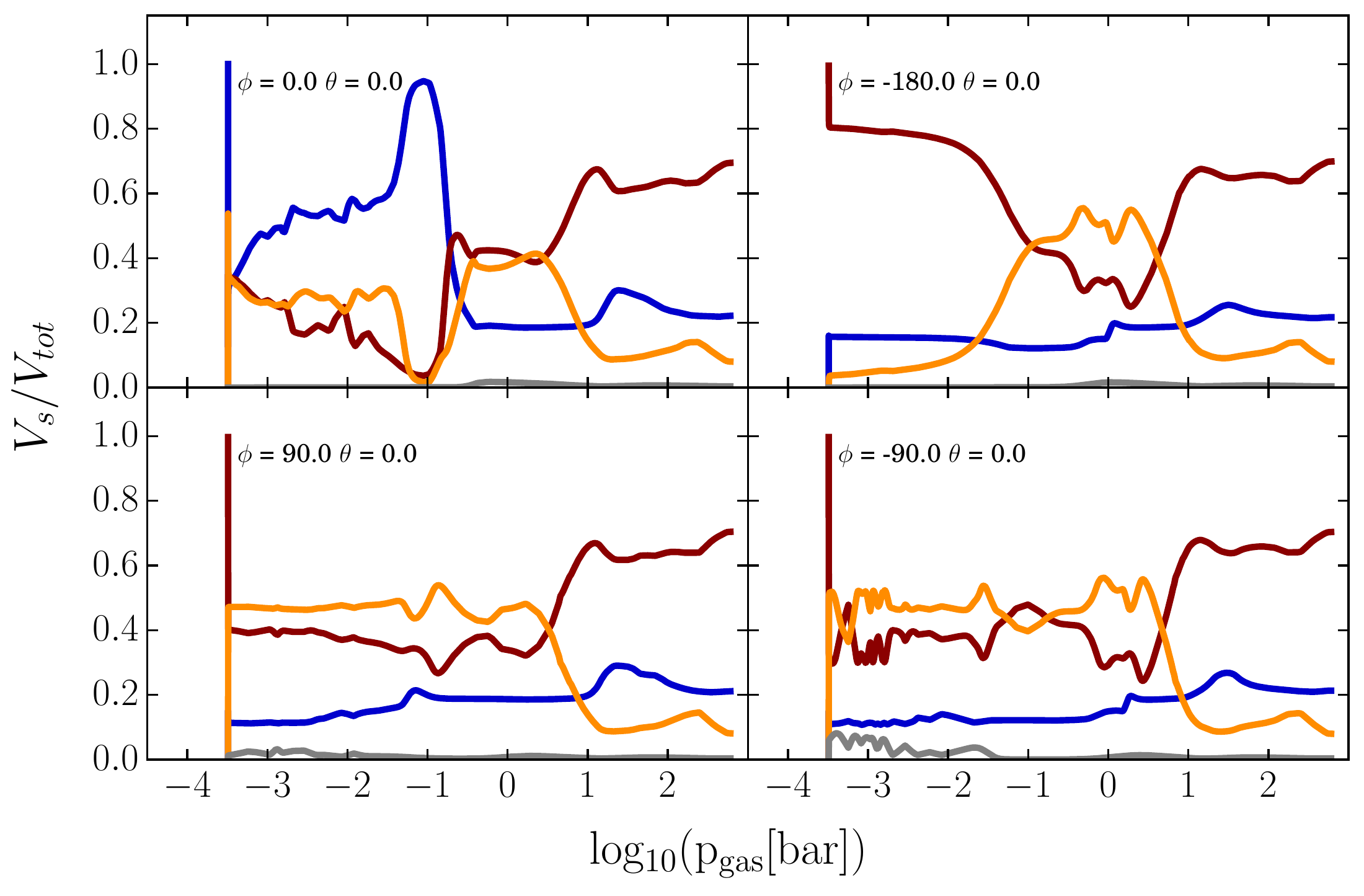}\\   
    \includegraphics[width=19pc]{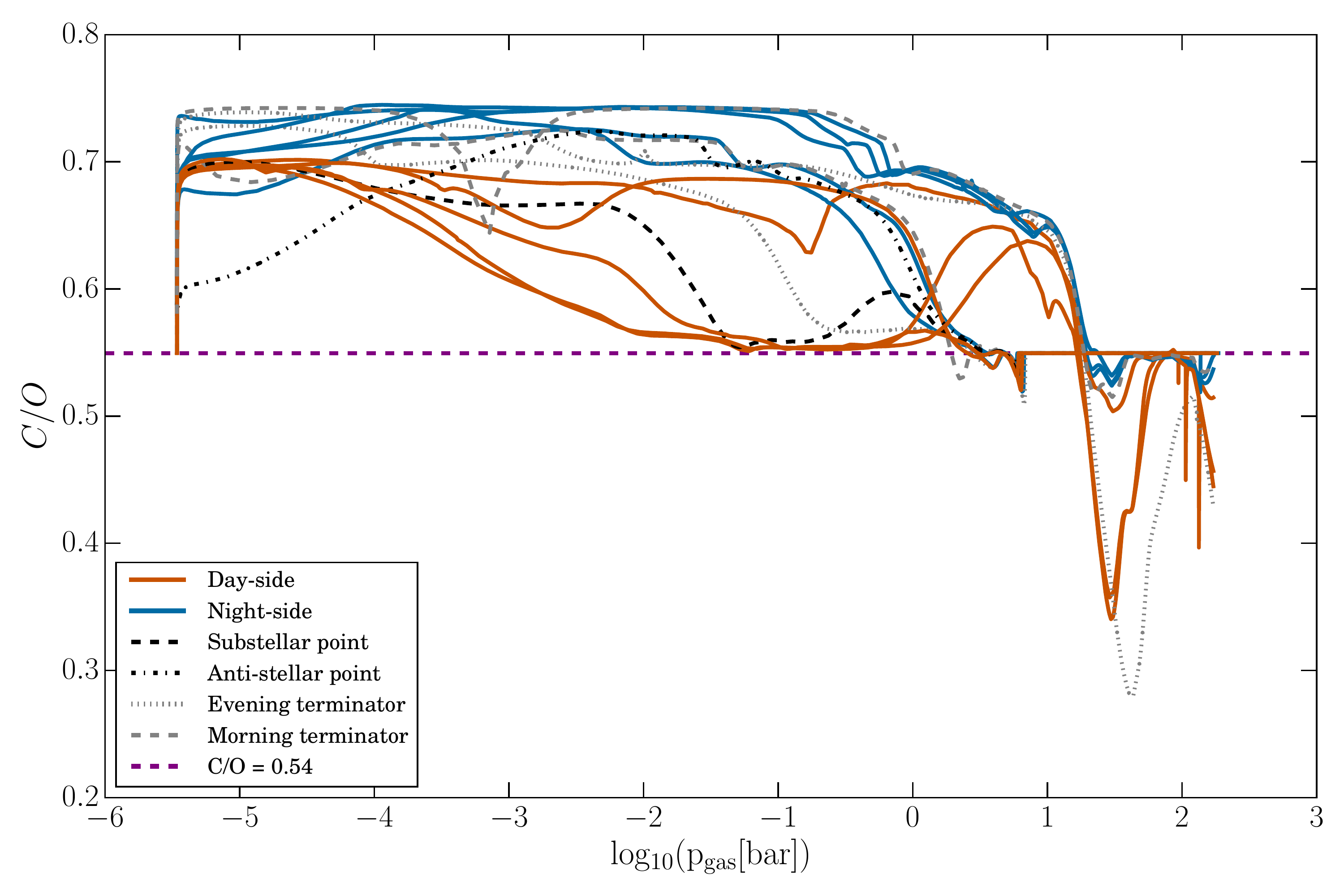}
    \includegraphics[width=19pc]{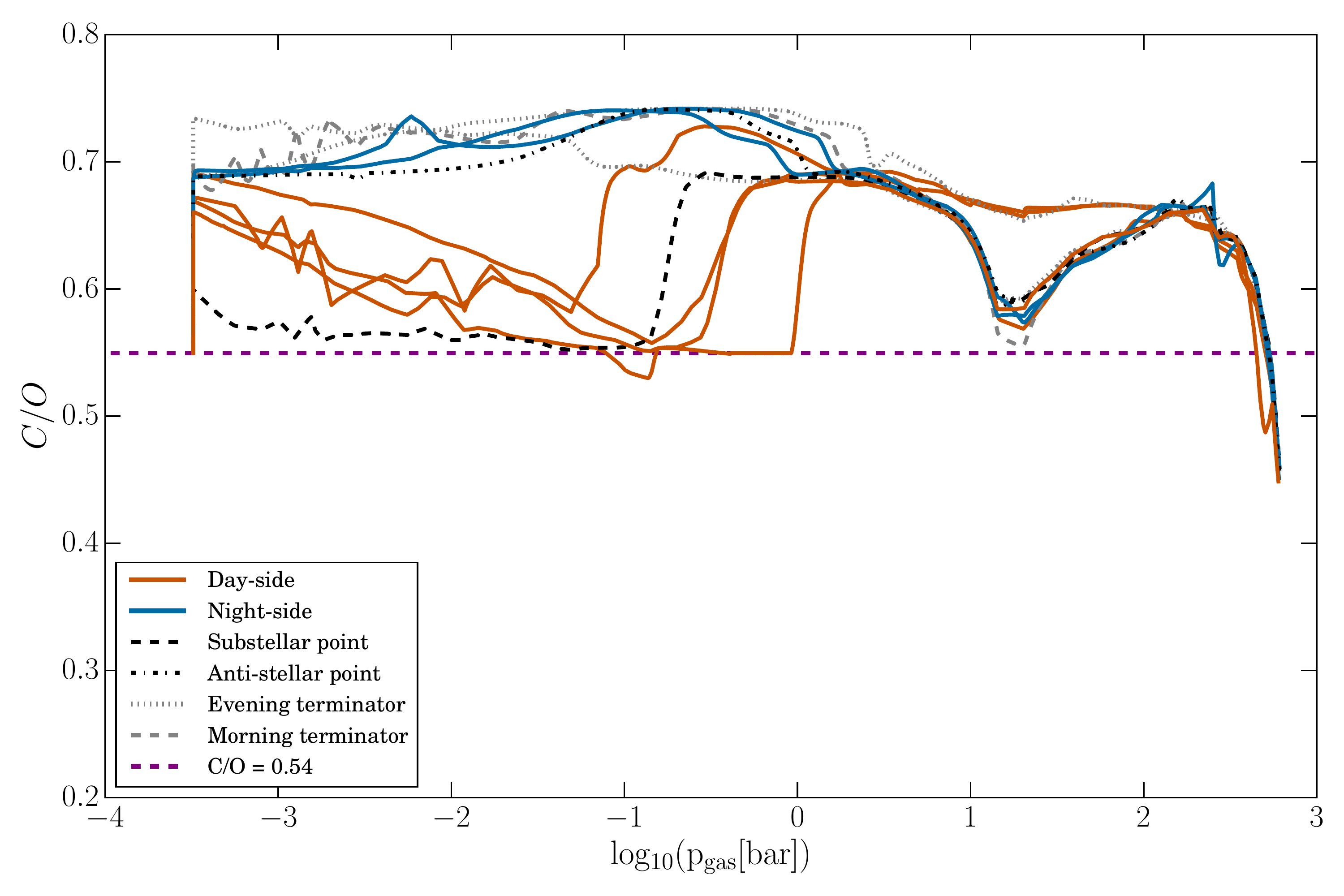}\\
    \includegraphics[width=19pc]{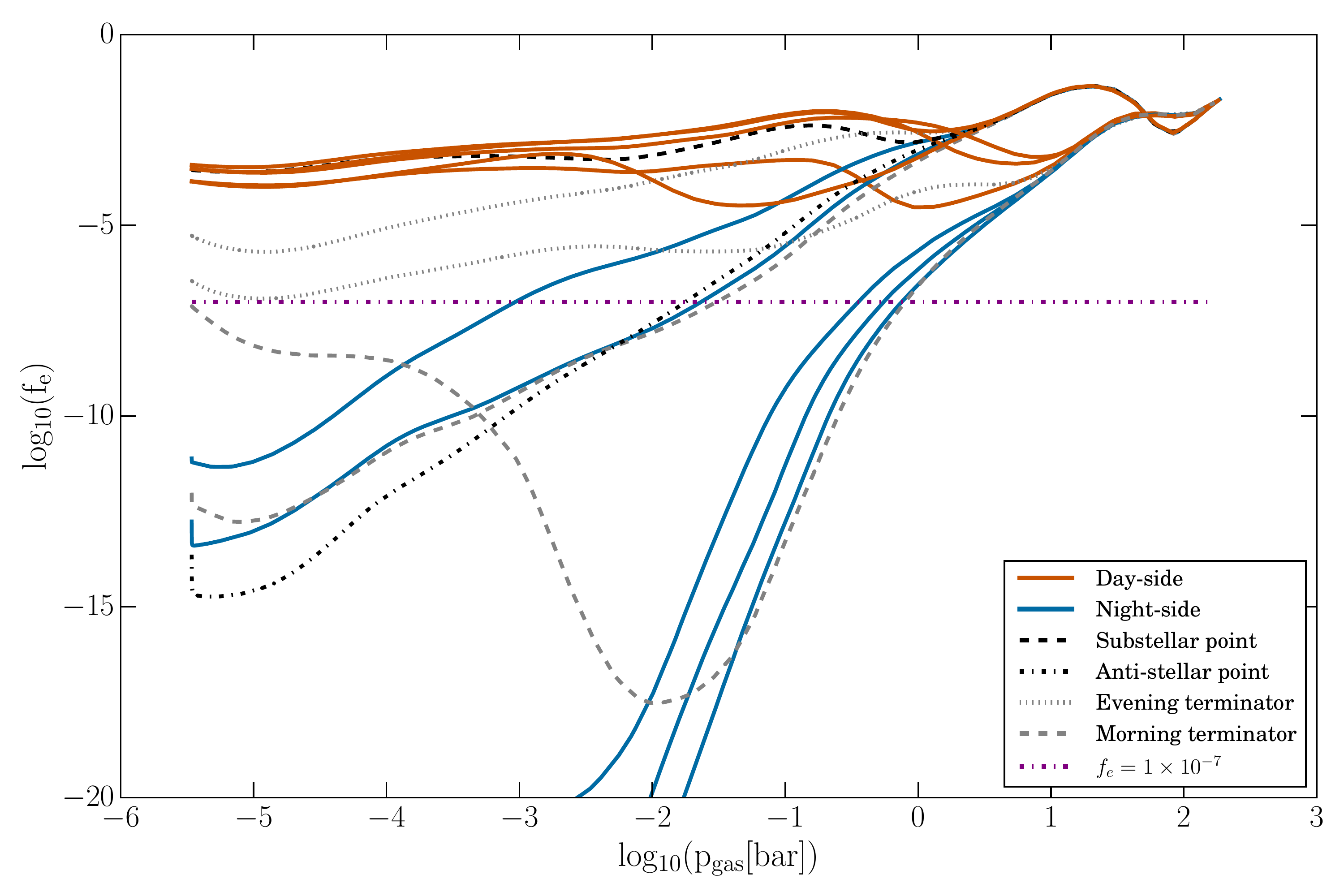}
    \includegraphics[width=19pc]{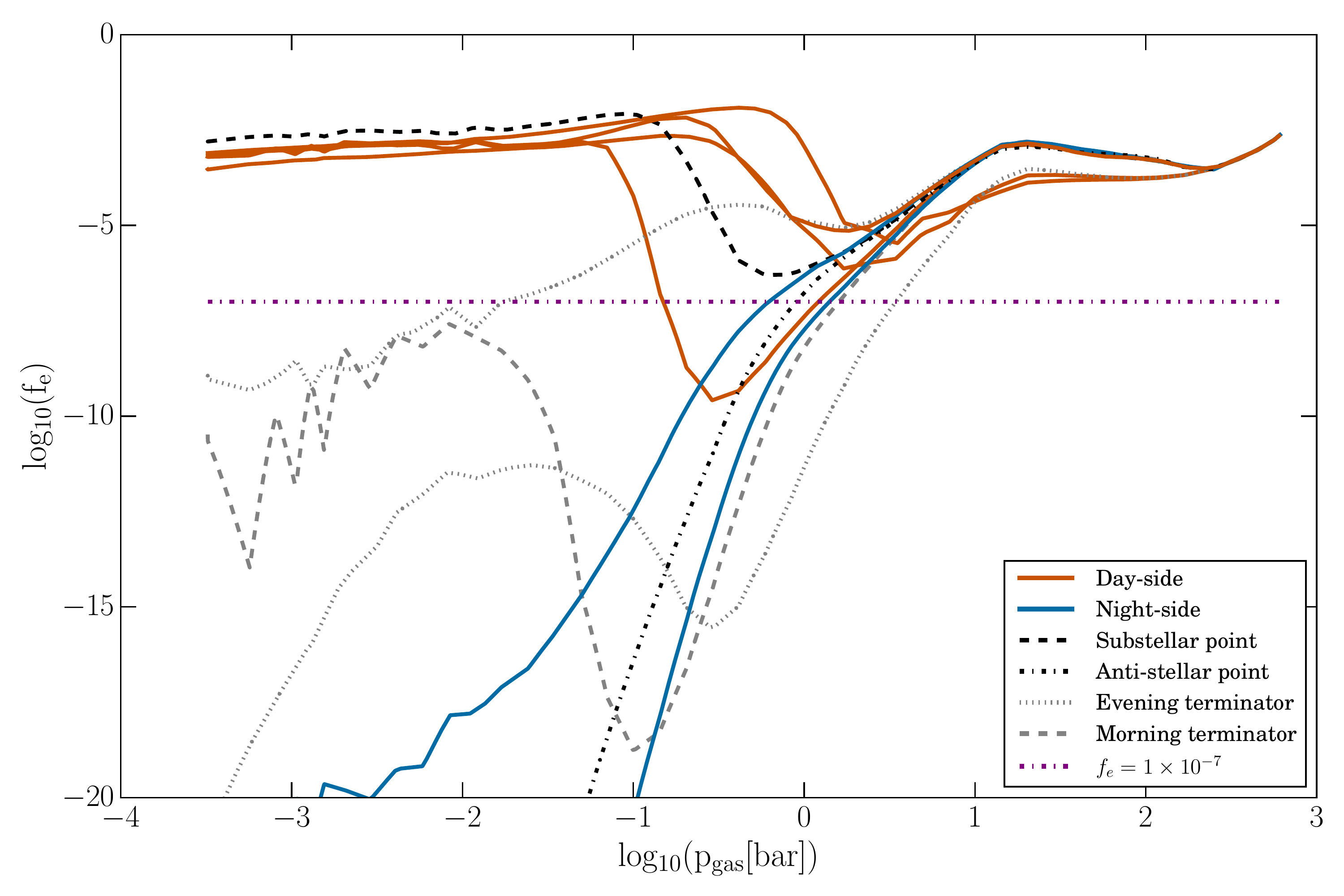}
    \caption{The effect of the inner boundary of the 3D GCM models on the grouped $V_{\rm s}/V_{\text{tot}}$, the local C/O, and the degree of ionisation  for the giant gas planet example WASP-43b. \textbf{Left: } Parmentier et al. \textbf{Right: }  Carone et al. The differences in the cloud particles material fractions result from the different local temperatures of the two GCM models, with the Carone model nightside being cooler than the Parmentier model nightside. }    \label{fig:wasp43b_LC_plots_b}
\end{figure*}

\subsection{Mean Molecular Weight}\label{ss:mu}

The mean molecular weight, $\mu$, defined as the mean mass of a particle in a gas, is an important quantity to know as it can be used to transform the local gas pressure into the local gas density via the ideal gas law. It, hence, enters calculations of transmission depth being expressed in terms of pressure scale heights ($H=k_{\rm B}T_{\rm eq}/(\mu g_{\rm P})$; e.g. \citealt{2020arXiv200707716A}) or for deriving a cloud top pressure for defining an isothermal transit radius (\citealt{2019MNRAS.490.3378H}). A constant value of $\mu$ is often assumed when running a 3D GCM (see introduction in \citealt{2018A&A...612A.105D})  as it is beneficial computationally. \cite{2017ApJ...836...73Z} show that a changing bulk composition  of the atmosphere (hence, a changing mean molecular weight) leads to a decreasing zonal wind velocity with increasing $\mu$, causing the planetary atmosphere to develop a more banded structure and a larger day/night temperature contrast. We demonstrate here that assuming a constant atmospheric bulk composition ($\mu$=const) globally may not be valid in all cases.
Figure~\ref{fig:CO_all1} (middle) summarises the results in terms of dayside and nightside averaged mean molecular weights (terminators excluded) for the hot  Jupiter WASP-43b, and for the ultra-hot Jupiters WASP-103b, WASP-121b, HAT-P-7b, and WASP-18b.  Figure~\ref{fig:mmw_all} provides the detailed results for each planet individually. 

For WASP-43b, the hot  giant gas panet in our sample, the mean molecular weight remains approximately constant ($\mu = 2.328\,\ldots\,2.337$) throughout the entire atmosphere on both the day and night side of the planet, consistent with a molecular hydrogen dominated atmosphere. 
The ultra-hot Jupiters present a very different story with the day- and nightside value of $\mu$ varying substantially across the global atmosphere. In the upper atmosphere, above 1 mbar, of the ultra-hot Jupiters the nightside value is $\mu \approx 2.3$, whereas the dayside has a value of $\mu \approx 1.3$. This difference is caused by the large temperature difference between the day and night sides of the planet seen in Fig.~\ref{Tp_all_new} which leads the dayside being highly ionised, in addition to molecular hydrogen being unabel to form.

The changing mean molecular weight which results from the local temperature effect on the gas-phase composition, results in substantial changes in the geometrical extension of the atmosphere around the globe. The details are summarised in  Appendix~\ref{ss:z} where the substantial effect of the changing mean molecular  weight on the hydrostatic pressure scale height is shown.




\subsection{Degree of Ionisation}


The degree of ionisation (Fig.~\ref{fig:CO_all1} (bottom) and Fig.~\ref{fig:deg_ion}), $f_{\rm e} = 
n_{\rm e}/n_{\rm tot}$ ($n_{\rm tot}$ - total gas number density, $n_{\rm e}$ - electron number density), 
provides a first insight into potential plasma behaviour within the atmospheres of exoplanets, including the possibility of forming an ionosphere and a magnetosphere in the presence of a potential magnetic field. A value of $f_{\rm e} > 10^{-7}$ is postulated as a threshold for transitioning from gas to plasma behaviour (\citealt{2015MNRAS.454.3977R}). Here, we consider thermal ionisation for the calculation of $f_{\rm e}$ only.

The high gas temperatures on the dayside of the ultra-hot Jupiters results in a highly ionised upper atmosphere with $f_{\rm e}$ approaching almost 1,  and a less ionised lower atmosphere with $f_{\rm e} = 10^{-2}\,\ldots\,10^{-1}$. The nightside gas temperatures of the ultra-hot Jupiters are sufficient for a partially ionised atmosphere where $f_{\rm e} \approx 10^{-6}\,\ldots\,10^{-2}$, where the increasing thermal ionisation is in line with the increased gas temperature deeper in the atmosphere.
The giant gas planet WASP-43b has dayside and nightside gas temperatures $\sim 1000-1500$ K and $\sim 500$ K less than at the same points of the ultra-hot Jupiters (Fig.~\ref{Tp_all_new}) and thus the atmosphere is much less ionised. The dayside has $f_{\rm e} \approx 10^{-4}$ throughout the entire atmosphere and the nightside has $f_{\rm e} = 10^{-13}-10^{-3}$ increasing with atmospheric depth.

The daysides of all the ultra-hot Jupiters (WASP-103b, WASP-121b, HAT-P-7b, WASP-18b) and the giant gas planet WASP-43b are sufficiently ionised by thermal processes such that an extended ionosphere is present.  Such an ionosphere is geometrically more extended for ultra-hot Jupiters compared to hot  giant gas planets according to the geometrical extension of the atmosphere (Appendix~\ref{ss:z}). The degree of ionisation will further be enhanced on the dayside  by the XUV radiation and stellar energetic particles of the host stars which will affect the outermost layers of the atmosphere. The nightside will not be affected by the stellar XUV and SEPs, but by the galactic cosmic rays. CRs have little effect in the atmospheric ionisation (\citealt{2013ApJ...774..108R}), but can open kinetic pathways to form complex hydrocarbon molecules (\citealt{2014IJAsB..13..173R,Barth2021}). 
\cite{2014ApJ...796...16K} demonstrate that the outer atmosphere of the hot gas giant HD\,209458b is magnetically coupled to a global magnetic field that may be present. The magnetic coupling amplifies with height in HD\,209458b due to the increasing effect of photoionisation. \cite{Barth2021} show that photochemistry amplifies the day/night asymmetry due to the high host star's radiation flux (XUV, FUV, SEPs) for tidally locked, close-in planets. Therefore, in the presence of a magnetic field (\citealt{2018ApJ...862...19Z,2019NatAs...3.1128C}), a  magnetosphere may affect the atmosphere of hot  giant gas planets globally (as on WASP-43b and  HD\,209733b), but it will have a strongly asymmetric affect on the atmosphere of ultra-hot Jupiters due to their strongly asymmetric ionosphere (as on  WASP-103b, WASP-121b, HAT-P-7b, WASP-18b). The asymmetry of an extended magnetosphere may be detectable as a bow shock as result of the interaction with the host-star wind (\citealt{2010ApJ...721..923L,2010ApJ...722L.168V,2011MNRAS.411L..46V}) or through  radio transit observation (\citealt{2020ApJ...895...62S}) in the future.

 The best candidates for detecting a magnetosphere could be  the ultra-hot Jupiters WASP-103b, WASP-121b and HAT-P-7b if using the hydrostatic scale height as a first guiding estimate (Fig.~\ref{fig:scaleheight}). Taking into account the interaction with the stellar wind, \cite{2011MNRAS.411L..46V} proposes WASP-18b (amongst others) as target for detecting a bow-shock. The coupling of the ionised part of a globally circulating atmosphere with a potentially existing exoplanet magnetic field may cause a current system to emerge that reduces the angular velocity at high latitudes and generate an auroral emission comparable to what has been suggested for brown dwarfs (\citealt{2012ApJ...760...59N}).  \cite{2017NatAs...1E.131R} present MHD simulations for a giant gas planet with a day/night temperature difference of $\Delta T=1000$ K, a prescribed temperature profile in order to solve the Saha equation,  and the planetary parameters of HAT-P7b. The arising Lorentz forces disrupt strong eastward atmospheric winds on such a dayside causing an oscillating pattern with a characteristics time scale (11.5 days for the chosen set up in \citealt{2017NatAs...1E.131R}).

\section{The effect of the inner boundary on GCM results for the example of WASP-43b}\label{s:wasp_gcm_comp}

Simulations using GCMs requires extensive computational resources, in particular if the radiation hydrodynamics is solved consistently with the gas chemistry and actual cloud formation modelling. Therefore, it is unsurprising that all exoplanet models apply a cloud parameterisation of some sort (\citealt{2013MNRAS.435.3159D,2018ApJ...854..172C,2018AJ....155..150M,2019MNRAS.488.1332L,2020arXiv201006936R,2021MNRAS.501...78P}). Grids of GCM simulations  are often run as completely cloud free or as opacity species only with cloud properties derived in post-processing (e.g., \citealt{2015ApJ...801...86K, 2018A&A...617A.110P}). Consequently, little time has been afforded to run extensive test on other assumptions, for example, the inner boundary.  Recently, \cite{2019arXiv190413334C} proposed that very deep layers (down to 700 bar) need to be considered to fully capture the emergence of waves that sculpt the observable climate in hot Jupiters that rotate faster than 1.5~days. This situation may be the case for WASP-43b but also generally for ultra-hot Jupiters (see Table~\ref{table:stpl}, most tidally locked ultra-hot Jupiter are expected to have rotation periods faster than 1.5 days). The possible importance to resolve deeper layers and a sufficiently long simulation times  has been further confirmed by \cite{2020ApJ...891....7W,2020SSRv..216..139S}. Here we take the opportunity to offer a first discussion on what effect the choice of the inner boundary may have on the cloud coverage of the hot  giant gas planets, WASP-43b.

In what follows, we compare the WASP-43b results from two 3D GCM simulations: Parmentier's version of SPARC/MITgcm (see Sect.~\ref{s:ap}) and  Carone 3D GCM. 
We follow the same approach as outlines in Sect~\ref{s:ap}, but additionally  utilize the input data from the Carone 3D GCM simulations (\citealt{2019arXiv190413334C}) for WASP-43b.  We will point out differences between the two 3D MITgcm  versions below. However,  an understanding of why the models differ in detail will require a more extensive comparison study which is outwith the scope of this paper. We demonstrate that shifting the inner boundary does extend the cloud layer into deeper, high pressure regimes. This raises the question about additional heating by backwarming from the cloud later which will, in turn, affect the thermal ionisation of the gas.

Also the Carone 3D GCM simulation for WASP-43b  is cloud-free. It uses simplified radiative transfer via Newtonanian cooling compared to a non-grey binned radiative transfer in \cite{2018A&A...617A.110P}. Thus, the temperature in the Carone model is less well constrained with a temperature uncertainty of  up to 100~K. \cite{2019arXiv190413334C}  has a dynamically active atmosphere that extends deeper downward to pressures of 700 bar, allowing for the formation of deep wind jets.   \cite{2019arXiv190413334C} uses several additional measures to stabilize the lower boundary  at  $p>10$~bar. These are: temperature convergence to the interior adiabat, a shorter convergence time scale $\tau_{\rm conv}=10^6$~s and drag between 550 and 700~bar \cite[][Section 2.3]{2019arXiv190413334C}. These measures are chosen ensure that the atmosphere is in a dynamical steady state from the top to the bottom. 

The SPARC/MITgcm runs of WASP-43b used here use a free slip, impermeable boundary condition situated at 200 bars and no drag other than numerical dissipation is added~\citep[see][for the detailed setup used]{Showman2009}. Because of the computational cost of running non-grey radiative transfer compared to the much faster newtonian cooling, the models are integrated for 300 days, which is shorter than the 2000 days of~\citet{2019arXiv190413334C}. With such a short integration timescale, the deep layers of the model are not yet equilibriated. The statistical steady-state reached at the photosphere is therefore dependent on the assumption that whatever circulation would develop in the deep layers of the planet is not strongly affecting the photospheric flow, an assumption recently challenged by~\citet{2019arXiv190413334C}. 

Different conditions in the deep atmosphere are proposed to induce a different climate regime compared to the thermal photosphere. Waves can travel upwards from the optically thick to the optically thin atmosphere regime. The different climate regime naturally leads to a larger day-to-night side temperature contrast and much colder (cloud free) night side temperatures compared to other GCMs.  Thus, despite the different level of approximations chosen in the  Paramentier SPARC/MITgcm and in the Carone GCM,  we  investigate if the proposed wave connection between the atmosphere at greater depth (>100 bar) affect cloud formation and chemistry higher up in the atmosphere where these effects maybe observable.

\begin{figure}
    \includegraphics[width=9.8cm]{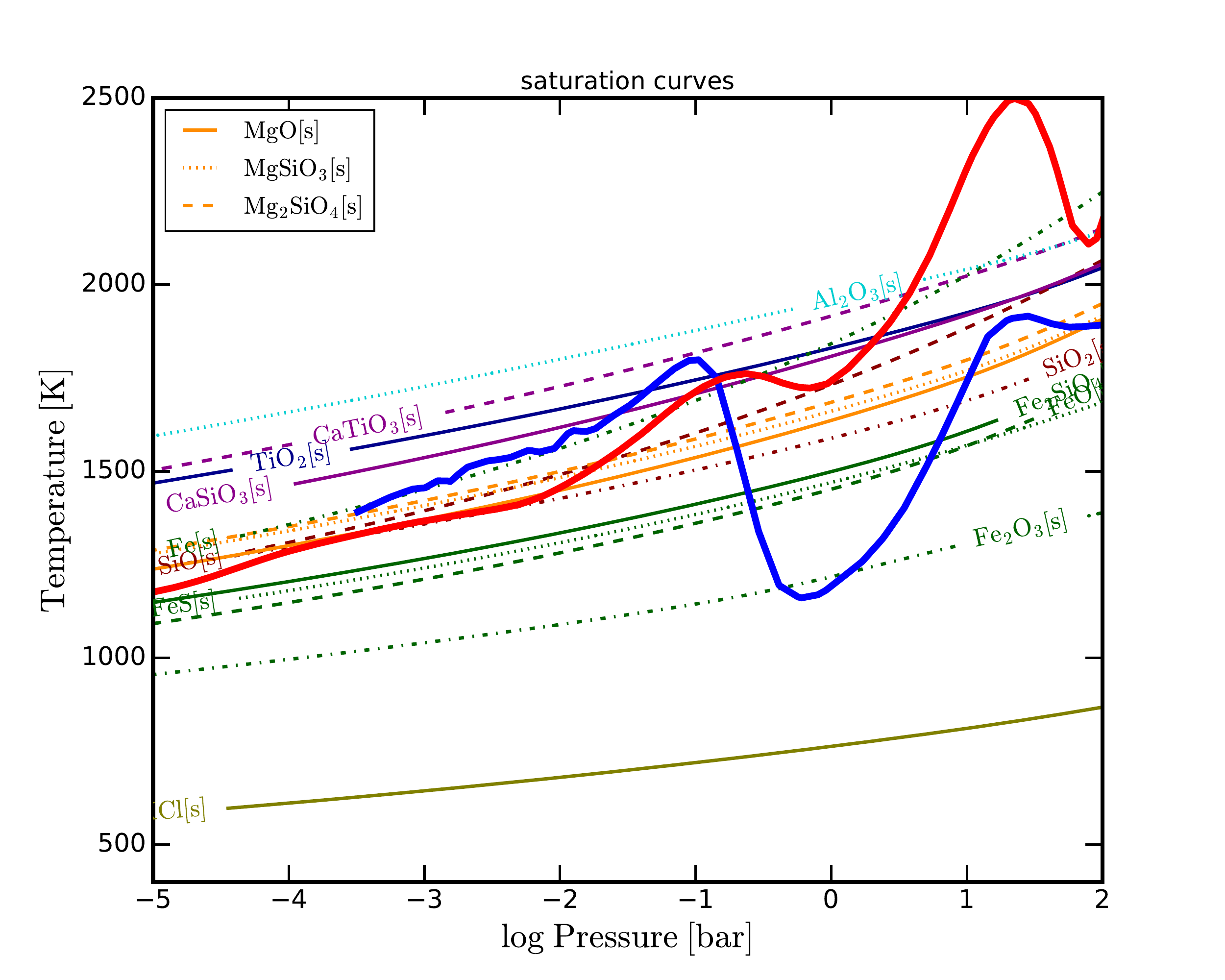}
    \caption{The sub-stellar $(T_{\rm gas}, p_{\rm gas})$-profiles from the Carone (blue, extended inner boundary) and the Parmentier (red, standard inner boundary) GCM runs for WASP-43b. The comparison to the thermal stability curves (supersaturation ratio S=1 for solar element abundances) of selected  solid materials shows that the local temperature differences at p$_{\rm gas}\approx 1$bar support our finding for the cloud particle material composition differences between the two models (Fig.~\ref{fig:wasp43b_LC_plots_b}). We note that the S=1 curves do not represent our full kinetic model approach and are provided here for the purpose of visualisation.
    }   
    \label{fig:S=1}
\end{figure}

\subsection{Does the inner boundary matter for cloud formation in atmospheres of  giant gas planets?}

We explore this question by comparing our cloud formation results for the Paramentier SPARC/MITgcm and the Carone GCM. We also explore the effect on C/O and on the local thermal ionisation as direct effects of cloud formation and thermodynamics.
Figure~\ref{fig:wasp43b_LC_plots} (top) shows the two sets of $(T_{\rm gas}$, $p_{\rm gas}$)-profiles, both showing temperature inversions at p$_{\rm gas}\sim 0.1$bar. Generally, the Carone models demonstrate that the cloud formation will extend deeper into the atmosphere due to the increasing thermal stability with increasing pressure for increasing temperature as the result of moving the inner boundary to higher pressures. The nucleation rate night profiles (Fig.~\ref{fig:wasp43b_LC_plots}, second row) have similar shape and values for both models, with the Carone model enabling seed formation  deeper into the lower atmosphere. 
Consequently, the grain size night profiles (Fig.~\ref{fig:wasp43b_LC_plots}, 3rd row)  have similar shapes and values for both models, and so does the cloud particle mass load in term of the dust-to-gas, $\rho_{d}/\rho$  (Fig.~\ref{fig:wasp43b_LC_plots}, 4th row).

The cloud formation on the dayside differs more between the two models than on the night side, reflecting the dayside differences in the $(T_{\rm gas}, p_{\rm gas})$ profiles. The inner dayside cloud (p$_{\rm gas}>0.1$bar) forms more efficiently in the Carone model, hence, more cloud particles form such that they remain smaller up to $p_{\rm gas}\approx 1$bar compared to the Parmentier model. At higher pressure, the surface growth of the gravitational settling cloud particles becomes more efficient in the Carone model (Fig.~\ref{fig:wasp43b_LC_plots}, right column) than in the Parmentier models (Fig.~\ref{fig:wasp43b_LC_plots}, left column) as the local densities are simply higher, hence the inner cloud has more and bigger cloud particles in the Carone model on the dayside and on the nightside.


Figure~\ref{fig:wasp43b_LC_plots_b} (top) shows that the thermodynamics of the atmosphere affects also  the material compositions of the cloud particles (for more details see Appendix~\ref{ss:pcapp}), suggesting that it is important to extend the 3D GCM models not only into lower atmospheric pressure regions where the stellar irradiation will affect the gas phase photo-chemically, but also toward higher pressures at the inner boundary. 
The cloud particle material composition will affect the element depletion of the gas phase locally, which we represent here in terms of the carbon-to-oxygen ratio (Fig.~\ref{fig:wasp43b_LC_plots_b}, 2nd row). Overall, the C/O values are comparable or even similar, but can differ in detail. For example is C/O<0.6 at $p_{\rm gas}\approx 10^{-2}$bar at the substellar point in the Carone model but C/O$\,\approx 0.68$ in the Parmentier model at the same pressure. The reason is that the element depletion is affected by the dynamics of the cloud particle formation which is determined by the cloud particle history in that a smaller particle will fall less fast into an atmosphere than a bigger particle.

Both models predict a partially ionised dayside through thermal ionisation (Fig.~\ref{fig:wasp43b_LC_plots_b}, 3rd row), but little ionisation on the nightside, hence, a magnetosphere should only be expected to form on the dayside from both models. 
We conclude this comparison by noting that the geometrical extension and hence the mean molecular weight are comparable in both models for a given pressure level (Fig.~\ref{fig:wasp43b_LC_plots_c}).

In this section, we studied how the treatment of the inner boundary and that of the inner atmosphere will effect the cloud properties, the C/O and the thermal degree of ionisation. {\it We conclude that the qualitative findings such as the presence of clouds, average C/O or degree of ionisation are in reasonable agreement between the two 3D atmosphere simulations.}

While the results discussed above remain qualitatively the same, the details of the material composition of the clouds (see top row of Fig. \ref{fig:wasp43b_LC_plots_c}) appears significantly differently for both the anti-stellar and sub-stellar  point.  
Here, the temperature and density differences are the largest between the Parmentier and Carone model (Fig. \ref{fig:wasp43b_LC_plots_b} top row). Thus, the night side clouds are composed of metal oxides and are geometrically more extended in the Carone model, whereas in the Parmentier model the clouds are composed of silicates and are thinner. 

There also appears to be a difference for the substellar points of the two models. For pressures greater than $\sim 10^{-1}$ bar this difference is easily explained by the differences between the thermal profiles of the models, with the Parmentier model being warmer in this region. However, for pressures between $10^{-4} - 10^{-1}$ bar the two models have temperatures within 100 K of one another for the same pressure. Thus it is surprising that in this region the Carone model produces clouds dominated by high-temperature condensates, whereas the clouds in the Parmentier model are mostly made of silicates and metal oxides. To explore this Figure  \ref{fig:S=1} shows the $S=1$ curves for all cloud condensates in our model with the substellar profiles for the two models (Carone in blue and Pamrmentier in red) over-plotted. This clearly shows that the slightly higher temperature of Carone models in this region puts it above the thermal stability curve for the magnesium bearing species \ce{MgO[s],MgSiO_3[s],Mg_2SiO_4[s]} as well as that of \ce{Fe} and \ce{SiO[s]}, which make up the majority of the cloud material in the Parmentier models (Fig. \ref{fig:mat_vol_WASP43b}). This leaves only the high-temperature condensates of \ce{TiO_2[s], CaSiO_3[s], CaTiO_3[s], Al_2O_3[s]} to be thermally stable. This explains the high-temperature condensate peak before  the drop in temperature at 0.1 bar in the Carone model, as here only \ce{CaTiO_3[s]} and \ce{Al_2O_3[s]}. Although we stress here the results shown in Fig~\ref{fig:S=1} are for solar abundances at all temperatures and pressures, and do not reflect our full kinetic model, where cloud formation depletes the gas phase of certain elements and hence changes the supersaturation ratios of condensate species bearing these elements.


\smallskip
As a summary, we note that generally the night sides of tidally locked exoplanets are the most susceptible to effects of the inner boundary. The nightside temperatures are set by the interior temperature and horizontal heat transport originating from the irradiated day side \citep{Thorngren2019}. Since horizontal heat transport is less efficient for ultra-hot Jupiters \citep{Komacek2016,2017ApJ...835..198K}, assumptions about the interior temperature and thus the lower boundary will become important for these planets. Recently, the temperature of the deep interior (at 200~bar) was invoked to explain observations of Fe and Mg in the ultra-hot WASP-121b, which led to constraints of the interior temperature $T_{\rm int}=500$~K \citep[][Fig.13]{Sing2019}.

An important next step  to shed further light on model differences would be to set up both 3D GCMs with the very same numerical parameters for the inner boundary and the radiative transfer treatment of the atmosphere. This is, however, outwith the scope of this paper.

\section{Observational implications}\label{s:obs}

We now take a look at the spectroscopic properties of the clouds for each of four ultra-hot Jupiters (WASP-18b, WASP-103b, WASP-121b, and HAT-P-7b) and one gas giant planet (WASP-43b), at four points around the equator ($\theta = 0.0\degree$), the sub-stellar and anti-stellar points ($\phi = 0.0\degree, 180.0\degree$), and the morning and evening terminators ($\phi = 90\degree,-90\degree$). The terminator profiles are indicative of what could be seen in emission for secondary eclipse, and in transmission spectroscopy. The anti-stellar point is difficult to observe, but is representative of nightside conditions, which is where cloud formation is very efficient and the most similarity in cloud structure between the planets occur. To investigate the atmosphere observable for both of these techniques (transmission and emission) we must know what pressure levels are optically thin, i.e. $\tau<1$; for example in transmission, the atmosphere deeper than this level is not visible to observers.

The optical depth along some path from $z_0$ to $z$ for a given wavelength is defined as
\begin{equation}
    \tau(\lambda,z-z_{0}) = \int_{z_0}^z \kappa(\lambda,z') \rho(z') {\rm d}z'
\end{equation}
where $\kappa$ is the extinction coefficient per unit atmospheric mass. For our atmospheres we use cloud spectral properties along vertical profiles, hence $z_{0}=0$ is the top of the atmosphere and z is the depth into the atmosphere. We then interpolate the pressure at the depth where $\tau = 1$ to get the pressure at which the clouds ceases to be optically thin. Extinction coefficients for the cloud particles are calculated using Mie theory \citep{Mie1908,Bohren1983} using the surface average particle radius from the moments as defined in Eq.~\ref{eq:surf_size} with corresponding number density as discussed in \cite{helling2020mineral}:
 
\begin{equation}
	n_{\rm d, A} = \frac{\rho L_{2}^{3}}{L_{3}^{2}}.
\end{equation} 
 
Mixed material refractive indices are determined using effective medium theory with the Bruggeman mixing rule \citep{Bruggeman1935}. Individual cloud species refractive indices are the same as in \cite{2019A&A...631A..79H}, with the addition of {\ce KCl} from \citep{1985hocs.book.....P} for all planets except WASP18-b and HATP-7b. To account for the effects of non-spherical cloud particles we include a Distribution of Hollow Spheres (DHS) \citep{Min2005,Samra20}. Hollow spheres are defined by a structure of a vacuous core and a mantle containing the material volume of a compact sphere of radius $\langle a \rangle_{\rm A}$, with volume fractions of materials as appropriate for that atmospheric layer. A distribution of these particles with different fractions of volume being the vacuum core are then averaged over. This  represents well the  distributions of irregularly shaped particles for protoplanetary disks, both in the Rayleigh regime and for larger particles \citep{Min2003,Min2008,Min2015}, and has now also been implemented in atmospheric models ATRES \citep{Stolker2017}, PetitCODE \citep{molliere2015model,molliere2017}, ARCiS \citep{Ormel2019,kchubb2019} and also in retrievals PetitRADTRANS \citep{molliere2019}.

 For ultra-hot Jupiter exoplanets there is significant difference in extension between the day and the nightside of the planet (see Appendix~\ref{ss:z}), furthermore as the stellar light passes through the atmosphere at a slant geometry \citep{Fortney2005}, there is a non-zero width of atmosphere probed around the terminator. The angle (in longitude) to which transit spectra are sensitive has been determined in recent works using both a parameterised estimation \citep{Caldas2019}, and by examining the impact of a full radiative transfer model \citep{lacy2020,pluriel2020}, both found the angle for these planets to vary between $10\degree - 40\degree$, i.e between $\pm 5\degree - 20\degree$ around the terminator. From our studied sample,  WASP-18b representing the lower end of this range and HAT-P-7b and WASP-103b representing the upper end (see \cite{lacy2020}, their Figure 3 top left).
As our approach produces 1D cloud profiles of selected longitude-latitude points, with a spacing in longitude of $45\degree$, our grid spacing is too wide to meaningfully integrate along line-of-sight trajectories through the atmosphere. We therefore chose to use vertically integrated optical depth of the clouds and apply a correction for the effect of slant geometries. In order to take into account the effect of slant geometry 
we use an correction factor calculated for a hydrostatic atmosphere, using the work of \citet{Fortney2005}, the slant geometry method of determining the optical depth adjusts the vertically integrated $\tau$ to

\begin{equation}
\label{eq:slant}
    \tau_{\rm s} = \tau\sqrt{\frac{2 \pi R_{\rm p}}{H_{\rm p}}},
\end{equation}

where $R_{\rm p}$ is the radius of the planet, and $H_{\rm p}$ is the hydrostatic pressure scale height, 
$    H_{\rm p} = (k T)/(\mu m_{H} \varg)$, 
with $k$ the Boltzmann constant,  $m_{H}$ the atomic mass unit, $\varg$ the gravitational acceleration of the planet, and $\mu$ the mean molecular weight, for which we use $\mu = 2.3$ for all substellar points. For the anti-stellar points and terminators we choose $\mu = 1.3$ (with the exception of WASP-43b which continues to use the sub-stellar value) as suggested by our results in Fig.~\ref{fig:CO_all1} (middle panel). This correction assumes that cloud properties along the vertical 1D profile are not substantially different from points probed in the optically thin atmosphere along the line-of-sight. This has been previously been investigated for more dense grids of longitude and latitudes (for example for HAT-P-7b in \cite{2019A&A...631A..79H} and for WASP-43b in \cite{helling2020mineral}), with especially rapid change in cloud properties around the terminators of ultra-hot Jupiters. Particularly at the morning terminator the pressure at which clouds form is a function of angle from the morning terminator, with the further day-ward points having clouds only at deeper levels and higher pressures, thus the optical depths of these profiles are likely affected by our assumption here. For calculations of transmission spectra it is clear that a fully three dimensional calculation is necessary as shown in \cite{pluriel2020} and \cite{lacy2020}, we leave such a full analysis to future works as previously noted, because of our wide grid spacing.

\subsection{The $p(\tau_{\rm s}(\lambda)=1)$-levels for a hot  gas giants and four ultra-hot Jupiters}

\begin{figure*}
     \includegraphics[width=1.00\textwidth]{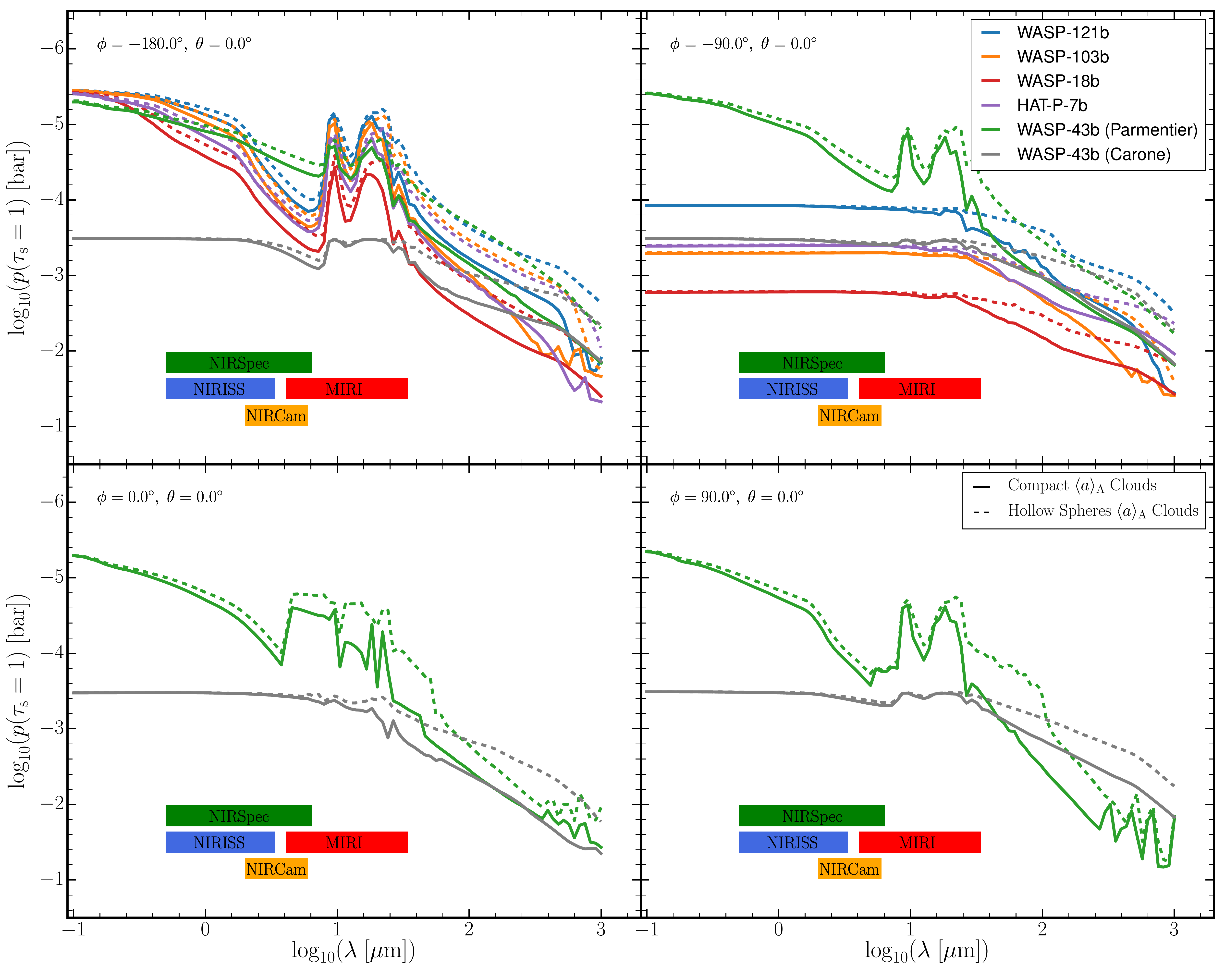}
     \caption{Wavelength-dependent pressure level where the giant gas planets WASP-43,  WASP-103b, WASP-121b, HAT-P-7b, and WASP-18b become optically thick due to cloud particles of different sizes and mixed materials forming inside these atmospheres, i.e. where $p_{\rm gas}=p(\tau_{\rm s}(\lambda)=1)$, in slant geometry. All results are based on the Parmentier GCM $(T_{\rm gas}, p_{\rm gas})$-structures, except for WASP-43b where we include both models (Parmentier GCM in green and Carone GCM in grey). When clouds remains optically thin such that $\tau_{\rm s}(\lambda)<1$, the pressure for the bottom of the atmosphere (p $\approx 10^{2.2}$ bar in the Parmentier GCMs) is returned (hence the lines for the sub-stellar points where there is no cloud are outside the plotted range).}     \label{fig:opitcal_depth4}
 \end{figure*}

Figures~\ref{fig:opitcal_depth4}  shows the results of optical thick pressure levels including the effect of slant geometry for compact (solid) and non-spherical (dashed) cloud particles. Full plots for vertically integrated optical depth alongside the slant values for each planet are found in Figure \ref{fig:opitcal_depth1_appendix}.
Only WASP-43b is sufficiently cool on the dayside to have clouds present at the sub-stellar point, and as it is substantially different from all of the other planets, it will be discussed separately in Sect.~\ref{ss:43b}. For the ultra-hot Jupiters, the inefficient heat re-distribution from the dayside to the nightside along with global circulation initiated by planetary rotation causes the evening terminators to be too hot for cloud formation, in addition to the nightside. Evidence for the effects of partial cloud coverage around the terminator has been found in retrievals \citep{Line2016,lacy2020}.
Clouds are present at the morning terminator ($(\phi,\theta) = (-90^o, 0^o)$) in all ultra-hot Jupiters (Fig.~\ref{fig:opitcal_depth4}, top left panels), but remain confined by the temperature inversion to deeper atmospheric levels of $p>10^{-3}$bar such that the atmosphere above would not be affected by cloud opacity. This results in cloud particles sizes of $10\,\ldots\,100\mu$m at $\tau_{\rm s}=1$ for $\lambda<10\mu$m in the ultra-hot Jupiters in comparison to $0.01\,\ldots\,0.1\,\mu$m  at $\tau_{\rm s}=1$ for $\lambda<10\mu$m for the hot  giant WASP-43b according to our present models. The material properties vary in accordance to the local temperatures. For WASP-103b, HAT-P-7b,  WASP-18b and WASP-121b, the optically thick pressure level appears wavelengths independent up to $\approx \,25\,{\rm \mu m}$ 
at the morning terminator. Whilst the optically thick pressure is constant at short wavelengths for each planet, the specific pressure varies between them from $10^{-4}\ldots\,10^{-3}$bar, due to the different geometric extensions. At the morning terminator ($(\phi, \theta) = (-90.0\degree, 0.0\degree$), top right in Figures~\ref{fig:opitcal_depth4}), $p(\tau_{\rm s}(\lambda)=1)$ differs considerably between the hot  gas giants and the ultra-hot Jupiters at $\lambda <30\mu$m. Beyond this, all planets show a trend of increasing transparency of clouds at longer wavelengths (i.e. increasing pressure at which $\tau_{\rm s}=1$), although the slope of this trend is affected by the details of the cloud micro-physics (see Sect \ref{ss:nonsph}).
%
%
For the anti-stellar points ($(\phi, \theta) = (-180.0\degree, 0.0\degree$)), all ultra-hot Jupiters show a consistent increase in pressure where $\tau_{\rm s}(\lambda)=1$  up to between $4-6\, {\rm \mu m}$. Hence, the atmospheric gas will be observable to greater depth and higher temperatures at these wavelengths. Strong silicate resonant features at $10\,\mu$m and $20\,\mu$m are prominent for all planets at the nightside, compared with only for WASP-43b at the morning terminator.
%
%
Figure~\ref{fig:opitcal_depth4} represent the maximum atmospheric depth that can be probed remotely. Additional gas opacity may cut the observable atmosphere to lower pressure levels than those depicted.

The distinct opacity difference between the morning and the evening  terminator of ultra-hot Jupiters may be probed by distinct asymmetries of the ingress and the egress in transmission light curves. Such an ingress/egress asymmetry effect due to clouds should be wavelength independent up to $\lambda\approx 20\mu$m according to Fig.~\ref{fig:opitcal_depth4}.  An observational signature of the evening terminator would furthermore be a clear atmosphere such that  the molecules in the atmosphere could be easily observable, and in the morning terminator one would see a much subdued spectral signature of gas phase molecules. The large difference in temperature between the two terminators will also play a role and will determine which gas species are present.


 \begin{figure}

\hspace*{-0.5cm}
    \includegraphics[width=1.15\columnwidth]{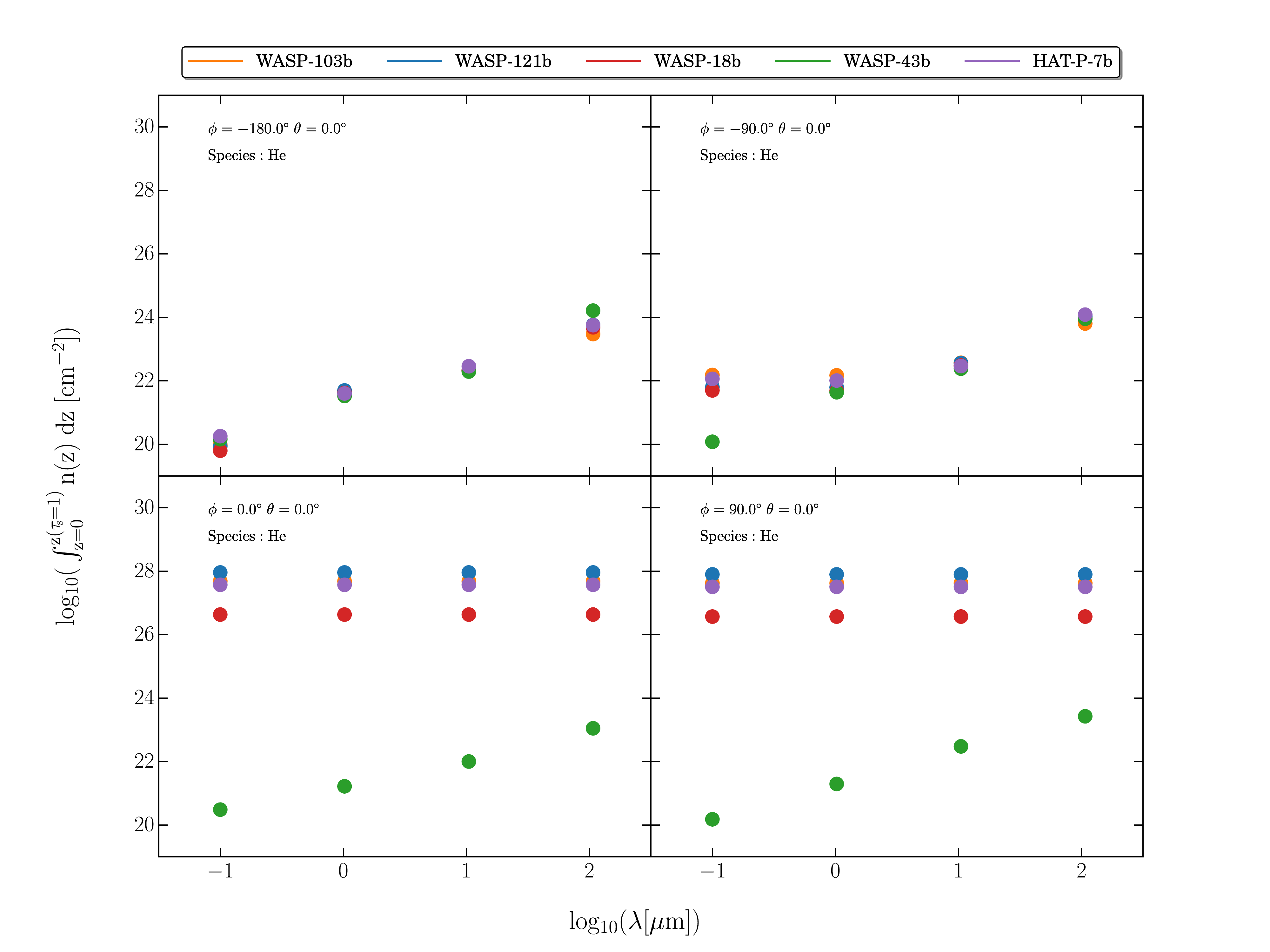}
    \caption{The cumulative, height-integrated number density of He, $\mathrm{\int^{z(\tau_{\rm s} =1)}_{z=0} n(z)~dz ~[cm^{-2}]}$,   in the optically thin region of the atmosphere $p<p(\tau_{\rm s}(\lambda)=1)$  for the sub-stellar, anti-stellar and equatorial morning and evening terminator points for $\lambda$ = 0.1 $\mu$m, 1.0 $\mu$m, 10.5 $\mu$m and 107 $\mu$m.} 
    \label{fig:He_number_density_vs_wavel}
\end{figure}
 
\subsection{The Case of WASP-43b}\label{ss:43b}
WASP-43b stands out amongst the selected planets as, in comparison to the  ultra-hot Jupiter results,  clouds form around the entire equator, thus observations both in transmission and emission will be significantly affected by cloud. It is worth noting that the sub-stellar point is not the hottest point on WASP-43b, due to the super-rotating equatorial jet, this is located at about $\phi = 30\degree$ (Fig.~2 in \citealt{helling2020mineral}).
For all profiles WASP-43b shows significant silicate spectral features at $\sim 5 - 25 {\rm \mu m}$ wavelengths, although the sub-stellar point displays a very different feature shape in this spectral region. The morning and evening terminators for WASP-43b, in contrast to the ultra-hot Jupiters, are both virtually identical to the anti-stellar point, with strong silicate features and a marginally lower pressure for optically thick clouds at all wavelengths. This provides a key difference for hot  giant gas planets vs ultra-hot Jupiters with total cloud cover of the terminators vs patchy cloud cover of only the morning terminator. A further notable difference for the sub-stellar point, is the total lack of \ce{Fe2SiO4}[s] which may explain the differentces in spectral  features. For the the sub-stellar point, using vertically integrated 
optical depth (Figure \ref{fig:opitcal_depth1_appendix}) results in almost none or very weak silicate features, and look much more like the morning terminator points for the ultra-hot Jupiters.

We previously investigated the optical depth of aerosols in WASP-43b in \citep{helling2020mineral}, for which the compact, vertically integrated clouds are identical. However, in \citep{helling2020mineral} we calculated tholin haze optical depths, we found that haze would be optically thin in the atmosphere of WASP-43b. Regardless of if compact or non-spherical shapes were considered for the haze, the clouds dominate the aerosol opacity. 

WASP-43b is also a key target with JWST, full phase curve observations are planned with MIRI as part of the community early release science (Program No. 1366; PI: N. Batalha and Co-PIs: J. Bean, K. Stevenson; \cite{Bean2018PASP..130k4402B}), followed by a NIRspec phase curve as part of GTO Program 1224  (PI: S. Birkmann). Previous observations of WASP-43b include full phase curves with Spitzer at $3.6\, \mu m$ and $4.5\, \mu m$ \citep{2017AJ....153...68S} and Hubble/WFC3 \citep{2014Sci...346..838S} across $1.1\,\ldots\,1.7\, \mu m$. Both sets of observations produced low nightside emission possibly due to poor heat redistribution \citep{2015ApJ...801...86K}, or disequilibrium chemistry and clouds \citep{2018AJ....155..150M}. However \cite{kchubb2019} did not find statistical evidence for inclusion of clouds in their retrievals, further \cite{venot2020} point out differentiating cloud scenarios is difficult using Hubble/WFC3 data. \cite{venot2020} included cloud microphysics in some of their modelling, where they assume that magnesium silicates are composed of \ce{Mg2SiO4}[s] (forsterite) over \ce{MgSiO3}[s]  (enstatite) and in addition including \ce{Fe}[s] , \ce{Cr}[s] ,
\ce{MnS}[s] , and \ce{Na2S}[s] . Overall they find that dayside magnesium silicate cloud species would be cold-trapped below 100 bar, and nightside cloud opacity dominated by \ce{MnS}[s]  and \ce{Na2S}[s]  shortward of $7\,\mu m$ and forsterite at longer wavelengths. However we find that clouds should be optically thick in the near-infrared across the planet at pressures as low as 0.1 mbar. At Spitzer wavelengths we only see minor differences between the profiles around the equator. As we do not include \ce{MnS} or \ce{Na2S} comparison of models is difficult, but NIRSpec phase curve observations will provide details requiring consistent cloud chemistry. In MIRI observations Fig.\ref{fig:opitcal_depth4} shows that (using the Parmentier model) we expect to see silicate features consistent around the equator of the planet, although variable cloud abundance with latitude would still affect the strength of these features in phase curve observations, particularly for dayside emission.

Every attempt to derive spectral information from models will depend on the computational domain for which the models are simulated. Utilising the two different GCM solution for WASP-43b we show the effect on the pressure level where $\tau_{\rm s}(\lambda)=1$ for completeness. Shifting the inner boundary does affect the location and the extension of the cloud layer resulting in some  lack of features to be explained by material composition of cloud, such as the sub-stellar point. The two models occupy different pressure domains: TOA(Parmentier) less than $10^{-5}$bar, TOA Carone only to $10^{-4}$bar. The clouds based for the  Carone GCM results have larger particles sizes in their upper cloud due the higher density supporting a higher surface growth efficiency,  which flatten the  wavelength-dependence of the optical depth substantially. The  morning terminator is comparable to that of the ultra-hot Jupiter in our sample, but  for a different reason: The large particle sizes in UHJs are cause by the low nucleation rate due to their locally higher gas temperatures. 

\subsection{Effects of Non-sphericity}\label{ss:nonsph}
The effects of non-spherical cloud particles are captured by a Distribution of Hollow Spheres (DHS) (Sect. \ref{s:obs}) and are not meant to be physical interpretations of the cloud particles, but instead by averaging over a distribution of these particles, the spectroscopic effects of a distribution of irregularly shaped cloud particles are well represented. 
The effects of non-sphericity are surprisingly limited for the case of ultra-hot Jupiters. In \cite{Samra20} we found that the wavelength at which clouds were no longer optically thick was increased with a DHS, however in these atmospheres (where they form) the clouds are never transparent at any wavelength. For profiles with strong silicate features, effects are largely limited to less than half an order of magnitude higher optically thick pressure levels for wavelengths longer than the silicate features, well outside the scope of what will be observable with JWST. 

For profiles with subdued features (i.e. WASP-43b sub-stellar point) a DHS does enhance the silicate features, marginally. For short wavelengths ($0.1 - 1\, {\rm \mu m}$) there is practically no difference between the spherical and DHS case for all profiles and planets. For profiles where the optically thick level is flat to $25\, {\rm \mu m}$ (e.g. morning terminators for HAT-P-7b and WASP-103b) the DHS increases the optical depth for all longer wavelengths, where compact particles become increasingly transparent. However, for all the Ultra-hot Jupiter morning terminators, in the slant geometry a DHS does not impact the optical depth for wavelengths observable by JWST, at these wavelengths the cloud deck is flat regardless of the micro-physics.
 
\subsection{A chemically inert global height asymmetry tracer?}

The changing day/night mean molecular weights (Fig.~\ref{fig:CO_all1}, ~\ref{fig:mmw_all}), being representative of a H/\ce{H2} dominated atmosphere gas in the cases studied here, respectively, leads to a $8\,\ldots\,10$ times more geometrically extended dayside compared to the nightside, if measured in hydrostatic pressure scale heights (Fig.~\ref{fig:scaleheight}). The vertical extension of the 3D GCM causes a factor of 2 (Fig.~\ref{fig:z}), which is not caused by the changing chemistry as the mean molecular weight is kept constant in these simulations.
The effect of geometrical asymmetry is also present in the terminator regions. Such geometrical effects may be traced by observing a chemically inert species, which is not affected by element depletion / enrichment by cloud formation, and possibly also not by changes of the ionisation state.

Helium (He) maybe such a species and  Fig.~\ref{fig:He_number_density_vs_wavel} shows the cumulative number density of He in the optically thin atmosphere (i.e. above the clouds) for the sub-stellar, anti-stellar and equatorial morning and evening terminator points at four selected wavelengths ($\lambda$ = 0.1 $\mu$m, 1.0 $\mu$m, 10.5 $\mu$m and 107 $\mu$m)  to match current and future observational capabilities. CARMENES can observe at 1.0 $\mu$m and JWST will be capable of observing at both 1.0 $\mu$m and 10.5 $\mu$m.  The column density of He changes with wavelength in Fig.~\ref{fig:He_number_density_vs_wavel} as the pressure level $p(\tau_{\rm s}(\lambda)=1)$ changes with wavelength, and hence, the geometrical extension of the optically thin atmosphere at $p<p(\tau_{\rm s}(\lambda)=1)$. The reason is the wavelength-dependent cloud opacity (see Fig.~\ref{fig:opitcal_depth4}).

The He column density (from the top of the atmosphere to where $p=p(\tau(_{\rm s}\lambda)=1)$) is largest on the dayside and the morning terminator for the ultra-hot Jupiters as no clouds form and these values therefore represent the whole atmosphere's He column density ($10^{27}\,\ldots\,10^{28}$cm$^{-2}$). On the nightside (Fig.~\ref{fig:He_number_density_vs_wavel}, top left), it follows the wavelength-dependent slope of $p=p(\tau_{\rm s}(\lambda)=1)$ for all sampled planets as shown in Fig.~\ref{fig:opitcal_depth4} (top left). The lowest He column density ($\approx 10^{20}$cm$^{-2}$) occurs in the optical, the highest in the IR on the nightside ($\approx 10^{23}\,\ldots\, 10^{24}$cm$^{-2}$).

\section{Discussion}\label{s:disc}

If two planets have a similar host star, similar orbital periods, radii, masses and similar undepeleted element abundances, the outcome of cloud formation should be largely similar, including certainly general trends such as clouds forming on the dayside or not.
\cite{2020AJ....160..109S} used HST/WCF3 data to study three (WASP-127b, WASP-79b, WASP-62b) hot  gas giants that are somewhat comparable to WASP-43b. Opaque clouds were retrieved at $\approx 10^{-3}$bar ($10^{2}$Pa) for WASP-127b, no clouds or at $p>10^{-2}$bar ($10^{3}$Pa) for WASP-79b, and at $\approx 10^{-1.5}$bar for WASP-62b, hence, at much greater depth than what our models predict. \cite{2020AJ....160...51A} use HST-WCF3 data for HAT-P-32Ab retrieveing a (isothermal) limp temperature of $\approx 1200$K, a cloud top pressure of $\approx 10^{-1.5}$bar and a C/O=0.12.
In our modelling, WASP-43b reaches $\tau_{\rm s}(\lambda=1\,\ldots\,2\mu$m$=1)$ at $p\approx 10^{-4}$bar at the terminators with $T_{\rm gas}\approx 600\,\ldots\,1000$K, and  with $\langle a \rangle_{\rm A} \approx 10^{-2}\mu$m but $\langle a \rangle_{\rm A} \approx 10^{-1}\mu$m at the dayside at the same pressure level (Fig.~\ref{fig:mean_particle_size_A}). The particles are made of a mix of metal oxides and silicates. The retrieved mean molecular weights are 2.34, 2.38, and 2.39, respectively. The values for WASP-127b, WASP-79b, WASP-62b, also within the unrealistically precise error bars, are consistent with an oxygen-depleted gas due to cloud formation.  The values suggest a higher oxygen depletion than what is derived from our model for the comparable hot  giant WASP-43b (Fig.~\ref{fig:mmw_all}, top left panel). This may be consistent with the formation of bigger cloud particles that can fall deeper into the atmosphere and consume more material. However, a quantitative comparison can only be made if similar undepleted element abundances and a similarly complete gas-phase chemistry is used. 
\cite{2020AJ....160...51A} derive a very low C/O ratio for HAT-P-32Ab, and a high metallicity for the host star.  Unless these values are plagued by the retrieval approach, the low C/O may point to either a carbon depletion and/or an initially high oxygen abundance as result of the planet formation and/or evolution processes.


The non-detection of  TiO and VO on WASP-121b (\citealt{2020A&A...636A.117M}) may be related to cloud formation at the evening terminator and to TiO/VO being thermally unstable, hence, Ti/V would be most abundant in their atomic (or ionic) form. \cite{2020AJ....160....8E} argue for hints of TiO and VO  in combination with a grey cloud layer based on HST/WFC3 data for the ultra-hot Jupiter WASP-76b which has a equilibrium temperature comparable to HAT-P-7b. This would suggest that the cloudy evening terminator dominates the transmission spectrum.
\cite{2020MNRAS.493.2215G} present  the VLT/UVES detection of Fe~I in WASP-121b and retrieve the presence of a cloud later at $p\approx 0.15\,\ldots\,0.4$ bar with a local temperature of 3000K$\,\ldots\,3700$K. Such a temperature-pressure combination renders cloud formation impossible, and may suggest that a continuum opacity source is missing in the retrieval.



As pointed out in previous works such as \cite{20VePaBl.wasp43b}, it is difficult to differentiate between a cloudy and cloud-free model by retrieving HST/WFC3 data alone. The influence on cloud modelling for retrievals of hot  Jupiters HD 189733~b and HD 209458~b has been studied in detail by \cite{20Barstow}. They find strong observational evidence that aerosols on HD 209458~b cover less than half of the terminator region, with unclear findings for HD~189733b. As demonstrated by the present work, there are differences between the morning and evening terminators, in terms of pressure-temperature structure, clouds, and chemistry. A 1D transmission retrieval assumes that the two terminator regions are identical. Works such as \cite{20McGoLe} point out the requirements for multi-dimensional retrieval techniques, which are starting to be developed, in particular for analysis of emission spectra (see, for example, \citealt{2019arXiv190903233I}). 


\section{Conclusion}\label{s:concl}

Our modelling work for a sample of ultra-hot Jupiters suggests that these exoplanets have a large day/night cloud and gas-phase asymmetry which  causes characteristic differences of their mean molecular weight, C/O (and other element ratios), and degree of ionisation. Similar properties differ less for hot giant gas planets. 

\smallskip\noindent
In conclusion, we identify the following trends:

\begin{itemize}
\item In ultra-hot Jupiter atmospheres where $p<1$ mbar, the nightside is molecule dominated with a mean molecular weight of $\mu \approx 2.3$ in case of H$_2$ dominating, whereas the dayside has a value representing an atomic gas like $\mu \approx 1.3$ for atomic hydrogen. This is caused by the large temperature difference between the day- and nightsides of these exoplanets. The dayside is therefore highly thermally ionised, in addition to molecular hydrogen being thermally unstable.
     \item A larger mean molecular weight of $\mu=2.3$ on the nightside than originally assumed in GCMs may decrease the zonal wind velocity (\citealt{2018arXiv180309149Z}).
     \item The day/night mean molecular weight differences cause a geometrically asymmetry between day- and nightsides in particular on tidally locked planets. We suggest a chemically inert species like He to probe this geometrical asymmetry.\\
    \item The immense atmospheric day/night temperature differences on ultra-hot Jupiters cause thermal ionisation to change substantially from the day- to the nightside. The thermal ionisation of the dayisde of  $>10^{-5}$ is sufficient to argue for the presence of very extended, thermally driven electrically conducting daysides, namely an atmospheric ionosphere.
    \item The atmospheric ionosphere suggests electromagnetic coupling to a potential planetary magnetic field which maybe observable through auroral emission or a bow-shock as result of the magnetosphere - stellar wind interaction.\\
    \item Ultra-hot Jupiters can be expected to have cloud-free daysides and cloudy nightsides, in comparison to hot  gas giants which have cloud-covered day- and nightsides.
        \item The detailed material composition is determined locally, but mineral silicate clouds made of Mg/Si/Fe/O may dominate the outer cloud layers, and high-temperature condensates will dominate the inner, warmer cloud layers. Photochemical hazes may also be present but will not affect the optical depth significantly on ultra-hot Jupiters.
        \item Exoplanet clouds may extend further inwards than previously assumed due to the increased thermal stability for increasing gas pressures. 
    \item The global atmosphere circulation results in cloud formation being more likely at the less-extended morning terminator, but not in the geometrically more extended, warmer evening terminator regions in ultra-hot Jupiters. \\
    \item Transmission spectra of ultra-hot Jupiters may be affected by cloud opacity at the morning terminator, but by atomic and/or ionic opacity sources at the evening terminator. This will affect the retrieval of mineral ratios like C/O/, Mg/Si etc.
    \item The different cloud properties at the morning terminator of  hot  giant gas planets like WASP-43b and ultra-hot Jupiters like WASP-103b,  WASP-121b,  WASP18b,  and  HAT-P-7b  cause their spectral features to differ characteristically.

\end{itemize}

\begin{acknowledgements}
 Ch.H. and M.M. acknowledge funding from the European Union H2020-MSCA-ITN-2019 under Grant Agreement no. 860470 (CHAMELEON).  D.S. acknowledges financial support from the Science and Technology Facilities Council (STFC), UK. for his PhD studentship (project reference 2093954), and  O.H. acknowledges the PhD stipend from the University of St Andrews' Centre for Exoplanet Science. V.G. acknowledges the hospitality of the School of Physics \& Astronomy at the University of St Andrews and the summer student funding from the Royal Astronomical Society.
\end{acknowledgements}

\bibliographystyle{aa}
\bibliography{reference.bib}

\appendix
\section{Cloud properties and gas-phase parameters (C/O, $\mu$, $f_{\rm e}$)}\label{s:ind_mat_fracs}
\begin{figure*}
   \centering
   \includegraphics[width=21pc]{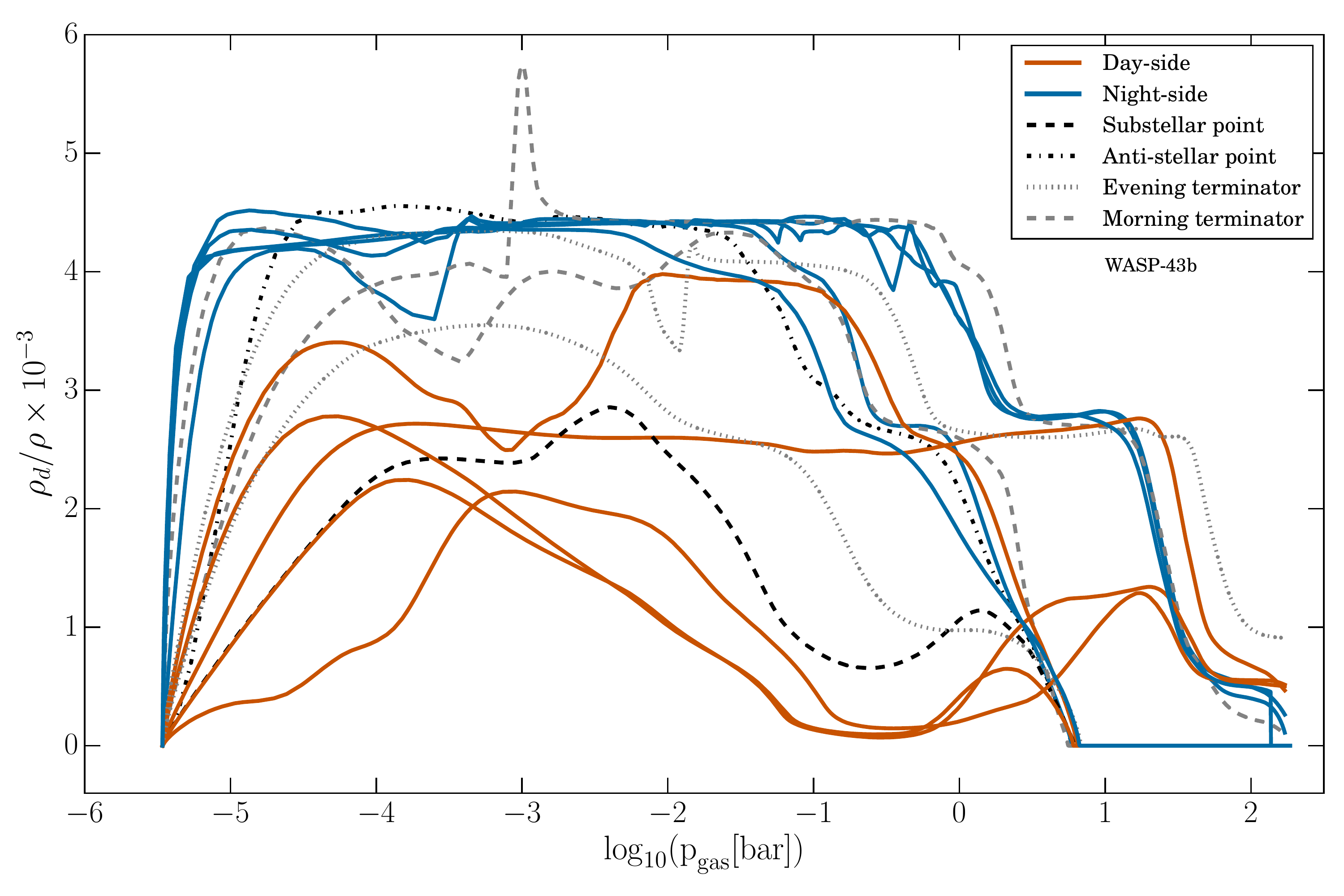}
   \includegraphics[width=21pc]{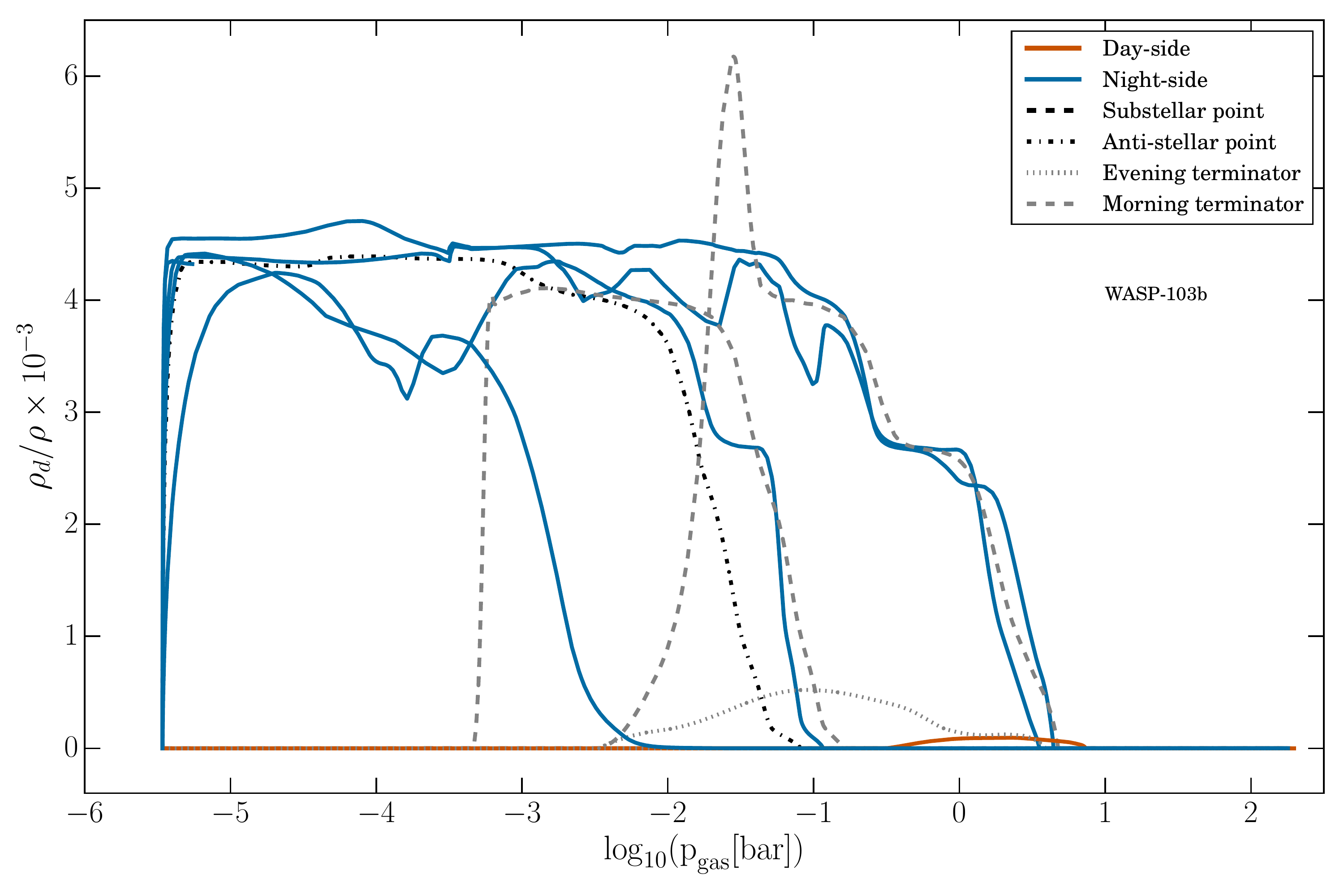}\\
   \includegraphics[width=21pc]{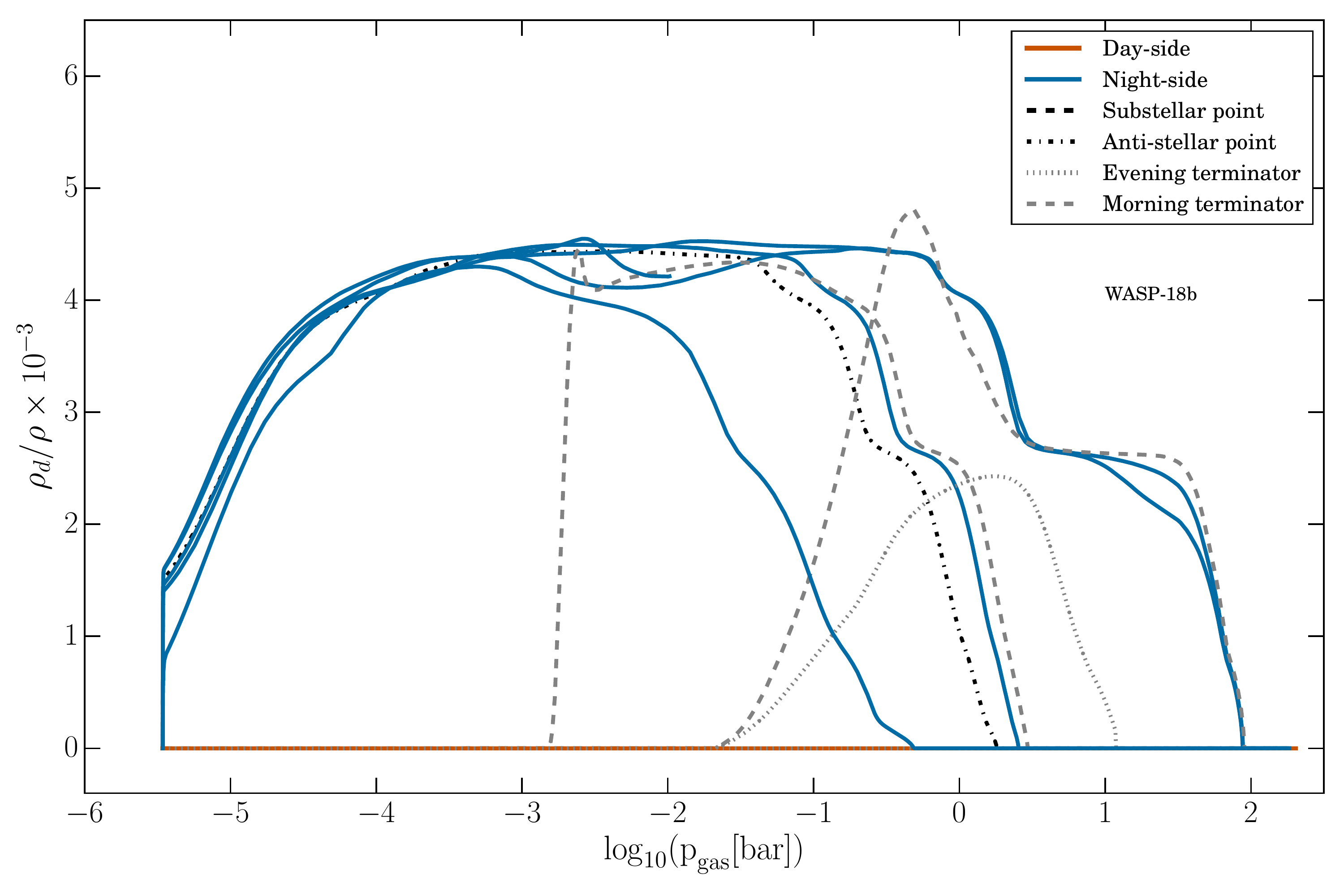}
   \includegraphics[width=21pc]{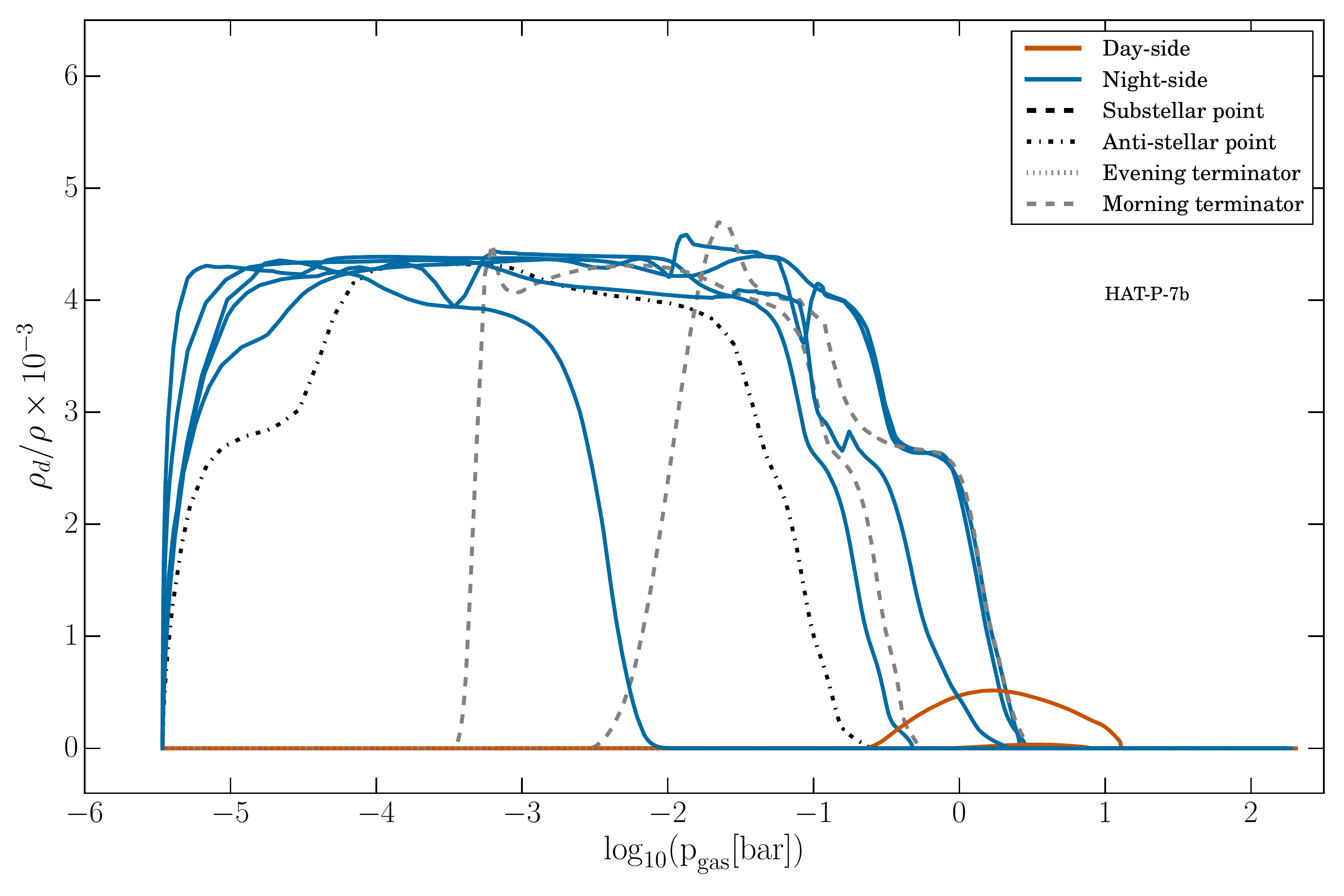}\\
   \includegraphics[width=21pc]{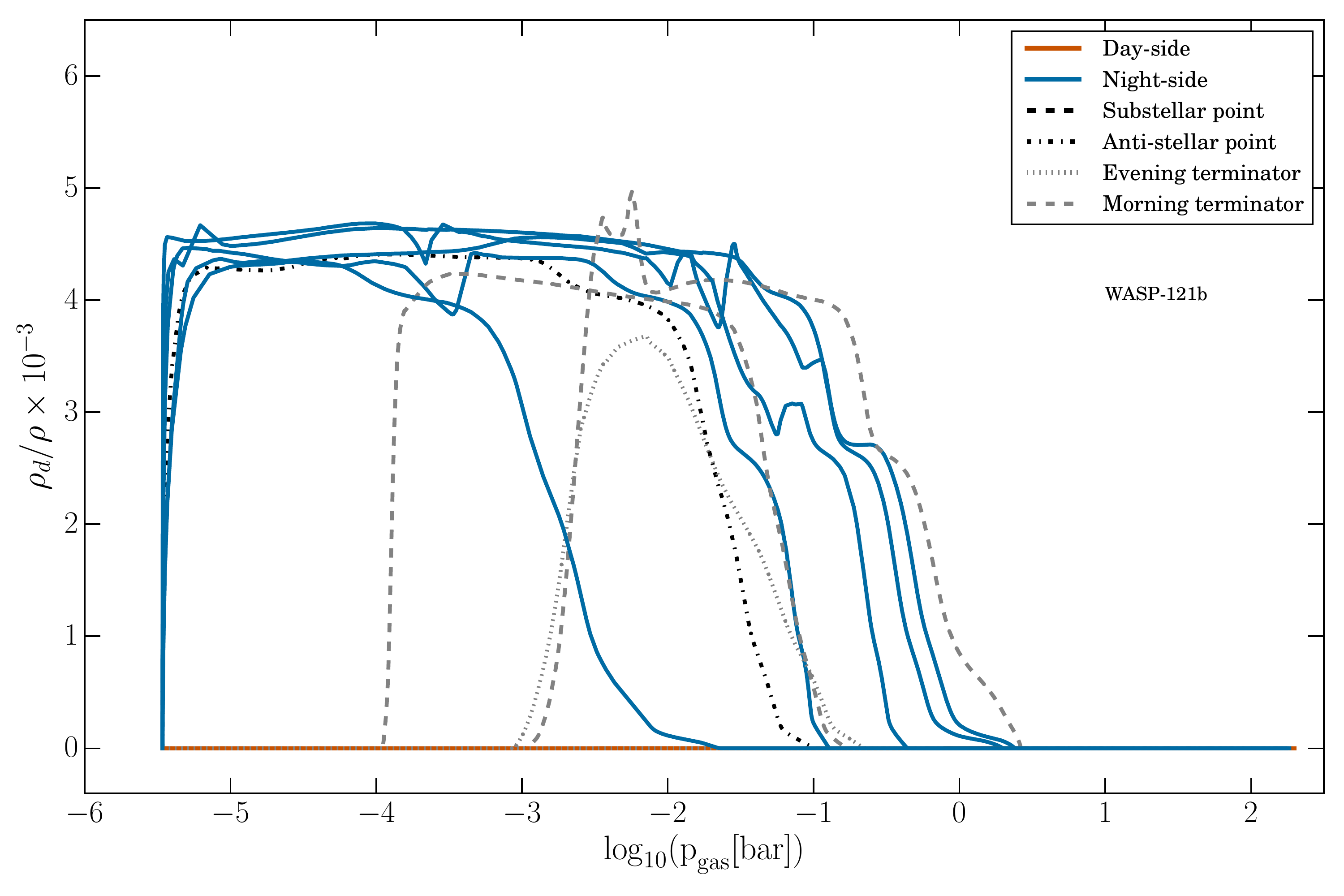}
   \includegraphics[width=21pc]{images/plots_wo_HD/avg_particle_size_rate_no_hd.pdf}
        \caption{Dust-to-gas ratio, $\rho_{d}/\rho$, for the giant gas planet WASP-43b, and the ultra-hot Jupiters WASP-18b, HAT-P-7b, WASP-103b, and WASP-121b.}
       \label{fig:rhod_rho_all}
\end{figure*}

\begin{figure*}
    \centering
    \includegraphics[width=21pc]{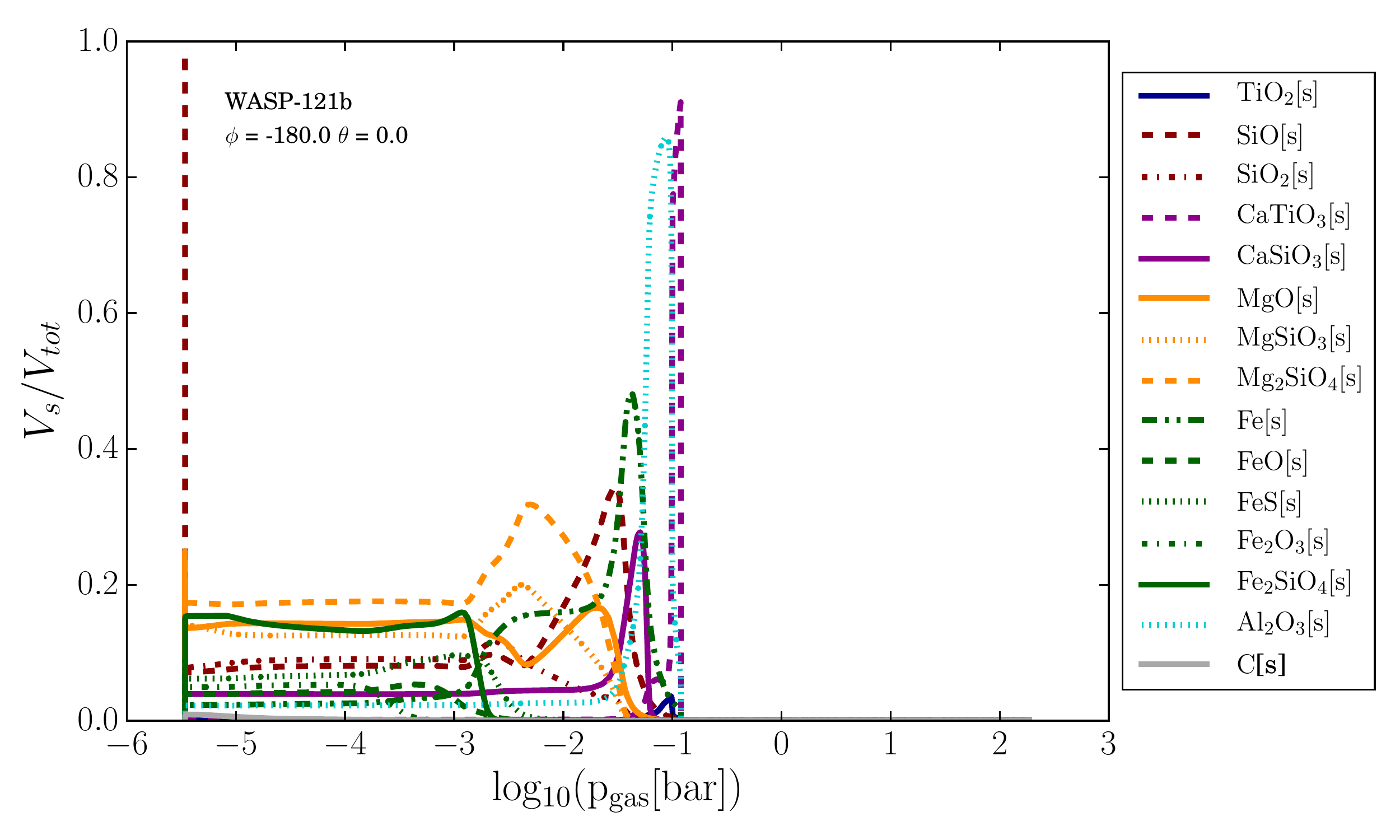}
    \includegraphics[width=21pc]{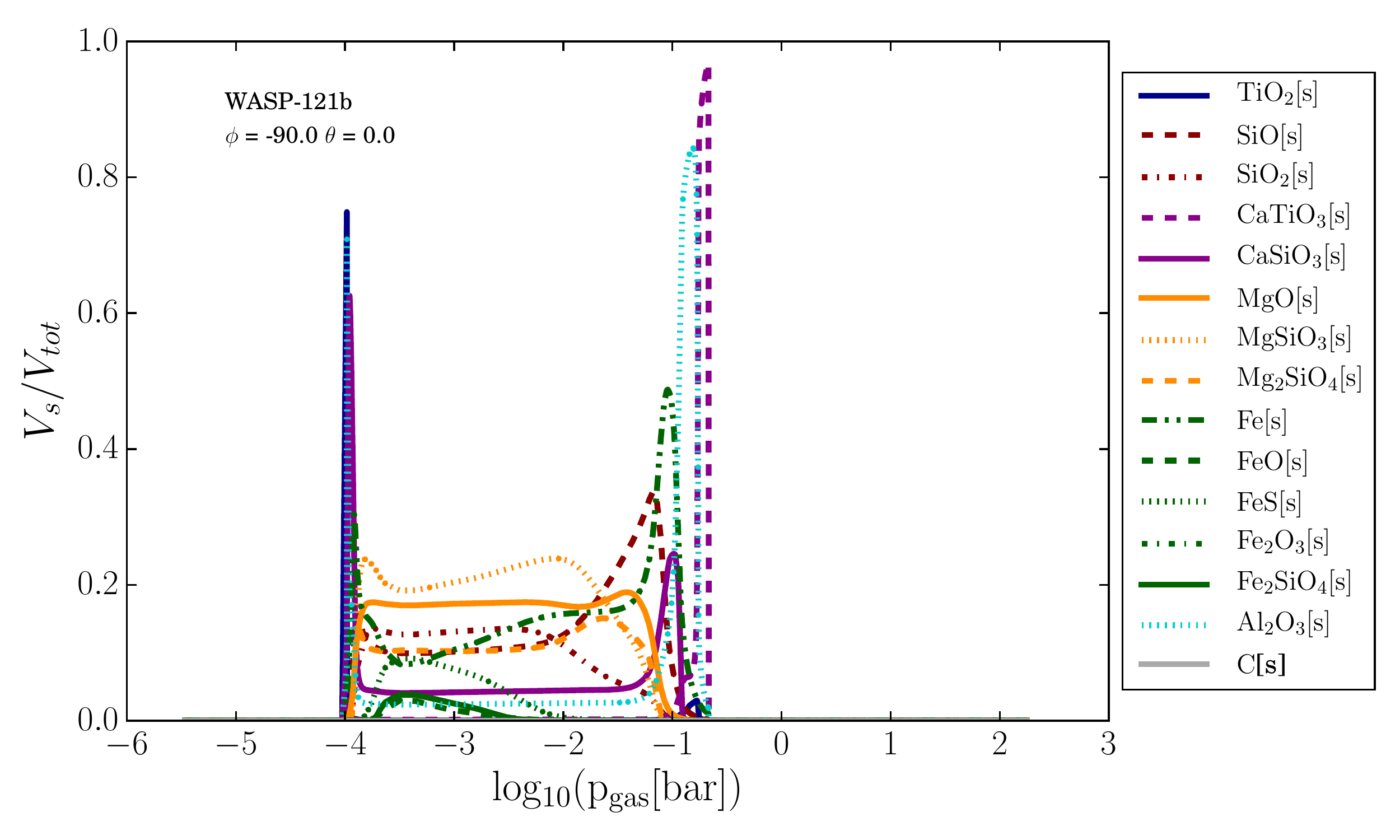}
    \caption{Individual bulk material volume fractions WASP-121b. Anti-stellar point: $(\phi, \theta)=(-180.0^o, 0.0^o)$ and Equatorial Morning Terminator: $(\phi, \theta)=(-90.0^o, 0.0^o)$.}
    \label{fig:mat_vol_WASP121b}
\end{figure*}

\begin{figure*}
    \centering
    \includegraphics[width=21pc]{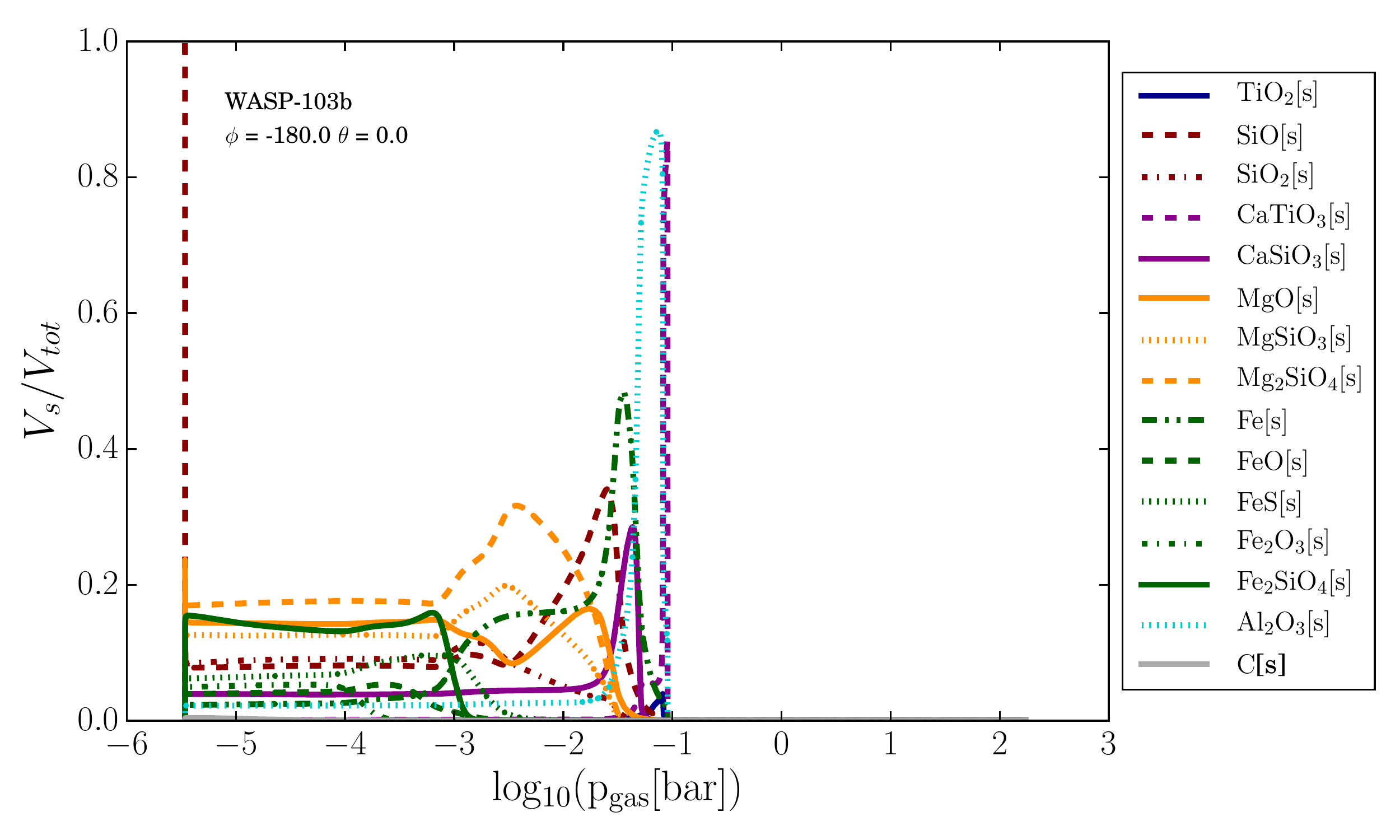}
    \includegraphics[width=21pc]{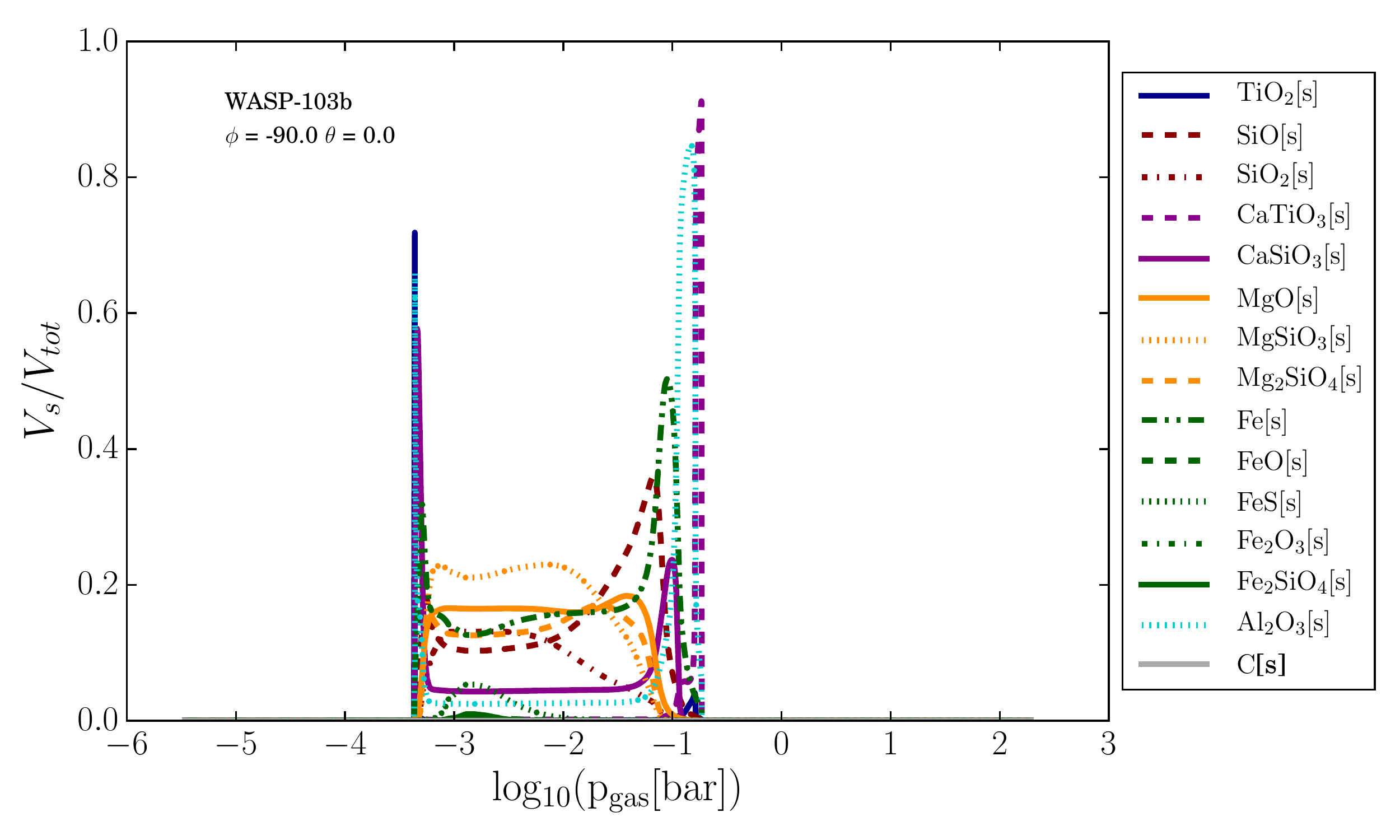}
    \caption{Individual bulk material volume fractions WASP-103b. Anti-stellar point: $(\phi, \theta)=(-180.0^o, 0.0^o)$ and Equatorial Morning Terminator: $(\phi, \theta)=(-90.0^o, 0.0^o)$.}
    \label{fig:mat_vol_WASP103b}
\end{figure*}

\begin{figure*}
    \centering
    \includegraphics[width=21pc]{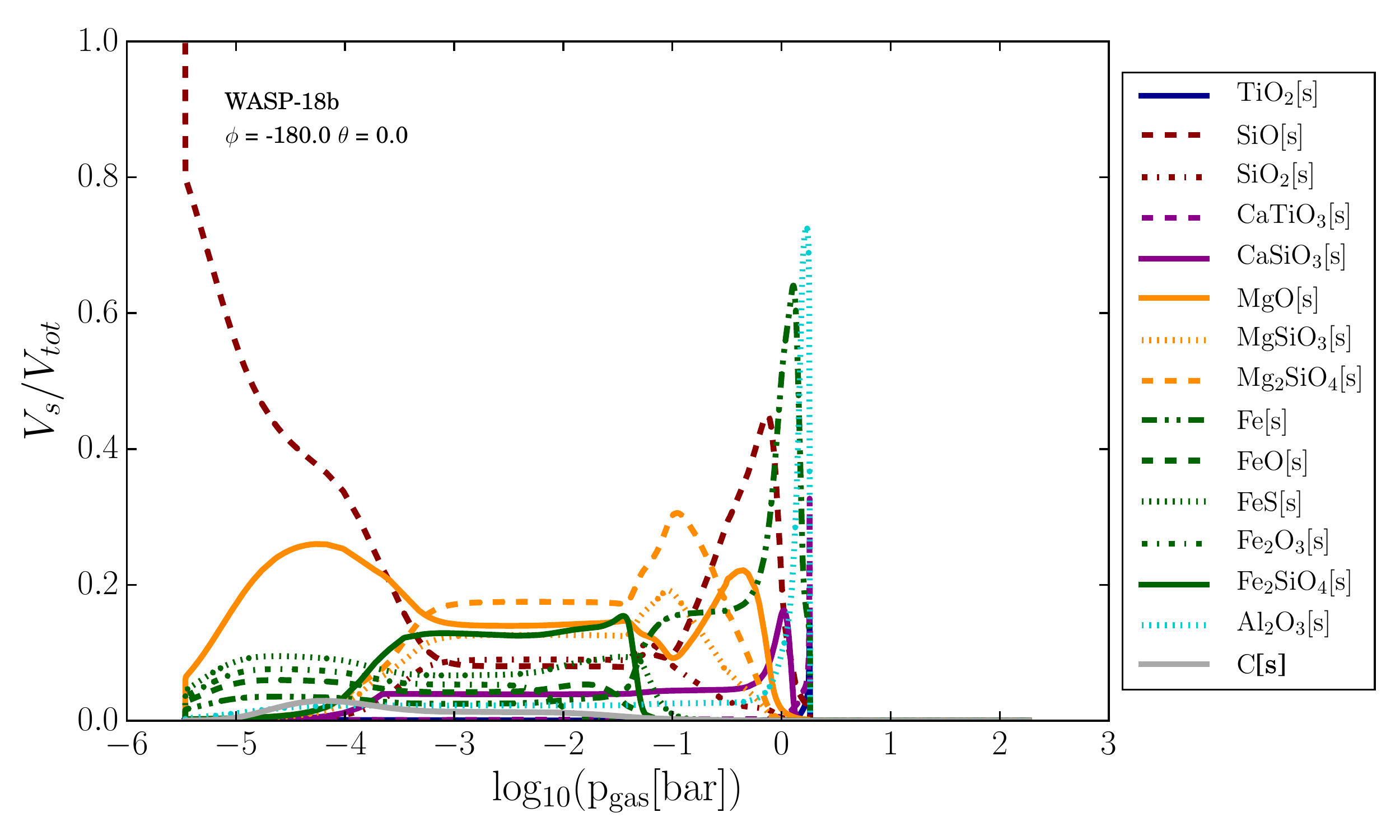}
    \includegraphics[width=21pc]{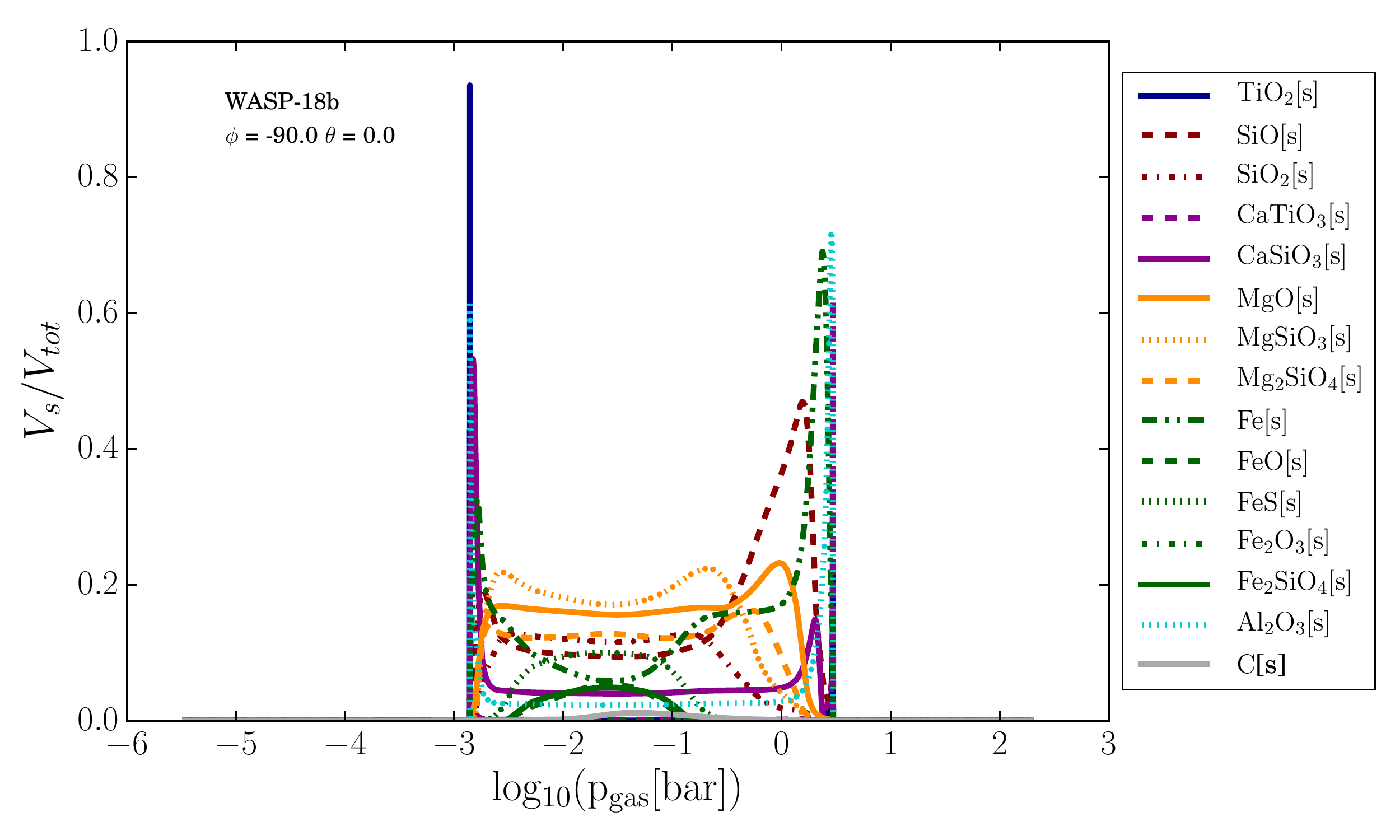}
    \caption{Individual bulk material volume fractions WASP-18b. Anti-stellar point: $(\phi, \theta)=(-180.0^o, 0.0^o)$ and Equatorial Morning Terminator: $(\phi, \theta)=(-90.0^o, 0.0^o)$.}
    \label{fig:mat_vol_WASP18b}
\end{figure*}

\begin{figure*}
    \centering
    \includegraphics[width=21pc]{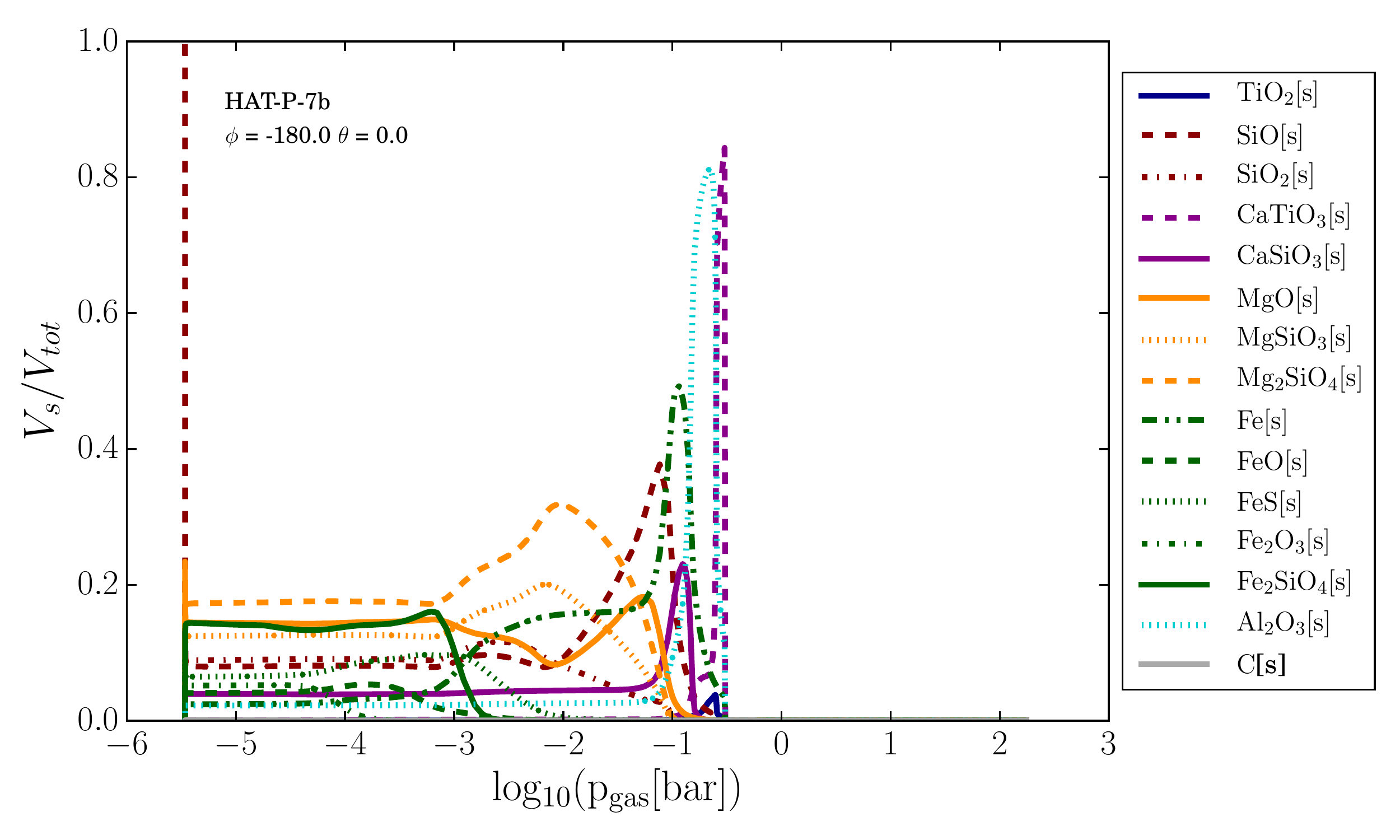}
    \includegraphics[width=21pc]{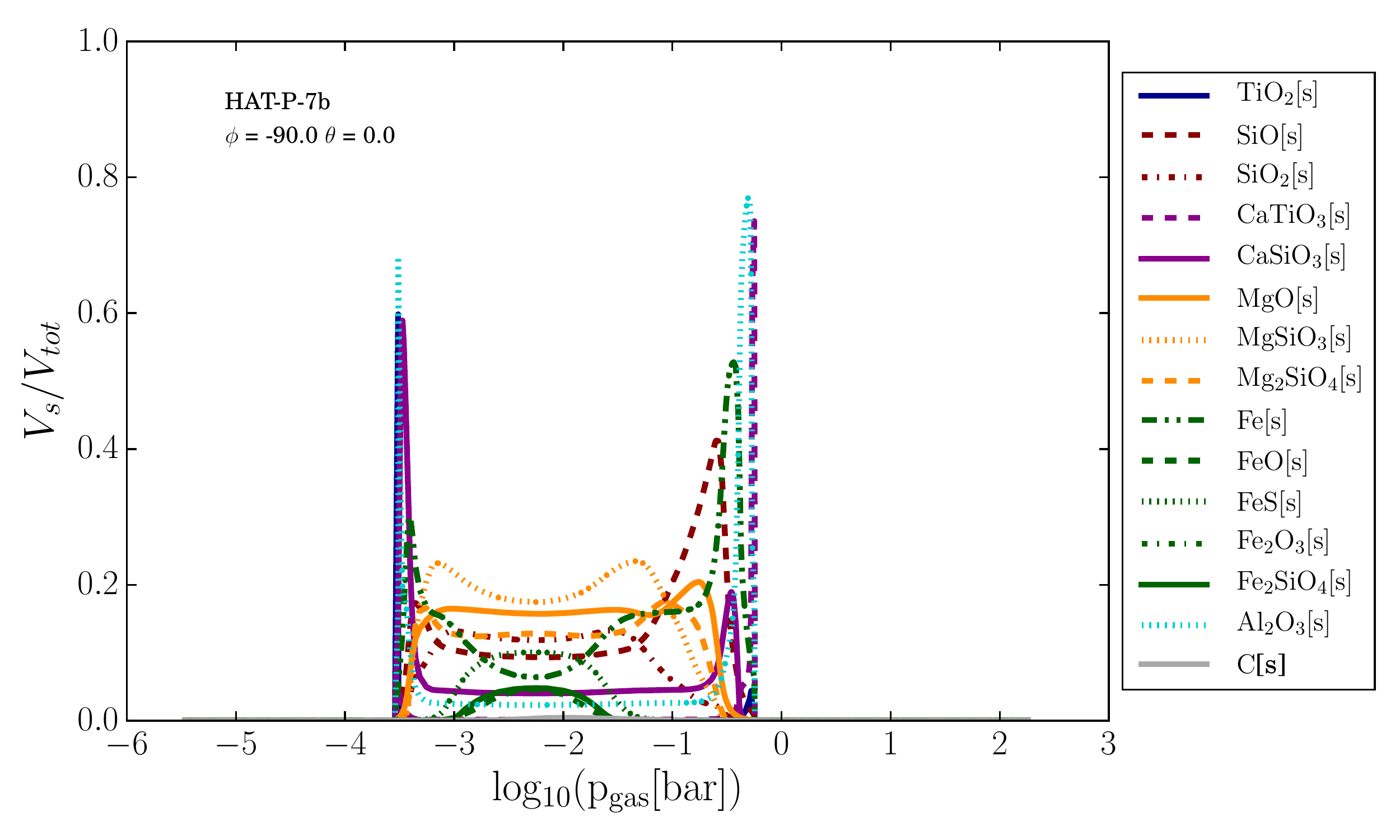}
    \caption{Individual bulk material volume fractions HAT-P-7b. Anti-stellar point: $(\phi, \theta)=(-180.0^o, 0.0^o)$ and Equatorial Morning Terminator: $(\phi, \theta)=(-90.0^o, 0.0^o)$.)}
    \label{fig:mat_vol_HATP7b}
\end{figure*}

\begin{figure*}
    \centering
    \includegraphics[width=21pc]{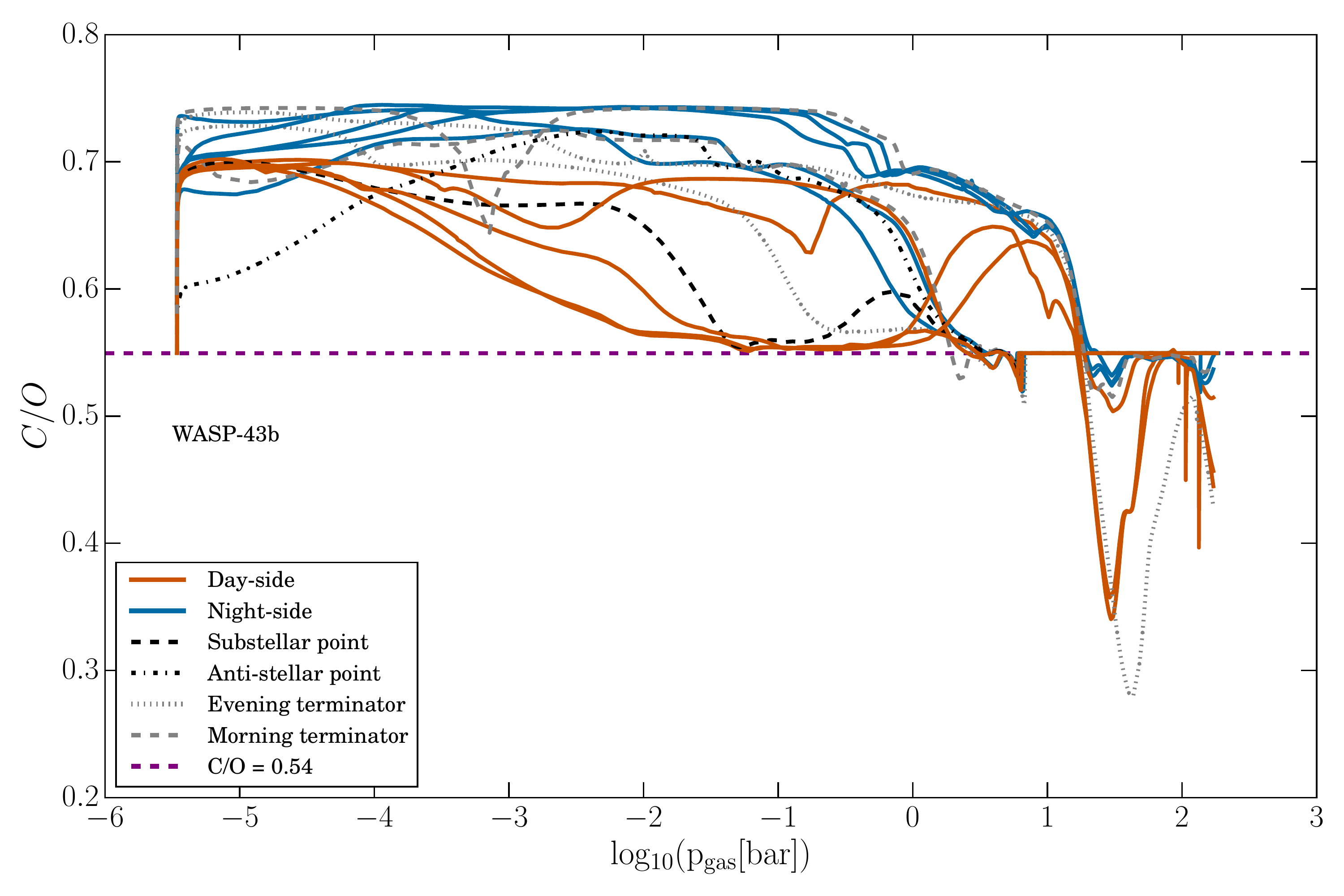}
    \includegraphics[width=21pc]{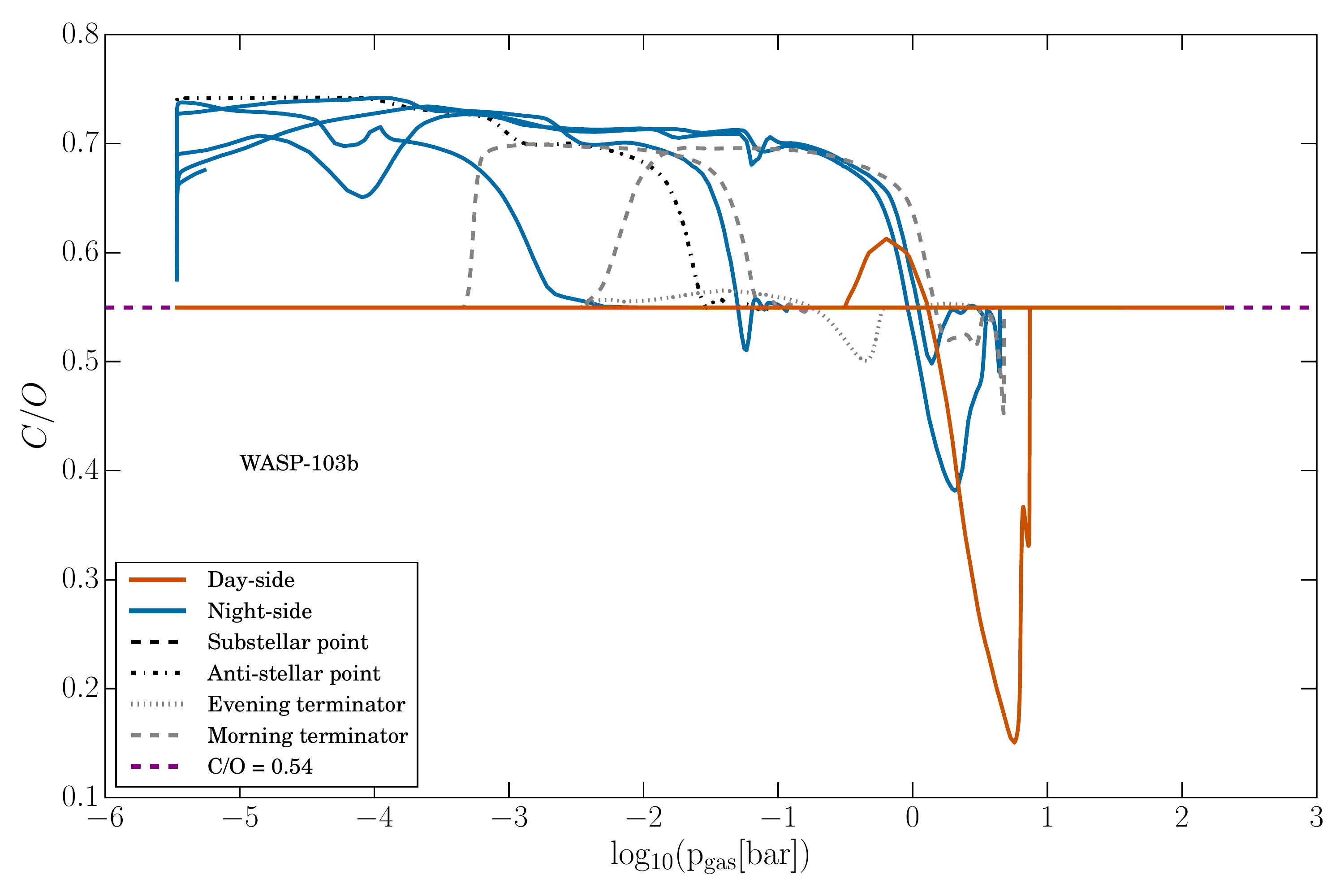}\\
    \includegraphics[width=21pc]{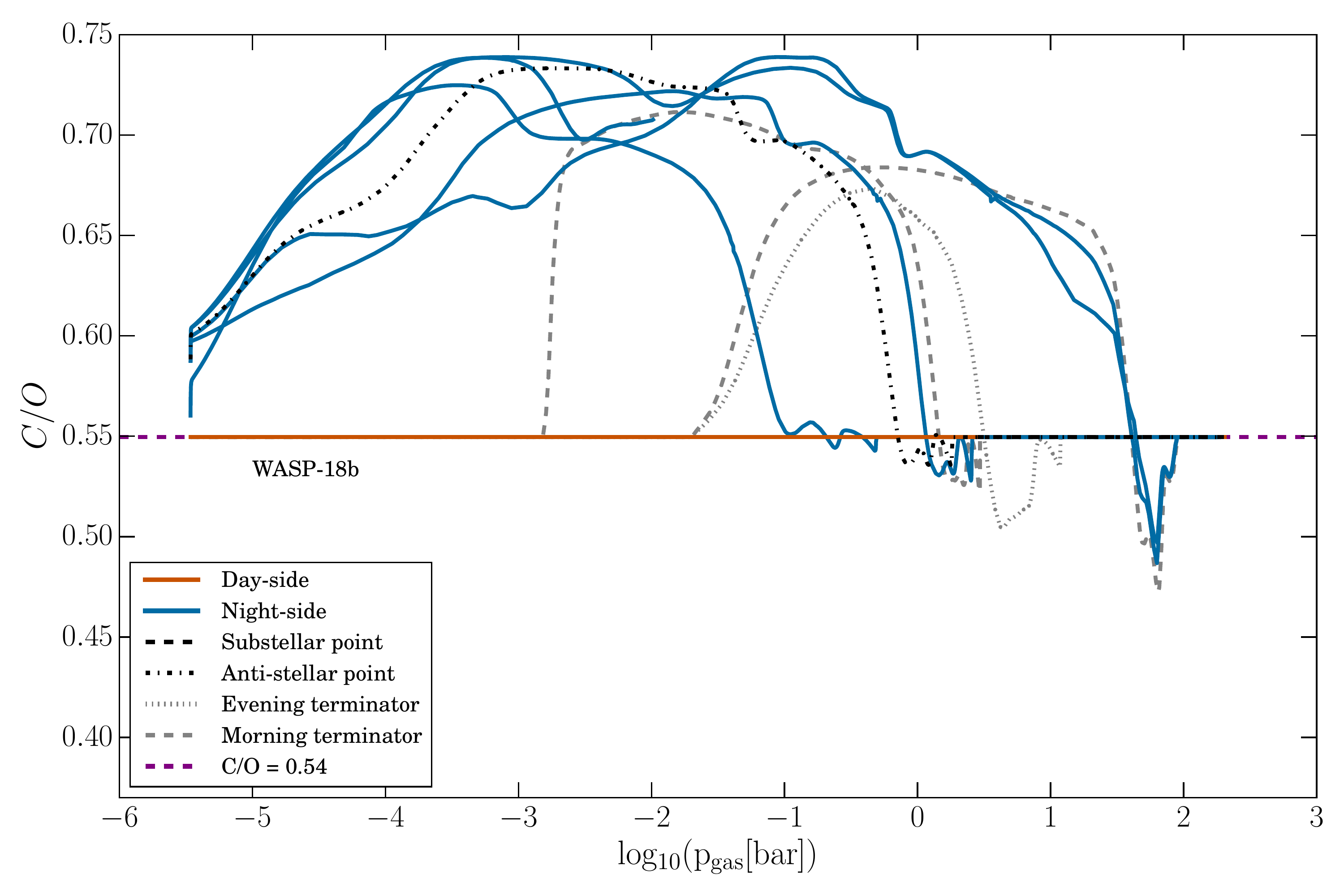}
    \includegraphics[width=21pc]{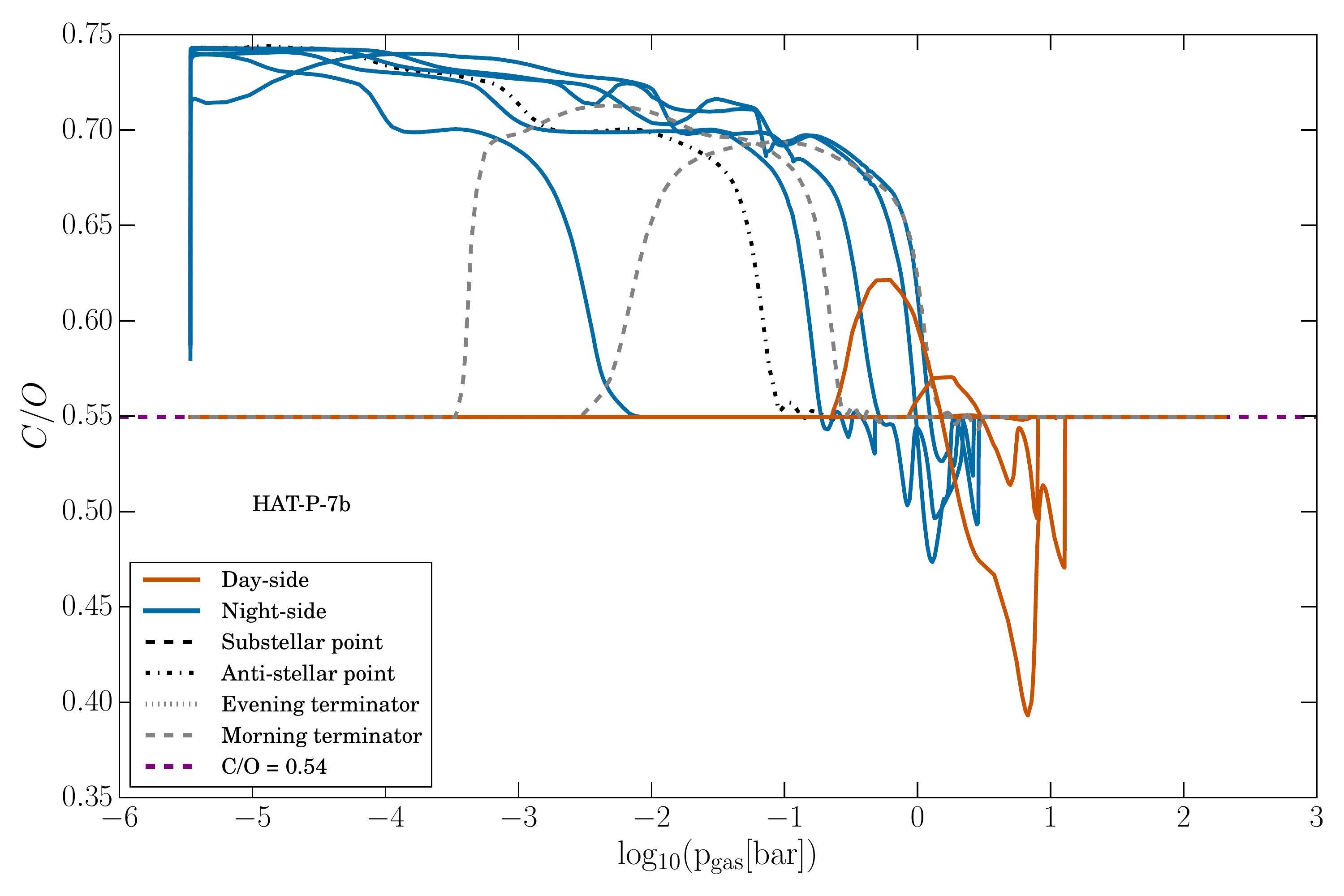}\\
    \includegraphics[width=21pc]{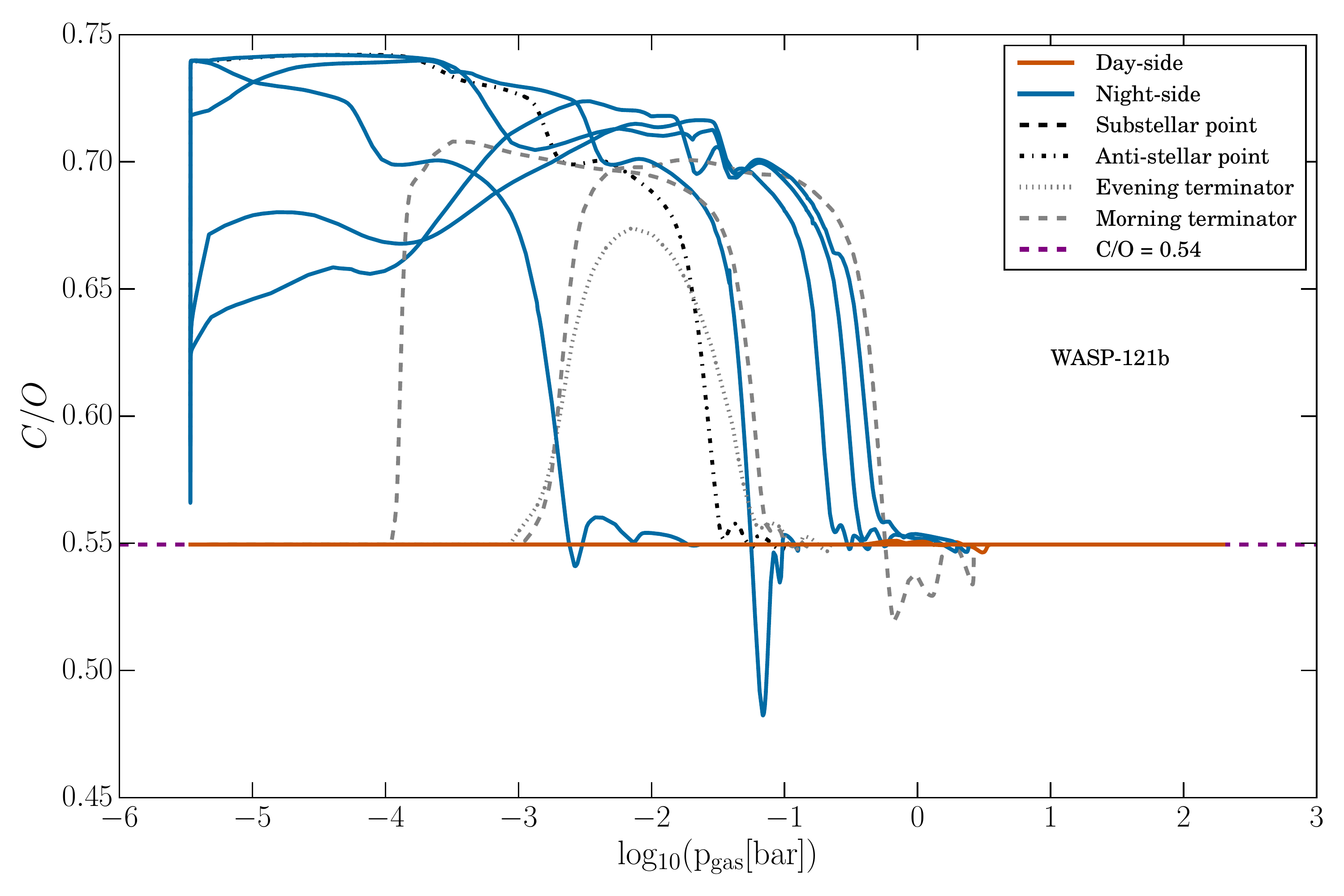}
    \includegraphics[width=21pc]{images/avg_c_o_all.pdf}
    \caption{The carbon-to-oxygen ratio (C/O) for the giant gas planet WASP-43b, and the ultra-hot Jupiters WASP-18b, HAT-P-7b, WASP-103b, and WASP-121b. The solar value C/O = 0.54 is plotted in dashed purple.}
    \label{fig:CO_all}
\end{figure*}

\begin{figure*}
    \centering
    \includegraphics[width=21pc]{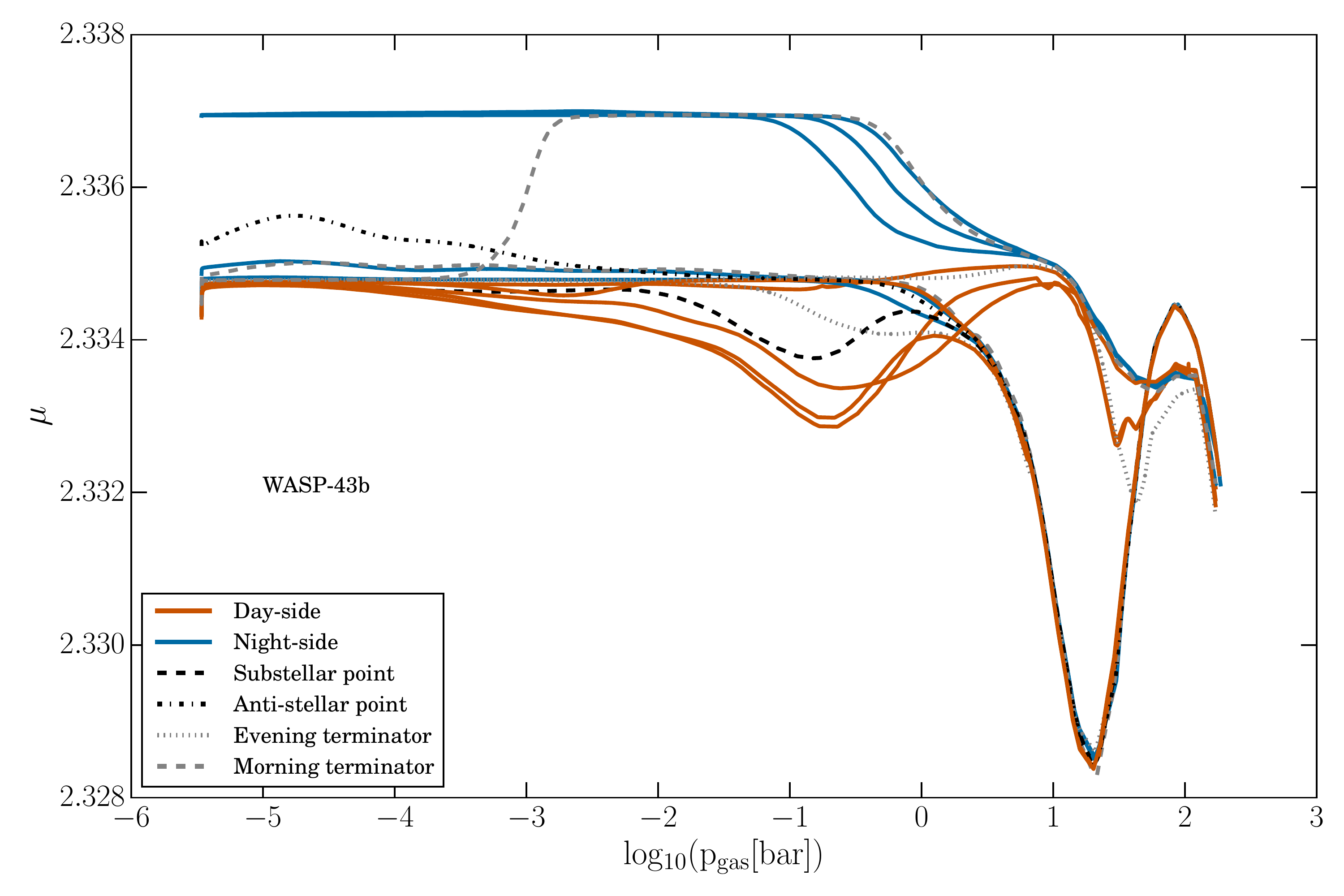}
    \includegraphics[width=21pc]{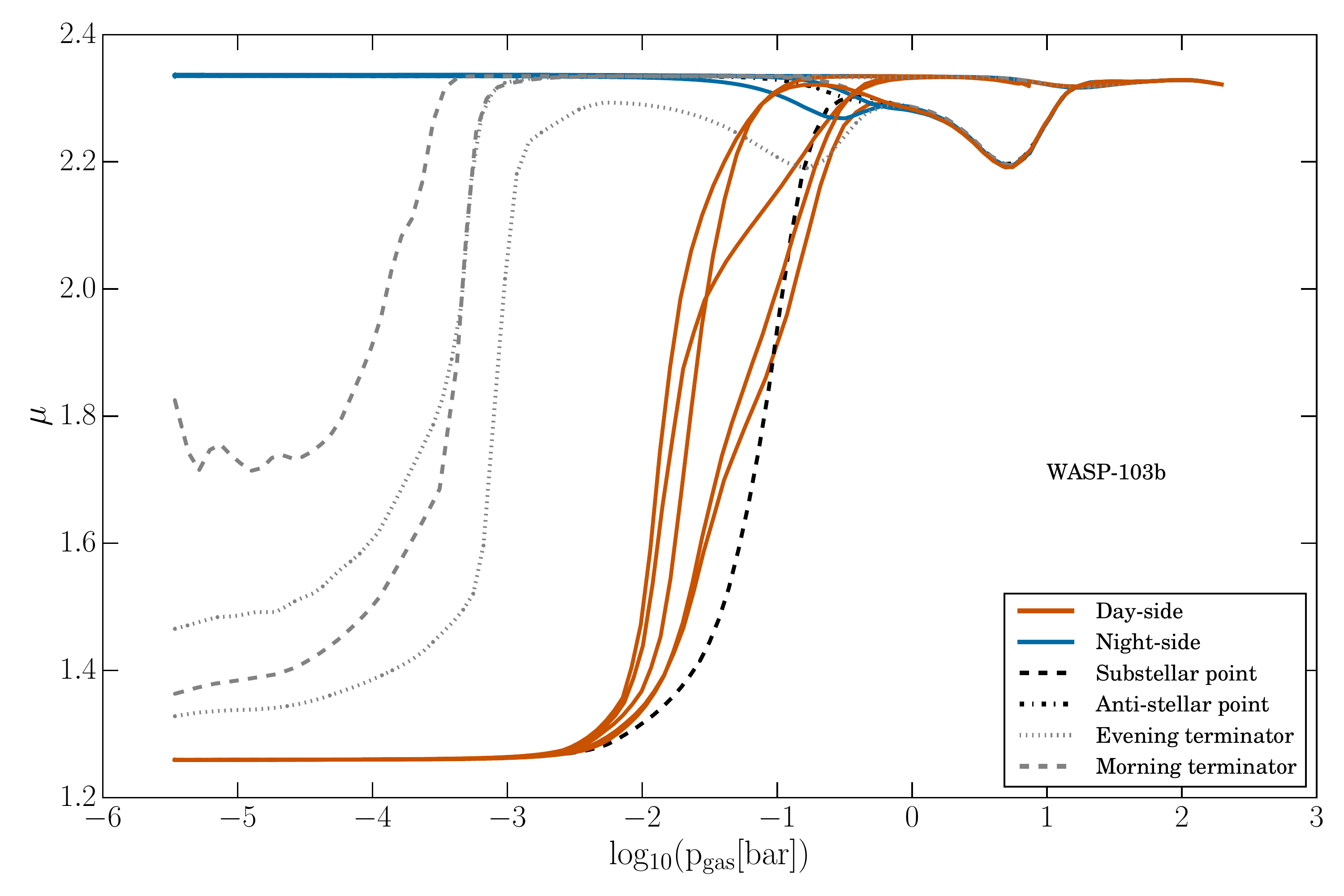}\\
    \includegraphics[width=21pc]{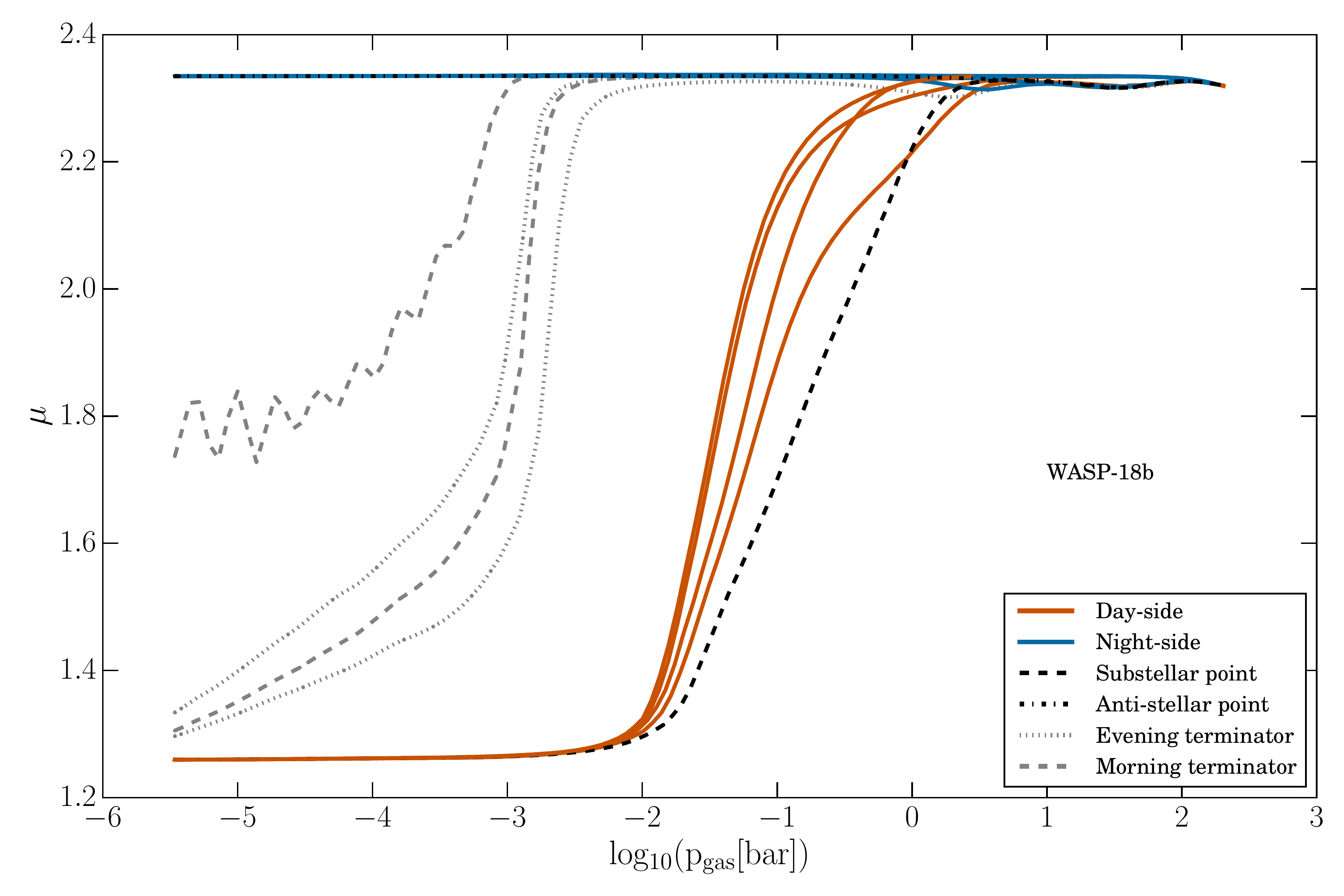}
    \includegraphics[width=21pc]{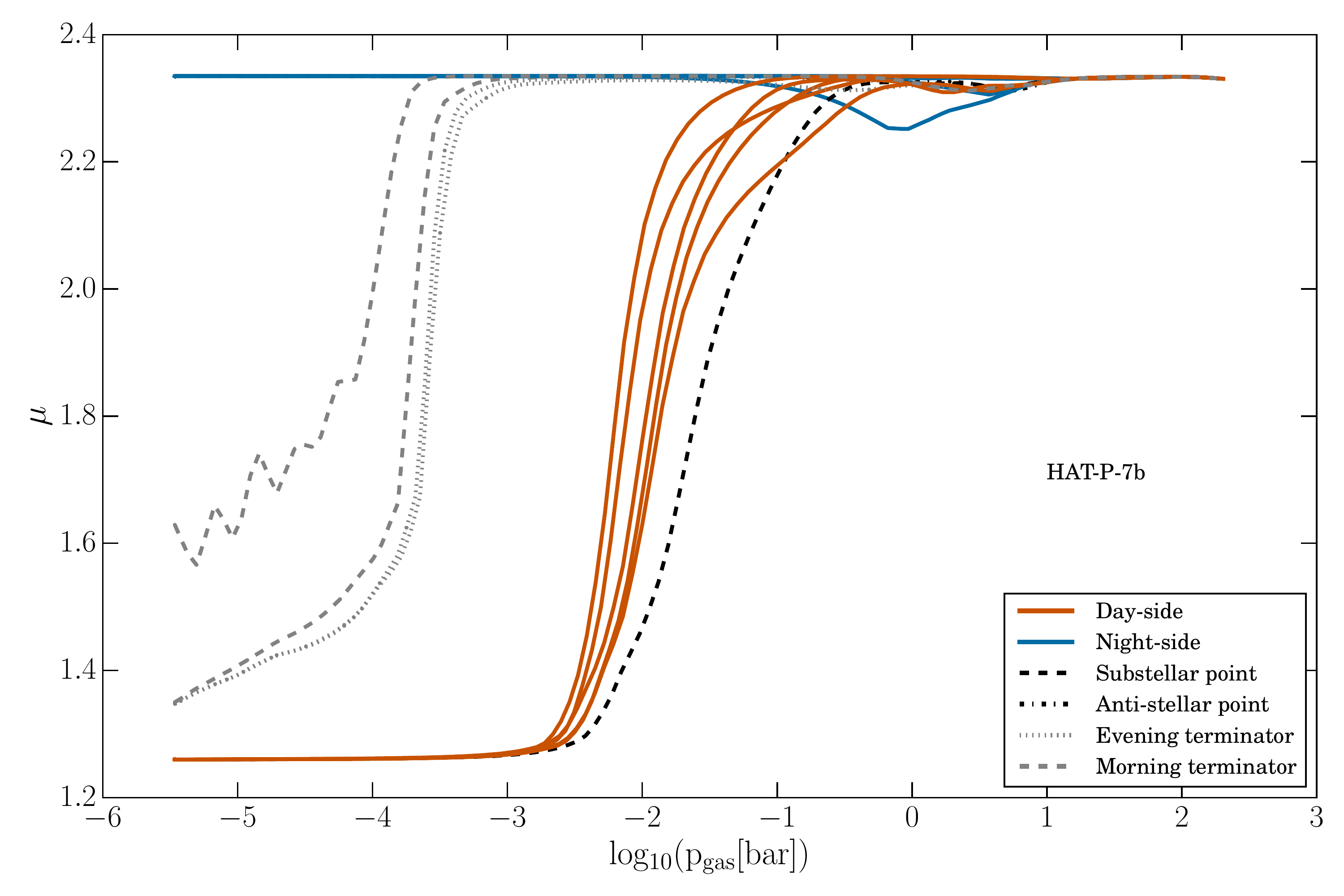}\\
    \includegraphics[width=21pc]{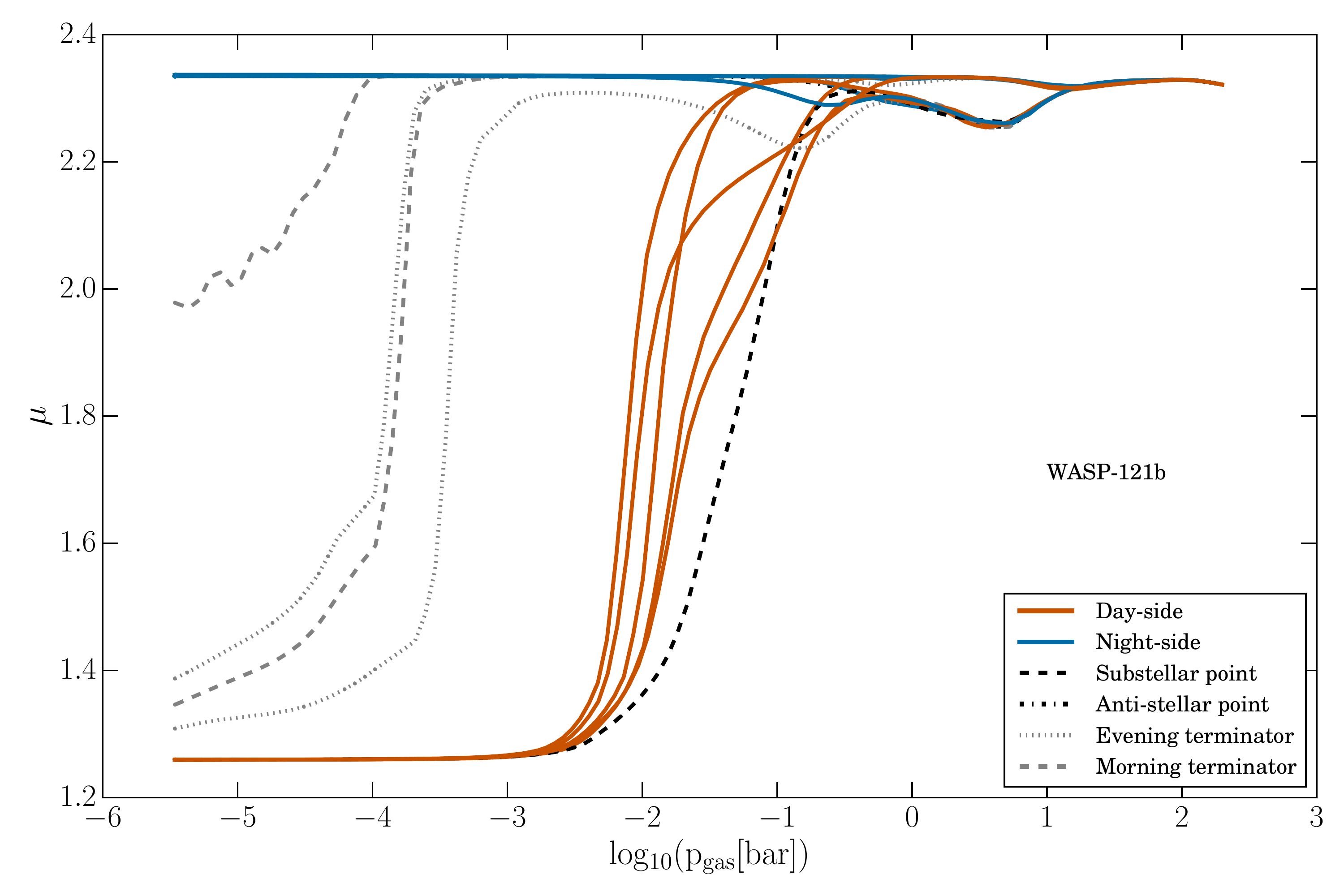}
    \includegraphics[width=21pc]{images/avg_mu_all.pdf}
    \caption{The atmospheric mean molecular weight, $\mu$, for the giant gas planet WASP-43b, and the ultra-hot Jupiters WASP-18b, HAT-P-7b, WASP-103b, and WASP-121b. The ultra-hot Jupiters show significant differences in $\mu$ between the dayside and nightside of the planets, whereas WASP-43b shows an approximately constant value around $\mu \approx 2.3$.}
    \label{fig:mmw_all}
\end{figure*}

\begin{figure*}
    \centering
    \includegraphics[width=21pc]{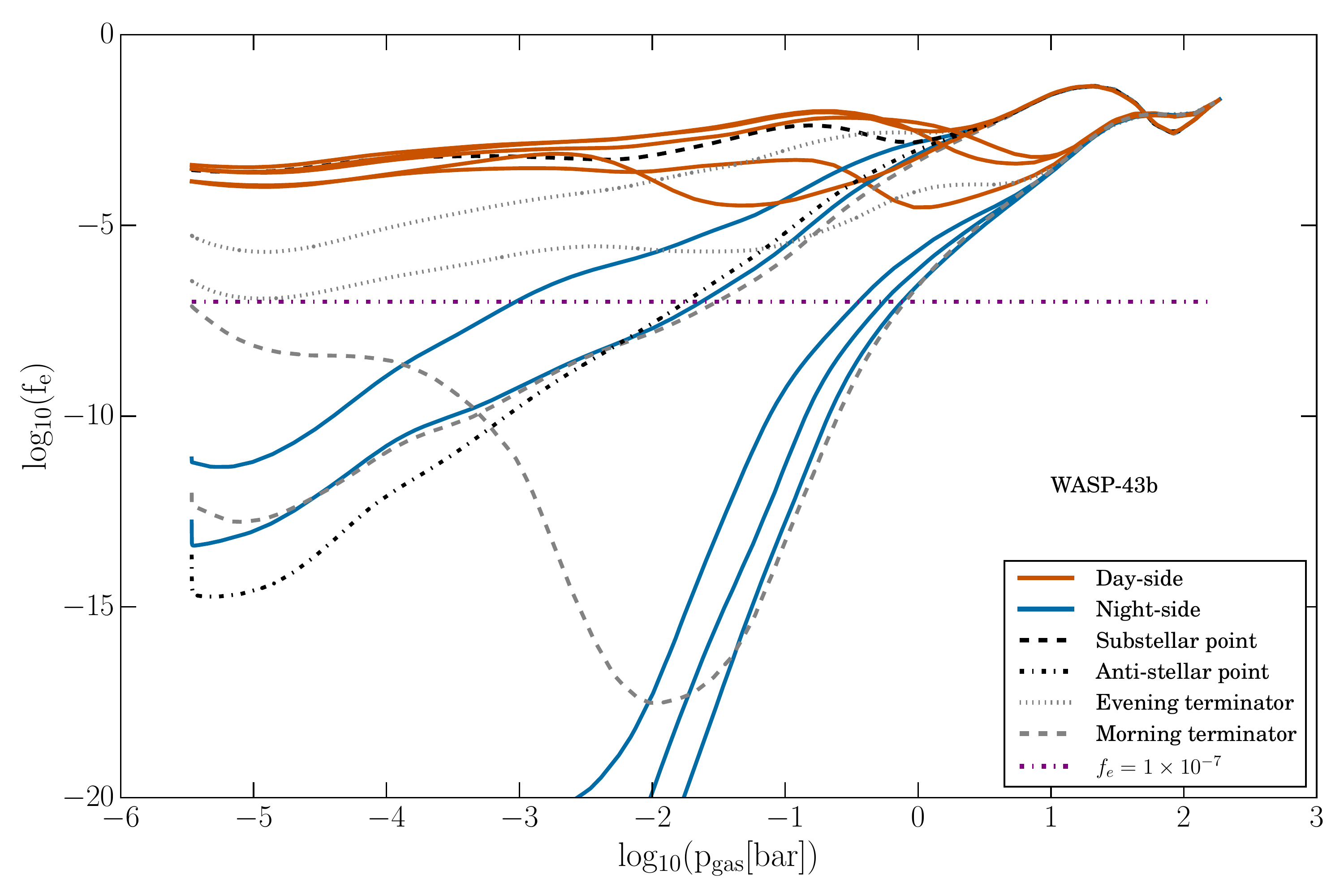}
    \includegraphics[width=21pc]{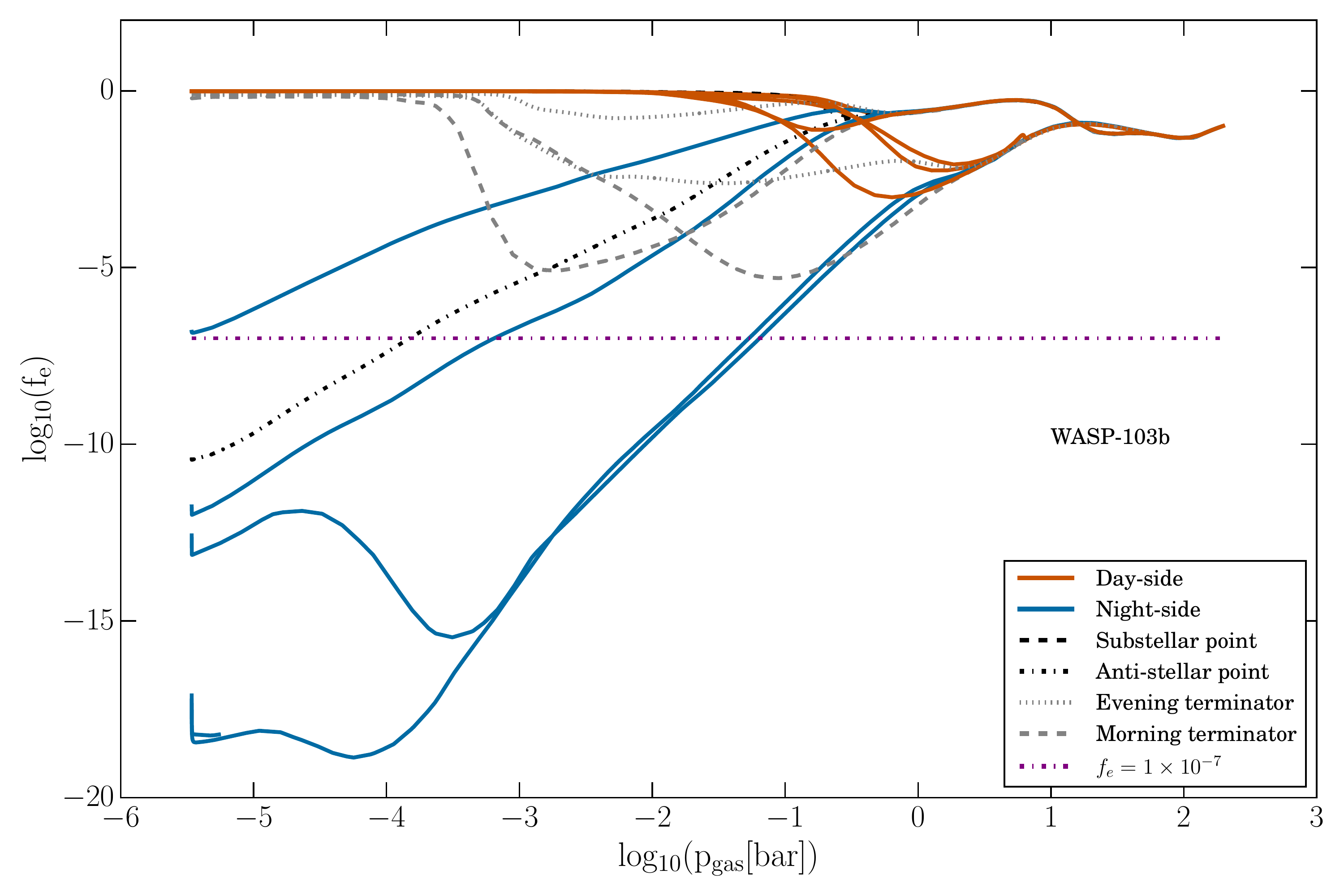}\\
    \includegraphics[width=21pc]{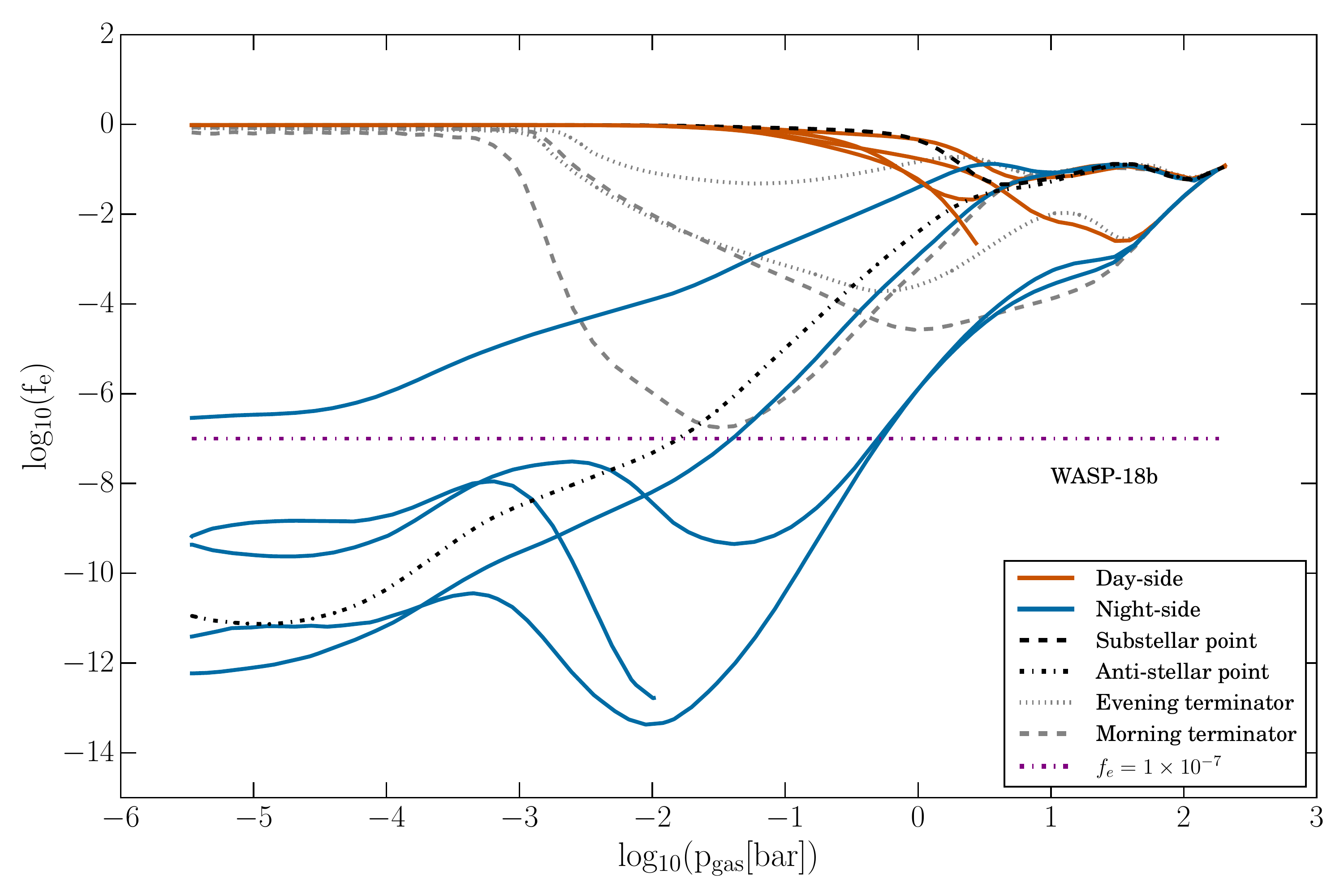}
    \includegraphics[width=21pc]{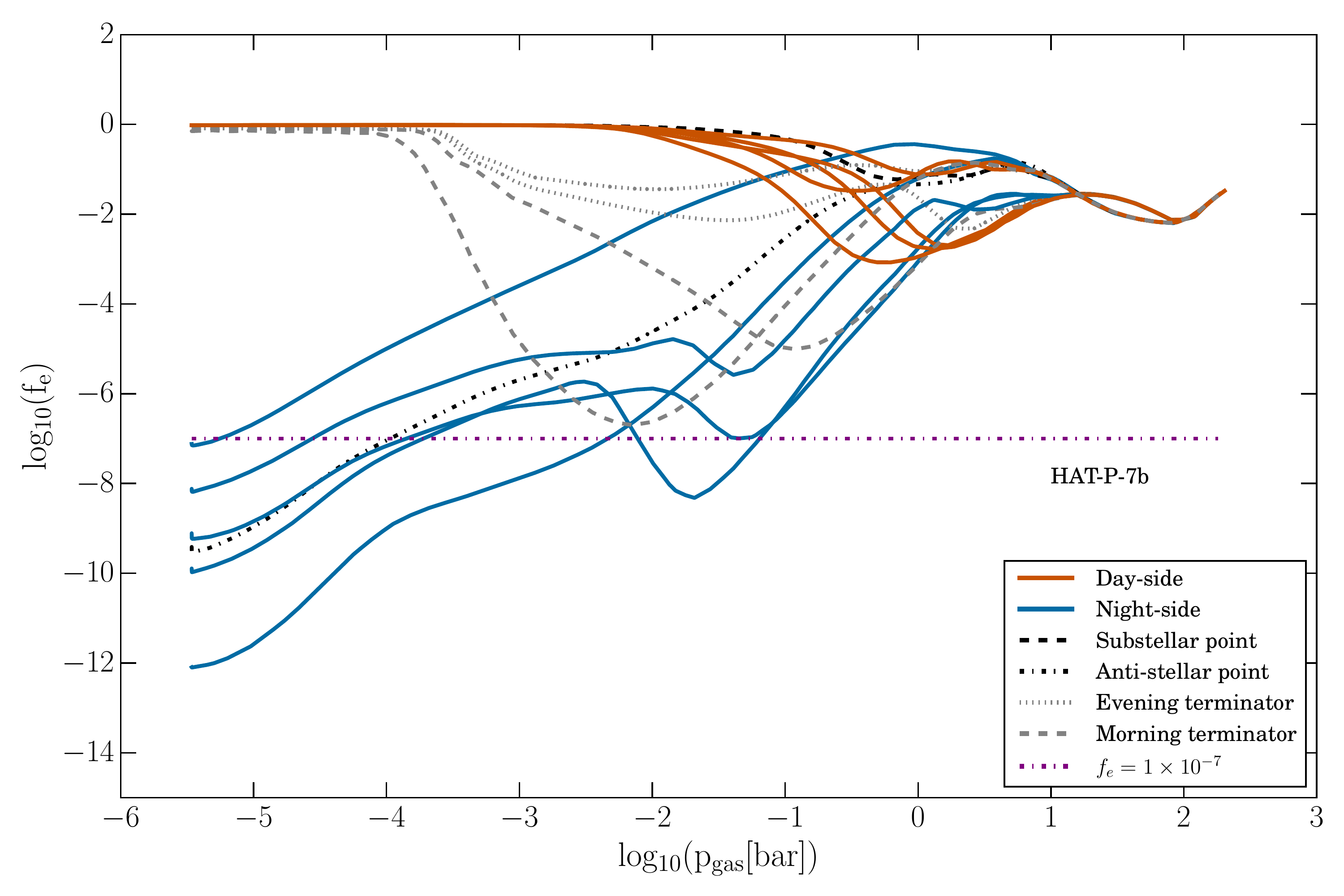}\\
    \includegraphics[width=21pc]{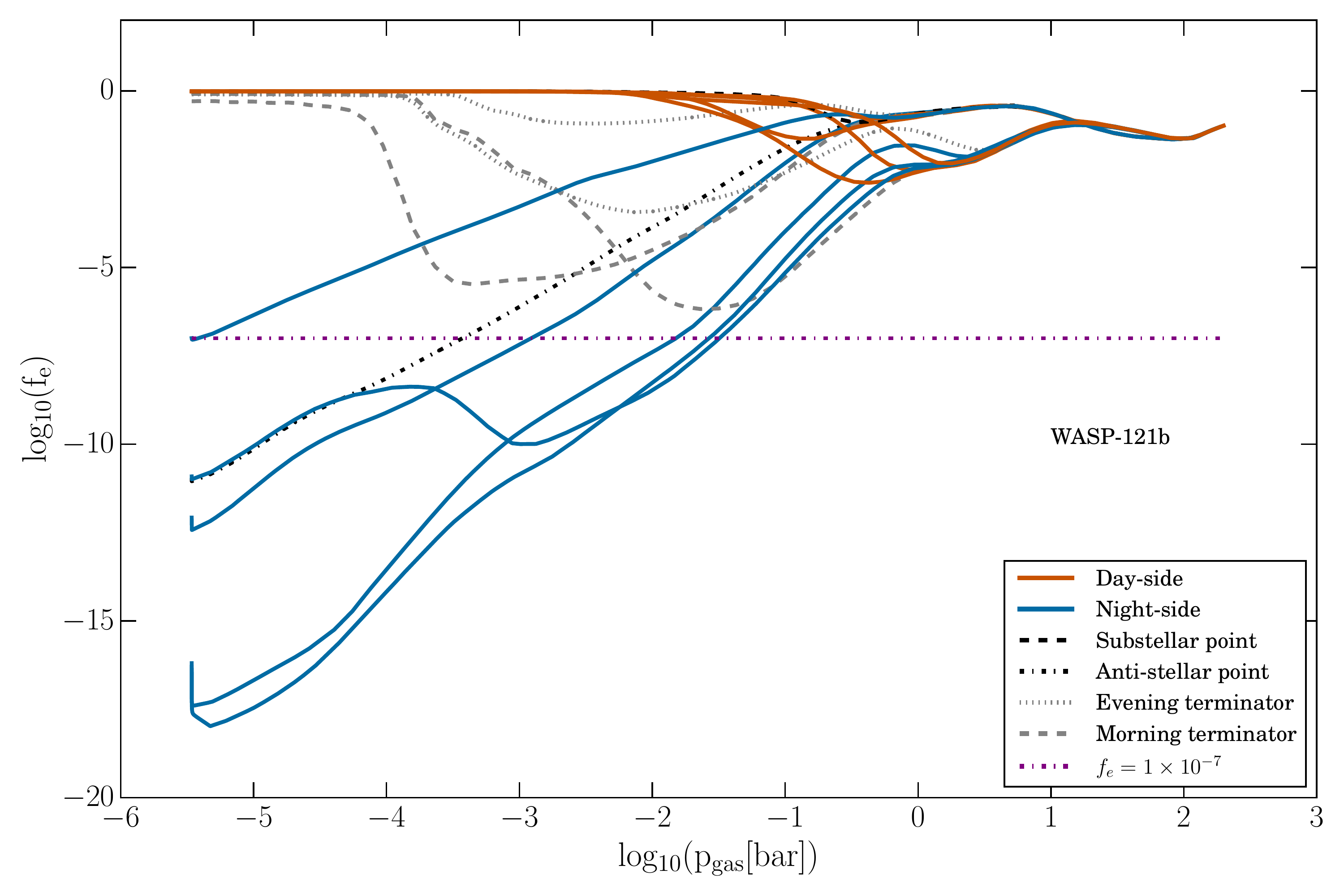}
    \includegraphics[width=21pc]{images/avg_deg_ion_all_new.pdf}
    \caption{Degree of thermal ionisation, $f_{\rm e} = p_{\rm e}/\left(p_{\rm gas}+p_{e}\right)$. 
    The dash-dot purple line shows $f_{e} = 1\times 10^{-7}$ as a threshold for plasma behaviour. All ultra-hot Jupiters have dayside  thermal ionisation $f_{\rm e}>10^{-4}$ suggesting an extended dayside ionosphere. Most of the cloud-forming nightsides are little affected by thermal ionisation.}
    \label{fig:deg_ion}
\end{figure*}

In order to enable model comparability, we provide here the detailed results of the cloud complex for the ultra-hot Jupiters which form the base for the more condensed representation within the main text body. 
Figure~\ref{fig:rhod_rho_all} provides all the dust-to- gas ratio, $\rho_{d}/\rho$, profiles clearly indicating where in these atmosphere most of the cloud particle mass is located according to our model. Figures~\ref{fig:mat_vol_WASP121b} --~\ref{fig:mat_vol_HATP7b} provide the details of the 16 considered bulk growth materials as part of our kinetic cloud formation model for the equatorial nightside and the morning terminator only. Mineral clouds do not form on the dayside and the evening terminator due to unfavourably high local gas temperatures.  Panels appear empty where no cloud particle formation occurs; this results from  profiles where no nucleation seeds form. Figure~\ref{fig:CO_all} --~\ref{fig:deg_ion} provide the detailed results of the carbon-to-oxygen ratio (C/O), the mean molecular weight, $\mu$, and the thermal degree of ionisation, $f_{\rm e}$ as data input for the averaged values shown  in previous sections. 

\begin{figure*}
     \centering
     \includegraphics[width=0.49\textwidth]{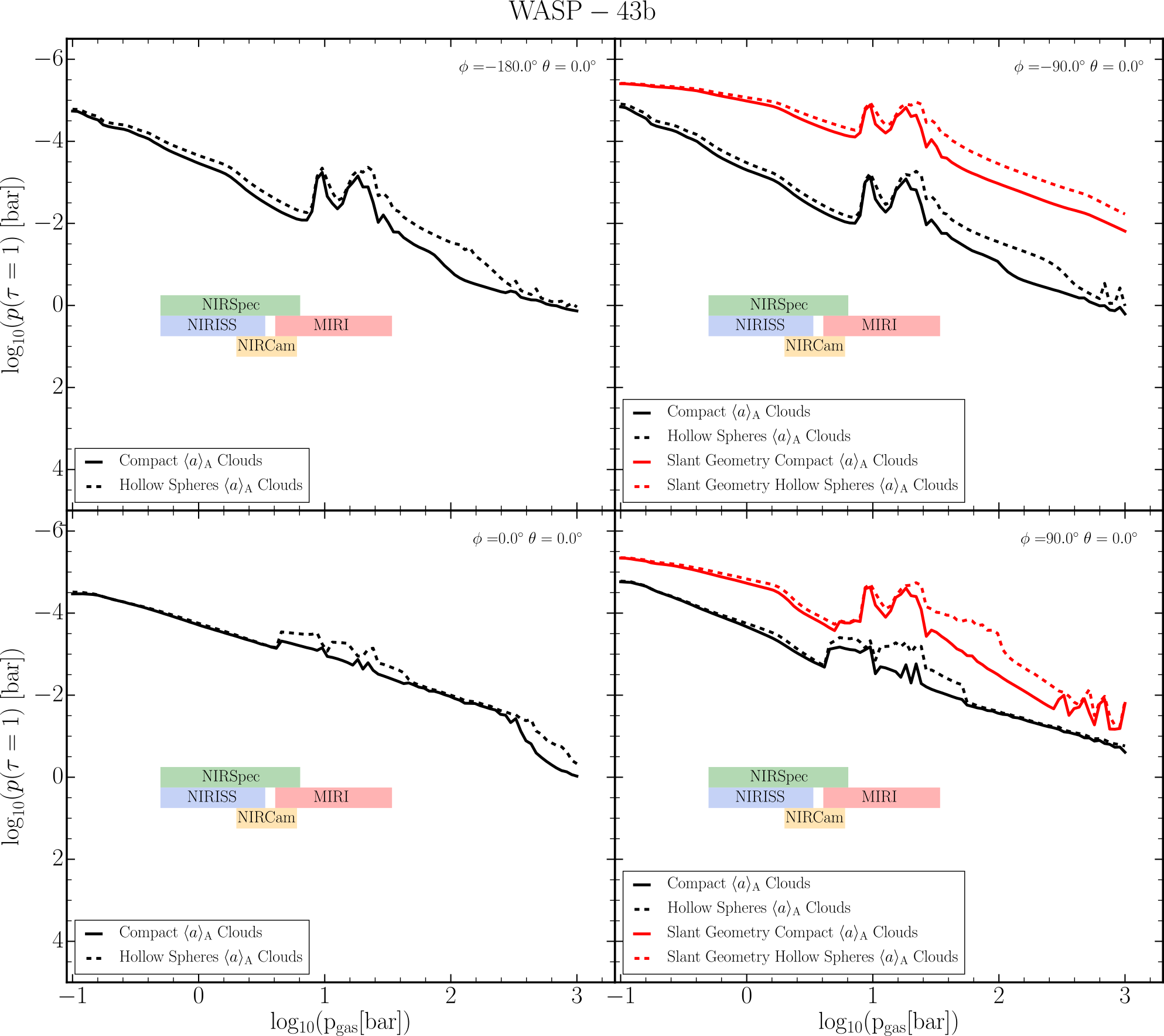}
     \includegraphics[width=0.49\textwidth]{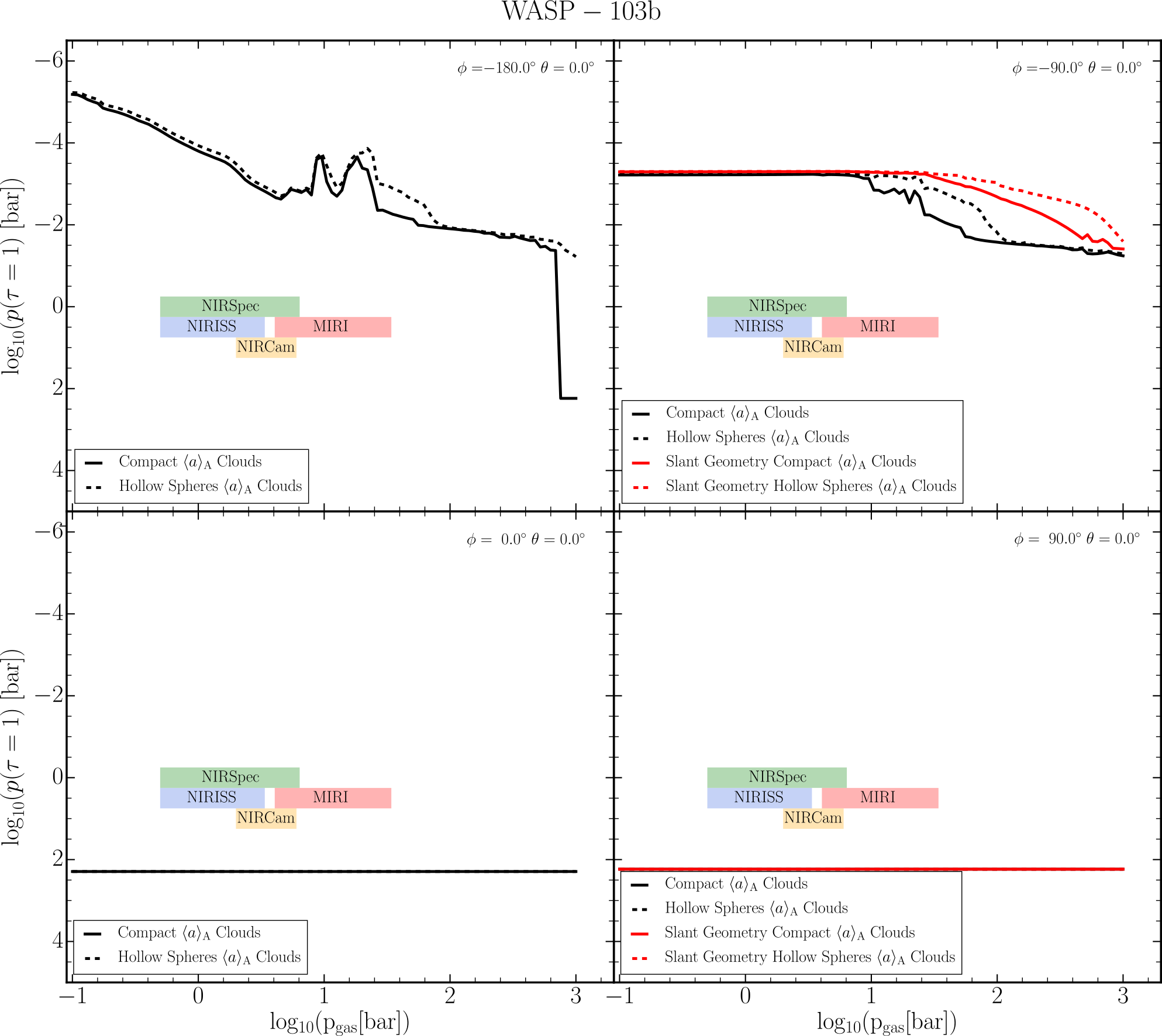}\\
     \includegraphics[width=0.49\textwidth]{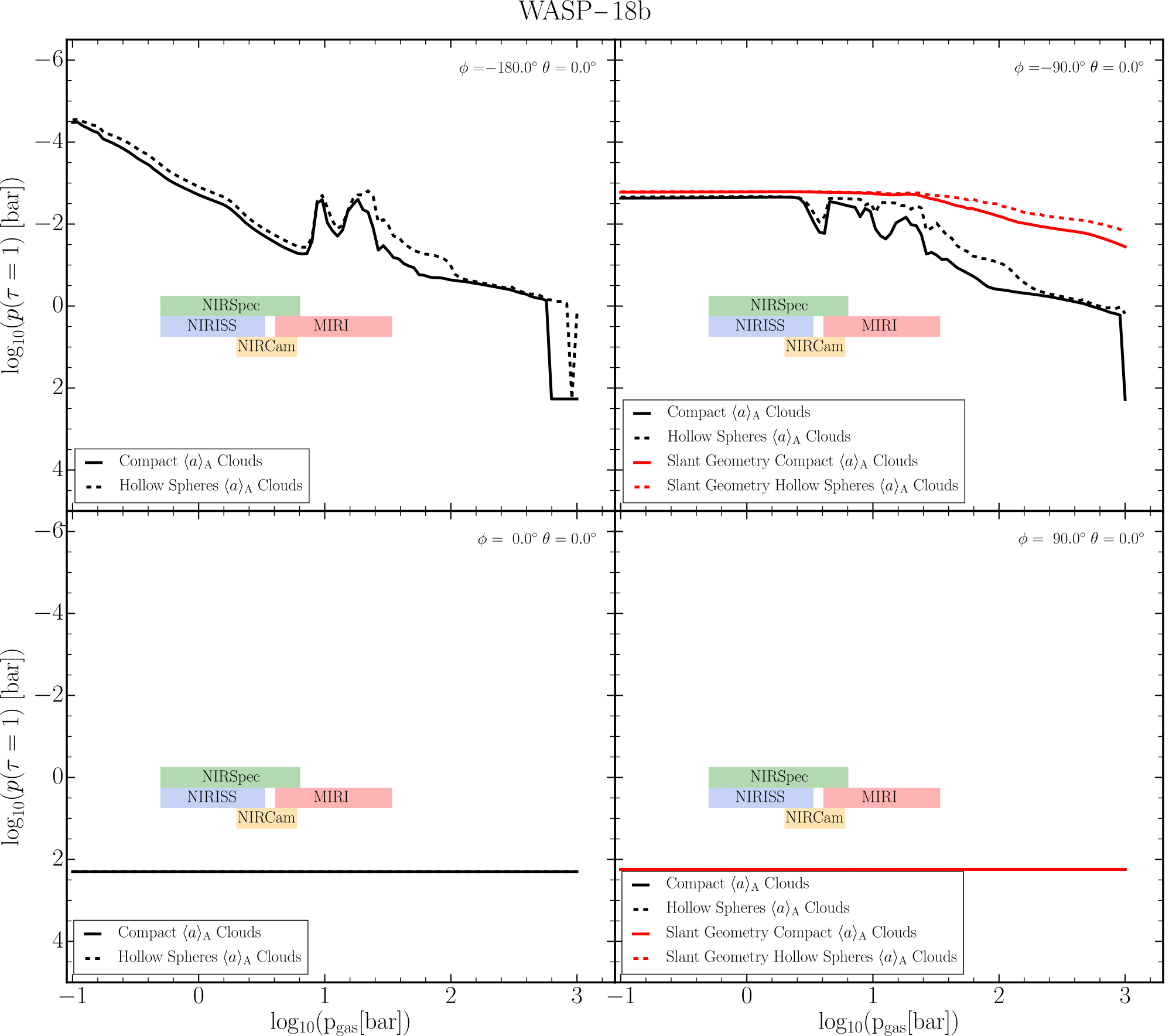}
          \includegraphics[width=0.49\textwidth]{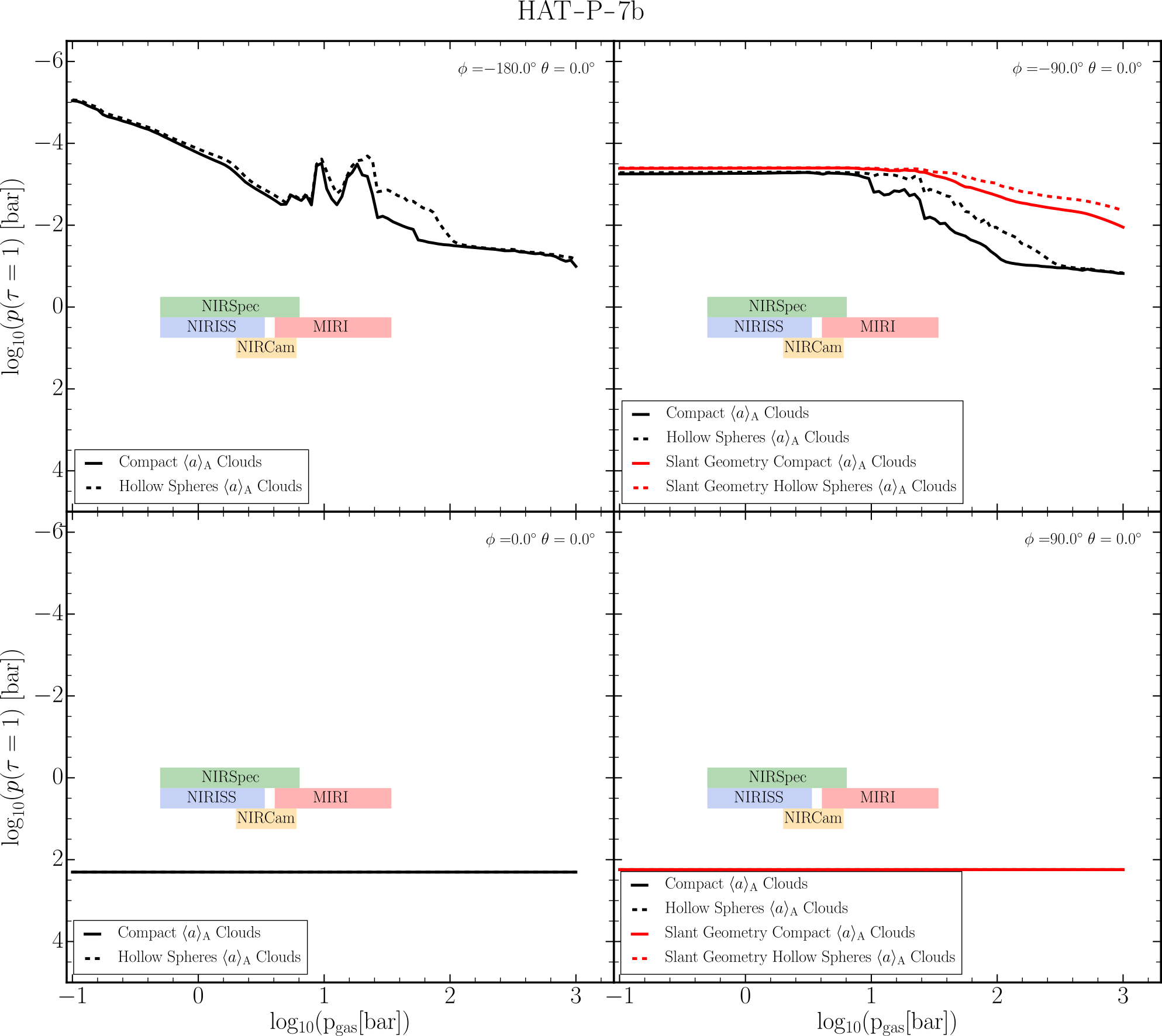}\\
     \includegraphics[width=0.49\textwidth]{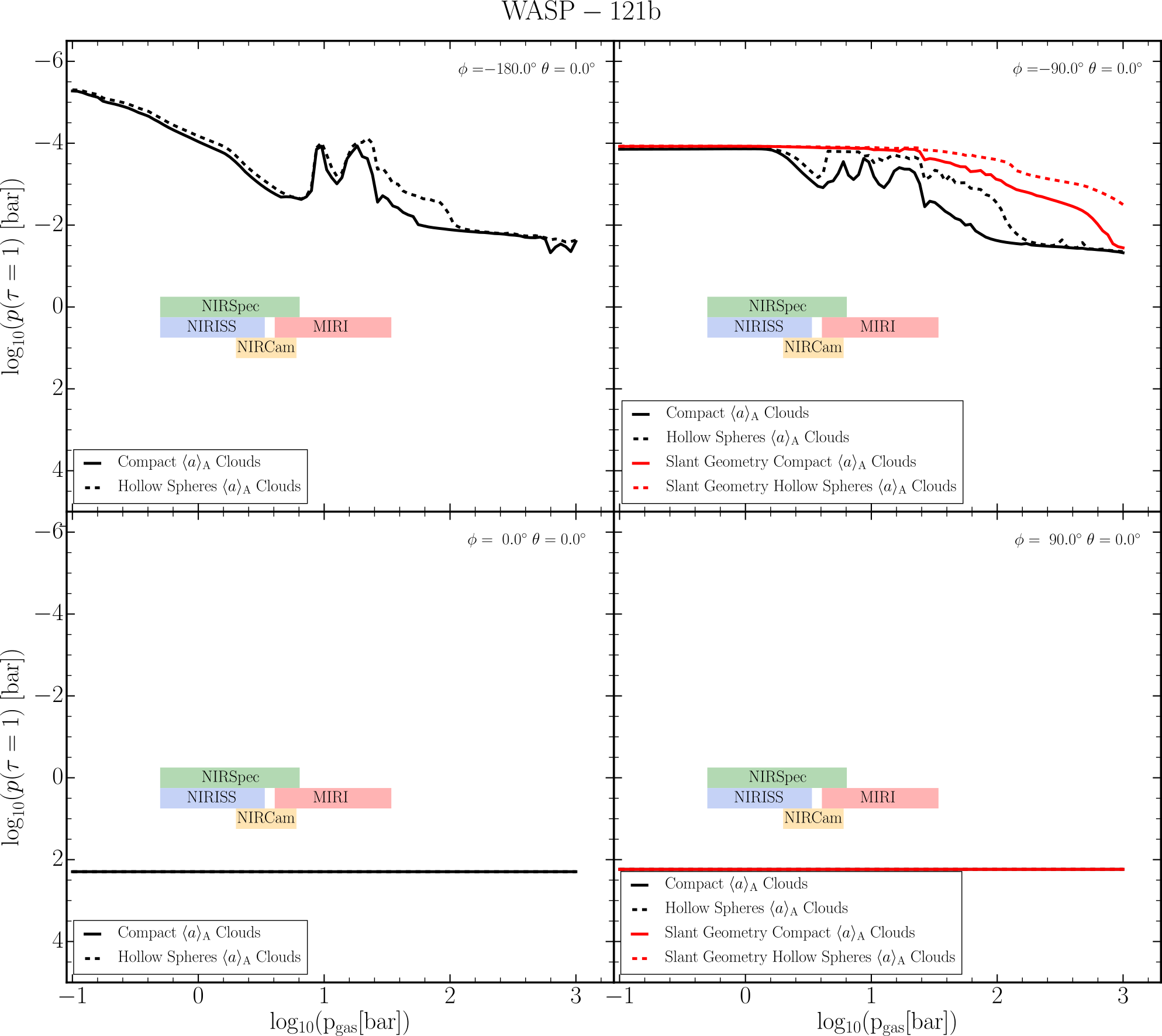}
     \caption{ Wavelength-dependent pressure level,  $p_{\rm gas}=p(\tau=1)$  where atmospheric gas above the clouds become optically thick (where $\tau=1$).  Included  is slant geometry (red curves,  using $\tau=\tau_{\rm s}$ Eq.~\ref{eq:slant}) and plane-parallel geometry (dashed) as comparison. When the optical depth of clouds never reaches 1, the pressure for the bottom of the atmosphere (here p $\approx 10^{2.2}$ bar) is returned (hence the lines for the sub-stellar points where there is no cloud).} 
     \label{fig:opitcal_depth1_appendix}
 \end{figure*}

Figure~\ref{fig:opitcal_depth1_appendix} provides the individual plots for the pressure where optical depth of unity (both vertical and slant geometry) is reached for the five sample planets, the  hot  giant gas planets like WASP-43b and ultra-hot Jupiters like WASP-103b,  WASP-121b,  WASP18b,  and  HAT-P-7b.

\clearpage
\section{Global atmosphere height asymmetries and hydrostatic pressure scale height}\label{ss:z}

\begin{figure*}
    \centering
    \includegraphics[width=21pc]{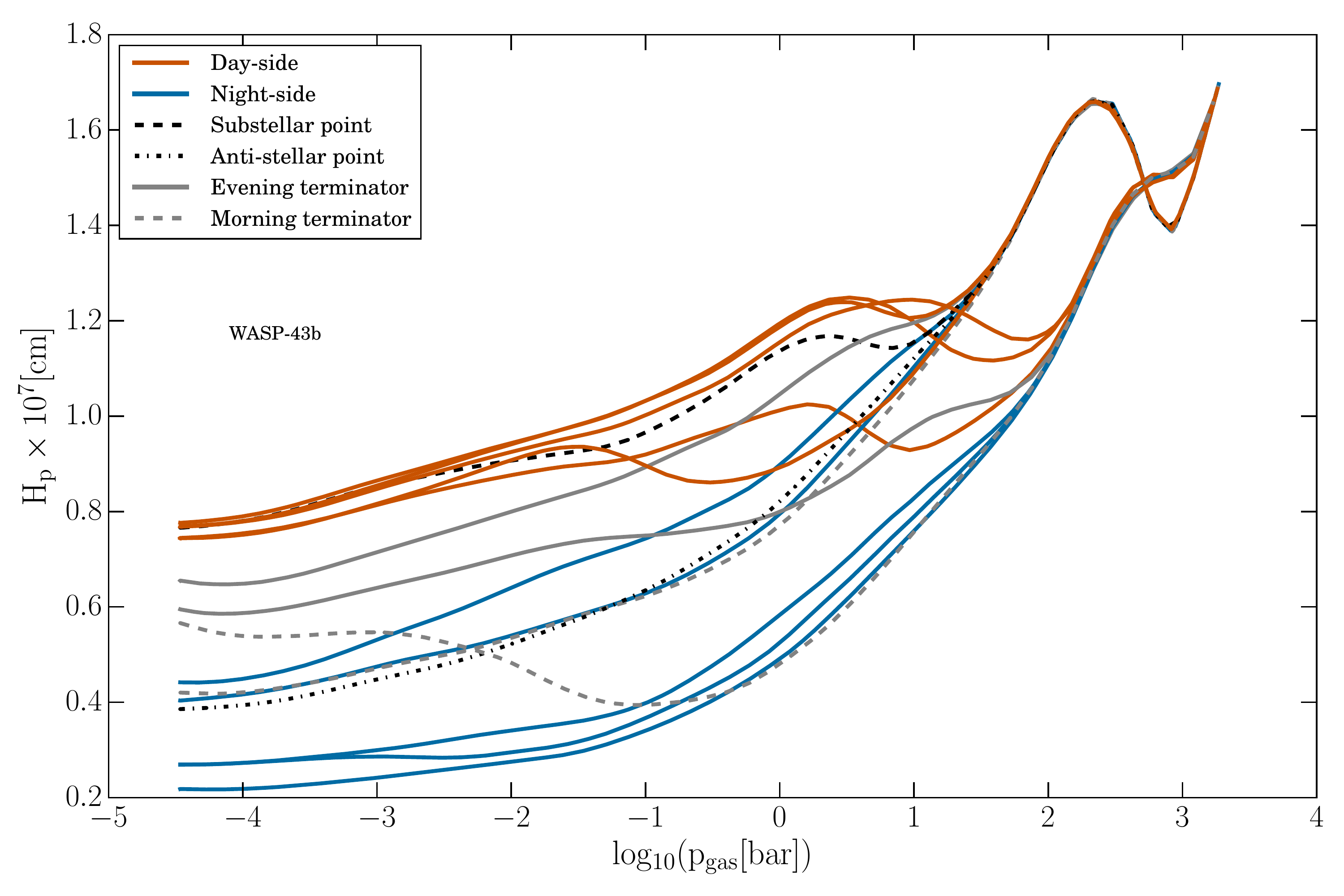}
    \includegraphics[width=21pc]{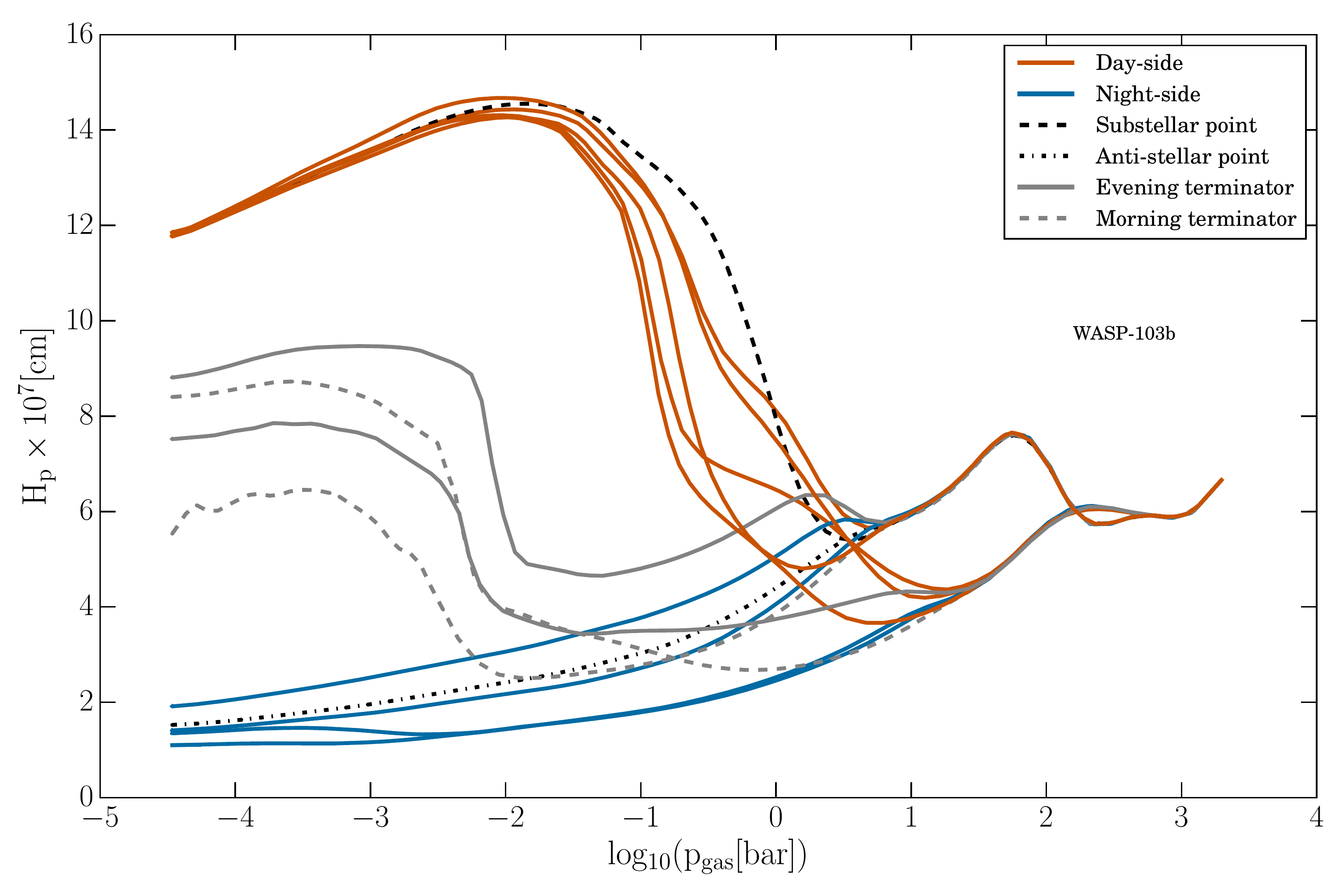}\\
    \includegraphics[width=21pc]{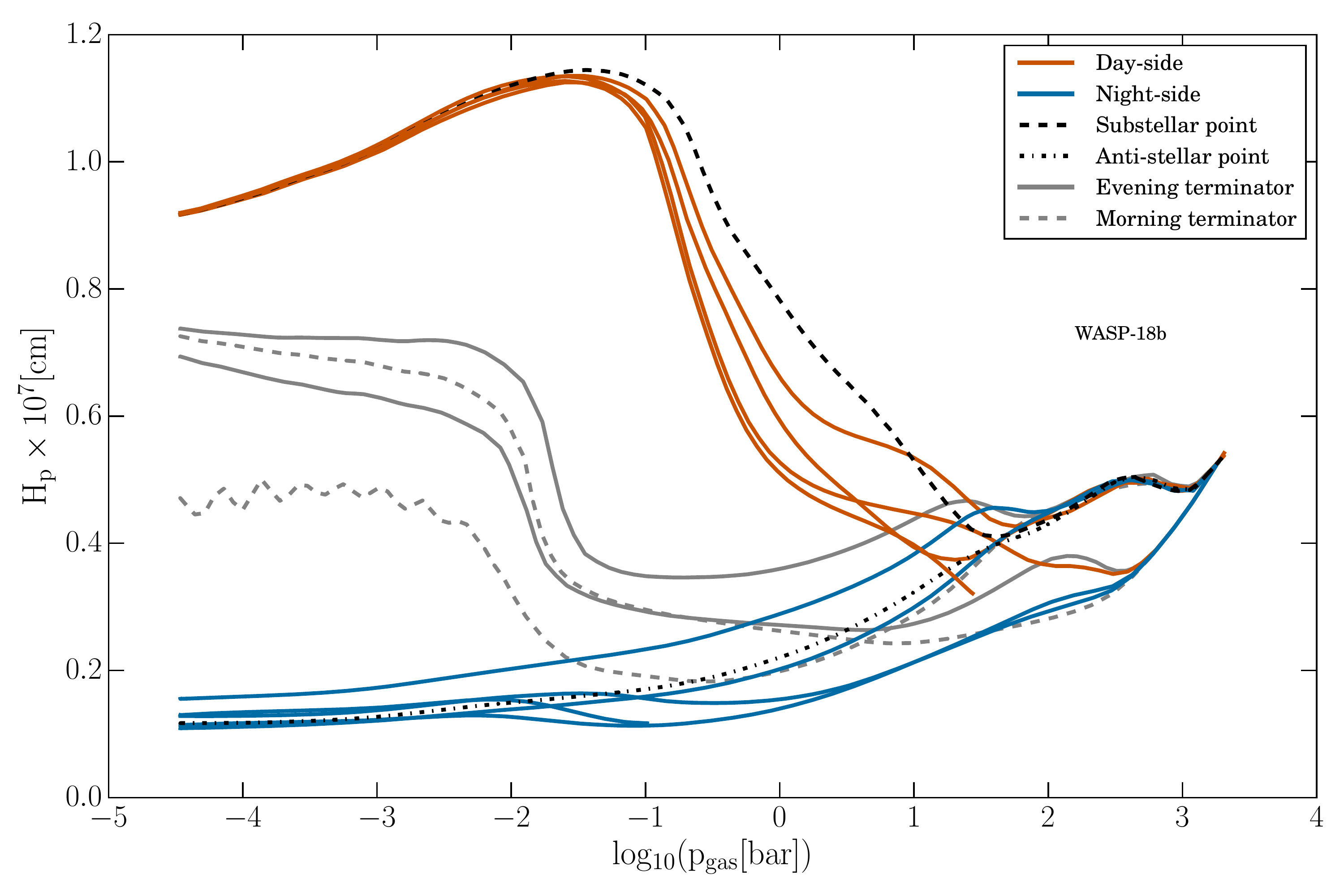}
    \includegraphics[width=21pc]{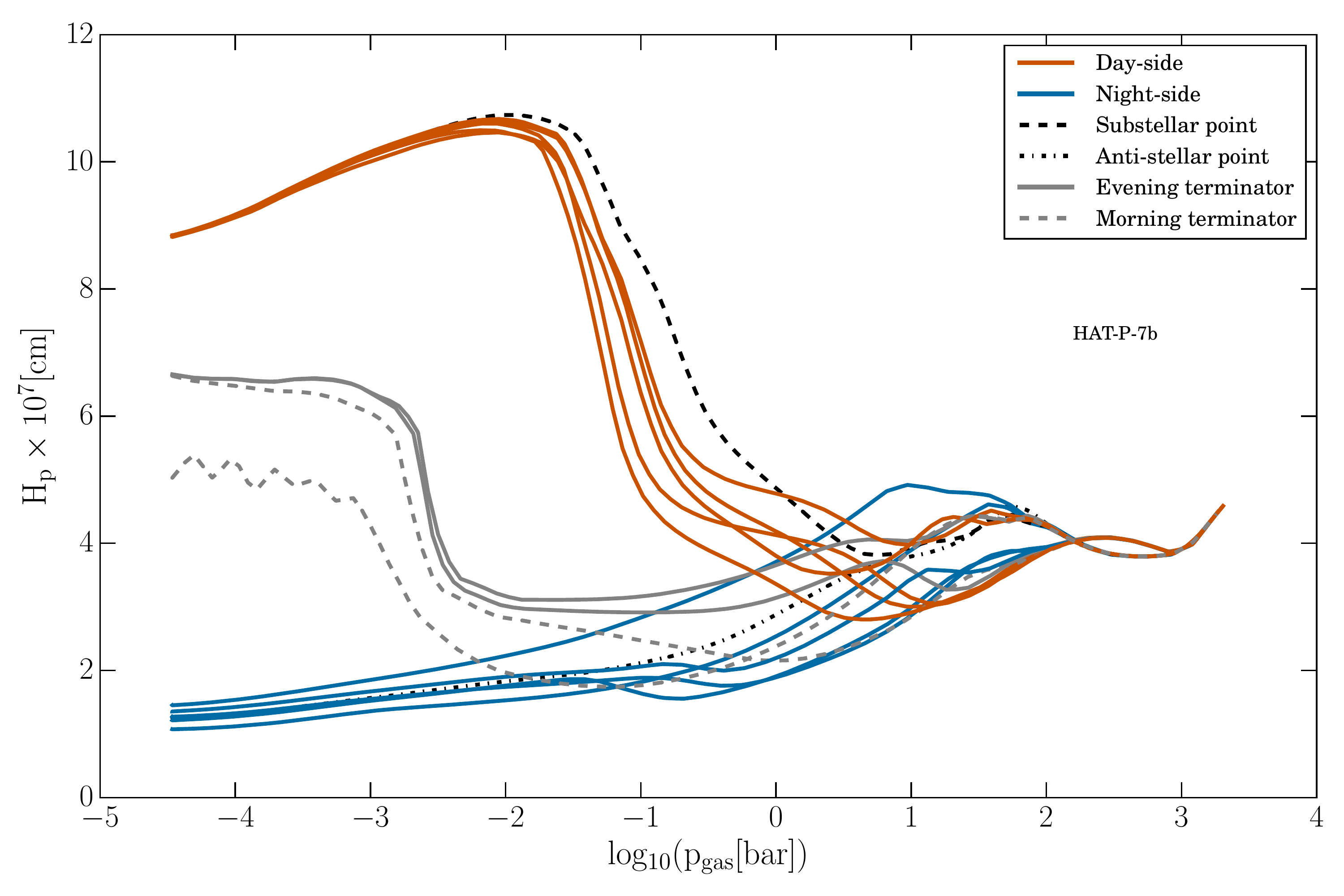}\\
    \includegraphics[width=21pc]{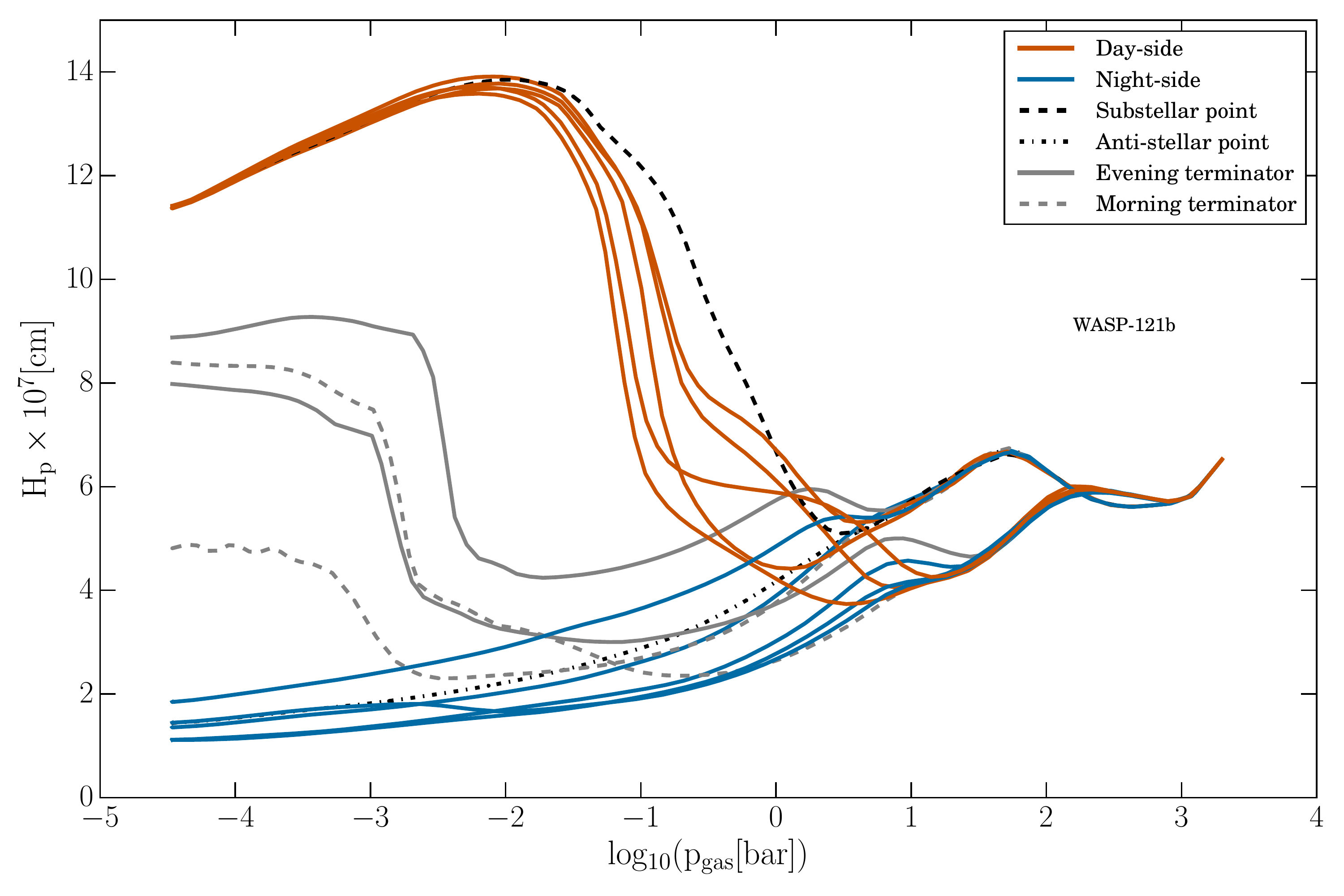}
    \includegraphics[width=21pc]{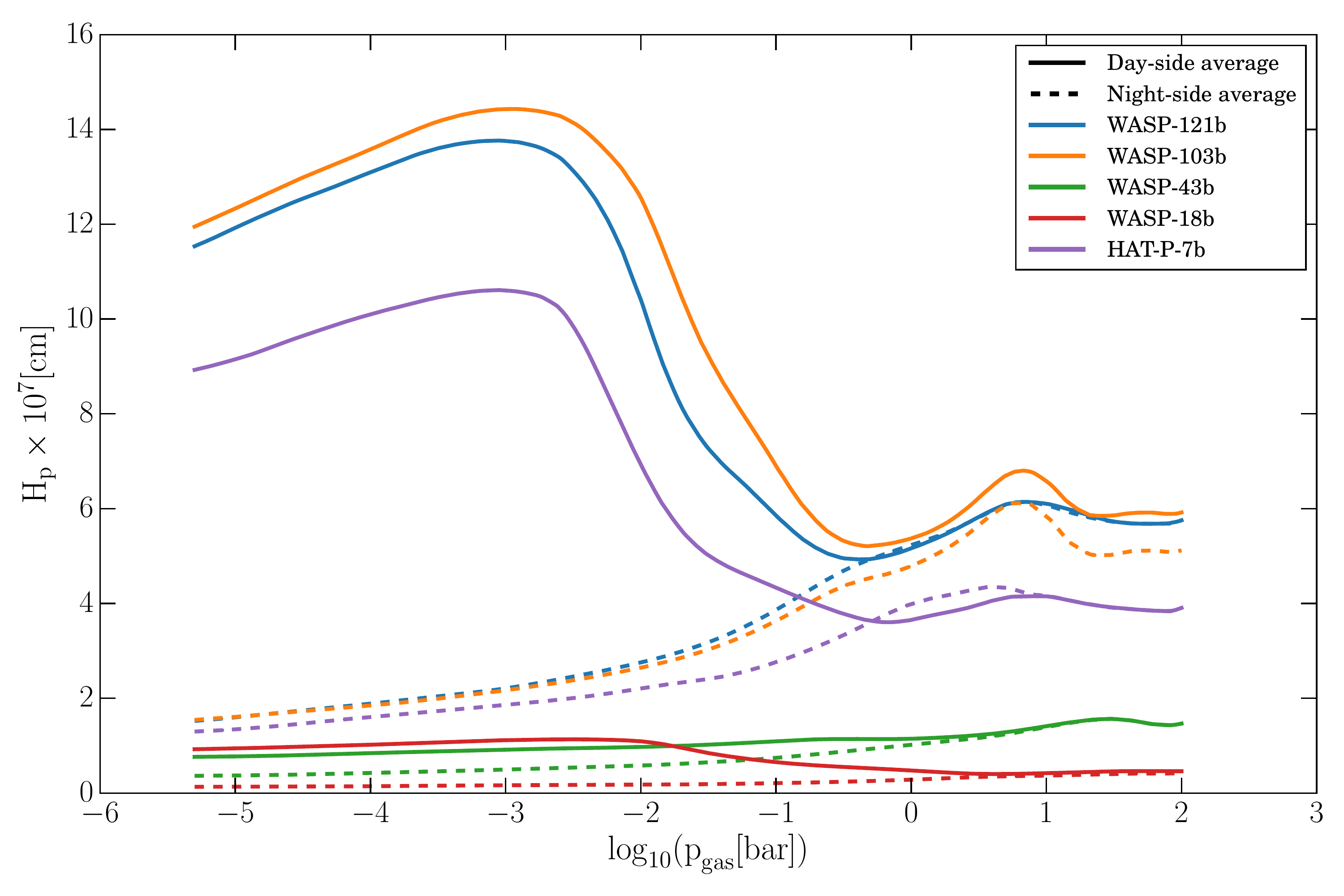}
    \caption{{Hydrostatic pressure scale height ($\rm H_{p} = (k T)(\mu m_{H} g)$) for the giant gas WASP-43b, and the ultra-hot Jupiters HAT-P-7b, WASP-18b, and WASP-103b, WASP-121b. The changing pressure scale height is caused by the temperature-dependent mean molecular weight, $\mu(T)$, that changes from the day- to the nightside due to the large differences in gas temperatures (see Fig.~\ref{fig:mmw_all}).}
    }
    \label{fig:scaleheight}
\end{figure*}

\begin{figure*}
    \centering
    \includegraphics[width=21pc]{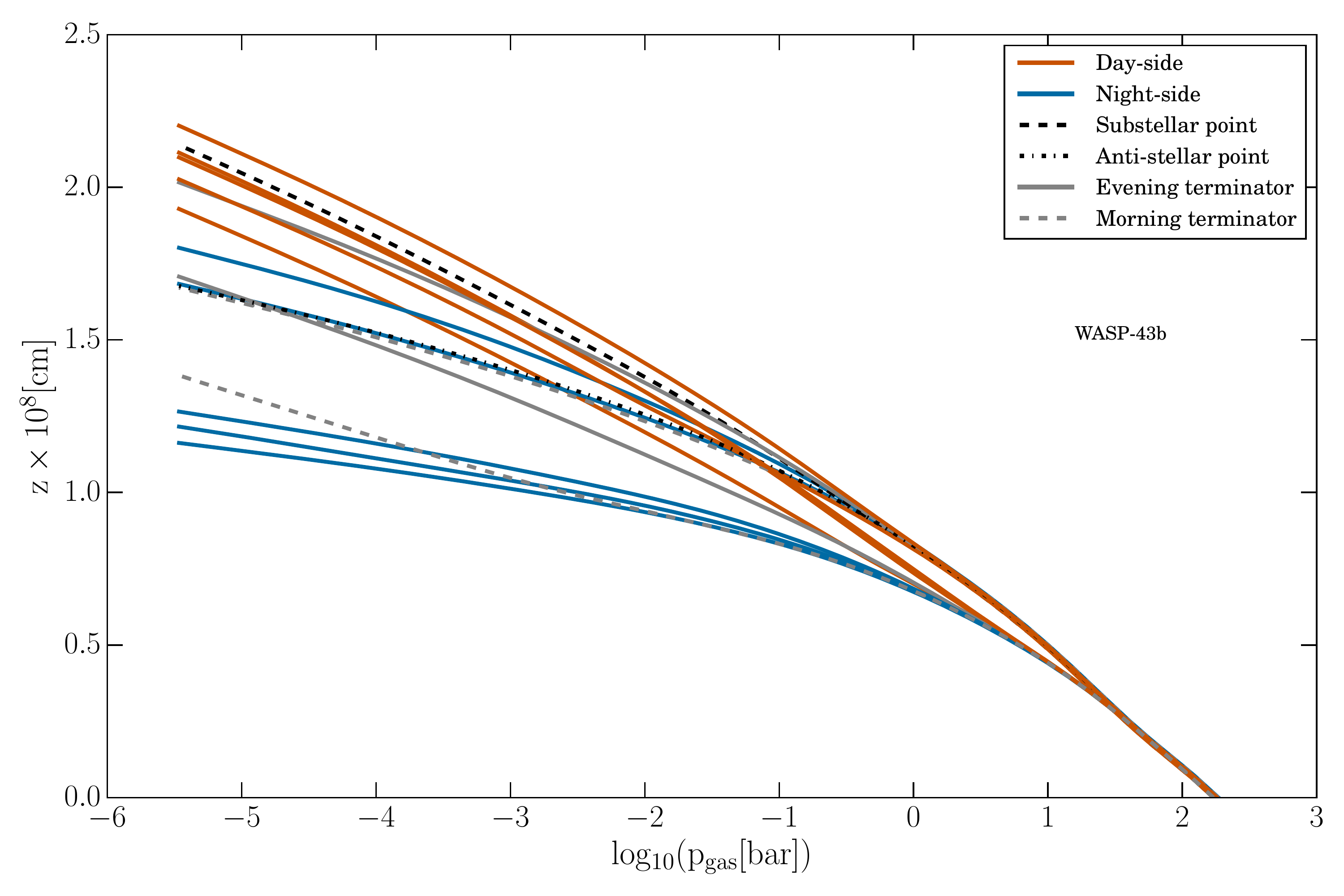}
    \includegraphics[width=21pc]{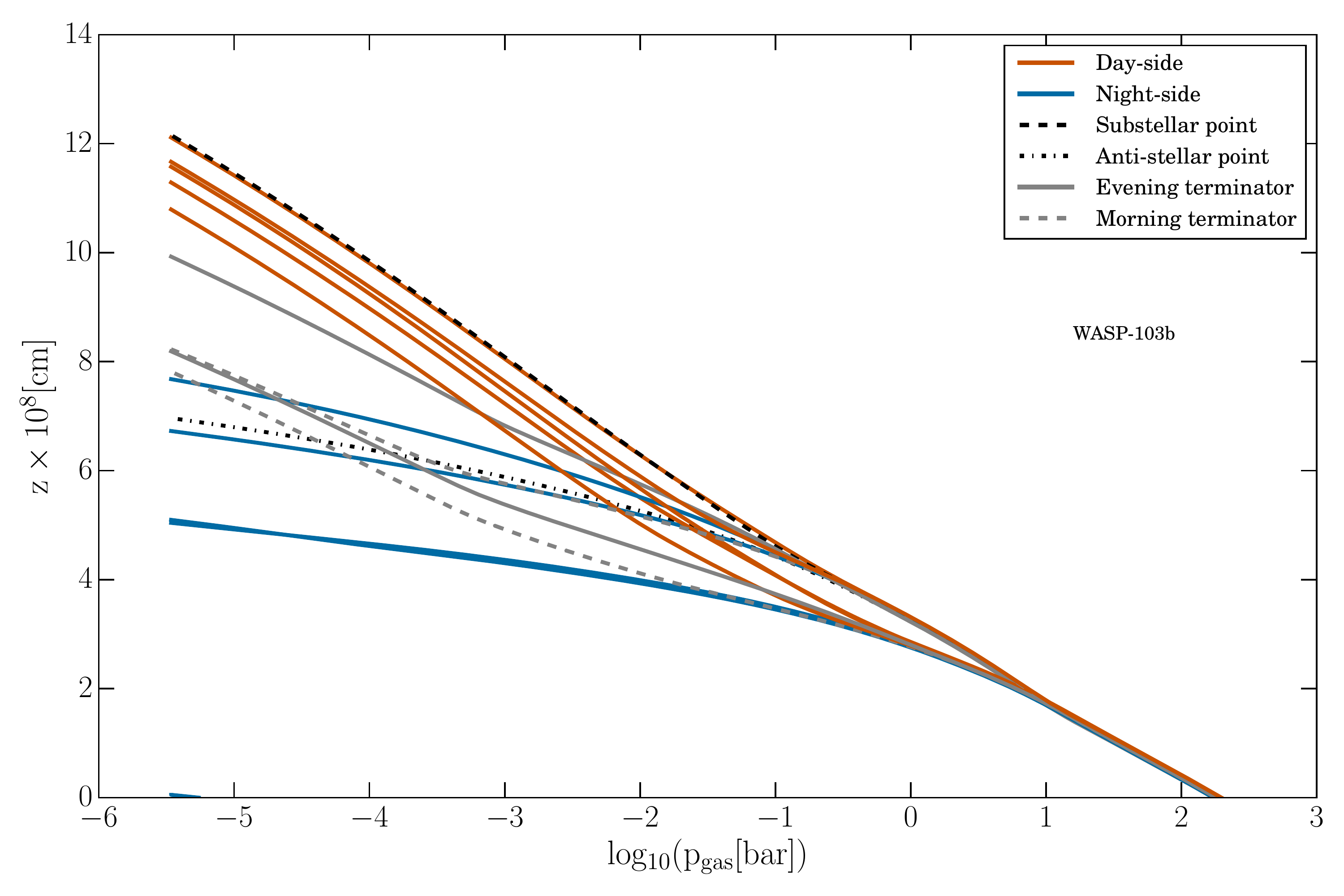}\\
    \includegraphics[width=21pc]{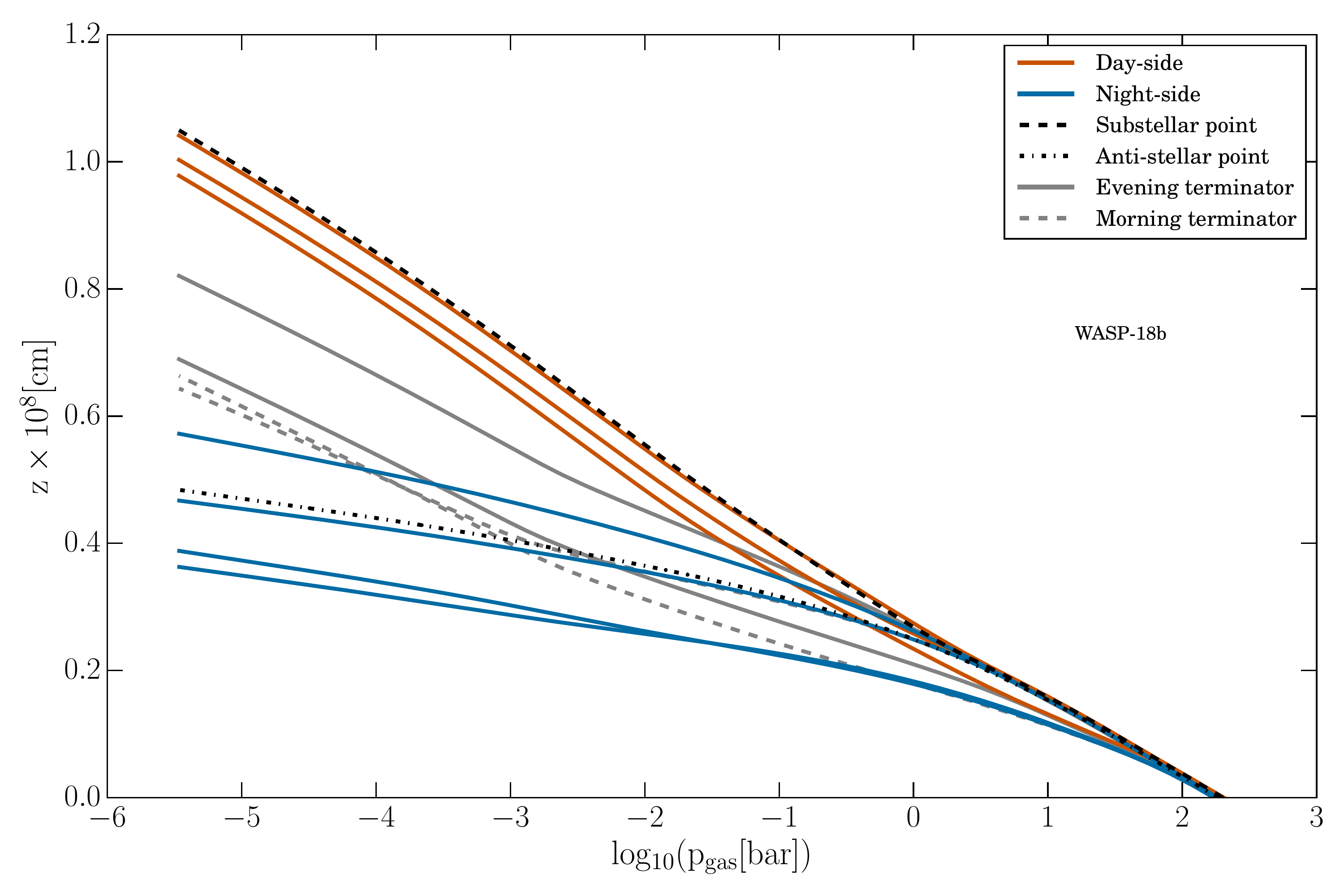}
    \includegraphics[width=21pc]{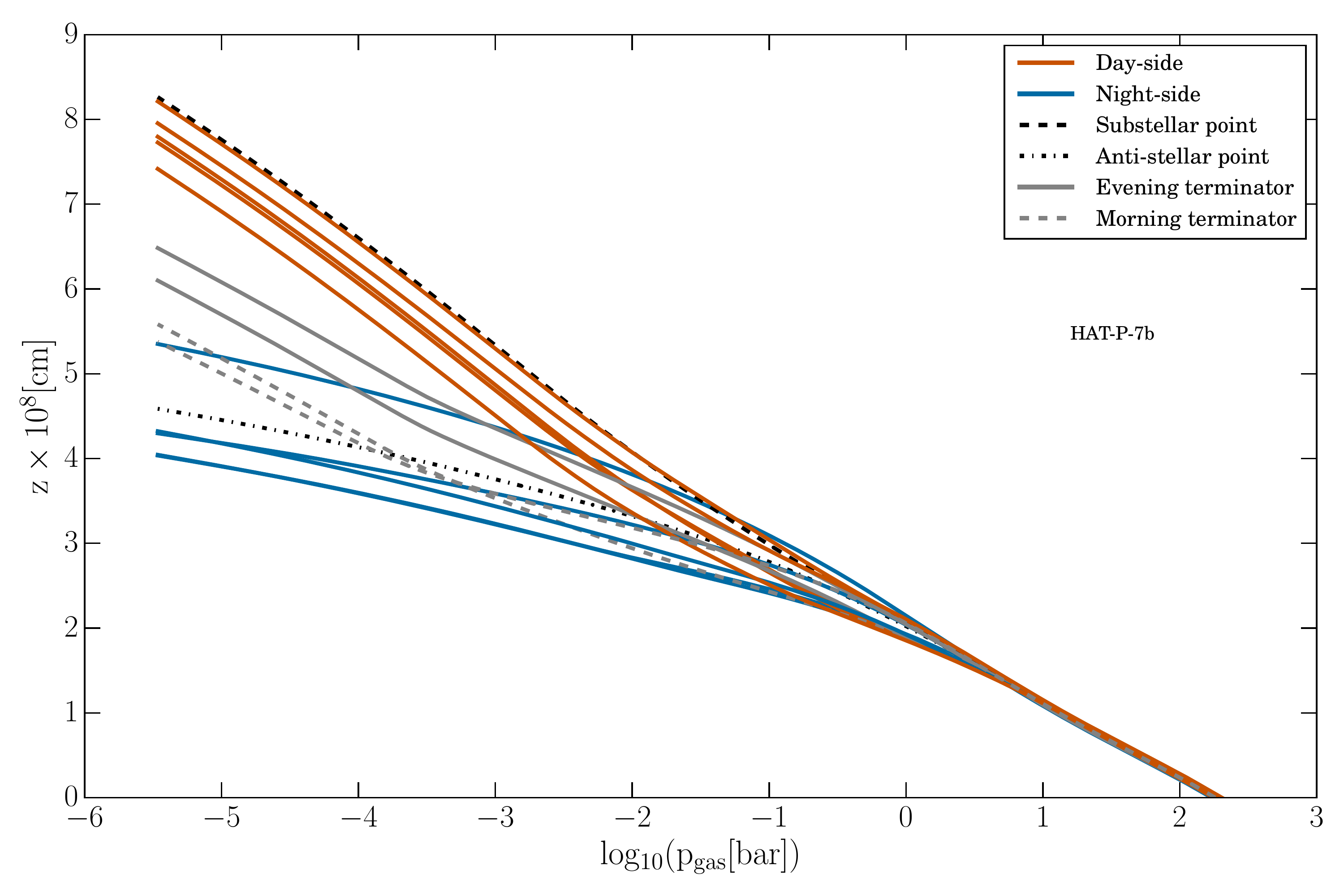}\\
    \includegraphics[width=21pc]{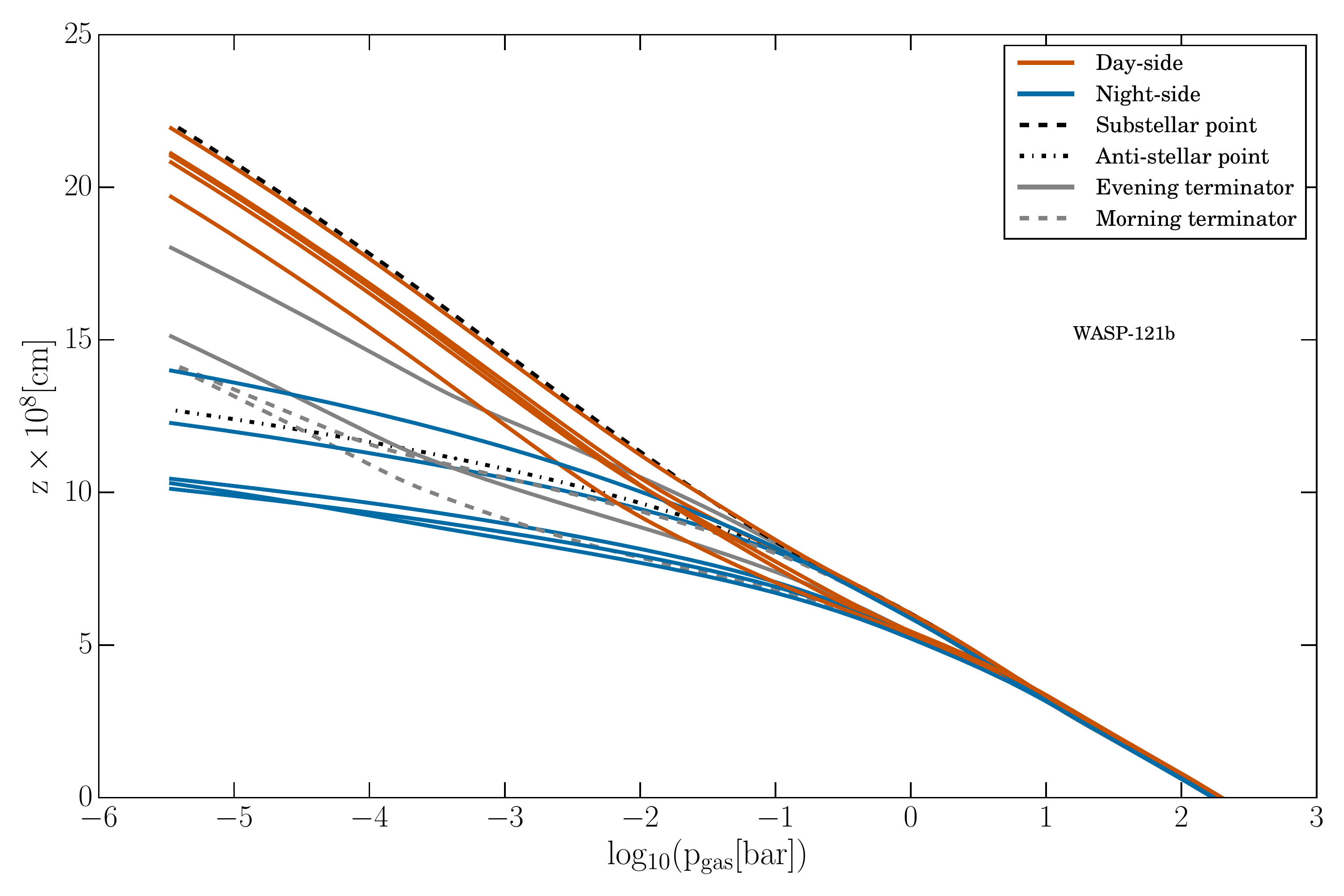}
    \includegraphics[width=21pc]{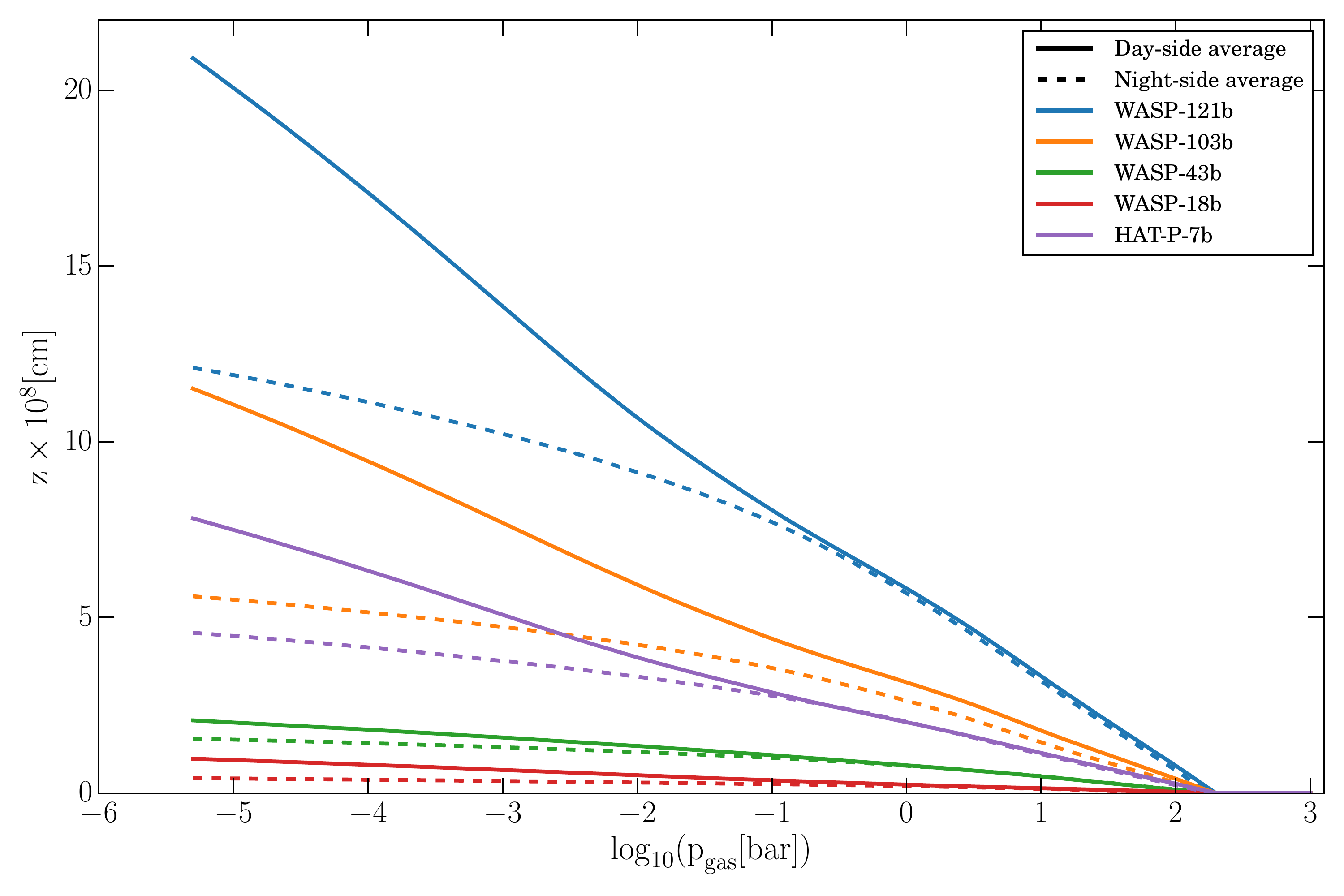}
    \caption{Geometric atmosphere height, $z$ [cm], calculated by summing the change in height dz between successive pressure layers starting from the inner boundary where $z=0$.}
    \label{fig:geo_atmos_all} 
    \label{fig:z}
\end{figure*}

We consider the (vertical) geometrical extension of the atmosphere (Fig.~\ref{fig:z}), and provide a comparison to the hydrostatic pressure scale height for completeness (Fig.~\ref{fig:scaleheight}). 
This geometric height is of interest as it gives  an indication for the asymmetry of the atmosphere. For example, have \cite{2018A&A...620A..97S} observe an asymmetric transit in the He~I line at 0.1083$\AA$   with CARMENES in conjunction with a net blue shift of $ - 3.5 \pm 0.4 $ km s$^{-1}$. One interpretation is a geometrical day/night asymmetry of 0.2 R$_{\rm P}$.
Figure~\ref{fig:z} demonstrates that the atmosphere of close-in planets are not spherical symmetric. The day/night geometric extension for the ultra-hot Jupiters in our sample is $\approx 2$,  but this is not a result of the changing mean molecular weight as a constant mean molecular weight is assumed in the 3D GCMs utilised here. Table~\ref{tab:geom_height} 
presents the effect of the changing geometrical extension in terms of a potential transit depth $\delta_{\rm transit} = ((R_{\rm P}+z)/ R_{\rm star})^{2}$ ($z$ - vertical extension of the atmosphere,Fig.~\ref{fig:z}).

The effect of the changing mean molecular weight is better seen in considering the hydostatic pressure scale height which was derived after the gas-phase chemistry was solved within our cloud formation model (Fig.~\ref{fig:scaleheight}).
Clearly visible in all plots is the onset of the thermal inversions, which produces an increase in the rate at which the vertical extent of the atmosphere grows at higher altitudes (moving left in the Fig.~\ref{fig:z}). This change is most noticeable for the ultra-hot Jupiters, for the terminators as they have the steepest inversions. The terminator profiles initially have atmospheric extensions similar to nightside profiles but around the millibar level they switch to a gradient parallel to the dayside profiles. The lower right of figure shows the difference between average the dayside and nightside profiles, with the general trend that in the deep atmosphere the extension is the same, but at higher altitudes diverges, this altitude similarly corresponds with the drop in mean molecular weight (see bottom right Fig.~\ref{fig:mmw_all}. It also shows that the effect is most prominent for planets with low surface gravities; WASP-18b shows little difference in extension despite being an ultra-hot Jupiter as it has a significantly higher surface gravity). Multiple studies have now investigated the affects of different extensions on the day- and nightsides in transmission spectra, \citep{Caldas2019,pluriel2020} found changes in temperature and compositional gradient across the terminator region can bias results of retrievals, and further that this is dependent on the gradient of the change as this affects the extent of the atmosphere and thus the amount of the dayside that the light ray passes through. 

Figure~\ref{fig:wasp43b_LC_plots_c} provides the detailed results on the mean molecular weight, $\mu$,  and geometric extension for the Parmentier/Carone  in order to enable comparison with future simulations.

\begin{table*}[bh]
    \centering
    \begin{tabular}{c|c c c c c}
    \hline
         \textbf{Planet} & \textbf{WASP-103b} & \textbf{WASP-18b} & \textbf{WASP-121b} & \textbf{HAT-P-7b} & \textbf{WASP-43b} \\[0.5ex]
         \hline
         \hline
         Average dayside maximum extension [R$_{\rm P}$]& 0.10777 & 0.0126  & 0.16558 & 0.07815 & 0.02856 \\
         Average nightside maximum extension [R$_{\rm P}$]& 0.05244 & 0.0055 & 0.09582 & 0.04561 & 0.00214 \\
         Day-/nightside ratio & 2.055 & 2.294 & 1.728 & 1.713 & 1.334 \\
         $\delta_{\rm transit}$ [\%] & 1.141 & 0.834 & 1.548 & 1.725 & 2.432 \\[0.5ex]
         \hline
    \end{tabular}
    \caption{The maximum average dayside and nightside extensions in terms of planetary radius, the day-to-nightside extension ratio and the expected transit depth calculated as $\delta_{\rm transit} = (( R_{\rm P}+z)/R_{\rm star})^{2}$. We use the $R_{\rm p}$ and $R_{\rm star}$ given in Table~\ref{table:stpl}, and the vertical extension of the atmosphere $z$ (Fig.~\ref{fig:z}).}
    \label{tab:geom_height}
\end{table*}

\clearpage
\subsection{Supplementary details on the WASP-43b simulations results by Parmentier and Carone}\label{ss:pcapp}

The detailed cloud modelling results regarding the cloud material volume fractions,$V_{\rm s}/V_{\rm tot}$, and the results for the mean molecular weight , $\mu$,  and the vertical, geometric extension, $z$, for the comparative study if the effect of the inner boundary for the example of WASP-32b in Sect.~\ref{s:wasp_gcm_comp} are provided.

\begin{figure*}
    \includegraphics[width=21pc]{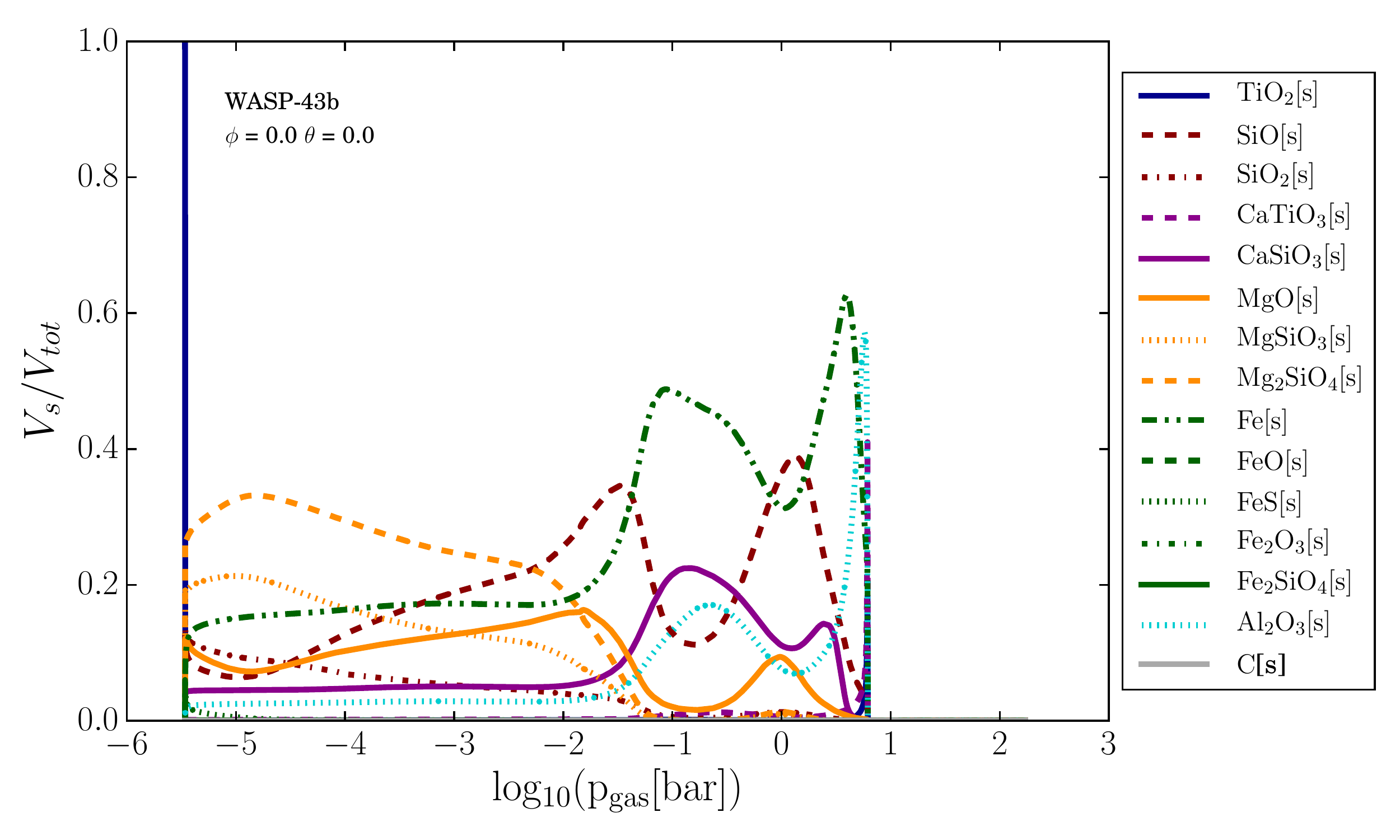}
    \includegraphics[width=21pc]{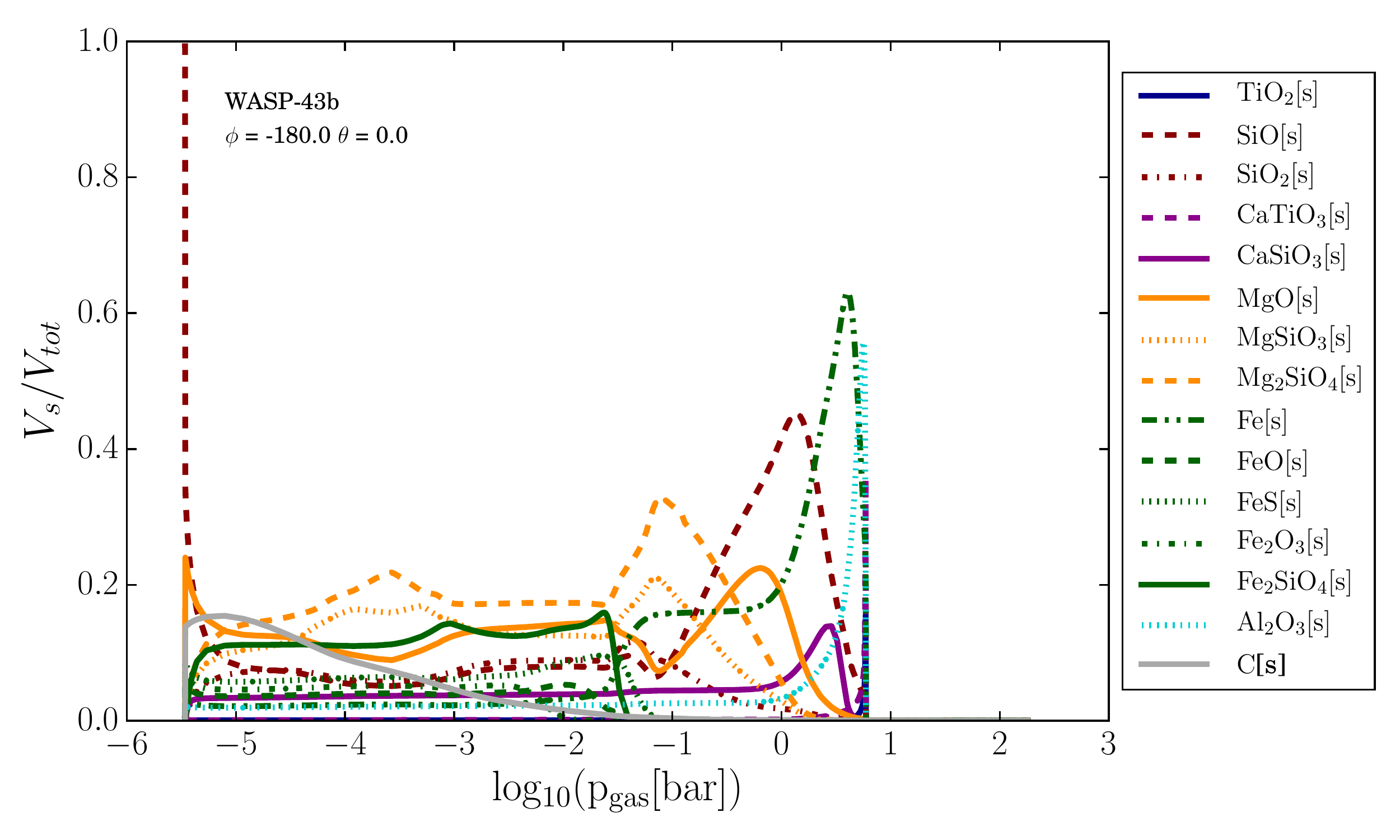}\\
    \includegraphics[width=21pc]{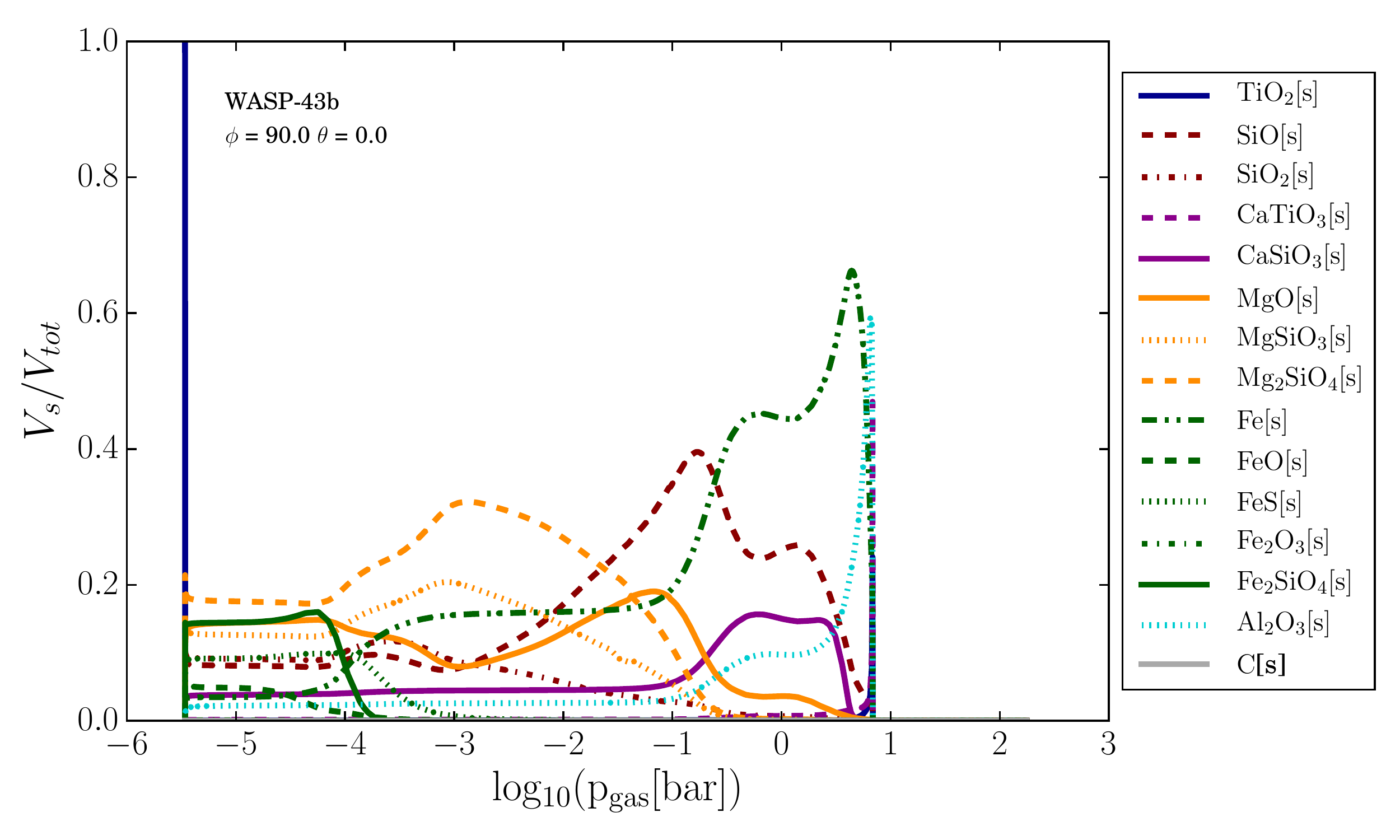}
    \includegraphics[width=21pc]{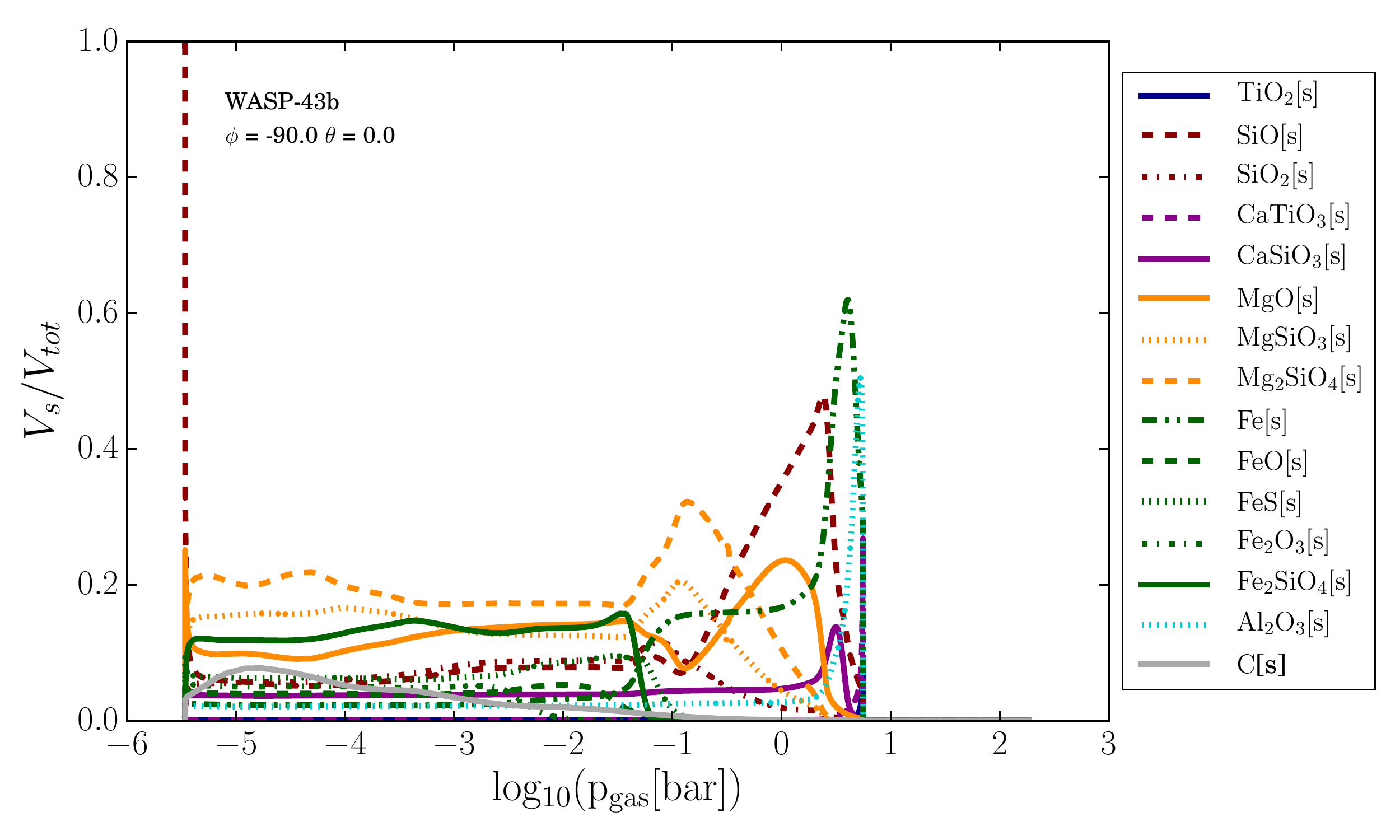}
    \caption{Individual bulk material volume fractions WASP-43b (based on the 1D thermodynamic profiles from Parmentier et al.). Sub-stellar point: $(\phi, \theta)=(0.0^o, 0.0^o)$, Anti-stellar point: $(\phi, \theta)=(-180.0^o, 0.0^o)$, Equatorial Morning Terminator: $(\phi, \theta)=(-90.0^o, 0.0^o)$, Equatorial Evening Terminator: $(\phi, \theta)=(90.0^o, 0.0^o)$.)}
    \label{fig:mat_vol_WASP43b}
\end{figure*}

\begin{figure*}
    \includegraphics[width=21pc]{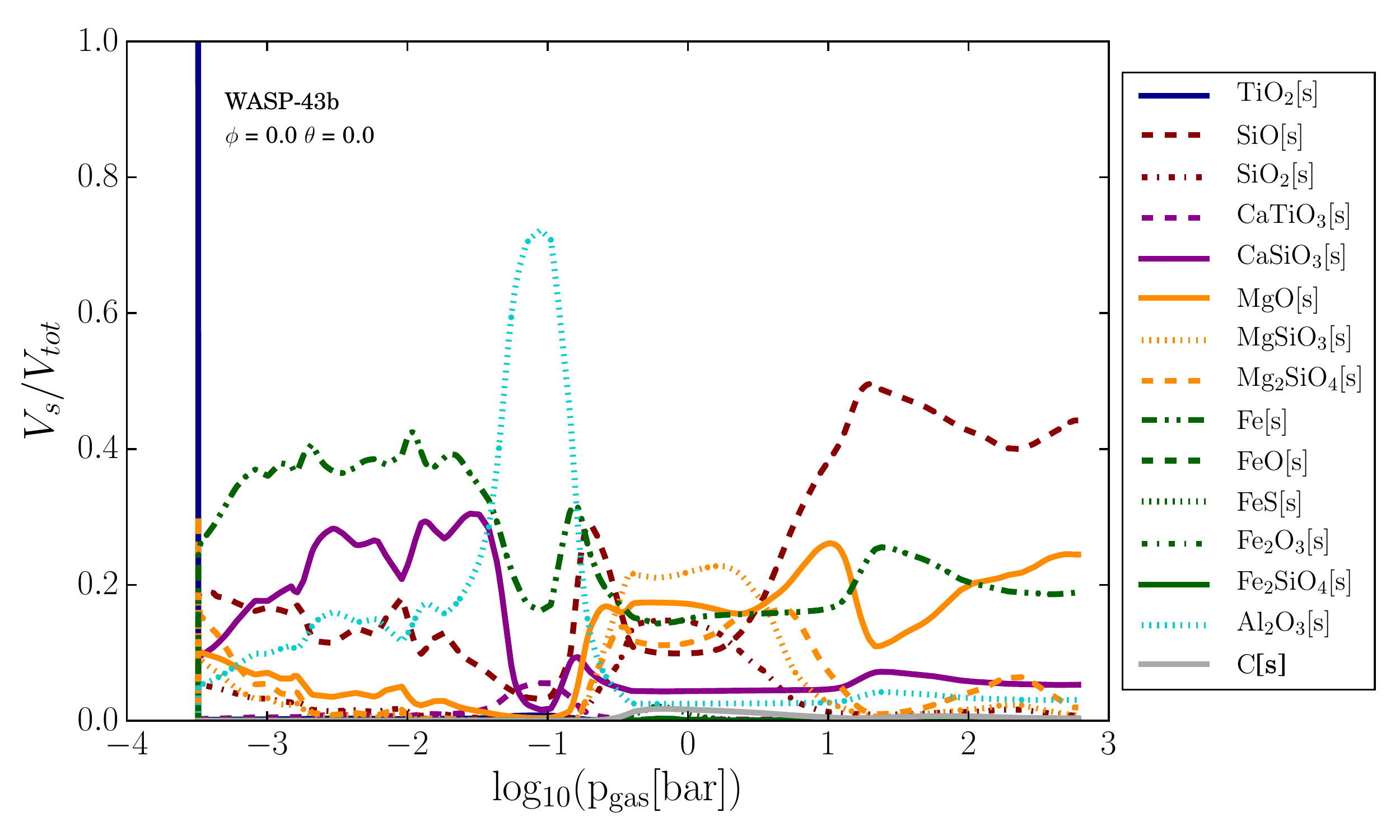}
    \includegraphics[width=21pc]{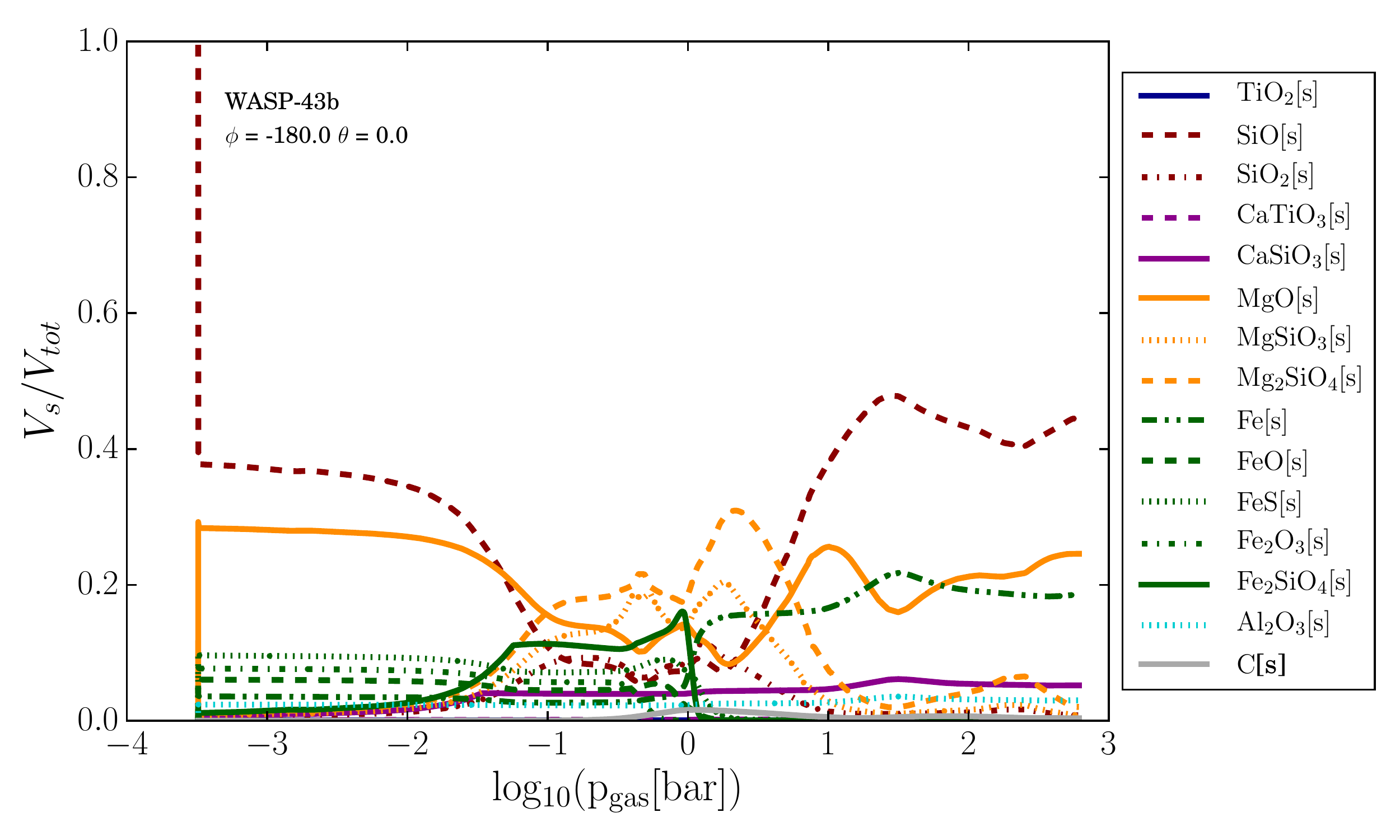}\\
    \includegraphics[width=21pc]{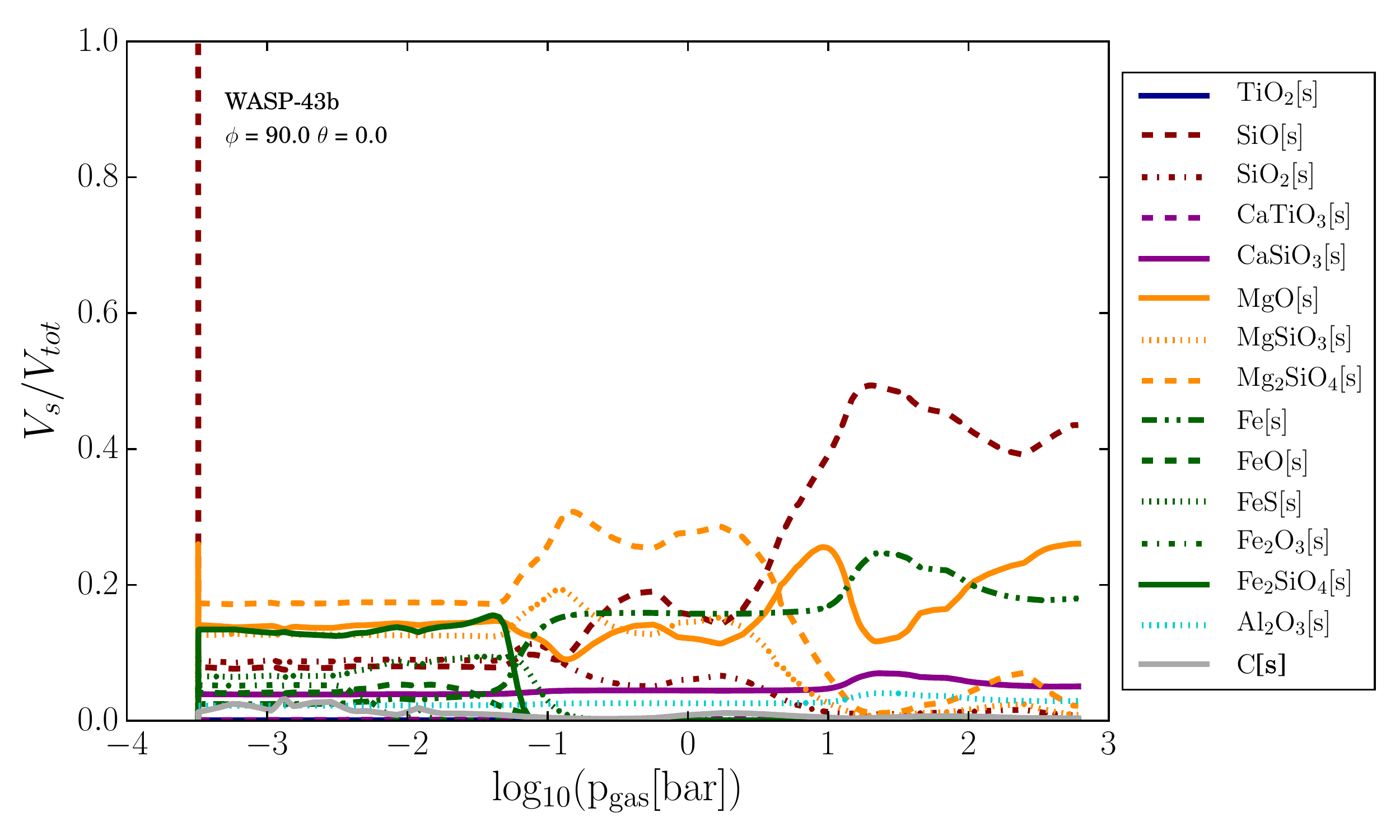}
    \includegraphics[width=21pc]{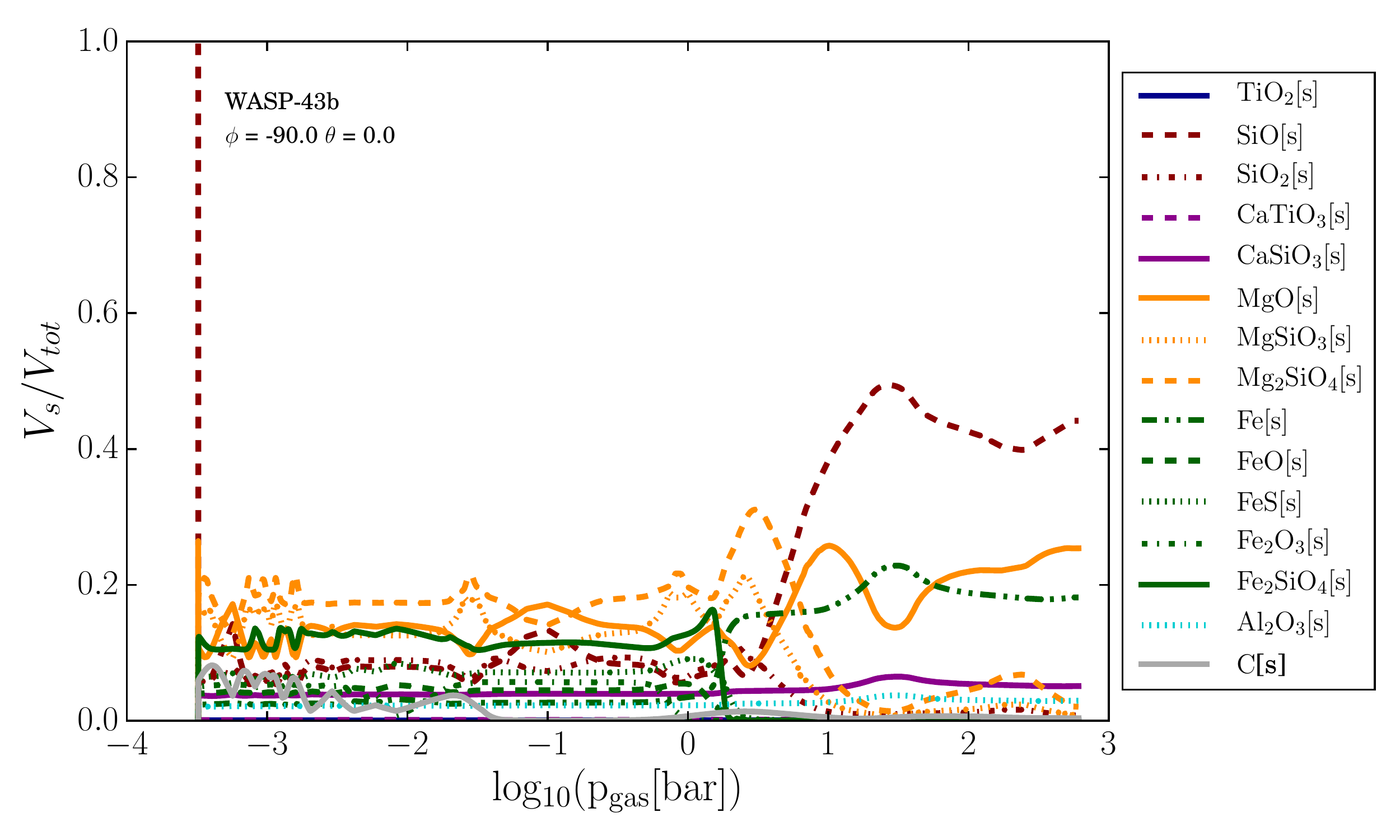}
    \caption{Individual bulk material volume fractions WASP-43b (based on the 1D thermodynamic profiles from Carone et al.). Sub-stellar point: $(\phi, \theta)=(0.0^o, 0.0^o)$, Anti-stellar point: $(\phi, \theta)=(-180.0^o, 0.0^o)$, Equatorial Morning Terminator: $(\phi, \theta)=(-90.0^o, 0.0^o)$, Equatorial Evening Terminator: $(\phi, \theta)=(90.0^o, 0.0^o)$.)}
    \label{fig:mat_vol_WASP43b_LC}
\end{figure*}

\begin{figure*}
    \centering
    \includegraphics[width=19pc]{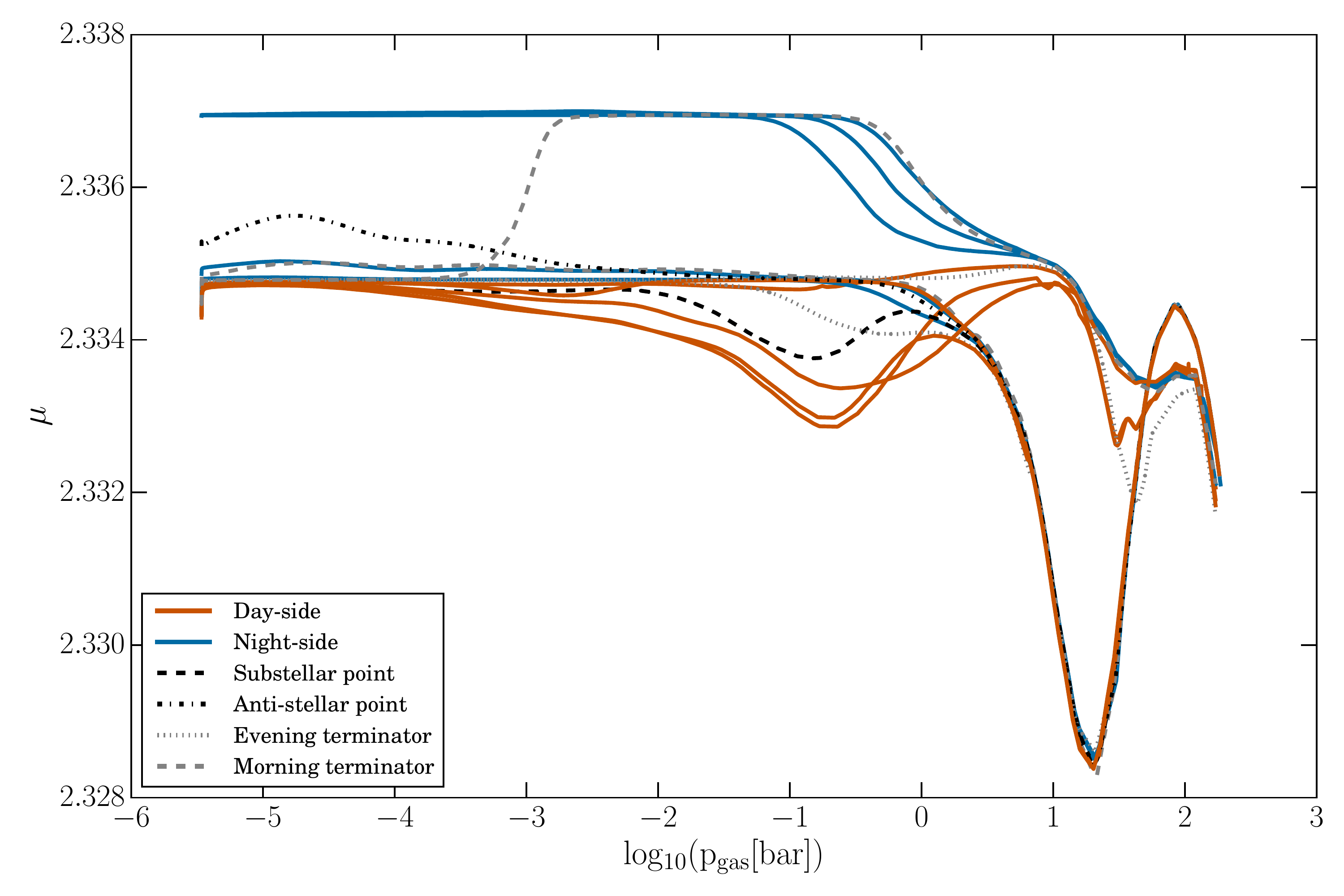}
    \includegraphics[width=19pc]{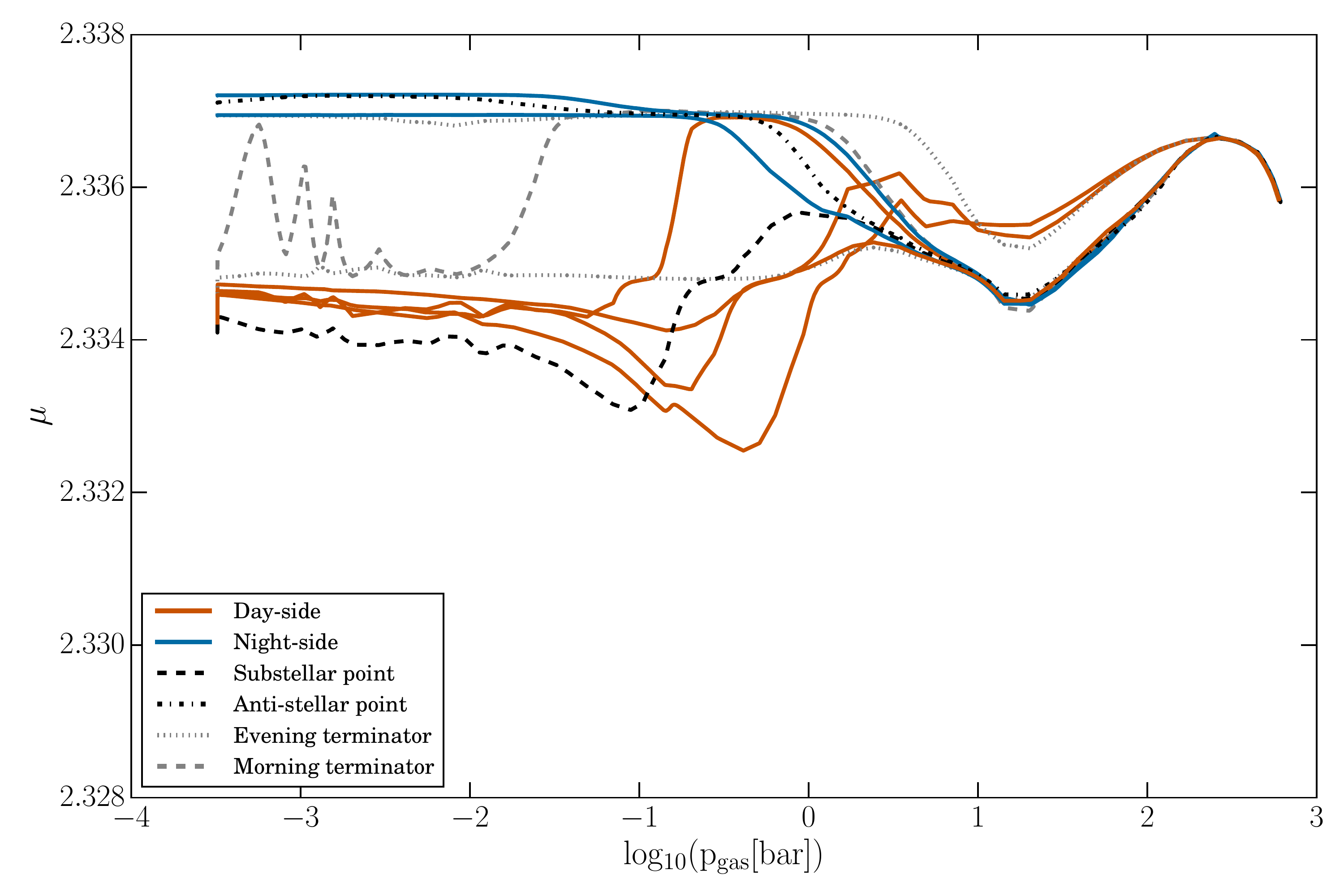}\\     
    \includegraphics[width=19pc]{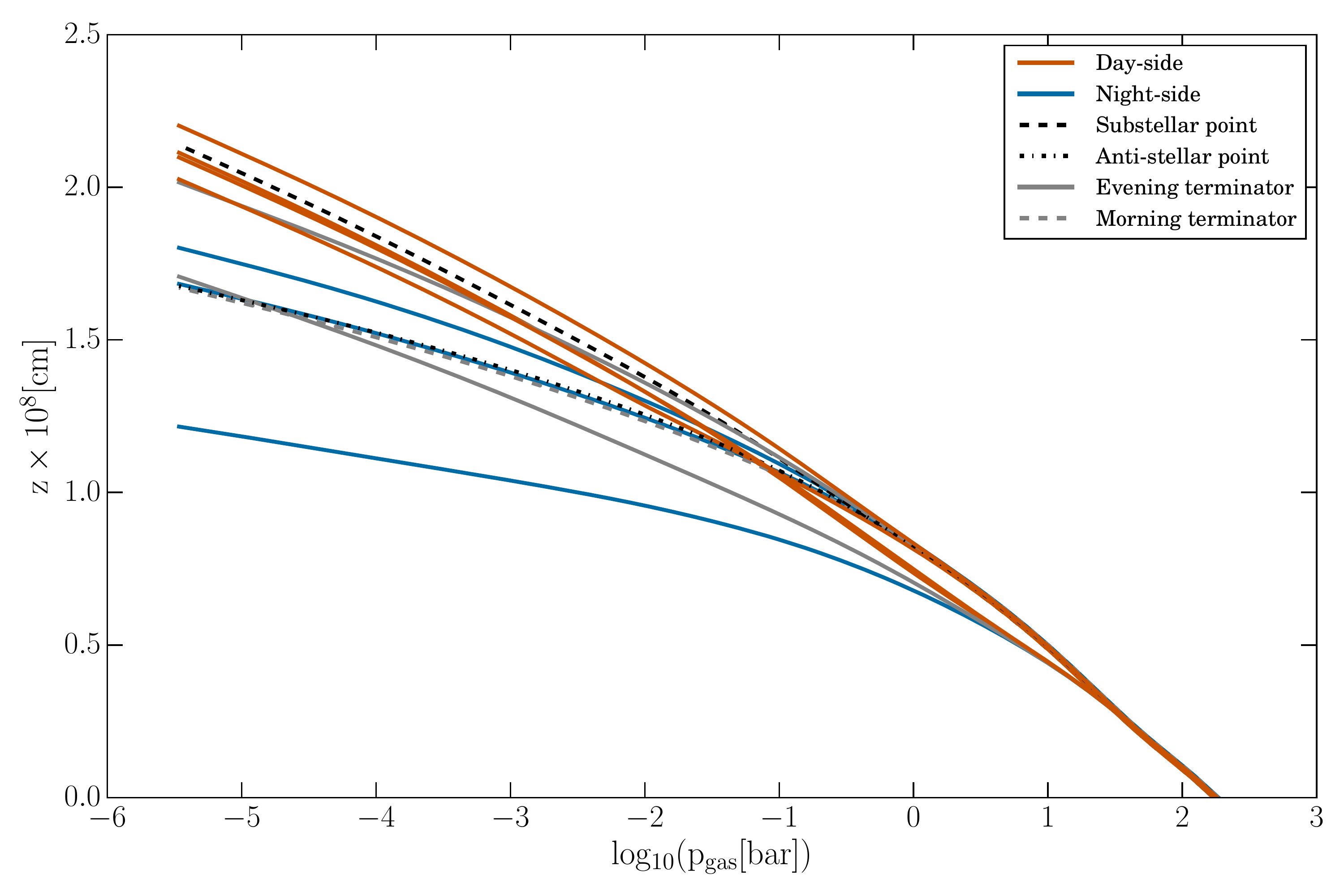}
    \includegraphics[width=19pc]{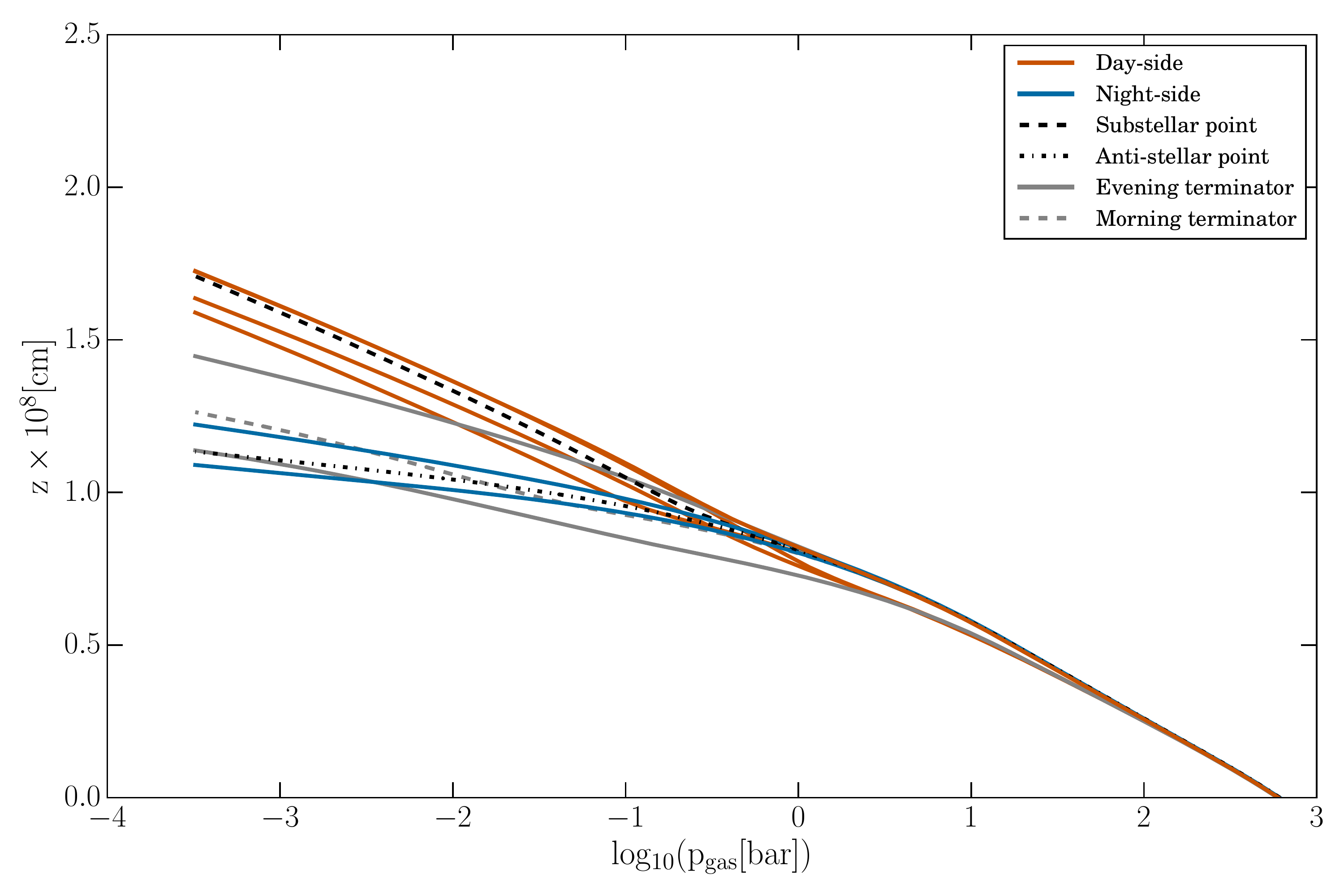}
    \caption{The effect of the inner boundary of the 3D GCM models on the mean molecular weight, $\mu$, and the geometric atmosphere extension for the giant gas planet example WASP-43b. \textbf{Right: } based on the 1D thermodynamic profiles from Parmentier et al. \textbf{Right: }  based on the 1D thermodynamic profiles from Carone et al. There is a difference of $\sim$0.5$\times 10^{8}$ cm between the day and night geometric extensions, of the same profiles, between the two models. }
    \label{fig:wasp43b_LC_plots_c}
\end{figure*}

\end{document}